\documentclass[%
 reprint,
superscriptaddress,
 amsmath,amssymb,
 aps,
pra,
]{revtex4-2}

\usepackage{graphicx}%
\usepackage{dcolumn}%
\usepackage{bm}%

\usepackage{graphicx}%
\usepackage{amsmath,amssymb,amsfonts}%
\usepackage{bm}%
\usepackage{braket}%
\usepackage{amsthm}%
\usepackage{mathrsfs}%
\usepackage[title]{appendix}%
\usepackage[dvipsnames]{xcolor}%
\usepackage{booktabs}%
\usepackage{hyperref}
\usepackage{subcaption}
\hypersetup{
    colorlinks=true,
    linkcolor=blue,
    filecolor=magenta,
    urlcolor=cyan,
    pdftitle={Overleaf Example},
    pdfpagemode=FullScreen,
    }

\usepackage{ragged2e}

\newtheorem{theorem}{Theorem}%
\newtheorem{proposition}[theorem]{Proposition}%
\newtheorem{remark}{Remark}%
\newtheorem{definition}{Definition}%

\makeatletter
\newtheorem*{rep@theorem}{\rep@title}
\newcommand{\newreptheorem}[2]{%
  \newenvironment{rep#1}[1]{%
    \def\rep@title{#2 \ref{##1} (restated)}%
    \begin{rep@theorem}}{\end{rep@theorem}}}
\makeatother

\newtheorem{corollary}[theorem]{Corollary}
\newtheorem{lemma}[theorem]{Lemma}

\newreptheorem{corollary}{Corollary}
\newreptheorem{lemma}{Lemma}
\newreptheorem{proposition}{Proposition}
\allowdisplaybreaks

\newcommand{\figMappingInner}{4}   %

\makeatletter
\AtBeginDocument{%
  \long\def\@makecaption#1#2{%
    \vskip\abovecaptionskip
    \begingroup
      \small
      \noindent
      \parbox{\linewidth}{%
        \setlength{\parindent}{0pt}%
        \setlength{\leftskip}{0pt}%
        \setlength{\rightskip}{0pt}%
        \setlength{\parfillskip}{0pt plus 1fil}%
        \justifying
        #1. #2%
      }%
      \par
    \endgroup
    \vskip\belowcaptionskip
  }%
}
\makeatother

\begin{document}

\title{Decoupling of the QAOA into independent spin-boson systems \\
and high-depth performance on pure and mixed spin glasses} %

\author{Sami Boulebnane}
\thanks{Equal contribution.}
\affiliation{Global Technology Applied Research, JPMorganChase, New York, NY 10017 USA}%
\author{Abid Khan}
\thanks{Equal contribution.}
\affiliation{Global Technology Applied Research, JPMorganChase, New York, NY 10017 USA}%
\author{Pragna Subrahmanya}
\affiliation{Global Technology Applied Research, JPMorganChase, New York, NY 10017 USA}%
\author{Dylan Herman}
\affiliation{Global Technology Applied Research, JPMorganChase, New York, NY 10017 USA}%
\author{Edward Farhi}
\affiliation{Google Quantum AI, Venice, CA 90291 USA}%
\author{Benjamin Villalonga}
\affiliation{Google Quantum AI, Venice, CA 90291 USA}%
\author{Ruslan Shaydulin}
\affiliation{Global Technology Applied Research, JPMorganChase, New York, NY 10017 USA}%

\date{\today}

\begin{abstract}
The Quantum Approximate Optimization Algorithm (QAOA) is regarded as a promising candidate for near-term quantum advantage in combinatorial optimization, yet our ability to study it at scale is limited.
Exact recursive formulas have been introduced for predicting QAOA performance on large spin glasses, but the computational cost of evaluating them grows exponentially with the number of QAOA layers, preventing the study of QAOA in the promising high-depth regime.
In this work, we show that for the task of computing QAOA energy on any mixed dense spin glass problem, spins approximately decouple into independent spin-boson systems. In the infinite-size limit, the decoupling becomes exact, establishing the spin-boson mapping as a natural framework for the many-body physics of QAOA.
This generalized spin-boson mapping gives a recursive procedure for computing QAOA energies that can be executed at modest cost using tensor networks. As a numerical application, we optimize QAOA on pure and mixed spin glasses at depths intractable for prior techniques, observing that higher-degree problems require more layers for a comparable approximation ratio while angle optimization becomes more challenging.
While decoupling enables the evaluations of QAOA energies at large size and depth, it does not enable strong simulation of QAOA; a quantum computer is required to sample the bitstring corresponding to the predicted energy.
\end{abstract}

\maketitle

\section{Introduction}
\begin{figure*}[t]
  \centering
    \includegraphics[width=\linewidth,page=\figMappingInner]{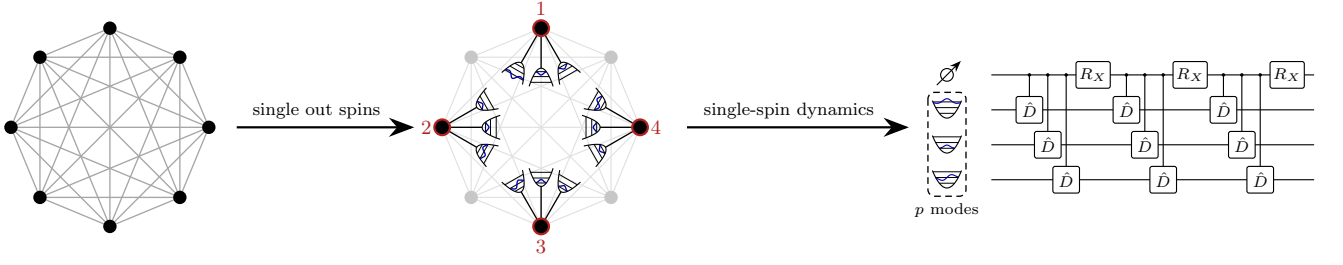}%
    \caption{Approximate decoupling of a subset $\{1, 2, 3, 4\}$ of spins in spin-glass QAOA. In the thermodynamic limit, these spins become independent, with each coupled to its own system of bosons. The evolution of each spin is described by the recursive spin-boson dynamics presented in Sec.~\ref{sec:mapping}.
    }
  \label{fig:mapping-inner}
\end{figure*}

Spin glasses constitute a paradigmatic random optimization problem, with their energy landscape and algorithmic complexity attracting considerable attention from the statistical mechanics and combinatorial optimization communities. Through Lindeberg-type replacement arguments~\cite{lindeberg1922} and Guerra--Toninelli interpolation~\cite{guerra_toninelli_thermodynamic_limit,guerra_toninelli_viana_bray}, mean-field spin glasses are closely related, in the large-degree limit, to random sparse constraint satisfaction problems such as MAX-$q$-XORSAT, both in their optimal values and in the algorithmic barriers they exhibit~\cite{extremal_cuts_sparse_random_graphs,random_max_csps_hardness_spin_glasses}.  This dense-sparse correspondence is what makes results on dense models reported here relevant to practically interesting sparse constraint satisfaction.

More recently, the power of quantum algorithms in spin glass optimization has been closely investigated in both approximate optimization~\cite{Farhi2022quantumapproximate,basso_et_al:LIPIcs.TQC.2022.7,qaoa_sparse_hypergraphs_spin_glass_models} and exact ground-state preparation settings~\cite{mind_the_gap,qaoa_polynomial_approximations}. Among other quantum combinatorial optimization paradigms (including short-path algorithms~\cite{mind_the_gap,markov_short_path_algorithms} and Decoded Quantum Interferometry~\cite{dqi}), the Quantum Approximate Optimization Algorithm (QAOA) is an attractive target for theory-grounded quantitative studies, as its analysis~\cite{Farhi2022quantumapproximate,basso_et_al:LIPIcs.TQC.2022.7,qaoa_sparse_hypergraphs_spin_glass_models,qaoa_spiked_tensor} produces both rigorous and sharp performance guarantees. The QAOA quantum circuit consists of a sequence of layers, and the performance of optimally tuned angles is non-decreasing as layers are added.
Recursive formulae for computing the performance of QAOA on spin glasses at a constant number of QAOA layers and in the infinite-size limit were introduced in Ref.~\cite{qaoa_sparse_hypergraphs_spin_glass_models}. In this regime and for most spin glass problems, the cost of solutions produced by QAOA is bounded away from the true optimum due to an intrinsic feature of the spin glass's optimization landscape known as the Overlap Gap Property (OGP)~\cite{Gamarnik2021}. The same limitation encompasses a variety of classical optimization algorithms, including low-degree polynomials~\cite{Gamarnik2020}, which were recently shown to include Approximate Message Passing~\cite{algorithmic_universality_ldp}; Simulated Annealing is also believed to be affected by the obstruction. A central question is therefore how QAOA approaches this threshold as its circuit depth increases.

We consider the performance of QAOA on various spin glass problems for a fixed but potentially large number of QAOA layers and in the infinite-size regime. This setting is already challenging, since classical evaluation of the formulae derived in Ref.~\cite{qaoa_sparse_hypergraphs_spin_glass_models} requires time exponential in the number of QAOA layers and spin order. We address this limitation by showing that the contribution of each spin to the quantities arising in these iterative formulae approximately decouples across space, making it possible to treat each spin as independently coupled to its own system of bosons (see Fig.~\ref{fig:mapping-inner}). This model provides a rigorous mean-field interpretation of spin-glass QAOA, with the important caveat that fields are quantum, i.e., non-commuting, at different layers. The correspondence is derived by a cavity calculation starting from the spin path integral representation of QAOA expectation values. Our results generalize the spin-boson mapping in Ref.~\cite{qaoa_spin_boson_mapping}, which derived the mapping for the quadratic Sherrington-Kirkpatrick (SK) model and interpreted it for quadratic problems on sparse graphs. As a byproduct of this analysis method in the context of QAOA, we obtain the first rigorous bounds on the performance of the algorithm at finite size for any $p > 1$ and mixed spin glass model. While these analytical bounds are loose, they hint at the power of these techniques to enable the analysis of QAOA at large but finite size.

The decoupling becomes exact in the infinite-size limit, enabling the evaluation of QAOA energy by simulating spin-boson systems using matrix product states (MPS). Surprisingly, the resource requirements of MPS grow only modestly with QAOA depth. Using this procedure, we optimized pure-spin models of orders $2\leq q\leq17$, at depths of several tens of layers, and extended the $p=120$ calculations of the 2-spin model reported in Ref.~\cite{qaoa_spin_boson_mapping} to $p=160$. These depths are far beyond the reach of the exact self-consistent equation evaluation of Ref.~\cite{qaoa_sparse_hypergraphs_spin_glass_models}, whose $\mathcal{O}\left(4^{pq}\right)$ cost is prohibitive.

Our numerical techniques reveal insights about QAOA performance on high-order spin glasses in high-depth regime. For pure spin glasses, we observe that QAOA approaches the OGP barrier slower as the spin order increases and that parameter optimization becomes more challenging with current techniques for higher spin order. Our results motivate further development of QAOA parameter optimization techniques targeting high-order regime. Additionally, we observe that for fixed constant depth values considered, QAOA gets closer to the algorithmic threshold for pure spin glasses than it does for mixed ones. This indicates pure spin glasses as the more promising regime for future study of QAOA as an approximation algorithm.

\section{Spin-boson mapping of spin-glass QAOA}\label{sec:mapping}

We begin by introducing an iterative procedure to evaluate the disorder-averaged, thermodynamic-limit QAOA energy density on any spin glass. This generalizes to the mixed case the $\bm{G}$ iteration derived in \cite{basso_et_al:LIPIcs.TQC.2022.7,qaoa_sparse_hypergraphs_spin_glass_models} for pure spin glasses. We then map the evaluation of this formula to computing expectation values of a recursively defined spin-boson system, generalizing to mixed spin glasses the procedure derived in Ref.~\cite{qaoa_spin_boson_mapping} for the quadratic SK model.

A spin glass problem over $n$ spins is defined by a random cost function of the form

\begin{align}
  C\left(\bm{\sigma}\right) & = \sum_{1 \leq q \leq k}\frac{c_q}{n^{(q-1)/2}}\sum_{j_1, \ldots, j_q \in [n]}J^{(q)}_{j_1, \ldots, j_q}\sigma_{j_1} \ldots \sigma_{j_q},\label{eq:random_cost_function}
\end{align}
where $\bm{\sigma} = \left(\sigma_1, \ldots, \sigma_n\right) \in \{1, -1\}^n$ is the spin configuration, the couplings $J^{(q)}_{j_1, \ldots, j_q} \overset{\mathrm{i.i.d}}{\sim} \mathcal{N}(0, 1)$ are independent standard Gaussian variables, and the $c_q \geq 0$ are non-negative coefficients. By convention, we may extend the sequence $c_1, \ldots, c_{k}$ to all positive integers by letting $c_q := 0$ for all $q > k$. This sequence entirely characterizes the random spin glass problem. From these coefficients, we define the spin glass mixture polynomial
\begin{align}
  \xi\left(x\right) & := \sum_{q \geq 0}c_q^2x^q.
\end{align}
The level-$p$ QAOA state associated with this random optimization problem, which depends on the Gaussian disorder $J^{(q)}_{j_1, \ldots, j_q}$, is denoted by
\begin{align}
    \ket{\bm{\gamma}, \bm{\beta}} & = \left(\overleftarrow{\prod_{t = 1}^p}e^{-i\beta_tB}e^{-i\gamma_tC}\right)\ket{+}^{\otimes n},\label{eq:qaoa_state}
\end{align}
with
\begin{align}
    B & := \sum_{j \in [n]}X_j
\end{align}
the transverse-field mixer Hamiltonian. We refer to $\bm{\gamma} = \left(\gamma_1, \ldots, \gamma_p\right)$ as phase separator parameters and $\bm{\beta} = \left(\beta_1, \ldots, \beta_p\right)$ as mixer angles. In the following, $\mathbb{E}_{\bm{J}}$ denotes averaging over the Gaussian disorder.

In this work, we prove a formula for the thermodynamic-limit, disorder-averaged energy density of QAOA on a problem of the above form. The formula provides a recursive way to construct a matrix
$\bm{G} \in \mathbb{C}^{2(p+1) \times 2(p+1)}$, indexed by

\begin{align}
    \mathcal{T}_p & := \left\{1, 2, \ldots, p, p + 1, -p - 1, -p, \ldots, -2, -1\right\}.\label{eq:layer_index_set}
\end{align}
The matrix is constructed in an iterative fashion, by increasing absolute value of its indices. That is, at iteration $m$, we fill entries $(r, s)$ of $\bm{G}$ such that $\max\left\{|r|, |s|\right\} = m$. We start the iteration at $m=1$ with:
\begin{align}
    G_{\pm 1,\,\pm 1} & = 1.
\end{align}
The other entries of the matrix need not be defined at this point, as they will be filled in with successive iterations.
Then, at iteration $m = 2, \ldots, p+1$, we follow:
\begin{align}
    G_{r, s} & = \sum_{\bm{a} \in \{1, -1\}^{2(m+1)}}a_ra_sf^{(m)}\left(\bm{a}\right)H^{(m)}\left(\bm{a}\right),\label{eq:g_iteration}\\
    H^{(m)}\left(\bm{a}\right) & := \exp\left(-\frac{1}{2}\sum_{t, u \in \mathcal{T}_{m - 1}}\Gamma_t\Gamma_u\xi'(G_{t, u})a_ta_u\right),
\label{eq:Hm}
\end{align}
where $\max\{|t|, |u|\} \leq m$ and $\max\{|r|, |s|\} = m + 1$.
Our results generalize the $\bm{G}$ iteration of Ref.~\cite{basso_et_al:LIPIcs.TQC.2022.7,qaoa_sparse_hypergraphs_spin_glass_models} to the mixed spin glass case.

The above equations involve summation over $\mathcal{T}_m$-indexed bitstrings
\begin{align}
    \bm{a} & = \left(a_1, \ldots, a_{m + 1}, a_{-m - 1}, \ldots, a_{-1}\right) \in \{1, -1\}^{2(m+1)}.
\end{align}
Tensor $f^{(m)}\left(\bm{a}\right)$ is defined from QAOA mixer angles by
\begin{align}
    f^{(m)}\left(\bm{a}\right) & = \frac{1}{2}\bm{1}\left[a_{m + 1} = a_{-m - 1}\right]\nonumber\\
    & \hspace*{15px} \times \prod_{1 \leq t \leq m}\bra{a_t}e^{i\beta_tX}\ket{a_{t + 1}}\bra{a_{-t - 1}}e^{-i\beta_tX}\ket{a_{-t}}.\label{eq:mixer_tensor}
\end{align}
In Eq.~\eqref{eq:Hm} we introduce:
\begin{align}
    \bm{\Gamma} & = \left(\Gamma_1, \ldots, \Gamma_p, \Gamma_{-p}, \ldots, \Gamma_{-1}\right)\\
    & := \left(\gamma_1, \ldots, \gamma_p, -\gamma_p, \ldots, -\gamma_1\right) \in \mathbb{R}^{2p}.
\end{align}
After constructing all $\bm{G}$ entries in this iterative fashion, the thermodynamic-limit, disorder-averaged QAOA energy density is given by
\begin{align}
    \lim_{n \to \infty}\mathbb{E}_{\bm{J}}\bra{\bm{\gamma}, \bm{\beta}}C/n\ket{\bm{\gamma}, \bm{\beta}} & = \sum_{t \in \mathcal{T}_{p - 1}}i\Gamma_t\xi\left(G_{p + 1, t}\right)
\end{align}
The above generalizes to mixed spin glasses a formula established for pure spin glasses in Refs.~\cite{Farhi2022quantumapproximate, basso_et_al:LIPIcs.TQC.2022.7,qaoa_sparse_hypergraphs_spin_glass_models}. In Supplementary Information, we provide two proofs of the formula: one starting from an alternative expression from Ref.~\cite{qaoa_sparse_hypergraphs_spin_glass_models}, and one from a new argument involving a cavity calculation. We then show the QAOA state maps to a recursively defined spin-boson state, allowing evaluation of $\bm{G}$ iteration Eq.~\ref{eq:g_iteration}. More specifically, we describe a joint update of a Hermitian matrix $\bm{G}^{\mathrm{herm}} \in \mathbb{C}^{(p + 1) \times (p + 1)}$, an upper-triangular matrix $\bm{\widetilde{L}} \in \mathbb{C}^{p \times p}$, and a spin-boson state $\ket{\Psi^{(m)}}$. $\bm{G}$ can then be deduced from $\bm{G}^{\mathrm{herm}}$ by symmetry. The spin-boson system consists of a single spin $1/2$ and $p$ bosonic modes, initially in state
\begin{align}
    \ket{\Psi^{(0)}} & := \ket{+}_{\mathrm{spin}} \otimes \ket{\overline{0}}_{\mathrm{boson}}^{\otimes p},\label{eq:spin_boson_initial_state}
\end{align}
with $\ket{\overline{0}}_{\mathrm{boson}}$ the bosonic vacuum. Initialize
\begin{align}
    G^{\mathrm{herm}}_{1, 1} = 1.
\end{align}
At the start of recursion step $m \in \{1, \ldots, p\}$, assume inductively that the top $m \times m$ block of $\bm{G}^{\mathrm{herm}}$, the top $(m - 1) \times (m - 1)$ block of $\bm{\widetilde{L}}$ and the state $\ket{\Psi^{(m - 1)}}$ have been constructed such that
\begin{align}
    \bm{\widetilde{L}}_{1:m - 1,\,1:m - 1}^{\dagger}\bm{\widetilde{L}}_{1:m - 1,\,1:m - 1} & = \xi'^{\odot}\left(\bm{G}_{1:m - 1,\,1:m - 1}^{\mathrm{herm}}\right),
\end{align}
with $\left( \xi'^{\odot}\left(\bm{G}_{1:m - 1,\,1:m - 1}^{\mathrm{herm}}\right) \right)_{r,s} := \xi'\left(G^{\mathrm{herm}}_{r, s}\right)$ for $1 \leq r, s \leq m-1$.
Update the first $m$ elements of column $m$ of $\bm{\widetilde{L}}$, $\bm{\ell}^{(m)} := \bm{\widetilde{L}}_{1:m,\,m}$, to extend the above Cholesky factorization to block $m \times m$:
\begin{align}
    \bm{\widetilde{L}}_{1:m,\,1:m}^{\dagger}\bm{\widetilde{L}}_{1:m,\,1:m} & = \xi'^{\odot}\left(\bm{G}^{\mathrm{herm}}_{1:m,\,1:m}\right)
\end{align}
The above only requires knowledge of top-left $m \times m$ block of $\bm{G}^{\mathrm{herm}}$, assumed already constructed at the start of the current recursion step $m$.
Explicitly,
\begin{align}
    \bm{\ell}^{(m)}_{1:m - 1} & = \bm{\widetilde{L}}_{1:m - 1,\,1:m - 1}^{\dagger{-1}}\xi'^{\odot}\left(\bm{G}^{\mathrm{herm}}_{1:m - 1,\,m}\right),\\
    \ell^{(m)}_m & = \sqrt{\xi'(1) - \bm{\ell}^{(m)\dagger}_{1:m - 1}\bm{\ell}^{(m)}_{1:m - 1}}.
\end{align}
Next, update spin-boson state as
\begin{align}
    \ket{\Psi^{(m)}} & := U^{(m)}\ket{\Psi^{(m - 1)}}\label{eq:spin_boson_state_update},\\
    U^{(m)} & := \exp\left(-i\beta_mX\right)D\left(-i\gamma_mZ\bm{\ell}^{(m)}\right),\label{eq:spin_boson_layer}
\end{align}
where
\begin{align}
    D\left(Z\bm{\alpha}\right) & := \sum_{\sigma \in \{1, -1\}}\ket{\sigma}\bra{\sigma}_{\mathrm{spin}} \otimes D\left(\sigma\bm{\alpha}\right)_{\mathrm{bosons}}
\end{align}
is the displacement operator of vector $\bm{\alpha}$, signed by the spin $Z$ projection. Explicitly,
\begin{align}
    D\left(\bm{\alpha}\right)_{\mathrm{bosons}} & := \exp\left(\bm{\alpha} \cdot \bm{\hat{c}}^{\dagger} - \overline{\bm{\alpha}} \cdot \bm{\hat{c}}\right), \quad \bm{\alpha} \in \mathbb{C}^p,
\end{align}
where $\bm{\hat{c}} = \left(\hat{c}_1, \ldots, \hat{c}_p\right)$ are the annihilation operators of bosonic modes. From $\ket{\Psi^{(m)}}$, construct vector $\bm{b}^{(m)} \in \mathbb{C}^m$ of $Z$-annihilation expectation values
\begin{align}
    b^{(m)}_l & := \bra{\Psi^{(m)}}Z\hat{c}_l\ket{\Psi^{(m)}}, \qquad 1 \leq l \leq m.\label{eq:z_annihilation_expectation_values}
\end{align}
Then, from now construct top $m \times m$ block of $\bm{\widetilde{L}}$, construct matrix $\bm{A}^{(m)} \in \mathbb{C}^{m \times m}$ as
\begin{align}
    A^{(m)}_{l, t} & := -i\gamma_t\widetilde{L}_{l, t}, \qquad 1 \leq l, t \leq m.
\end{align}
Then, define row $(m + 1)$ of $\bm{G}^{\mathrm{herm}}$ as:
\begin{align}
    \bm{G}^{\mathrm{herm}}_{m + 1,\,1:m} = \bm{A}^{(m)-1}\bm{b}^{(m)}, \quad G^{\mathrm{herm}}_{m + 1,\,m + 1} = 1,\label{eq:g_iteration_spin_boson_column_update}
\end{align}
and deduce column $(m + 1)$ by Hermiticity. This completes the construction of $\bm{G}^{\mathrm{herm}}_{1:m + 1,\,1:m + 1}, \bm{\widetilde{L}}_{1:m,\,1:m}, \ket{\Psi^{(m)}}$ given $\bm{G}^{\mathrm{herm}}_{1:m,\,1:m}, \bm{\widetilde{L}}_{1:m - 1,\,1:m - 1}, \ket{\Psi^{(m - 1)}}$. As step $m = p$, the procedure terminates and we define for all $1 \leq r \leq s \leq p + 1$,
\begin{align}
    G_{-r, s} & = G_{s, -r} = G_{-r, -s} = G_{-s, -r} = G^{\mathrm{herm}}_{s, r},\\
    G_{r, s} & = G_{s, r} = G_{r, -s} = G_{-s, r} = \overline{G^{\mathrm{herm}}_{s, r}}.
\end{align}
We claim the above procedure based on a spin-boson state evolution yields the same $\bm{G}$ matrix as exact iteration Eq.~\ref{eq:g_iteration}.

\section{Physical interpretation of the mapping}
\label{sec:physical_interpretation_mapping}

\subsection{The spin path integral formalism}

In this paragraph, we introduce the spin path integral formalism for expressing expectation values of unitary quantum circuits. This formalism was the starting point of many analytic formulae for QAOA performance metrics~\cite{Farhi2022quantumapproximate,qaoa_sparse_hypergraphs_spin_glass_models,qaoa_spiked_tensor}, and also forms the basis of our cavity argument showing approximate decoupling of spins and spin-boson equivalence in the thermodynamic limit. The spin path integral is a sum over the computational basis trajectory of all spins, weighted by the path integral weight. The spin path integral can be defined for any $n$-qubit quantum circuit that is partitioned into $p$ unitary layers $U_1, \ldots, U_p$ and starts in an initial state $\ket{\psi_0}$.

The path integral weight associated with a spin trajectory $\bm{z} = \left(\bm{z}^{[1]}, \ldots, \bm{z}^{[p + 1]}, \bm{z}^{[-p - 1]}, \ldots, \bm{z}^{[-1]}\right)$ is
\begin{align}
    \mu\left(\bm{z}\right) & := \prod_{1 \leq t \leq p}\bra{\bm{z}^{[t]}}U_t^{\dagger}\ket{\bm{z}^{[t + 1]}}\bra{\bm{z}^{[-t - 1]}}U_t\ket{\bm{z}^{[-t]}}\nonumber\\
    & \hspace*{15px} \times \bm{1}\left[\bm{z}^{[p + 1]} = \bm{z}^{[-p - 1]}\right]\braket{\psi_0|\bm{z}^{[1]}}\braket{\bm{z}^{[-1]}|\psi_0}.\label{eq:path_integral_measure_generic_layered_circuit_p2}
\end{align}
In the above scalar expression, $\bm{z}^{[t]} = \left(z_j^{[t]}\right)_{j \in [n]}$ collects the computational basis states of all spins at layer $t$. For $j \in [n]$, we also denote by $\bm{z}_j = \left(z_j^{[t]}\right)_{t \in \mathcal{T}_p}$ the computational basis states assumed by spin $j$ across all layers.
$\bm{z}$ is an $n \times 2(p+1)$ bit matrix with columns indexed by $[t]$ and rows indexed by $j$.
The above path integral weight
is associated with the layered circuit that prepares the state $\ket{\psi} = U_p \ldots U_1\ket{\psi_0}$. By the computational basis completeness relation, any expectation value involving $Z$ operators can be evaluated as a (quasiprobability) expectation under the path integral weights. For instance,
\begin{align}
    \bra{\psi}Z_1Z_2\ket{\psi} & = \sum_{\bm{z} \in \{1, -1\}^{n \times 2(p + 1)}}\mu\left(\bm{z}\right)z_1^{[p + 1]}z_2^{[p + 1]}\\
    & =: \left\langle z_1^{[p + 1]}z_2^{[p + 1]} \right\rangle_{\mu\left(\bm{z}\right)}.
\end{align}

Besides fixed-time observables, the path integral can express spatio-temporal correlators of $Z$ operators under the quantum circuit. For instance,

\begin{align}
    \bra{\psi_0}Z_1(u)Z_2(t)\ket{\psi_0} & = \left\langle z_1^{[-u]}z_2^{[-t]} \right\rangle_{\mu\left(\bm{z}\right)},
\end{align}

where $Z_1(u) := U_1^{\dagger} \ldots U_{u - 1}^{\dagger} Z_1 U_{u - 1} \ldots U_1$ and $Z_2(t) := U_1^{\dagger} \ldots U_{t - 1}^{\dagger} Z_2 U_{t - 1} \ldots U_1$ ($u \geq t$) are Heisenberg-evolved operators.

\subsection{Spin decoupling in the thermodynamic limit}

We now explain the physical mechanism enabling the spin-boson formula. Our central insight is that,
in the thermodynamic limit and after disorder averaging, the spins in any constant-size set approximately behave independently, with each coupled to its own $p$ bosonic modes, as illustrated in Fig.~\ref{fig:mapping-inner}. What factorizes, concretely, is the disorder-averaged path integral weights restricted to that set: expectations of products of the spins' computational-basis values across layers reduce to a product over the set, each factor being the single-spin dynamics defined in Eq.~\ref{eq:spin_boson_layer}. Physically, the bosonic couplings emerge as an effective interaction with the integrated-out spin bulk. Hence, the correspondence is a system-bath interaction in a discrete setting whose bath parameters are given explicitly by familiar QAOA quantities. The approximate decoupling results can be conveniently framed in spin path integral form.

\begin{proposition}[Spin decoupling in thermodynamic limit]
\label{prop:spin_decoupling}
Let $\mu^{[p]}\left(\bm{z}; \bm{J}\right)$ be the path integral weight of Eq.~\ref{eq:path_integral_measure_generic_layered_circuit_p2} applied to the spin-glass depth-$p$ QAOA state of Eq.~\ref{eq:qaoa_state}, with Gaussian disorder $\bm{J}$. Let $\mu^{[p]}_{SB}\left(\bm{a}\right)$ be the path integral weight of the spin in the spin-boson circuit defined by the iterative update in Eq.~\ref{eq:spin_boson_state_update}~and~\ref{eq:spin_boson_layer}. The path integral weight is defined for a spin computational basis trajectory $\bm{a} = \left(a_t\right)_{t \in \mathcal{T}_p}$. Isolate a constant number, $n'$, of spins and let $\left(\tau_j\right)_{j \in \left[n'\right]}, \tau_j \subset \mathcal{T}_p$ layer index sets for each spin $\{1, \ldots, n'\}$. Then,
\begin{align}
    \lim_{n \to \infty}\mathbb{E}_{\bm{J}}\left\langle \prod_{j \in \left[n'\right]}z_j^{\left[\tau_j\right]} \right\rangle_{\mu^{[p]}\left(\bm{z}; \bm{J}\right)} & = \prod_{j \in \left[n'\right]}\left\langle a_{\tau_j} \right\rangle_{\mu_{SB}^{[p]}\left(\bm{a}\right)},
\end{align}
where path integral expectations are denoted by brackets, and we used shorthand
\begin{align}
    z_j^{\left[\tau_j\right]} := \prod_{t \in \tau_j}z_j^{[t]}, \qquad a_{\tau_j} := \prod_{t \in \tau_j}a_t.
\end{align}
for Boolean monomials in spin variables.
\end{proposition}

Technically, Proposition~\ref{prop:spin_decoupling} asserts approximate spatial independence of qubits in spin-glass QAOA, while their independent temporal $Z$ correlators are approximately reproduced by a spin-boson system. Proposition~\ref{prop:spin_decoupling} elucidates two aspects of QAOA.

First, it clarifies the relationship between QAOA and the mean-field classical algorithms like those of Refs~\cite{mean_field_approximate_optimization_algorithm,2604.01512,2607.08708} Our results show that in QAOA, the effective field experienced by each spin is non-commuting, i.e. quantum. Consequently, the correspondence we derive does not say anything about efficient classical simulation of spin-glass QAOA, including in the average-instance case and for constant-size subsystems.

Second, the derivation can be interpreted as a reduction: any disorder-averaged expectation of $Z$ operators on a constant-size set (at a single layer, or as a time autocorrelation across layers) becomes a statement about the corresponding independent spin-boson dynamics. This makes the spin-boson mapping as a natural framework for studying the spin-glass QAOA dynamics in the thermodynamic limit and average-instance setting.

The full derivation of the mapping is deferred to Supplementary Information. At a high level, the mapping is derived from the spin path integral weights of the QAOA circuit, the starting point of many analytic formulae for QAOA expectation values~\cite{Farhi2022quantumapproximate,basso_et_al:LIPIcs.TQC.2022.7,qaoa_sparse_hypergraphs_spin_glass_models,qaoa_spiked_tensor}. We derive a cavity decomposition of this path integral weight associated with a partition of spins between a small subsystem $\{1, \ldots, n''\}$ and its environment, the bulk system $\{n'' + 1, \ldots, n\}$. We then find that, after integrating out the environment $\{n'' + 1, \ldots, n\}$ and taking the thermodynamic limit $n' := n - n'' \longrightarrow \infty$, this bulk system acts on the small subsystem $\{1, \ldots, n''\}$ as a bosonic environment, and that the path integral weight further factorizes according to particles $\{1, \ldots, n''\}$. Unlike applications of the cavity method in classical statistical mechanics, the effective field created by the bulk system on the small system is quantum and is described by non-commuting displacement operators. Additionally, the bulk's overlaps stop fluctuating. These overlaps are site-averaged correlations between two layers of the same trajectory, and under the bulk path integral weights they converge to the entries of $\bm{G}$ in the infinite-size limit. Once they are deterministic, the bulk's influence on the small subsystem enters only as a single sum over spins $\{1, \ldots, n''\}$, so the path integral weight factorizes spin by spin (Eq.~\ref{eq:cavity_factorization}). Because the overlaps entering a given layer involve only earlier ones, these two steps chain into an induction over the number of QAOA layers, whose full proof is given in Supplementary Information.

\section{Numerical Results}\label{sec:numerics}

The spin-boson mapping gives a recursive procedure for evaluating QAOA energies that can be accelerated by simulating the (recursive) quantum evolution of spin-boson states approximately using matrix product states (see Methods).
This gives direct access to thermodynamic-limit QAOA energies at previously inaccessible depths. We use this access to optimize pure and mixed spin models, and arrive at three main numerical results.

\subsection{Convergence with depth on pure spin glasses}

\begin{figure*}[htbp]
    \centering
    \includegraphics[width=\linewidth]{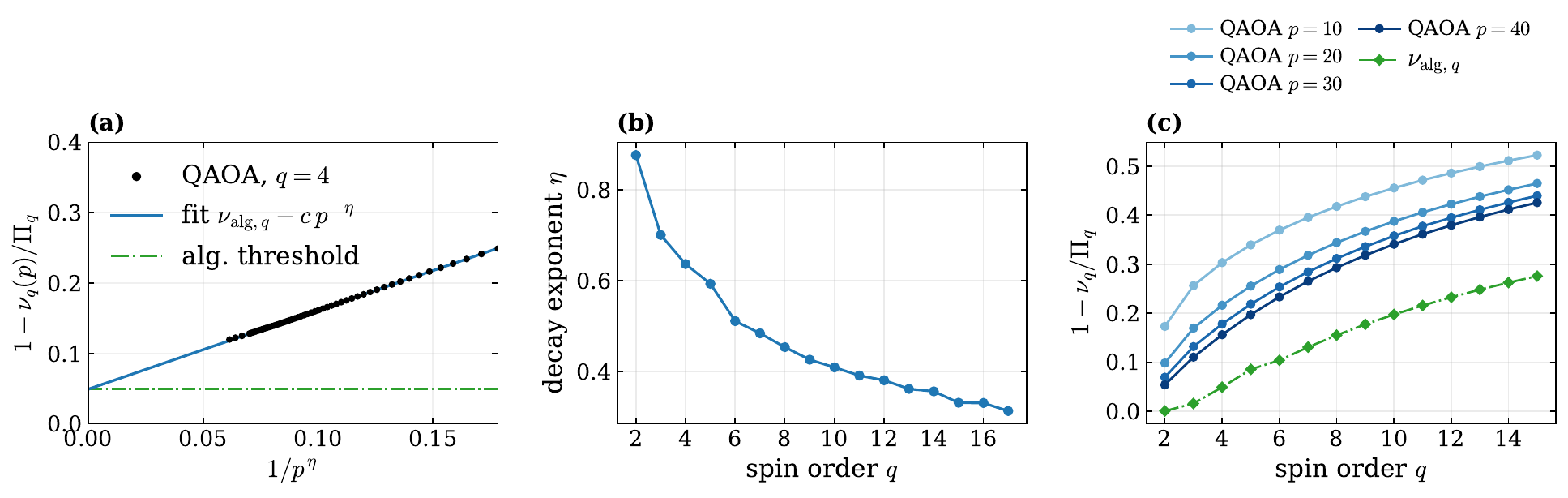}
    \caption{Optimized QAOA on pure $q$-spin models. \textbf{(a)} Relative gap to the Parisi value, $1-\nu_q(p)/\Pi_q$, for the pure 4-spin model against $1/p^\eta$, with the fit $\nu_q(p)=\nu_{\mathrm{alg},q}-c\,p^{-\eta}$ (blue), whose $p\to\infty$ asymptote is pinned to the algorithmic threshold (green dash-dot); $c=0.669$ and $\eta=0.648$, from the $54$ points with $15\leq p\leq80$. \textbf{(b)} Decay exponent $\eta$ (left axis) and the algorithmic threshold as a fraction of the Parisi optimum, $\nu_{\mathrm{alg},q}/\Pi_q$ (right axis, dash-dot), against spin order; curve colors match their axes, which scale independently. \textbf{(c)} Relative gap $1-\nu_q/\Pi_q$ at fixed depths $p=10,20,30,40$ (blue shades) and at the algorithmic threshold (green dash-dot). }
    \label{fig:pure-spin-overview}
\end{figure*}

First, we find that QAOA approaches the algorithmic barrier slower as the spin order $q$ increases. We optimize pure-spin models with orders $2\leq q\leq17$, reaching maximum depths between $p=40$ and $p=80$, depending on the order. Additionally, for the 2-spin model, we extend the depth-$120$ results of Ref.~\cite{qaoa_spin_boson_mapping} to $p=160$. For each pure model, we compare the optimized QAOA energy density $\nu_q(p)$ with the order-dependent algorithmic threshold $\nu_{\mathrm{alg},q}$ obtained from the variational principle of Ref.~\cite{optimization_mean_field_spin_glasses}. Energies are reported relative to $\Pi_q$, the optimal energy density of the pure $q$-spin model in the thermodynamic limit, given by the Parisi formula~\cite{PhysRevLett.43.1754,talagrand_parisi_formula,Panchenko2013}. The quantity $1-\nu/\Pi_q$ is then the relative gap to the optimum, and $\nu_{\mathrm{alg},q}\leq\Pi_q$.

To quantify the large-depth trend without introducing a free asymptote, we test the two-parameter form
\begin{align}
    \nu_q(p) = \nu_{\mathrm{alg},q} - c p^{-\eta},
\end{align}
with $\nu_{\mathrm{alg},q}$ fixed. The fits begin at $p=15$ and extend to the largest available depths. We emphasize that pinning the asymptote makes this a consistency test of convergence to the algorithmic threshold, rather than evidence that such convergence must occur. As a representative example, Fig.~\ref{fig:pure-spin-overview}a shows the 4-spin data and fit over the 54 available depths $15\leq p\leq80$, giving $c=0.669$ and $\eta=0.648$. Repeating the fit across spin orders gives Fig.~\ref{fig:pure-spin-overview}b. The fitted exponent $\eta$ decreases monotonically with $q$, from $\eta=0.88$ at $q=2$ to $\eta=0.39$ at $q=17$, indicating that QAOA needs more layers on problems with higher spin order to reach the same achievable solution quality. Appendix~\ref{app:per-order-fits} shows the underlying fits separately for $q=3$ through $8$. While Fig.~\ref{fig:pure-spin-overview}b rests on the consistency test fits, the same conclusion follows without any fit and extrapolation. At fixed depth, the relative gap $1-\nu_q(p)/\Pi_q$ widens monotonically with spin order, as Fig.~\ref{fig:pure-spin-overview}c shows for $p=10,20,30,40$. Overall, our results suggest that for pure spin glasses, and under the optimistic assumption that infinite-size QAOA converges to the algorithmic threshold as $p \to \infty$, the convergence slows down as the order of the spin glass increases.

\subsection{Mixed order spin glasses}

\begin{figure*}[htbp]
    \centering
    \includegraphics[width=\textwidth]{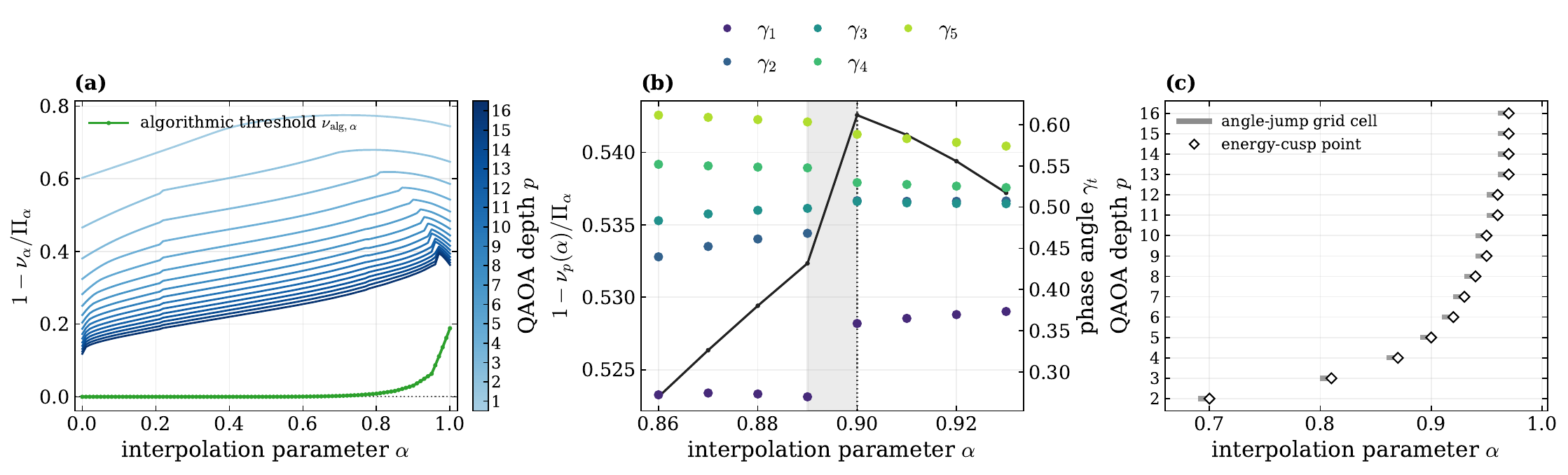}
    \caption{QAOA along the 2-spin to 8-spin path $\xi(t)=(1-\alpha)t^2/2+\alpha t^8/8$ (pure 2-spin at $\alpha=0$, pure 8-spin at $\alpha=1$). \textbf{(a)} Algorithmic threshold and optimized QAOA energy densities for different values of $p$. Energies are reported as the relative gap to the optimum $1-\nu_\alpha/\Pi_\alpha$; each blue curve is a fixed QAOA depth $p$ (darker is deeper), and the green curve is the algorithmic threshold $\nu_{\mathrm{alg},\alpha}$. The optimized energy exhibits points of non-differentiability (``cusp"). The rightmost cusp occurs at an increasingly higher value of $\alpha$ as $p$ increases. \textbf{(b)} Optimized $\bm{\gamma}$ angles and energy at $p= 5$ around the rightmost cusp, located at a threshold value $\alpha^* \in [0.89, 0.90]$. Discontinuity of optimized $\bm{\gamma}$ angles is visible around $\alpha^*$ and more pronounced for the earliest angles of the schedule; in contrast, the optimized energy remains continuous, but non-differentiable, at $\alpha^*$. \textbf{(c)} Location of the rightmost cusp $\alpha^*$ as a function of QAOA depth $p$. Grey horizontal bars denote the magnitude of the largest $\bm{\gamma}$ jump across the threshold value. The location of the cusp appears to get closer to $1$ as $p$ increases. It remains unclear whether $\alpha^*$ converges to $1$ as $p \to \infty$}
    \label{fig:gap-vs-alpha-p8}
\end{figure*}

Second, we consider a spin glass mixture of 2-spin and 8-spin orders, and find that the performance gap between QAOA and the algorithmic barrier is smallest for pure spin and varies nontrivially with the level of mixture of the spin glass. Additionally, we find that the optimized angles do not continuously transform from pure 2-spin to pure 8-spin optimized angles.
The pure models isolate the effect of spin order, but the mapping applies to arbitrary mixtures. To demonstrate this broader scope, we consider the simplest nontrivial interpolation, with cost function
\begin{align}
    C_{\alpha}\left(\bm{\sigma}\right) & := \sqrt{\frac{1 - \alpha}{2}}\sum_{j_1, j_2 \in [n]}n^{-1/2}J_{j_1, j_2}\sigma_{j_1}\sigma_{j_2}\nonumber\\
    & \hspace*{15px} + \sqrt{\frac{\alpha}{8}}\sum_{j_1,\ldots,j_8 \in [n]}n^{-7/2}J_{j_1,\ldots,j_8}\sigma_{j_1}\cdots\sigma_{j_8}.
\end{align}
The parameter $\alpha \in [0, 1]$ interpolates continuously between the pure 2-spin glass ($\alpha = 0$) and the pure 8-spin glass ($\alpha = 1$), while preserving $\xi'(1)=1$. The corresponding mixture polynomial is
\begin{align}
    \xi\left(x\right) & = \xi_2x^2 + \xi_8x^8\\
    & = \frac{1 - \alpha}{2}x^2 + \frac{\alpha}{8}x^8.
\end{align}

The left-hand panel of Figure~\ref{fig:gap-vs-alpha-p8} shows the optimized QAOA energy density across this path for depths $1\leq p\leq20$, normalized by the corresponding Parisi value $\Pi_\alpha$. Interestingly, the gap between algorithmic threshold and QAOA value depends on interpolation parameter $\alpha$.

We note that along the interpolation path, optimized angles vary discontinuously, as exemplified in the middle panel at depth $p = 5$. The right-hand panel shows the location of the discontinuity of largest interpolation parameter $\alpha$ as a function of $p$. The corresponding $\alpha = \alpha^*\left(p\right)$ value appears to increase with $p$, though a numerical resolution of a potential $p \to \infty$ limit challenging. However, assuming for the sake of the argument that $\alpha^*\left(p\right) \to 1$ as $p \to \infty$, the observed phase transition implies that for any mixed spin glass with $\alpha$ close to $1$, parameter extrapolation in $p$ will ultimately fail at sufficiently large $p$: for sufficiently small $p$, $\alpha^*\left(p\right) < \alpha$ and parameters belong to the phase on the right of the cusp, while for sufficiently large $p$, $\alpha^*\left(p\right) > \alpha$ and parameters belong to the phase on the left of the cusp. In Appendix~\ref{sec:angles_smoothness}, we present theoretical evidence that these discontinuities are essential features of the optimization landscape, rather than accidental artifacts of our optimization procedure.

\subsection{The angle-optimization landscape}

\begin{figure*}[htbp]
    \centering
    \includegraphics[width=\linewidth]{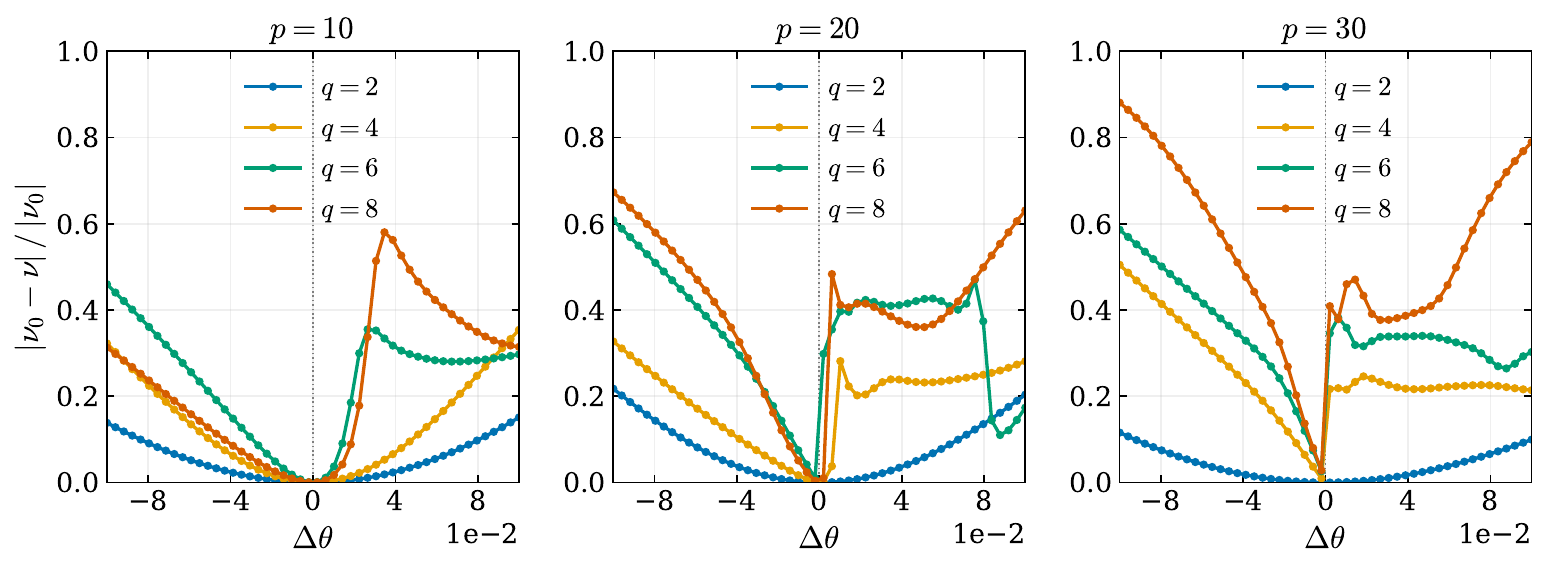}
    \caption{Relative energy change $|\nu_0-\nu|/|\nu_0|$ vs.\ signed angle displacement
    $\Delta\theta$ along the residual gradient direction at the optimized angles, where
    $\left|\Delta\theta\right|$ is the largest displacement applied to any single angle. The panels correspond to QAOA depths
    $p=10, 20, 30$; within each panel, the spin orders $q=2,4,6,8$ are shown, with markers
    at the sampled displacements (50 per curve). All panels share the fixed ranges
    $\Delta\theta\in[-10^{-1},10^{-1}]$ and $|\nu_0-\nu|/|\nu_0|\in[0,1]$. The basin narrows as $p$ grows for every order except $q=2$, and at each depth the largest relative change over the window increases with spin order; the individual curves are non-monotonic in $\Delta\theta$. Due to this narrow gorge phenomenon, optimization of QAOA parameters at $q > 2$ becomes increasingly challenging as $p$ increases, unless a very accurate initialization can be inferred.}
    \label{fig:sensitivity-k4-k8}
\end{figure*}

Finally, we find that, for high depth $p$ and spin order larger than $q=2$, finding optimal angles proves rather difficult, motivating future work on better warm starting QAOA angle optimization.
To understand this, we examine the local geometry around the obtained optimal parameters. Starting from each optimized angle vector $\bm{\theta}_0 =(\bm{\gamma},\bm{\beta})$ for a given configuration $(p,q)$, we displace all $2p$ angles along the residual finite-difference gradient $\nabla_{\bm{\theta}}\nu_q(p;\bm{\theta}_0)$ and record the relative energy change $|\nu_0-\nu|/|\nu_0|$. We note that the gradient does not completely vanish at $\bm{\theta}_0$ due to the stopping criteria set by the optimizer. We parameterize the displacement by its largest component, $\left|\Delta\theta\right| = \Delta\lVert\nabla_{\bm{\theta}}\nu\rVert_\infty$, so that $\left|\Delta\theta\right|$ is the sup-norm of the angle displacement and no individual angle moves further than it.

For $p = 10,20,30$ and $q=2,4,6,8$, we scan the window $\Delta\theta\in\left[-10^{-1}, 10^{-1}\right]$ and show the energy lanscape in Fig.~\ref{fig:sensitivity-k4-k8}.
We find that for  $q=2$ (the SK model), across all three depths, the basin formed by this slice along parameter space remains convex, and one would expect an optimizer to reach the optimum starting from a initial point $\Delta\theta = 10^{-1}$ away from the optimal parameters, independent of depth $p$. On the other hand, for $q>2$, the landscape sharpens markedly with $p$ and $q$. For example, at the smallest displacement we sample, $\left|\Delta\theta\right|=2\times10^{-3}$, the relative energy change of the 4-spin model grows from $0.02\%$ to $22\%$ as $p$ grows from $10$ to $30$. The sharpening with increasing $p$ becomes more drastic as as $q$ increases. The extreme case is the 8-spin model at $p=30$, where a displacement of $\left|\Delta\theta\right|=2\times10^{-3}$ changes the energy by $41\%$.

While these observations are made by examining only a 1D slice of the parameter landscape,
they suggest highly accurate initialization of parameters is needed for optimization in the high-$p$, high-$q$ regime. We supply our techniques in Section~\ref{sec:methods}, while Appendix~\ref{app:angle-schedules} shows the optimized schedules at $p=40$ for all $q=2,\ldots,17$.

All in all, the presence of narrowing gorges around optimal angles at large $q$ motivates further research into parameter setting in the high-$p$, high-$q$ regime, previously inaccessible numerically due to the exponential cost in $p$.

\section{Discussion}\label{sec:discussion}

We proved a sharp correspondence between constant-size reduced subsystems of spin-glass QAOA and independent spin-boson systems, valid after disorder averaging. This generalizes the mapping established for the quadratic model in Ref.~\cite{qaoa_spin_boson_mapping} to the dense case and a much broader variety of optimization problems. Constant-size subsystems are natural objects for studying QAOA in the constant-angles, infinite-size regime. Indeed, in this regime, the statevector overlap between consecutive layers is expected to decay exponentially in $n$, ruling out the statevector as a physically informative description of the system's dynamics. In contrast, low-degree observables, notably the energy density, are expected to vary smoothly between consecutive layers. Our result reduces statements about constant-size subsystems to corresponding statements about the spin-boson mapping, establishing the latter as a natural framework, rather than a mere computational artifact, for the fundamental understanding of the QAOA dynamics in the constant-angles, infinite-size regime.

From a path integral point of view, the approximate decoupling of spins in the thermodynamic limit manifests as a convergence of the path integral weight to that of decoupled spin-boson systems. By expressing the finite-size path integral weight relative to this limiting weight, we were able to derive explicit bounds on the performance of QAOA at finite size. To the best of our knowledge, no such bounds were previously derived at $p > 1$ and general (potentially, mixed) spin glasses. For all finite $p$, the bounds imply that the QAOA energy-density error decays as $n^{-\alpha_p}$, up to logarithmic factors, with $\alpha_p=1/(2a_p)$; for fixed $\gamma$ and $x$, this exponent decreases at an inverse-factorial scale with $p$. We believe this estimate to be extremely loose and expect, based on the $p = 1$ case, that the correct scaling is instead $1/n$. Moreover, the prefactor in our sufficient-size threshold grows superexponentially with $p$, whereas our numerics suggest that the finite-size correction has only polynomial dependence on $p$. We hope the detailed derivation of our finite-size bounds may help identify the current fundamental obstructions to more accurate estimates, and support their gradual improvement.

Irrespective of the limitations of our mathematical techniques, we empirically demonstrated efficient simulation of the spin-boson mapping using matrix product states, following and improving upon the methods of Ref.~\cite{qaoa_spin_boson_mapping}. While the theoretical proof of spin decoupling left open the amount of entanglement between bosonic modes, the efficiency of the MPS simulation suggests only moderate entanglement growth. Understanding the structural origin of this limited growth and why angle sensitivity increases with spin order is an interesting direction for future research.

Our numerical simulations reached depths of approximately $p=40$--$50$ for most pure models with $2\leq q\leq17$. For the 2-spin model, we extended the previous $p=120$ result~\cite{qaoa_spin_boson_mapping} to $p=160$, a regime intractable for the exact $\mathcal{O}(4^p)$ evaluation. From optimized energy densities computed at large $p$, we inferred a scaling for this quantity as a function of $p$, based on a frugal, two-parameter power-law fit. Irrespective of the functional form of the fit, the available data at large $p$ indicate decreasing efficiency of QAOA convergence to the algorithmic threshold as spin order increases. While our numerical studies focused on pure spin glasses for definiteness, our methods generalize to mixed spin glasses, and hence, by sparse-dense equivalence, to a broad variety of random constraint satisfaction problems. Exploring this much broader space of problems in a principled way is an interesting direction for future research. This may help uncover a more substantial quantum advantage and lead to a better understanding of problem features leveraged by QAOA.

\section{Methods}\label{sec:methods}

\subsection{Sketch of derivation of spin decoupling}

We now sketch the cavity calculation argument yielding approximate decoupling of spins into independent spin-boson systems, with full proofs deferred to Supplementary Information. The cavity calculation is based on the spin path integral formulation of QAOA expectation values. According to the general formula Eq.~\ref{eq:path_integral_measure_generic_layered_circuit_p2}, for any spin glass instance characterized by Gaussian disorder $\bm{J}$, the corresponding QAOA admits a disorder-dependent path integral weight:

\begin{align}
    \mu\left(\bm{z}; \bm{J}\right) & = \braket{+|\bm{z}^{[1]}}\braket{\bm{z}^{[-1]}|+}\bm{1}\left[\bm{z}^{[p + 1]} = \bm{z}^{[-p - 1]}\right]\nonumber\\
    & \hspace*{15px} \times \prod_{1 \leq t \leq p}\bra{\bm{z}^{[t]}}U_t^{\dagger}\ket{\bm{z}^{[t + 1]}}\nonumber\\
    & \hspace*{15px} \times \prod_{1 \leq t \leq p}\bra{\bm{z}^{[-t - 1]}}U_t\ket{\bm{z}^{[-t]}},
\end{align}

where the dependence on the Gaussian disorder enters through the unitaries $U_t$. In Supplementary Information, we first show the disorder-averaged energy density $\nu := \mathbb{E}_{\bm{J}}\bra{\bm{\gamma}, \bm{\beta}}C/n\ket{\bm{\gamma}, \bm{\beta}}$ can be expressed exactly in terms of the disorder-averaged path integral weight
\begin{align}
    \mu\left(\bm{z}\right) & := \mathbb{E}_{\bm{J}}\mu\left(\bm{z}; \bm{J}\right).\label{eq:disorder_averaged_path_integral_measure}
\end{align}
While the calculation is straightforward, this statement does require a justification since both the loss function $C/n$ and the state $\ket{\bm{\gamma}, \bm{\beta}}$ in the disorder-averaged energy density depend on $\bm{J}$. We obtain the final expression
\begin{align}
    \nu & = \sum_{t \in \mathcal{T}_{p - 1}}i\Gamma_t\left\langle \xi\left(R^{[p + 1, t]}\left(\bm{z}\right)\right) \right\rangle_{\mu\left(\bm{z}\right)},\label{eq:energy_density_exact_expression}
\end{align}
where the disorder-averaged path integral weight
is given by
\begin{align}
    \mu\left(\bm{z}\right) & = \mu_B\left(\bm{z}\right)\mu_C\left(\bm{z}\right).
\end{align}
In the above
\begin{align}
    \mu_B\left(\bm{z}\right) & := \prod_{j \in [n]}f\left(\bm{z}_j\right),
\end{align}
where $f\left(\bm{z}_j\right) := f^{[p]}\left(\bm{z}_j\right)$, with tensor $f^{[p]}$ defined in Eq.~\ref{eq:mixer_tensor}, comes from QAOA mixer unitaries. On the other hand,
\begin{align}
    \mu_C\left(\bm{z}\right) & := \exp\left(-\frac{n}{2}\sum_{t, u \in \mathcal{T}_{p - 1}}\Gamma_t\Gamma_u\xi\left(R^{[t, u]}\left(\bm{z}\right)\right)\right),\label{eq:disorder_averaged_path_integral_measure_cost_contribution}
\end{align}
where we introduced normalized overlaps
\begin{align}
    R^{[t, u]}\left(\bm{z}\right) & := \frac{1}{n}\left\langle \bm{z}^{[t]}, \bm{z}^{[u]} \right\rangle\\
    & = \frac{1}{n}\sum_{j \in [n]}z_j^{[t]}z_j^{[u]},
\end{align}
comes from disorder-averaged cost unitaries.

The previous formulae are only rephrasings of identities proven in Ref.~\cite{qaoa_sparse_hypergraphs_spin_glass_models}, adapted to the current work's formalism. Since $\xi$ is a polynomial, the energy density Eq.~\ref{eq:energy_density_exact_expression} expands into expectations of monomials in $z_j^{[t]}$ involving up to a constant number of spins $j$. One may further exploit permutation invariance in spin indices of the disorder-averaged path integral weight Eq.~\ref{eq:disorder_averaged_path_integral_measure}. From these considerations, it is sufficient to determine the joint distribution of a fixed constant-size set of spins under the path integral weights.

Without loss of generality, we then fix small spin set $\{1, \ldots, n''\}$ of constant size $n''$. We call this small set the cavity set and its complement $\{n'' + 1, \ldots, n'' + n'\}$ of size $n' := n - n''$ the bulk set. For any spin trajectories $\bm{z} \in \{1, -1\}^{n \times 2(p + 1)}$, we let $\bm{z}_{\text{c}} := \left(\bm{z}_j\right)_{j \in \{1, \ldots, n''\}}$ the trajectories of cavity spins and $\bm{z}_{\text{b}} := \left(\bm{z}_j\right)_{j \in \{n'' + 1, \ldots, n'' + n'\}}$ the trajectories of bulk spins. From classical probability intuition, the distribution of cavity spins can be understood from the distribution of bulk spins and the conditional distribution of cavity spins given bulk spins trajectories. Specifically, from classical formula $\mathbb{P}\left(Y = y \,|\,X = x\right) = \mathbb{P}(X = x, Y = y)/\mathbb{P}(X = x)$, the following ratio:
\begin{align}
    \mu_{\text{c}\,|\,\text{b}}\left(\bm{z}_{\text{c}}\,|\,\bm{z}_{\text{b}}\right) & := \frac{\mu\left(\bm{z}\right)}{\mu_{\text{b}}\left(\bm{z}_{\text{b}}\right)}
\end{align}
gives the conditional distribution of cavity trajectories $\bm{z}_c$ given bulk trajectories $\bm{z}_c$. $\mu_{\text{b}}\left(\bm{z}_{\text{b}}\right)$ is a path integral weight for bulk trajectories, defined similarly to full path integral weight $\mu\left(\bm{z}\right)$. After a relatively straightforward calculation, we arrive at the following approximation for this pseudo-conditional probability:
\begin{align}
    \mu_{\text{c}\,|\,\text{b}}\left(\bm{z}_{\text{c}}\,|\,\bm{z}_{\text{b}}\right) & \approx \mu_{B}\left(\bm{z}_{\text{c}}\right)\exp\left(-\frac{1}{2}\hspace*{-8px}\sum\limits_{\smash{\raisebox{-0.7ex}{$\scriptstyle t, u \in \mathcal{T}_{p - 1}$}}}\hspace*{-8px}\Gamma_t\Gamma_u\xi'\left(R^{[t, u]}\left(\bm{z}_{\text{b}}\right)\right)\right.\nonumber\\
    & \hspace*{110px} \times \left\langle \bm{z}^{[t]}_{\text{c}}, \bm{z}^{[u]}_{\text{c}} \right\rangle\Bigg).\label{eq:approximate_cavity_decomposition}
\end{align}
The approximation in the above calculation only results from dropping factors that are manifestly equivalent to $1$ as $n \to \infty$. The above calculation is rigorous as expressing ratio $\mu\left(\bm{z}\right)/\mu_{\text{b}}\left(\bm{z}_{\text{b}}\right)$ and does not require the non-rigorous conditional probability interpretation. We now invoke this interpretation for intuition only, with the full rigorous reasoning deferred to Supplementary Information. Eq.~\ref{eq:approximate_cavity_decomposition} then claims that conditioned on bulk overlaps $R^{[t, u]}\left(\bm{z}_{\text{b}}\right)$, the distribution of cavity spins factorizes since
\begin{align}
    & \mu_B\left(\bm{z}_{\text{c}}\right)\exp\left(-\frac{1}{2}\sum_{t, u}\Gamma_t\Gamma_u\xi'\left(R^{[t, u]}\left(\bm{z}_{\text{b}}\right)\right)\left\langle \bm{z}^{[t]}_{\text{c}}, \bm{z}^{[u]}_{\text{c}} \right\rangle\right)\nonumber\\
    & = \mu_B\left(\bm{z}_{\text{c}}\right)\exp\left(-\frac{1}{2}\sum_{t, u}\Gamma_t\Gamma_u\xi'\left(R^{[t, u]}\left(\bm{z}_{\text{b}}\right)\right)\sum_{j \in \left[n''\right]}\bm{z}^{[t]}_j\bm{z}^{[u]}_j\right)\nonumber\\
    & = \prod_{j \in \left[n''\right]}f\left(\bm{z}_j\right)\exp\left(-\frac{1}{2}\sum_{t, u}\Gamma_t\Gamma_u\xi'\left(R^{[t, u]}\left(\bm{z}_{\text{b}}\right)\right)\bm{z}^{[t]}_j\bm{z}^{[u]}_j\right).\label{eq:cavity_conditional_distribution_approximate_factorization}
\end{align}
More specifically, in Supplementary Information we show that under the bulk path integral weight $\mu_{\text{b}}\left(\bm{z}_{\text{b}}\right)$, normalized bulk overlaps $R^{[t, u]}\left(\bm{z}_{\text{b}}\right)$ converge to deterministic variables $G_{t, u}$ in the infinite-size limit.
This convergence of normalized bulk overlaps and the factorization across cavity spins are established together, by induction over the QAOA level $p$. The overlaps appearing in Eq.~\ref{eq:approximate_cavity_decomposition} carry layer indices in $\mathcal{T}_{p - 1}$, whereas the expectations under construction at level $p$ carry indices in $\mathcal{T}_p$, so the induction hypothesis applies to the overlaps at the level below. Granting it, $R^{[t, u]}\left(\bm{z}_{\text{b}}\right)$ is the average of $n'$ bounded, approximately independent site terms, and the law of large numbers makes it converge to a deterministic value, namely $G_{t, u}$ as $n' \to \infty$, equivalently $n \to \infty$. Replacing $R^{[t, u]}\left(\bm{z}_{\text{b}}\right)$ by that limit in Eq.~\ref{eq:cavity_conditional_distribution_approximate_factorization},

\begin{align}
    & \mu_{\text{c}\,|\,\text{b}}\left(\bm{z}_{\text{c}}\,|\,\bm{z}_{\text{b}}\right)\nonumber\\
    & \approx \prod_{j \in \left[n''\right]}f\left(\bm{z}_j\right)\exp\left(-\frac{1}{2}\sum_{t, u}\Gamma_t\Gamma_u\xi'\left(G_{t, u}\right)z_j^{[t]}z_j^{[u]}\right).\label{eq:cavity_factorization}
\end{align}

Informally, the above asserts conditional independence of cavity spins given any configuration of bulk spins. The conditional distribution is the same for all bulk spin configurations, which follows from convergence of bulk overlaps to deterministic variables. Hence, cavity spins are unconditionally independent. While this reasoning is based on a non-rigorous classical probability analogy, it can be precisely conceptualized and quantified when working with complex path integral weights. Details are deferred to Supplementary Information.

\subsection{MPS simulation and QAOA angle optimization}

For numerical evaluation, we represent the spin-boson state by two MPS, $\ket{\psi_+}$ and $\ket{\psi_-}$, corresponding to the two spin-$Z$ sectors, over a chain of $p$ bosonic modes initially in their vacuum states.
For spin-boson evolution step $m$, we then apply unitary layer $U^{(m)}$(Eq.~\ref{eq:spin_boson_layer}), consisting of a spin-conditioned displacement operator and an $X$ rotation over the spin, and measure $Z$-annihilation expectations $\bm{b}^{(m)}$ (Eq.~\ref{eq:z_annihilation_expectation_values}). Spin-conditioned displacement operator $D\left(-i\gamma_tZ\bm{\ell}^{(m)}\right)$ acts as an ordinary displacement on each spin-$Z$ sector: $D\left(-i\gamma_t\bm{\ell}^{(m)}\right)$ on $\ket{\psi_+}$, $D\left(i\gamma_t\bm{\ell}^{(m)}\right)$ on $\ket{\psi_-}$. Since multimode displacements factorize over modes: $D\left(-i\gamma_t\sigma\bm{\ell}^{(m)}\right) = \bigotimes_{l \in [p]}D_l\left(-i\gamma_t\sigma\ell^{(m)}_l\right)$, this amounts to applying a product operator on each of $\ket{\psi_+}$, $\ket{\psi_-}$, which does not grow bond dimension.
Applying an $X$ rotation of angle $\beta$ amounts to a linear operation over the bosonic MPS's representing each spin sector: $\left(\ket{\psi_+}, \ket{\psi_-}\right) \longrightarrow \left(\cos\beta\ket{\psi_+} - i\sin\beta\ket{\psi_-}, \cos\beta\ket{\psi_-} - i\sin\beta\ket{\psi_+}\right)$.
The sum of two MPS's can be represented as a MPS, but the resulting bond dimension before truncation may be as large as the sum of bond dimensions of original MPS's. $Z$-annihilation expectation values reduce to $1$-local expectation values under $\ket{\psi_+}, \ket{\psi_-}$: $\bra{\Psi}Z\hat{c}_l\ket{\Psi} = \bra{\psi_+}\hat{c}_l\ket{\psi_+} - \bra{\psi_-}\hat{c}_l\ket{\psi_-}$ and are therefore efficient to compute. Unlike the implementation of Ref.~\cite{qaoa_spin_boson_mapping}, our procedure only requires equal-time expectation values rather than two-times correlators. As a result, it is sufficient to evolve one state, rather than $p$, across the entire computation. This results in a factor $p$ saving of memory and sequential computation time.
All tensors use complex double precision. For the reported calculations, each bosonic mode was truncated to 40 Fock states; the SVD truncation cutoff was $10^{-13}$ and no explicit maximum MPS bond dimension was imposed.

Calculations were performed on Polaris~\cite{polaris} and Perlumutter~\cite{perlmutter}, both supercomputers whose nodes contain four NVIDIA A100 GPUs. The implementation of Ref.~\cite{qaoa_spin_boson_mapping} distributed the temporal-correlation calculations over multiple GPUs. In the present work, all annihilation-operator expectations required at a given layer are extracted from a single pair of MPS, so one GPU performs the complete depth-$p$ spin-boson evolution.

The depth-one angles were obtained from the exact scalar energy expression. At each subsequent depth, the initial angle schedules were constructed from optimized lower-depth schedules by scaled interpolation and Chebyshev extrapolation~\cite{apte2026iterativeinterpolationschedulesquantum}. The angles were optimized using L-BFGS-B in an adaptive Chebyshev parameterization, retaining the smallest number of coefficients that reconstructed each seed schedule with a maximum error of $10^{-5}$. Gradients were evaluated by central finite differences with a step size of $10^{-6}$. The relative-function and projected-gradient tolerances were $10^{-8}$ and $10^{-6}$, respectively, and physical angles were restricted to $10^{-6}\leq\gamma_l,\beta_l\leq0.99$. The best angles found at each depth supplied the warm start for the next depth.

For the sensitivity diagnostic in Fig.~\ref{fig:sensitivity-k4-k8}, we recomputed the complete angle gradient with the same finite-difference step and evaluated the energy at $\bm\theta_0+\Delta\nabla_{\bm\theta}\nu(\bm\theta_0)$. We report the signed angle displacement $\Delta\theta=\Delta\lVert\nabla_{\bm\theta}\nu\left(\bm{\theta}_0\right)\rVert_\infty$, the largest displacement applied to any single angle, and normalize the energy change by $|\nu(\bm\theta_0)|$. Each scan uses 50 values of $\Delta$ chosen so that $\Delta\theta$ is equally spaced over $[-10^{-1},10^{-1}]$. The figure reports $q\in\{2,4,6,8\}$ at $p\in\{10,20,30\}$, the depths for which scans are available at every one of these orders.

\begin{acknowledgements}
We thank Jeffrey Larson and Anuj Apte for helpful discussions of parameter optimization methods.
We thank Rob Otter and the technical staff at JPMorganChase’s Global Technology Applied Research for their support of this work.
This research used resources of the National Energy Research Scientific Computing Center (NERSC), a Department of Energy Office of Science User Facility using NERSC award ERCAP0036355.
An award of computer time was provided by the U.S. Department of Energy’s (DOE) Innovative and Novel Computational Impact on Theory and Experiment (INCITE) Program. This research used resources from the Argonne Leadership Computing Facility, a U.S. DOE Office of Science user facility at Argonne National Laboratory, which is supported by the Office of Science of the U.S. DOE under Contract No. DE-AC02-06CH11357.
\end{acknowledgements}

\section*{Author contributions}
R.S., A.K. and S.B. conceptualized the project. S.B. derived the $\bm{G}$ iteration for a mixed spin glass and its corresponding spin boson mapping and produced the cavity calculation analysis. A.K. developed the matrix product states simulation code and numerical experiments infrastructure. A.K. and B.V. developed the classical angle optimization protocol. R.S., A.K. and S.B. designed the numerical experiments. A.K. and P.S. ran numerical experiments. D.H. implemented numerical schemes to evaluate the optimal and algorithmic values of spin glasses. R.S. and E.F. provided valuable guidance on the overall conduct of the project and the presentation of our results.

\bibliography{bibliography}

@InProceedings{basso_et_al:LIPIcs.TQC.2022.7,
  author =	{Basso, Joao and Farhi, Edward and Marwaha, Kunal and Villalonga, Benjamin and Zhou, Leo},
  title =	{{The Quantum Approximate Optimization Algorithm at High Depth for MaxCut on Large-Girth Regular Graphs and the Sherrington-Kirkpatrick Model}},
  booktitle =	{17th Conference on the Theory of Quantum Computation, Communication and Cryptography (TQC 2022)},
  pages =	{7:1--7:21},
  series =	{Leibniz International Proceedings in Informatics (LIPIcs)},
  ISBN =	{978-3-95977-237-2},
  ISSN =	{1868-8969},
  year =	{2022},
  volume =	{232},
  editor =	{Le Gall, Fran\c{c}ois and Morimae, Tomoyuki},
  publisher =	{Schloss Dagstuhl -- Leibniz-Zentrum f{\"u}r Informatik},
  address =	{Dagstuhl, Germany},
  URL =		{https://drops.dagstuhl.de/entities/document/10.4230/LIPIcs.TQC.2022.7},
  URN =		{urn:nbn:de:0030-drops-165144},
  doi =		{10.4230/LIPIcs.TQC.2022.7}
}

@article{Farhi2022quantumapproximate,
  doi = {10.22331/q-2022-07-07-759},
  url = {https://doi.org/10.22331/q-2022-07-07-759},
  title = {The {Q}uantum {A}pproximate {O}ptimization {A}lgorithm and the {S}herrington-{K}irkpatrick {M}odel at {I}nfinite {S}ize},
  author = {Farhi, Edward and Goldstone, Jeffrey and Gutmann, Sam and Zhou, Leo},
  journal = {{Quantum}},
  issn = {2521-327X},
  publisher = {{Verein zur F{\"{o}}rderung des Open Access Publizierens in den Quantenwissenschaften}},
  volume = {6},
  pages = {759},
  month = jul,
  year = {2022}
}

@INPROCEEDINGS{qaoa_sparse_hypergraphs_spin_glass_models,
  author={Basso, Joao and Gamarnik, David and Mei, Song and Zhou, Leo},
  booktitle={2022 IEEE 63rd Annual Symposium on Foundations of Computer Science (FOCS)},
  title={Performance and limitations of the QAOA at constant levels on large sparse hypergraphs and spin glass models},
  year={2022},
  volume={},
  number={},
  pages={335-343},
  doi={10.1109/FOCS54457.2022.00039}}

@article{qaoa_spin_boson_mapping,
  title = {Spin-Boson Mapping of the Quantum Approximate Optimization Algorithm},
  author = {Boulebnane, Sami and Khan, Abid and Liu, Minzhao and Larson, Jeffrey and Herman, Dylan and Shaydulin, Ruslan and Pistoia, Marco},
  journal = {Phys. Rev. Lett.},
  volume = {136},
  issue = {24},
  pages = {240601},
  numpages = {7},
  year = {2026},
  month = {Jun},
  publisher = {American Physical Society},
  doi = {10.1103/2w94-rymn},
  url = {https://link.aps.org/doi/10.1103/2w94-rymn}
}

@article{talagrand_parisi_formula,
 ISSN = {0003486X},
 URL = {http://www.jstor.org/stable/20159953},
 author = {Michel Talagrand},
 journal = {Annals of Mathematics},
 number = {1},
 pages = {221--263},
 publisher = {Annals of Mathematics},
 title = {The Parisi Formula},
 urldate = {2026-07-04},
 volume = {163},
 year = {2006}
}

@article{Gamarnik2021,
  title = {The overlap gap property and approximate message passing algorithms for $p$-spin models},
  volume = {49},
  ISSN = {0091-1798},
  url = {http://dx.doi.org/10.1214/20-AOP1448},
  DOI = {10.1214/20-aop1448},
  number = {1},
  journal = {The Annals of Probability},
  publisher = {Institute of Mathematical Statistics},
  author = {Gamarnik,  David and Jagannath,  Aukosh},
  year = {2021},
  month = Jan
}

@inproceedings{Gamarnik2020,
  title = {Low-Degree Hardness of Random Optimization Problems},
  url = {http://dx.doi.org/10.1109/FOCS46700.2020.00021},
  DOI = {10.1109/focs46700.2020.00021},
  booktitle = {2020 IEEE 61st Annual Symposium on Foundations of Computer Science (FOCS)},
  publisher = {IEEE},
  author = {Gamarnik,  David and Jagannath,  Aukosh and Wein,  Alexander S.},
  year = {2020},
  month = Nov,
  pages = {131–140}
}

@misc{algorithmic_universality_ldp,
Author = {Houssam El Cheairi and David Gamarnik},
Title = {Algorithmic Universality, Low-Degree Polynomials, and Max-Cut in Sparse Random Graphs},
Year = {2024},
Eprint = {arXiv:2412.18014},
}

@InProceedings{limitations_local_algorithms_max_k_xor,
  author =	{Chou, Chi-Ning and Love, Peter J. and Sandhu, Juspreet Singh and Shi, Jonathan},
  title =	{{Limitations of Local Quantum Algorithms on Random MAX-k-XOR and Beyond}},
  booktitle =	{49th International Colloquium on Automata, Languages, and Programming (ICALP 2022)},
  pages =	{41:1--41:20},
  series =	{Leibniz International Proceedings in Informatics (LIPIcs)},
  ISBN =	{978-3-95977-235-8},
  ISSN =	{1868-8969},
  year =	{2022},
  volume =	{229},
  editor =	{Boja\'{n}czyk, Miko{\l}aj and Merelli, Emanuela and Woodruff, David P.},
  publisher =	{Schloss Dagstuhl -- Leibniz-Zentrum f{\"u}r Informatik},
  address =	{Dagstuhl, Germany},
  URL =		{https://drops.dagstuhl.de/entities/document/10.4230/LIPIcs.ICALP.2022.41},
  URN =		{urn:nbn:de:0030-drops-163822},
  doi =		{10.4230/LIPIcs.ICALP.2022.41}
}

@misc{qaoa_needs_to_see_whole_graph_typical_case,
Author = {Edward Farhi and David Gamarnik and Sam Gutmann},
Title = {The Quantum Approximate Optimization Algorithm Needs to See the Whole Graph: A Typical Case},
Year = {2020},
Eprint = {arXiv:2004.09002},
}

@InProceedings{random_max_csps_hardness_spin_glasses,
  author =	{Jones, Chris and Marwaha, Kunal and Sandhu, Juspreet Singh and Shi, Jonathan},
  title =	{{Random Max-CSPs Inherit Algorithmic Hardness from Spin Glasses}},
  booktitle =	{14th Innovations in Theoretical Computer Science Conference (ITCS 2023)},
  pages =	{77:1--77:26},
  series =	{Leibniz International Proceedings in Informatics (LIPIcs)},
  ISBN =	{978-3-95977-263-1},
  ISSN =	{1868-8969},
  year =	{2023},
  volume =	{251},
  editor =	{Tauman Kalai, Yael},
  publisher =	{Schloss Dagstuhl -- Leibniz-Zentrum f{\"u}r Informatik},
  address =	{Dagstuhl, Germany},
  URL =		{https://drops.dagstuhl.de/entities/document/10.4230/LIPIcs.ITCS.2023.77},
  URN =		{urn:nbn:de:0030-drops-175804},
  doi =		{10.4230/LIPIcs.ITCS.2023.77}
}

@article{PhysRevLett.43.1754,
  title = {Infinite Number of Order Parameters for Spin-Glasses},
  author = {Parisi, G.},
  journal = {Phys. Rev. Lett.},
  volume = {43},
  issue = {23},
  pages = {1754--1756},
  numpages = {0},
  year = {1979},
  month = {Dec},
  publisher = {American Physical Society},
  doi = {10.1103/PhysRevLett.43.1754},
  url = {https://link.aps.org/doi/10.1103/PhysRevLett.43.1754}
}

@article{PhysRevE.65.046137,
  title = {Analysis of the $\ensuremath{\infty}$-replica symmetry breaking solution of the Sherrington-Kirkpatrick model},
  author = {Crisanti, A. and Rizzo, T.},
  journal = {Phys. Rev. E},
  volume = {65},
  issue = {4},
  pages = {046137},
  numpages = {9},
  year = {2002},
  month = {Apr},
  publisher = {American Physical Society},
  doi = {10.1103/PhysRevE.65.046137},
  url = {https://link.aps.org/doi/10.1103/PhysRevE.65.046137}
}

@article{Marwaha2022boundsapproximating,
  doi = {10.22331/q-2022-07-07-757},
  url = {https://doi.org/10.22331/q-2022-07-07-757},
  title = {Bounds on approximating {M}ax {$k$}{XOR} with quantum and classical local algorithms},
  author = {Marwaha, Kunal and Hadfield, Stuart},
  journal = {{Quantum}},
  issn = {2521-327X},
  publisher = {{Verein zur F{\"{o}}rderung des Open Access Publizierens in den Quantenwissenschaften}},
  volume = {6},
  pages = {757},
  month = jul,
  year = {2022}
}

@book{Panchenko2013,
  title = {The Sherrington-Kirkpatrick Model},
  ISBN = {9781461462897},
  ISSN = {1439-7382},
  url = {http://dx.doi.org/10.1007/978-1-4614-6289-7},
  DOI = {10.1007/978-1-4614-6289-7},
  journal = {Springer Monographs in Mathematics},
  publisher = {Springer New York},
  author = {Panchenko,  Dmitry},
  year = {2013}
}

@misc{polaris,
    title       = {Argonne Leadership Computing Facility},
    url         = {https://docs.alcf.anl.gov/polaris/getting-started/},
}

@misc{perlmutter,
    title       = {National Energy Research Scientific Computing},
    url         = {https://docs.nersc.gov/},
}

@article{optimization_sk,
author = {Montanari, Andrea},
title = {Optimization of the Sherrington–Kirkpatrick Hamiltonian},
journal = {SIAM Journal on Computing},
volume = {54},
number = {4},
pages = {FOCS19-1-FOCS19-38},
year = {2025},
doi = {10.1137/20M132016X},
URL = {https://doi.org/10.1137/20M132016X},
eprint = {https://doi.org/10.1137/20M132016X}
}

@article{optimization_mean_field_spin_glasses,
author = {Ahmed El Alaoui and Andrea Montanari and Mark Sellke},
title = {{Optimization of mean-field spin glasses}},
volume = {49},
journal = {The Annals of Probability},
number = {6},
publisher = {Institute of Mathematical Statistics},
pages = {2922 -- 2960},
year = {2021},
doi = {10.1214/21-AOP1519},
URL = {https://doi.org/10.1214/21-AOP1519}
}

@misc{apte2026iterativeinterpolationschedulesquantum,
      title={Iterative Interpolation Schedules for Quantum Approximate Optimization Algorithm},
      author={Anuj Apte and Shree Hari Sureshbabu and Ruslan Shaydulin and Sami Boulebnane and Zichang He and Dylan Herman and James Sud and Marco Pistoia},
      year={2026},
      eprint={2504.01694},
      archivePrefix={arXiv},
      primaryClass={quant-ph},
      url={https://arxiv.org/abs/2504.01694},
}

@inbook{potential_hessian_ascent,
author = {David Jekel and Juspreet Singh Sandhu and Jonathan Shi},
title = {Potential Hessian Ascent: The Sherrington-Kirkpatrick Model},
booktitle = {Proceedings of the 2025 Annual ACM-SIAM Symposium on Discrete Algorithms (SODA)},
year = {2025},
publisher = {SIAM},
chapter = {},
pages = {5307-5387},
doi = {10.1137/1.9781611978322.182},
URL = {https://epubs.siam.org/doi/abs/10.1137/1.9781611978322.182},
eprint = {https://epubs.siam.org/doi/pdf/10.1137/1.9781611978322.182}
}

@article{local_algorithms_sparse_maxcut,
author = {El Alaoui, Ahmed and Montanari, Andrea and Sellke, Mark},
title = {Local algorithms for maximum cut and minimum bisection on locally treelike regular graphs of large degree},
journal = {Random Structures \& Algorithms},
volume = {63},
number = {3},
pages = {689-715},
doi = {https://doi.org/10.1002/rsa.21149},
url = {https://onlinelibrary.wiley.com/doi/abs/10.1002/rsa.21149},
eprint = {https://onlinelibrary.wiley.com/doi/pdf/10.1002/rsa.21149},
year = {2023}
}

@article{extremal_cuts_sparse_random_graphs,
author = {Amir Dembo and Andrea Montanari and Subhabrata Sen},
title = {{Extremal cuts of sparse random graphs}},
volume = {45},
journal = {The Annals of Probability},
number = {2},
publisher = {Institute of Mathematical Statistics},
pages = {1190 -- 1217},
year = {2017},
doi = {10.1214/15-AOP1084},
URL = {https://doi.org/10.1214/15-AOP1084}
}

@inproceedings{qaoa_spiked_tensor,
 author = {Zhou, Leo and Basso, Joao and Mei, Song},
 booktitle = {Advances in Neural Information Processing Systems},
 doi = {10.52202/079017-0896},
 editor = {A. Globerson and L. Mackey and D. Belgrave and A. Fan and U. Paquet and J. Tomczak and C. Zhang},
 pages = {28537--28588},
 publisher = {Curran Associates, Inc.},
 title = {Statistical Estimation in the Spiked Tensor Model via the Quantum Approximate Optimization Algorithm},
 url = {https://proceedings.neurips.cc/paper_files/paper/2024/file/32133a6a24d6554263d3584e3ac10faa-Paper-Conference.pdf},
 volume = {37},
 year = {2024}
}

@inproceedings{mind_the_gap,
author = {Dalzell, Alexander M. and Pancotti, Nicola and Campbell, Earl T. and Brand{\~a}o, Fernando G.S.L.},
title = {Mind the Gap: Achieving a Super-Grover Quantum Speedup by Jumping to the End},
year = {2023},
isbn = {9781450399135},
publisher = {Association for Computing Machinery},
address = {New York, NY, USA},
url = {https://doi.org/10.1145/3564246.3585203},
doi = {10.1145/3564246.3585203},
booktitle = {Proceedings of the 55th Annual ACM Symposium on Theory of Computing},
pages = {1131–1144},
numpages = {14},
location = {Orlando, FL, USA},
series = {STOC 2023}
}

@article{qaoa_polynomial_approximations,
  doi = {10.22331/q-2023-05-11-999},
  url = {https://doi.org/10.22331/q-2023-05-11-999},
  title = {Concentration bounds for quantum states and limitations on the {QAOA} from polynomial approximations},
  author = {Anshu, Anurag and Metger, Tony},
  journal = {{Quantum}},
  issn = {2521-327X},
  publisher = {{Verein zur F{\"{o}}rderung des Open Access Publizierens in den Quantenwissenschaften}},
  volume = {7},
  pages = {999},
  month = may,
  year = {2023}
}

@article{dqi,
	author = {Jordan, Stephen P. and Shutty, Noah and Wootters, Mary and Zalcman, Adam and Schmidhuber, Alexander and King, Robbie and Isakov, Sergei V. and Khattar, Tanuj and Babbush, Ryan},
	date = {2025/10/01},
	doi = {10.1038/s41586-025-09527-5},
	id = {Jordan2025},
	isbn = {1476-4687},
	journal = {Nature},
	number = {8086},
	pages = {831--836},
	title = {Optimization by decoded quantum interferometry},
	url = {https://doi.org/10.1038/s41586-025-09527-5},
	volume = {646},
	year = {2025}}

@misc{markov_short_path_algorithms,
Author = {Shouvanik Chakrabarti and Dylan Herman and Guneykan Ozgul and Shuchen Zhu and Brandon Augustino and Tianyi Hao and Zichang He and Ruslan Shaydulin and Marco Pistoia},
Title = {Generalized Short Path Algorithms: Towards Super-Quadratic Speedup over Markov Chain Search for Combinatorial Optimization},
Year = {2024},
Eprint = {arXiv:2410.23270},
}

@article{mean_field_approximate_optimization_algorithm,
  title = {Mean-Field Approximate Optimization Algorithm},
  author = {Misra-Spieldenner, Aditi and Bode, Tim and Schuhmacher, Peter K. and Stollenwerk, Tobias and Bagrets, Dmitry and Wilhelm, Frank K.},
  journal = {PRX Quantum},
  volume = {4},
  issue = {3},
  pages = {030335},
  numpages = {19},
  year = {2023},
  month = {Sep},
  publisher = {American Physical Society},
  doi = {10.1103/PRXQuantum.4.030335},
  url = {https://link.aps.org/doi/10.1103/PRXQuantum.4.030335}
}

@article{lindeberg1922,
  author = {Lindeberg, J. W.},
  title = {Eine neue Herleitung des Exponentialgesetzes in der Wahrscheinlichkeitsrechnung},
  journal = {Mathematische Zeitschrift},
  volume = {15},
  number = {1},
  pages = {211--225},
  year = {1922},
  doi = {10.1007/BF01494395}
}

@article{guerra_toninelli_thermodynamic_limit,
  author = {Guerra, Francesco and Toninelli, Fabio Lucio},
  title = {The Thermodynamic Limit in Mean Field Spin Glass Models},
  journal = {Communications in Mathematical Physics},
  volume = {230},
  number = {1},
  pages = {71--79},
  year = {2002},
  doi = {10.1007/s00220-002-0699-y}
}

@article{guerra_toninelli_viana_bray,
  author = {Guerra, Francesco and Toninelli, Fabio Lucio},
  title = {The High Temperature Region of the {Viana--Bray} Diluted Spin Glass Model},
  journal = {Journal of Statistical Physics},
  volume = {115},
  number = {1--2},
  pages = {531--555},
  year = {2004},
  doi = {10.1023/B:JOSS.0000019815.11115.54}
}

@article{Jagannath_2015,
   title={A dynamic programming approach to the Parisi functional},
   volume={144},
   ISSN={0002-9939},
   url={http://dx.doi.org/10.1090/proc/12968},
   DOI={10.1090/proc/12968},
   number={7},
   journal={Proceedings of the American Mathematical Society},
   publisher={American Mathematical Society (AMS)},
   author={Jagannath, Aukosh and Tobasco, Ian},
   year={2015},
   month=Dec, pages={3135–3150} }

@article{Gamarnik_2021_ogp_survey,
   title={The overlap gap property: A topological barrier to optimizing over random structures},
   volume={118},
   ISSN={1091-6490},
   url={http://dx.doi.org/10.1073/pnas.2108492118},
   DOI={10.1073/pnas.2108492118},
   number={41},
   journal={Proceedings of the National Academy of Sciences},
   publisher={National Academy of Sciences},
   author={Gamarnik, David},
   year={2021},
   month=Oct }

@article{2604.01512,
Author = {Flaviano Morone and Andrew D. Kent and Dries Sels},
Title = {Variational Iterative Rotation Algorithm: Combinatorial Optimization with Classical Kicked Tops},
Year = {2026},
journal = {arXiv:2604.01512},
}

@article{2607.08708,
Author = {Dries Sels and Flaviano Morone},
Title = {Absence of quantum advantage for approximate spin glass optimization},
Year = {2026},
journal = {arXiv:2607.08708},
}

\section*{Disclaimer}

This paper was prepared for informational purposes with contributions from the Global Technology Applied Research center of JPMorgan Chase \& Co. This paper is not a product of the Research Department of JPMorgan Chase \& Co. or its affiliates. Neither JPMorgan Chase \& Co. nor any of its affiliates makes any explicit or implied representation or warranty and none of them accept any liability in connection with this paper, including, without limitation, with respect to the completeness, accuracy, or reliability of the information contained herein and the potential legal, compliance, tax, or accounting effects thereof. This document is not intended as investment research or investment advice, or as a recommendation, offer, or solicitation for the purchase or sale of any security, financial instrument, financial product or service, or to be used in any way for evaluating the merits of participating in any transaction.

\appendix

\makeatletter
\@addtoreset{figure}{section}
\@addtoreset{table}{section}
\makeatother
\renewcommand{\thefigure}{\thesection\arabic{figure}}
\renewcommand{\thetable}{\thesection\arabic{table}}

\onecolumngrid

\clearpage

\section{Per-order pinned fits}\label{app:per-order-fits}

\begin{figure*}[!htbp]
    \centering
    \includegraphics[width=\linewidth]{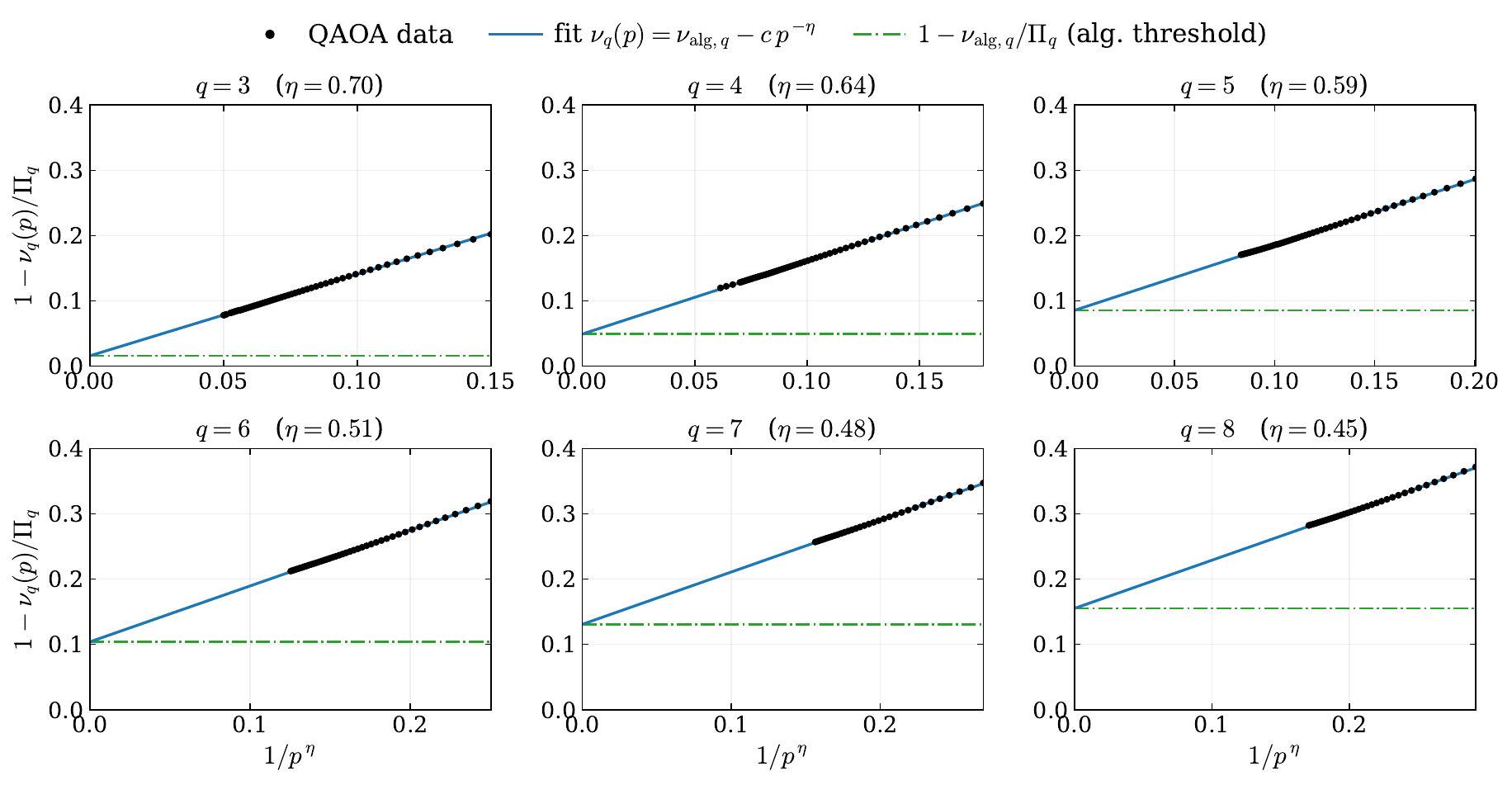}
    \caption{Optimized QAOA energy density on pure $q$-spin models, for $q=3\ldots8$. Each panel shows $1-\nu_q(p)/\Pi_q$
    vs.\ $1/p^\eta$ with the pinned fit $\nu_q(p)=\nu_{\mathrm{alg},q}-c\,p^{-\eta}$
    (solid blue) and algorithmic threshold (green dash-dot). The fitted exponent $\eta$ is given in each panel title and is plotted for a larger range of $q$ values in Fig.~\ref{fig:pure-spin-overview}b.}
    \label{fig:per-order-fits}
\end{figure*}

Figure~\ref{fig:per-order-fits} reports the pinned fits of Sec.~\ref{sec:numerics} separately for each spin order, using the same fit window and pinned asymptote as the main text. Presented this way, the quality of the two-parameter form can be judged order by order, which the single exponent per order in Fig.~\ref{fig:pure-spin-overview}b necessarily obscures.

\section{Optimized angles and energies}\label{app:angle-schedules}

\begin{figure*}[htbp]
    \centering
    \includegraphics[width=\linewidth]{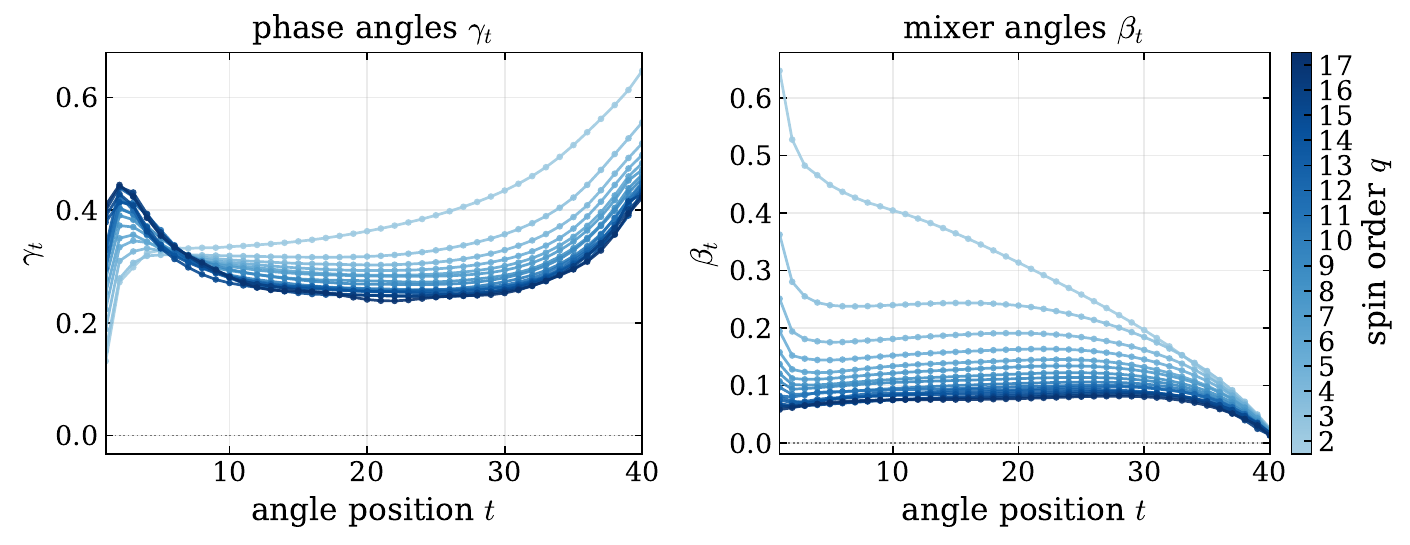}
    \caption{Optimized QAOA angle schedules at fixed depth $p=40$, one curve per spin order $q=2\ldots17$ (color encodes $q$). Left: phase angles $\gamma_i$; right: mixer angles $\beta_i$, both as functions of layer index $i=1\ldots p$.}
    \label{fig:angle-schedules}
\end{figure*}

Tables~\ref{tab:pure-energies} and~\ref{tab:pure-energies-tail} list the optimized energy densities $\nu_q(p)$ that underlie Figs.~\ref{fig:pure-spin-overview} and~\ref{fig:per-order-fits}, to four decimal places, together with the algorithmic threshold $\nu_{\mathrm{alg},q}$ and the Parisi optimum $\Pi_q$ against which they are normalized. Every fit, gap and exponent reported for the pure models can therefore be recomputed from the printed numbers. The depth coverage is uneven because the maximum depth reached depends on the spin order: all sixteen orders are tabulated through $p=50$, while beyond that depth only $q=2,\ldots,6$ and $q=9$ were optimized, and those entries are collected in the continuation table. The 2-spin model extends furthest, to $p=160$. Figure~\ref{fig:angle-schedules} shows the optimized angles at $p=40$ for all orders $2\leq q\leq17$. The schedules vary smoothly with layer index throughout, by an amount of order $1/p$ per layer, but although $\bm{\gamma}$ increases and $\bm{\beta}$ decreases monotonically at $q=2$, that monotonicity is lost at large $q$; the limiting behaviour as $q\to\infty$, if any, remains open.

\begin{table*}[htbp]
    \centering
    \scriptsize
    \setlength{\tabcolsep}{3.5pt}
    \begin{tabular}{r|rrrrrrrrrrrrrrrr}
        \hline\hline
         & \multicolumn{16}{c}{spin order $q$} \\
        \cline{2-17}
        $p$ & $2$ & $3$ & $4$ & $5$ & $6$ & $7$ & $8$ & $9$ & $10$ & $11$ & $12$ & $13$ & $14$ & $15$ & $16$ & $17$ \\
        \hline
        $1$ & 0.3033 & 0.2210 & 0.1780 & 0.1507 & 0.1315 & 0.1172 & 0.1060 & 0.0970 & 0.0896 & 0.0833 & 0.0780 & 0.0734 & 0.0693 & 0.0657 & 0.0625 & 0.0596 \\
        $2$ & 0.4075 & 0.3036 & 0.2461 & 0.2087 & 0.1822 & 0.1622 & 0.1465 & 0.1339 & 0.1234 & 0.1146 & 0.1070 & 0.1005 & 0.0948 & 0.0897 & 0.0852 & 0.0812 \\
        $3$ & 0.4726 & 0.3554 & 0.2889 & 0.2453 & 0.2140 & 0.1904 & 0.1718 & 0.1568 & 0.1444 & 0.1339 & 0.1250 & 0.1172 & 0.1105 & 0.1045 & 0.0992 & 0.0944 \\
        $4$ & 0.5157 & 0.3913 & 0.3190 & 0.2710 & 0.2364 & 0.2103 & 0.1897 & 0.1730 & 0.1592 & 0.1476 & 0.1376 & 0.1290 & 0.1215 & 0.1149 & 0.1089 & 0.1036 \\
        $5$ & 0.5476 & 0.4184 & 0.3419 & 0.2906 & 0.2536 & 0.2255 & 0.2033 & 0.1853 & 0.1705 & 0.1580 & 0.1473 & 0.1380 & 0.1299 & 0.1227 & 0.1164 & 0.1107 \\
        $6$ & 0.5721 & 0.4397 & 0.3600 & 0.3062 & 0.2672 & 0.2375 & 0.2141 & 0.1952 & 0.1795 & 0.1662 & 0.1549 & 0.1451 & 0.1365 & 0.1290 & 0.1223 & 0.1162 \\
        $7$ & 0.5915 & 0.4570 & 0.3748 & 0.3190 & 0.2784 & 0.2475 & 0.2230 & 0.2032 & 0.1868 & 0.1730 & 0.1612 & 0.1509 & 0.1420 & 0.1341 & 0.1271 & 0.1208 \\
        $8$ & 0.6073 & 0.4714 & 0.3872 & 0.3297 & 0.2877 & 0.2558 & 0.2305 & 0.2100 & 0.1930 & 0.1787 & 0.1664 & 0.1558 & 0.1466 & 0.1384 & 0.1311 & 0.1246 \\
        $9$ & 0.6203 & 0.4835 & 0.3977 & 0.3388 & 0.2957 & 0.2628 & 0.2368 & 0.2157 & 0.1982 & 0.1835 & 0.1709 & 0.1600 & 0.1504 & 0.1420 & 0.1345 & 0.1279 \\
        $10$ & 0.6314 & 0.4940 & 0.4067 & 0.3466 & 0.3026 & 0.2689 & 0.2423 & 0.2207 & 0.2027 & 0.1876 & 0.1747 & 0.1636 & 0.1538 & 0.1452 & 0.1375 & 0.1306 \\
        $11$ & 0.6408 & 0.5030 & 0.4146 & 0.3535 & 0.3086 & 0.2743 & 0.2471 & 0.2250 & 0.2067 & 0.1913 & 0.1781 & 0.1667 & 0.1567 & 0.1479 & 0.1401 & 0.1331 \\
        $12$ & 0.6490 & 0.5110 & 0.4216 & 0.3595 & 0.3139 & 0.2790 & 0.2513 & 0.2288 & 0.2102 & 0.1945 & 0.1811 & 0.1694 & 0.1593 & 0.1503 & 0.1424 & 0.1352 \\
        $13$ & 0.6561 & 0.5181 & 0.4278 & 0.3649 & 0.3186 & 0.2831 & 0.2550 & 0.2322 & 0.2133 & 0.1973 & 0.1837 & 0.1719 & 0.1616 & 0.1525 & 0.1444 & 0.1371 \\
        $14$ & 0.6623 & 0.5244 & 0.4333 & 0.3697 & 0.3228 & 0.2869 & 0.2584 & 0.2352 & 0.2160 & 0.1999 & 0.1860 & 0.1741 & 0.1636 & 0.1544 & 0.1462 & 0.1389 \\
        $15$ & 0.6679 & 0.5300 & 0.4383 & 0.3741 & 0.3267 & 0.2903 & 0.2614 & 0.2380 & 0.2185 & 0.2022 & 0.1882 & 0.1760 & 0.1655 & 0.1561 & 0.1478 & 0.1404 \\
        $16$ & 0.6729 & 0.5352 & 0.4429 & 0.3780 & 0.3301 & 0.2933 & 0.2641 & 0.2405 & 0.2208 & 0.2042 & 0.1901 & 0.1778 & 0.1671 & 0.1577 & 0.1493 & 0.1418 \\
        $17$ & 0.6773 & 0.5398 & 0.4470 & 0.3816 & 0.3333 & 0.2961 & 0.2666 & 0.2427 & 0.2229 & 0.2061 & 0.1918 & 0.1795 & 0.1686 & 0.1591 & 0.1506 & 0.1431 \\
        $18$ & 0.6814 & 0.5441 & 0.4508 & 0.3849 & 0.3361 & 0.2986 & 0.2689 & 0.2448 & 0.2248 & 0.2079 & 0.1934 & 0.1810 & 0.1700 & 0.1604 & 0.1519 & 0.1442 \\
        $19$ & 0.6850 & 0.5480 & 0.4542 & 0.3879 & 0.3388 & 0.3010 & 0.2710 & 0.2467 & 0.2265 & 0.2095 & 0.1949 & 0.1823 & 0.1713 & 0.1616 & 0.1530 & 0.1453 \\
        $20$ & 0.6884 & 0.5516 & 0.4574 & 0.3907 & 0.3412 & 0.3032 & 0.2730 & 0.2484 & 0.2281 & 0.2109 & 0.1963 & 0.1836 & 0.1725 & 0.1627 & 0.1540 & 0.1463 \\
        $21$ & 0.6914 & 0.5549 & 0.4604 & 0.3933 & 0.3435 & 0.3052 & 0.2748 & 0.2501 & 0.2296 & 0.2123 & 0.1975 & 0.1847 & 0.1736 & 0.1638 & 0.1550 & 0.1472 \\
        $22$ & 0.6943 & 0.5580 & 0.4632 & 0.3957 & 0.3456 & 0.3070 & 0.2764 & 0.2516 & 0.2309 & 0.2135 & 0.1987 & 0.1858 & 0.1746 & 0.1647 & 0.1559 & 0.1480 \\
        $23$ & 0.6969 & 0.5609 & 0.4657 & 0.3979 & 0.3475 & 0.3087 & 0.2780 & 0.2530 & 0.2322 & 0.2147 & 0.1998 & 0.1868 & 0.1755 & 0.1656 & 0.1567 & 0.1488 \\
        $24$ & 0.6993 & 0.5636 & 0.4681 & 0.4000 & 0.3494 & 0.3104 & 0.2794 & 0.2543 & 0.2334 & 0.2158 & 0.2008 & 0.1878 & 0.1764 & 0.1664 & 0.1575 & 0.1495 \\
        $25$ & 0.7015 & 0.5661 & 0.4704 & 0.4019 & 0.3511 & 0.3119 & 0.2808 & 0.2555 & 0.2345 & 0.2169 & 0.2017 & 0.1887 & 0.1772 & 0.1672 & 0.1582 & 0.1502 \\
        $26$ & 0.7036 & 0.5684 & 0.4725 & 0.4038 & 0.3527 & 0.3133 & 0.2820 & 0.2566 & 0.2356 & 0.2178 & 0.2026 & 0.1895 & 0.1780 & 0.1679 & 0.1589 & 0.1508 \\
        $27$ & 0.7055 & 0.5707 & 0.4744 & 0.4055 & 0.3542 & 0.3146 & 0.2832 & 0.2577 & 0.2366 & 0.2187 & 0.2034 & 0.1903 & 0.1787 & 0.1686 & 0.1595 & 0.1514 \\
        $28$ & 0.7073 & 0.5727 & 0.4763 & 0.4071 & 0.3556 & 0.3159 & 0.2844 & 0.2587 & 0.2375 & 0.2195 & 0.2042 & 0.1910 & 0.1793 & 0.1692 & 0.1601 & 0.1520 \\
        $29$ & 0.7090 & 0.5747 & 0.4781 & 0.4086 & 0.3569 & 0.3171 & 0.2854 & 0.2597 & 0.2384 & 0.2202 & 0.2050 & 0.1917 & 0.1800 & 0.1698 & 0.1607 & 0.1525 \\
        $30$ & 0.7106 & 0.5766 & 0.4797 & 0.4101 & 0.3582 & 0.3182 & 0.2864 & 0.2606 & 0.2391 & 0.2209 & 0.2057 & 0.1923 & 0.1806 & 0.1704 & 0.1612 & 0.1530 \\
        $31$ & 0.7121 & 0.5783 & 0.4813 & 0.4115 & 0.3594 & 0.3192 & 0.2874 & 0.2615 & 0.2400 & 0.2216 & 0.2064 & 0.1929 & 0.1812 & 0.1709 & 0.1617 & 0.1535 \\
        $32$ & 0.7135 & 0.5800 & 0.4828 & 0.4128 & 0.3606 & 0.3203 & 0.2883 & 0.2623 & 0.2407 & 0.2223 & 0.2070 & 0.1935 & 0.1817 & 0.1714 & 0.1621 & 0.1539 \\
        $33$ & 0.7148 & 0.5816 & 0.4842 & 0.4140 & 0.3616 & 0.3212 & 0.2892 & 0.2630 & 0.2414 & 0.2229 & 0.2076 & 0.1940 & 0.1822 & 0.1718 & 0.1625 & 0.1544 \\
        $34$ & 0.7161 & 0.5831 & 0.4856 & 0.4152 & 0.3627 & 0.3222 & 0.2900 & 0.2638 & 0.2421 & 0.2235 & 0.2082 & 0.1945 & 0.1827 & 0.1723 & 0.1629 & 0.1548 \\
        $35$ & 0.7173 & 0.5845 & 0.4869 & 0.4163 & 0.3636 & 0.3230 & 0.2908 & 0.2645 & 0.2427 & 0.2240 & 0.2087 & 0.1950 & 0.1832 & 0.1728 & 0.1633 & 0.1552 \\
        $36$ & 0.7184 & 0.5859 & 0.4881 & 0.4174 & 0.3646 & 0.3238 & 0.2915 & 0.2652 & 0.2433 & 0.2245 & 0.2092 & 0.1955 & 0.1836 & 0.1732 & 0.1636 & 0.1555 \\
        $37$ & 0.7195 & 0.5872 & 0.4893 & 0.4184 & 0.3655 & 0.3247 & 0.2922 & 0.2659 & 0.2439 & 0.2253 & 0.2097 & 0.1960 & 0.1841 & 0.1736 & 0.1641 & 0.1558 \\
        $38$ & 0.7205 & 0.5885 & 0.4904 & 0.4194 & 0.3664 & 0.3254 & 0.2929 & 0.2665 & 0.2445 & 0.2258 & 0.2101 & 0.1964 & 0.1845 & 0.1740 & 0.1643 & 0.1560 \\
        $39$ & 0.7215 & 0.5897 & 0.4915 & 0.4203 & 0.3672 & 0.3261 & 0.2935 & 0.2670 & 0.2450 & 0.2263 & 0.2105 & 0.1968 & 0.1850 & 0.1743 & 0.1645 & 0.1563 \\
        $40$ & 0.7224 & 0.5908 & 0.4925 & 0.4212 & 0.3680 & 0.3268 & 0.2942 & 0.2675 & 0.2454 & 0.2267 & 0.2109 & 0.1971 & 0.1851 & 0.1746 & 0.1646 & 0.1564 \\
        $41$ & 0.7233 & 0.5919 & 0.4935 & 0.4221 & 0.3687 & 0.3275 & 0.2948 & 0.2681 & 0.2460 & 0.2271 & 0.2114 & 0.1975 & 0.1856 & 0.1749 & -- & 0.1566 \\
        $42$ & 0.7242 & 0.5930 & 0.4945 & 0.4229 & 0.3694 & 0.3281 & 0.2953 & 0.2686 & 0.2464 & 0.2275 & 0.2117 & 0.1977 & 0.1858 & 0.1751 & -- & 0.1567 \\
        $43$ & 0.7250 & 0.5940 & 0.4954 & 0.4237 & 0.3701 & 0.3287 & 0.2958 & 0.2691 & 0.2468 & 0.2281 & 0.2118 & 0.1981 & -- & 0.1754 & -- & 0.1568 \\
        $44$ & 0.7257 & 0.5950 & 0.4963 & 0.4245 & 0.3708 & 0.3293 & 0.2964 & 0.2696 & 0.2474 & 0.2285 & -- & 0.1983 & -- & 0.1756 & -- & 0.1569 \\
        $45$ & 0.7265 & 0.5959 & 0.4971 & 0.4252 & 0.3714 & 0.3299 & 0.2968 & 0.2701 & 0.2478 & -- & -- & 0.1986 & -- & 0.1757 & -- & -- \\
        $46$ & 0.7272 & 0.5968 & 0.4980 & 0.4259 & 0.3720 & 0.3305 & 0.2974 & 0.2705 & 0.2481 & -- & -- & -- & -- & 0.1759 & -- & -- \\
        $47$ & 0.7279 & 0.5977 & 0.4987 & 0.4266 & 0.3726 & -- & 0.2979 & 0.2709 & 0.2485 & -- & -- & -- & -- & 0.1760 & -- & -- \\
        $48$ & 0.7285 & 0.5985 & 0.4995 & 0.4270 & 0.3732 & -- & 0.2982 & 0.2713 & 0.2489 & -- & -- & -- & -- & 0.1761 & -- & -- \\
        $49$ & 0.7292 & 0.5993 & 0.5002 & 0.4277 & 0.3737 & -- & 0.2987 & 0.2716 & 0.2493 & -- & -- & -- & -- & -- & -- & -- \\
        $50$ & 0.7298 & 0.6001 & 0.5009 & 0.4283 & 0.3743 & -- & -- & 0.2719 & -- & -- & -- & -- & -- & -- & -- & -- \\
        \hline
        $\nu_{\mathrm{alg},q}$ & 0.7632 & 0.6535 & 0.5538 & 0.4775 & 0.4195 & 0.3744 & 0.3386 & 0.3095 & 0.2854 & 0.2653 & 0.2481 & 0.2332 & 0.2203 & 0.2090 & 0.1989 & 0.1900 \\
        $\Pi_q$ & 0.7632 & 0.6639 & 0.5832 & 0.5238 & 0.4787 & 0.4433 & 0.4146 & 0.3908 & 0.3706 & 0.3533 & 0.3382 & 0.3248 & 0.3130 & 0.3023 & 0.2927 & 0.2839 \\
        \hline\hline
    \end{tabular}
    \caption{Optimized QAOA energy densities $\nu_q(p)$ in the thermodynamic limit, to 4 decimal places. An en dash marks a depth that was not optimized for that spin order. The final two rows give the algorithmic threshold $\nu_{\mathrm{alg},q}$ and the Parisi optimum $\Pi_q$ that the figures normalize against, in the same units. Depths $p>50$ continue in Table~\ref{tab:pure-energies-tail}.}
    \label{tab:pure-energies}
\end{table*}

\begin{table*}[htbp]
    \centering
    \footnotesize
    \setlength{\tabcolsep}{3.5pt}
    \begin{tabular}{r|rrrrrr}
        \hline\hline
         & \multicolumn{6}{c}{spin order $q$} \\
        \cline{2-7}
        $p$ & $2$ & $3$ & $4$ & $5$ & $6$ & $9$ \\
        \hline
        $51$ & 0.7304 & 0.6009 & 0.5016 & 0.4289 & 0.3747 & 0.2722 \\
        $52$ & 0.7309 & 0.6016 & 0.5023 & 0.4294 & 0.3754 & 0.2725 \\
        $53$ & 0.7315 & 0.6023 & 0.5027 & 0.4299 & 0.3757 & -- \\
        $54$ & 0.7320 & 0.6030 & 0.5033 & 0.4304 & 0.3763 & -- \\
        $55$ & 0.7325 & 0.6037 & 0.5039 & 0.4309 & 0.3767 & -- \\
        $56$ & 0.7329 & 0.6043 & 0.5044 & 0.4314 & 0.3772 & -- \\
        $57$ & 0.7334 & 0.6050 & 0.5049 & 0.4318 & 0.3775 & -- \\
        $58$ & 0.7339 & 0.6056 & 0.5055 & 0.4322 & 0.3781 & -- \\
        $59$ & 0.7343 & 0.6062 & 0.5060 & 0.4326 & -- & -- \\
        $60$ & 0.7347 & 0.6067 & 0.5065 & 0.4330 & -- & -- \\
        $61$ & 0.7352 & 0.6073 & 0.5070 & 0.4334 & -- & -- \\
        $62$ & 0.7356 & 0.6075 & 0.5075 & 0.4338 & -- & -- \\
        $63$ & 0.7360 & 0.6080 & 0.5079 & 0.4341 & -- & -- \\
        $64$ & 0.7363 & 0.6085 & 0.5084 & 0.4345 & -- & -- \\
        $65$ & 0.7367 & 0.6091 & 0.5088 & 0.4348 & -- & -- \\
        $66$ & 0.7371 & 0.6096 & -- & 0.4351 & -- & -- \\
        $67$ & 0.7374 & 0.6100 & -- & -- & -- & -- \\
        $68$ & 0.7378 & -- & -- & -- & -- & -- \\
        $69$ & 0.7381 & -- & -- & -- & -- & -- \\
        $70$ & 0.7384 & 0.6115 & 0.5108 & -- & -- & -- \\
        $71$ & 0.7387 & 0.6120 & -- & -- & -- & -- \\
        $72$ & 0.7390 & 0.6124 & -- & -- & -- & -- \\
        $73$ & 0.7393 & -- & -- & -- & -- & -- \\
        $74$ & 0.7396 & -- & -- & -- & -- & -- \\
        $75$ & -- & -- & 0.5123 & -- & -- & -- \\
        $80$ & 0.7412 & -- & 0.5138 & -- & -- & -- \\
        $85$ & 0.7423 & -- & -- & -- & -- & -- \\
        $90$ & 0.7432 & -- & -- & -- & -- & -- \\
        $95$ & 0.7441 & -- & -- & -- & -- & -- \\
        $100$ & 0.7449 & -- & -- & -- & -- & -- \\
        $110$ & 0.7462 & -- & -- & -- & -- & -- \\
        $120$ & 0.7473 & -- & -- & -- & -- & -- \\
        $140$ & 0.7478 & -- & -- & -- & -- & -- \\
        $160$ & 0.7483 & -- & -- & -- & -- & -- \\
        \hline\hline
    \end{tabular}
    \caption{Optimized QAOA energy densities $\nu_q(p)$ in the thermodynamic limit, to 4 decimal places. An en dash marks a depth that was not optimized for that spin order, continued from Table~\ref{tab:pure-energies} for $p>50$. Only $q={}$2, 3, 4, 5, 6 and 9 were optimized beyond that depth, so the remaining orders are omitted rather than shown as empty columns.}
    \label{tab:pure-energies-tail}
\end{table*}

\clearpage
\twocolumngrid

\section{Smoothness of angles and energies in interpolated mixed spin}
\label{sec:angles_smoothness}

In this Appendix, we give closer consideration to the discontinuity of angles and non-differentiability of energy in the $2 \to 8$ spin interpolation experiment reported on Fig.~\ref{fig:gap-vs-alpha-p8}.

\begin{figure*}[!htbp]
    \includegraphics[width=\linewidth]{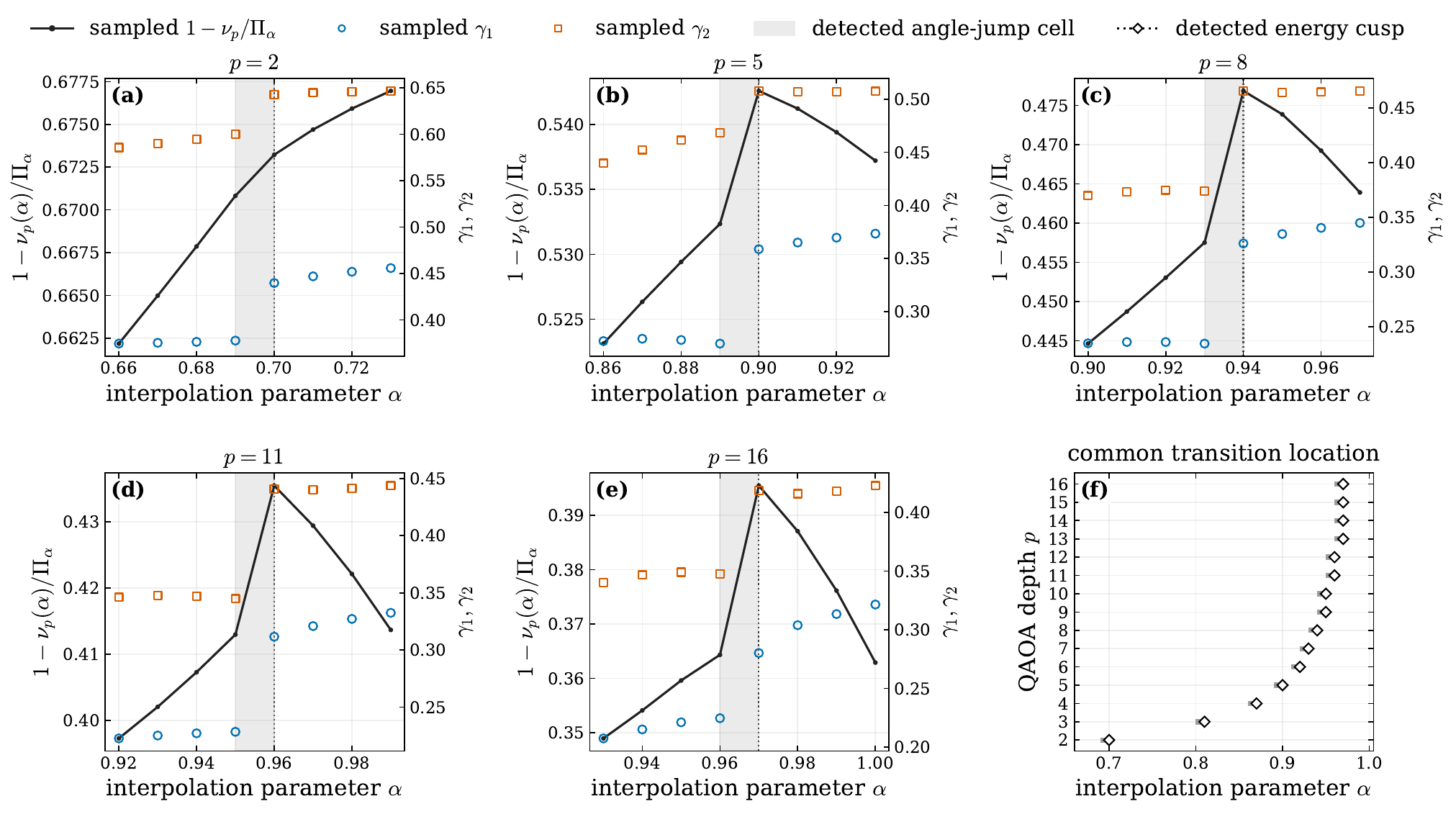}
    \caption{$\bm{\gamma}$ angles and energy around optimized angle discontinuity closest to $\alpha = 1$. \textbf{(a) - (e)} Discontinuity of earliest two $\bm{\gamma}$ angles and non-differentiability of energy around rightmost cusp for increasing values of $p$. Quantitatively, the discontinuity manifests at all covered $p$ by a jump of magnitude at least $0.05$. \textbf{(f)} Location of cusp as a function of depth $p$, inferred from large $\bm{\gamma}$ jumps visible on previous figures.}
\end{figure*}

\begin{figure*}
    \includegraphics[width=0.6\linewidth]{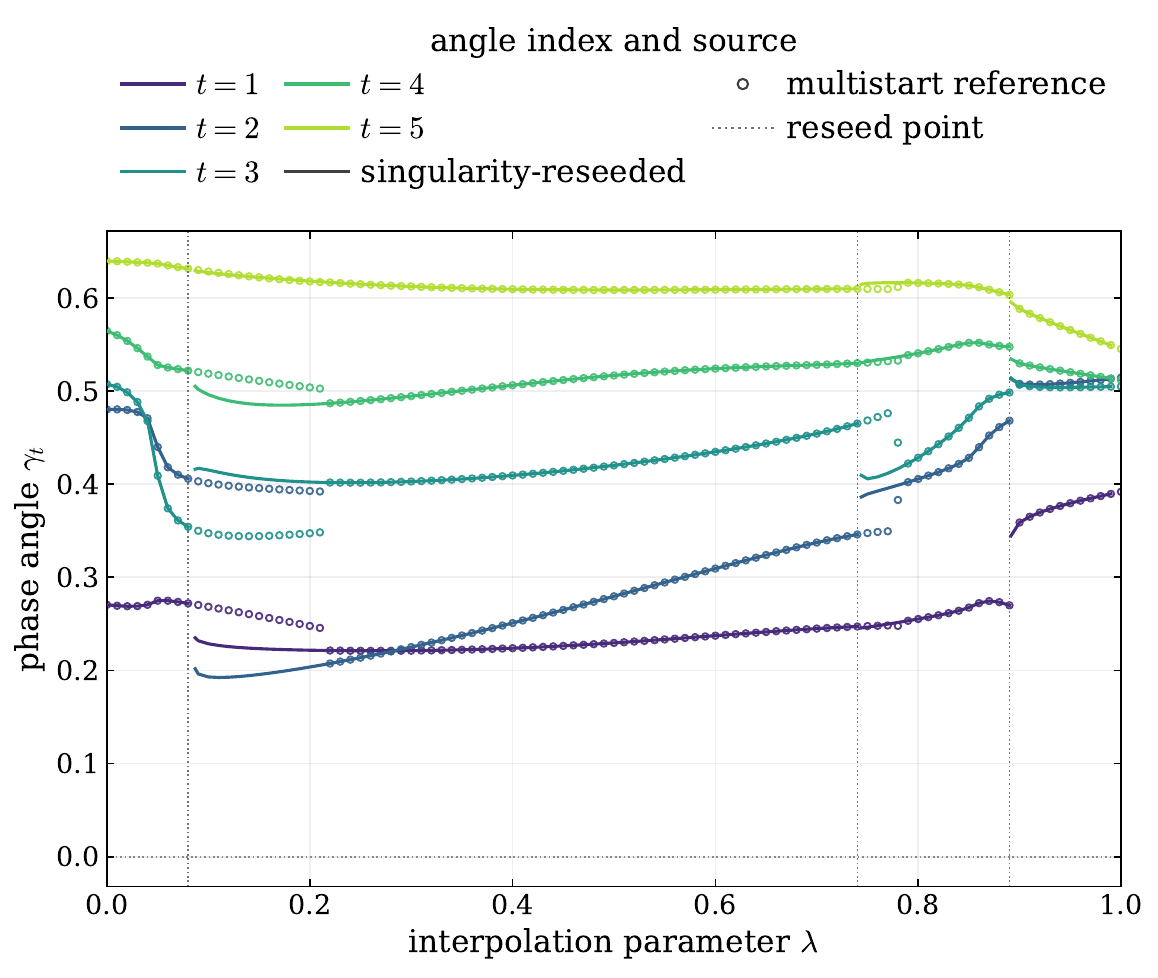}
    \caption{Tracking a local optimum using its analytic derivative along the interpolation path at $p = 5$. The interpolation path runs from right to left. Vertical dashed lines correspond to curvature singularities in the analytic derivatives of optimal angles (curvature vanishing in one direction). Upon encountering a singularity, the tracking is reseeded with a multistart optimization, thereby catching up with the multistart reference.}
    \label{fig:analytic_differentiation_optimized_angles_singularity_reseeded}
\end{figure*}

To rule out an optimizer artifact as the source of this discontinuity, Fig.~\ref{fig:analytic_differentiation_optimized_angles_singularity_reseeded} compares angles obtained by multistart optimization with those obtained by analytically tracking the local optimum along the interpolation path. This comparison reveals that at certain interpolation parameters $\alpha$, the discontinuity is associated with a singularity (vanishing curvature) in the analytically tracked local optimum. In other cases, however, the local optimum continues to exist as an analytic function of the interpolation parameter, but stops being global. From these observations, angle discontinuities in the interpolation parameter do not rule out the optimality of our optimized angles. However, they reveal limitations in optimized parameter transfer between mixed spin glasses with close coefficients, suggesting transfer is efficient only locally in problem space.

We now introduce our method of computing analytic derivatives of optimized angles. The idea is to regard the QAOA energy density as a parametrized function to optimized, parametrized by mixture polynomial coefficients $\bm{\xi}$. We then regard optimized angles $\left(\bm{\gamma}^*\left(\bm{\xi}\right), \bm{\beta}^*\left(\bm{\xi}\right)\right)$ as a critical point of this parametric function, and themselves locally differentiable functions of $\bm{\xi}$. Our calculation of analytic derivatives $\left(\bm{\nabla}_{\bm{\xi}}\bm{\gamma}^*\left(\bm{\xi}\right), \bm{\nabla}_{\bm{\xi}}\bm{\beta}^*\left(\bm{\xi}\right)\right)$ then follows from general formula for the parametric derivative of critical points of parametrized functions. We keep the exposition informal, focusing on the operational aspects of the formulae. Consider a function
\begin{align}
    \varphi: \left\{\begin{array}{ccc}
         \mathbb{R}^n \times \mathbb{R}^m & \longrightarrow & \mathbb{R}\\
         \left(\bm{x}, \bm{\eta}\right) & \longrightarrow & \varphi\left(\bm{x}, \bm{\eta}\right)
    \end{array}\right.,
\end{align}
where the first argument $\bm{x}$ is regarded as an optimized variable, and the second argument $\bm{\eta}$ as a parameter defining the optimization problem. For each parameter $\bm{\eta}$ in an appropriate neighbourhood, let $\bm{x}^*\left(\bm{\eta}\right) \in \mathbb{R}^n$ be an extremum for the problem $\bm{\eta}$:
\begin{align}
    \bm{\nabla}\varphi\left(\bm{x}^*\left(\bm{\eta}\right), \bm{\eta}\right) & = \bm{0}_n,\label{eq:parametrized_critical_point_equation}
\end{align}
and further assume that $\bm{x}^*\left(\bm{\eta}\right)$ is differentiable in $\bm{\eta}$ in the appropriate $\bm{\eta}$ neighbourhood. The above critical point equation is a special case of the following root equation:
\begin{align}
    \bm{\Phi}\left(\bm{x}^*\left(\bm{\eta}\right), \bm{\eta}\right) & = \bm{0}_n,\label{eq:parametrized_root_equation}
\end{align}
where
\begin{align}
    \bm{\Phi}: \left\{\begin{array}{ccc}
         \mathbb{R}^n \times \mathbb{R}^m & \longrightarrow & \mathbb{R}^n\\
         \left(\bm{x}, \bm{\eta}\right) & \longmapsto & \bm{\Phi}\left(\bm{x}, \bm{\eta}\right).
    \end{array}\right..
\end{align}
The critical point equation Eq.~\ref{eq:parametrized_critical_point_equation} reduces to this form by letting $\bm{\Phi} := \bm{\nabla}_{\bm{x}}\varphi$. Taking the total derivative of the parametrized root equation Eq.~\ref{eq:parametrized_root_equation} with respect to $\bm{\eta}$ yields:
\begin{align}
    \bm{\nabla}_{\bm{x}}\bm{\Phi}\left(\bm{x}^*\left(\bm{\eta}\right), \bm{\eta}\right)\bm{\nabla}_{\bm{\eta}}\bm{x}^*\left(\bm{\eta}\right) + \bm{\nabla}_{\bm{\eta}}\bm{\Phi}\left(\bm{x}^*\left(\bm{\eta}\right), \bm{\eta}\right) & = 0,
\end{align}
with the convention of matrix columns iterating over matrix coordinates. Assuming invertibility, the above determines the parametric derivative of the root as:
\begin{align}
    \bm{\nabla}_{\bm{\eta}}\bm{x}^*\left(\bm{\eta}\right) & = -\bm{\nabla}_{\bm{x}}\bm{\Phi}\left(\bm{x}^*\left(\bm{\eta}\right), \bm{\eta}\right)^{-1}\bm{\nabla}_{\bm{\eta}}\bm{\Phi}\left(\bm{x}^*\left(\bm{\eta}\right), \bm{\eta}\right).
\end{align}
Specializing to the critical point equation Eq.~\ref{eq:parametrized_critical_point_equation}, the above becomes:
\begin{align}
    \bm{\nabla}_{\bm{\eta}}\bm{x}^*\left(\bm{\eta}\right) & = -\bm{\nabla}_{\bm{x}\bm{x}}\varphi\left(\bm{x}^*\left(\bm{\eta}\right), \bm{\eta}\right)^{-1}\bm{\nabla}_{\bm{x}\bm{\eta}}\varphi\left(\bm{x}^*\left(\bm{\eta}\right), \bm{\eta}\right),\label{eq:parametrized_critical_point_solution_derivative}
\end{align}
where $\bm{\nabla}_{\bm{x}\bm{x}}\varphi$ is the diagonal Hessian block of $\bm{x}$ derivatives and $\bm{\nabla}_{\bm{x}\bm{\eta}}\varphi$ the Hessian block with $\bm{x}$ derivatives in rows and $\bm{\eta}$ derivatives in columns. We now specify the application of this formalism to local differentiation of optimal QAOA angles with respect to problem parameters. In this case, the optimized variables are the QAOA angles: $\bm{x} = \left(\bm{\gamma}, \bm{\beta}\right) = \left(\gamma_1, \ldots, \gamma_p, \beta_1, \ldots, \beta_p\right)$ and the parameters are the spin glass polynomial coefficients: $\bm{\eta} = \bm{\xi} = \left(\xi_1, \xi_2, \ldots\right)$. The scalar function $\varphi$ is the thermodynamic energy density of QAOA for angles $\bm{\gamma}, \bm{\beta}$ over the problem $\bm{\xi}$:
\begin{align}
    \varphi\left(\bm{x}, \bm{\eta}\right) & := \nu_p\left(\bm{\gamma}, \bm{\beta};\,\bm{\xi}\right)\nonumber\\
    & := \sum_{t \in \mathcal{T}_{p - 1}}i\Gamma_t\xi\left(G_{p + 1, t}\left(\bm{\gamma}, \bm{\beta};\,\bm{\xi}\right)\right),
\end{align}
where we explicitly indicated the dependence of $\bm{G}$ on angles $\bm{\gamma}, \bm{\beta}$ and the spin glass mixture polynomial coefficients $\bm{\xi}$. According to Eq.~\ref{eq:parametrized_critical_point_solution_derivative}, to compute the derivative (assuming differentiability) of extremal angles $\left(\bm{\gamma}^*\left(\bm{\xi}\right), \bm{\beta}^*\left(\bm{\xi}\right)\right)$ as a function of mixture coefficients $\bm{\xi}$, we require the derivative of $\bm{G}$ up to second order in the angles and first mixed order in the angles and mixture coefficients. $\bm{G}$ can be expressed from a $p$-step iteration of elementary functions (arithmetic operations and exponentials), so these derivatives can be efficiently computed by an automatic differentiation engine. This systematic method is only applicable when working with the sum-over-bitstrings representation of the iteration, and is therefore restricted to small numbers of layers. We believe these derivatives may be recast as spin-boson observables amenable to larger depth simulations, but leave this direction for future research.

\clearpage
\onecolumngrid
\setcounter{section}{0}
\setcounter{equation}{0}
\setcounter{theorem}{0}
\setcounter{definition}{0}
\setcounter{remark}{0}
\renewcommand{\thesection}{S\arabic{section}}
\numberwithin{equation}{section}
\numberwithin{theorem}{section}
\numberwithin{definition}{section}
\numberwithin{remark}{section}
\section*{Supplementary Information: Spin-boson mapping and cavity-method analysis of spin-glass QAOA}
\section{Background}

\subsection{Spin glasses and random sparse constraint satisfaction problems}
\label{si:sec:spin_glasses_csps}

Spin glasses constitute a major family of benchmark optimization problems. They are defined by a random cost function of the form
\begin{align}
    C_{\textrm{spin glass}}\left(\bm{\sigma}\right) & := \sum_{q \geq 1}\frac{c_q}{n^{(q - 1)/2}}\sum_{j_1, \ldots, j_q \in [n]}J^{(q)}_{j_1, \ldots, j_q}\sigma_{j_1} \ldots \sigma_{j_q},\label{si:eq:spin_glass_cost_function}
\end{align}
where $\left(c_q\right)_{q \geq 1}$ is a sequence of non-negative numbers vanishing above a certain term and all $J^{(q)}_{j_1, \ldots, j_q}$ are i.i.d $\mathcal{N}(0, 1)$.

In the thermodynamic limit ($n \to \infty$), the optimal energy density of a spin glass converges to a constant known as the Parisi value. A procedure to evaluate this constant was first derived by heuristic means~\cite{PhysRevLett.43.1754,PhysRevE.65.046137} and more recently confirmed by a fully rigorous argument~\cite{talagrand_parisi_formula}. The original argument applied to the quadratic spin glass model, defined by cost function,
\begin{align}
    C\left(\bm{\sigma}\right) & := \frac{1}{\sqrt{2n}}\sum_{j, k \in [n]}J_{j, k}\sigma_j\sigma_k,
\end{align}
and with approximate Parisi value
\begin{align}
    P_*\left(2\right) & \approx 0.763166\ldots
\end{align}
The argument was later generalized to arbitrary mixed spin glasses. We refer to \cite{Panchenko2013} for a modern account, and to \cite{Marwaha2022boundsapproximating} for a concrete evaluation of Parisi constants on higher-order pure spin glasses.

Random sparse constraint satisfaction problems are defined over a random hypergraph. Following Ref.~\cite{random_max_csps_hardness_spin_glasses}, given a $k$-hypergraph $G = \left(V, E\right)$ with vertex set $V$ and hyperedge multiset $E$ (we allow repetition of hyperedges), and a truth table $T: \{1, -1\}^k \longrightarrow \mathbb{R}$, a random constraint satisfaction problem is defined by cost function:
\begin{align}
    C_{\textrm{CSP}}\left(G, \bm{\sigma}\right) & := \frac{1}{\sqrt{D}}\sum_{\{\{j_1, \ldots, j_k\}\} \in E}T\left(\varepsilon_{e, 1}\sigma_{j_1}, \ldots, \varepsilon_{e, k}\sigma_{j_k}\right), \qquad G = \left(V, E\right) \textrm{(multi)hypergraph},\label{si:eq:random_csp_general_form}
\end{align}
with $\varepsilon_{e, k} \overset{\textrm{i.i.d}}{\sim} \mathrm{Unif}(\{1, -1\})$ are independent random signs. For maximum generality, we accommodated the possibility of variable repetition within hyperedges, and denoted a multiset of $k$ variable $j_1, \ldots, j_k$ by $\{\{j_1, \ldots, j_k\}\}$. Randomizing hypergraphs $G$ allows to study this optimization problem in the average case. Ref.~\cite{extremal_cuts_sparse_random_graphs} focuses on MAXCUT, characterized by cost function
\begin{align}
    C_{\mathrm{MAXCUT}}\left(G, \bm{\sigma}\right) & := \frac{1}{\sqrt{D}}\sum_{\{j, k\} \in E}\sigma_j\sigma_k,
\end{align}
 on sparse random graphs, including regular and Poisson-random hypergraphs  They show that in the large degree limit, the energy density is approximated by
\begin{align}
    \frac{P_*\left(2\right)}{\sqrt{D}} + o\left(D^{-1/2}\right),
\end{align}
i.e. by a simple rescaling of the Parisi constant of the quadratic dense spin glass. This dense-sparse correspondence was generalized to problems of arbitrary degree in Ref.~\cite{random_max_csps_hardness_spin_glasses}. The authors show that for all CSP of the form of Eq.~\ref{si:eq:random_csp_general_form}, there is an associated dense spin glass of identical degree such that in the infinite-degree limit, the optimal energy density is $P_*/\sqrt{D}$ up to a negligible degree correction, with $P_*$ the Parisi value of the dense spin glass. Beyond optimal energy values, their methods show equivalent algorithmic hardness for a broad variety of classical and quantum algorithms between sparse and dense problems. These algorithms include in particular classical low-degree polynomial algorithms and QAOA.  While the analysis restricts to Poisson-random hypergraphs for convenience, it is reasonably expected to generalize to other standard sparse hypergraph ensembles, including random regular hypergraphs.

\subsection{Classical and quantum algorithms}

Several algorithms with or without rigorous performance guarantees have been developed to approximately optimize spin glasses and sparse random constraint satisfaction problems.

Approximate Message Passing (AMP) \cite{optimization_sk,optimization_mean_field_spin_glasses} and Potential Hessian Ascent \cite{potential_hessian_ascent} are two prominent examples of rigorously analyzed classical algorithms for optimizing spin glasses.

Both can approximate the optimal energy of the quadratic Sherrington-Kirkpatrick model to arbitrary approximation ratio under a widely believed full replica symmetry breaking assumption. AMP was also analyzed for sparse constraint satisfaction problems, for instance MAXCUT on locally treelike, large-degree graphs~\cite{local_algorithms_sparse_maxcut}. In this setting, and under the same full replica symmetry breaking assumption, AMP was shown to approximate the optimal energy to arbitrary constant approximation ratio. The result holds asymptotically in the degree, where the optimal energy is related to the Parisi constant of some spin glass according to the general dense-sparse correspondence introduced in Section~\ref{si:sec:spin_glasses_csps}. Interestingly, to the best of our knowledge, no statement can be made on the optimality of AMP or any other algorithm at any fixed finite degree.

Unlike the quadratic Sherrington-Kirkpatrick model, higher-order spin glasses typically exhibit an overlap gap property \cite{Gamarnik2021} obstructing a broad family of algorithms from approximating the optimum. These include classical low-degree polynomial algorithms \cite{Gamarnik2020}, which were recently proven to include AMP~\cite{algorithmic_universality_ldp}, as well as quantum algorithms like QAOA~\cite{qaoa_needs_to_see_whole_graph_typical_case,limitations_local_algorithms_max_k_xor,qaoa_sparse_hypergraphs_spin_glass_models}. By the dense-sparse correspondence principle, these limitations also apply to higher-order sparse-dense constraint satisfaction problems. The overlap gap property was rigorously established~\cite{Gamarnik2021} for the pure $q$-spin model at $q \geq 4$.

\section{Results}

The main operational result enabling quantitative investigations in this work is a spin-boson mapping of QAOA applied to a mixed spin glass problem. This generalizes the mapping derived for the quadratic Sherrington-Kirkpatrick model in Ref.~\cite{qaoa_spin_boson_mapping}. By the universality result from Ref.~\cite{qaoa_sparse_hypergraphs_spin_glass_models}, the energy density achieved by QAOA on dense spin glass models coincides with the satisfaction fraction it achieves on sparse constraint satisfaction problems, in the large size and degree limit. A classical simulation of the mapping was empirically observed~\cite{qaoa_spin_boson_mapping} to enable evaluation of QAOA energies for the quadratic Sherrington-Kirkpatrick model up to a depth $p = 160$, out of reach for the exact evaluation, whose $\mathcal{O}\left(p^24^p\right)$ cost is prohibitive at that depth. The generalization of the spin-boson mapping developed in this work enables evaluation and optimization of QAOA satisfaction fractions for a broad variety of random constraint satisfaction problems. Beyond the operational description of the mapping, we provide its physical interpretation in the dense spin glass setting. This complements the insight from Ref.~\cite{qaoa_spin_boson_mapping}, analyzing the correspondence for quadratic problems on a sparse graph. Our interpretation of the mapping for dense spin glasses relies on a cavity calculation directly derived from the spin path integral representation of QAOA expectation values, a starting point of many earlier analytic studies of the algorithm~\cite{Farhi2022quantumapproximate,basso_et_al:LIPIcs.TQC.2022.7,qaoa_spiked_tensor}. Besides this physical interpretation, our cavity calculation yields concrete finite-size bounds on the QAOA energy. While we deem these bounds extremely loose, they represent, to the best of our knowledge, the first documented analytic attempt at quantifying the finite-size performance of QAOA. More generally, we believe our techniques may provide a framework to move beyond the constant depth, infinite-size limit restricting prior analyses of QAOA~\cite{Farhi2022quantumapproximate,qaoa_sparse_hypergraphs_spin_glass_models,qaoa_spiked_tensor}. We hope that despite its current quantitative limitations, our method may serve as a baseline for future more sophisticated attempts at understanding QAOA at finite size.

In Section~\ref{si:sec:spin_glass_qaoa_g_iteration_spin_boson_mapping}, we start by introducing the spin-boson mapping enabling evaluation of QAOA energies on arbitrary mixed spin glasses. This mapping is in turn based on an exact formula for the thermodynamic-limit QAOA energy of time complexity $\mathcal{O}\left(p^24^p\right)$. The classical simulation of the spin-boson mapping using Matrix Product States, enabling optimization of the QAOA energy on various spin glasses at depths intractable for exact evaluation, together with the resulting numerical comparison to Simulated Annealing, is presented in the accompanying main text. Finally, Section~\ref{si:sec:qaoa_cavity_method_analysis_heuristic} outlines the cavity argument enabling an alternative derivation of the exact formula for the energy of spin-glass QAOA, together with a physical interpretation of the spin-boson mapping in the dense case.

\subsection{The spin-glass QAOA iteration and its spin-boson mapping}
\label{si:sec:spin_glass_qaoa_g_iteration_spin_boson_mapping}

This Section introduces the exact formula for evaluating the energy of QAOA on an arbitrary mixed spin glass, as well as the spin-boson mapping of this formula.

Section~\ref{si:sec:spin_glass_qaoa_g_iteration}, specifically Theorem~\ref{si:th:spin_glass_qaoa_energy_from_g_matrix}, introduces the exact evaluation formula for the QAOA energy. The formula generalizes to mixed spin glasses the procedure derived in earlier work for the quadratic (Ref.~\cite{basso_et_al:LIPIcs.TQC.2022.7}) and higher-order (Ref.~\cite{qaoa_sparse_hypergraphs_spin_glass_models}) pure spin glass models. An expression for the energy in the mixed case was previously derived in a different form in Ref.~\cite{qaoa_sparse_hypergraphs_spin_glass_models}. The evaluation complexity was reported as exponential in the maximum arity $q_{\mathrm{max}}$ of the spin glass, but was observed to be independent of $q_{\mathrm{max}}$ in the special case of a pure spin glass. The present work generalizes this simplification, ultimately extending the iteration formula initially derived in Ref.~\cite{basso_et_al:LIPIcs.TQC.2022.7}, and of time complexity $\mathcal{O}\left(p^24^p\right)$, to an arbitrary spin glass. In this context, the evaluation formula is proved by direct calculation, starting from the original form derived in Ref.~\cite{qaoa_sparse_hypergraphs_spin_glass_models} and showing that our new proposed form satisfies the self-consistent equation characterizing the QAOA energy on a generic mixed spin-glass. Section~\ref{si:sec:unitary_circuit_path_integrals_and_spin_boson_systems} lays out background about quantum circuit path integrals and spin-boson systems, required both for the description of the spin-boson mapping and the sketch of the cavity argument. Finally, Section~\ref{si:sec:spin_boson_mapping_qaoa_iteration} introduces a spin-boson mapping of the QAOA energy evaluation procedure for an arbitrary spin glass. This generalizes the correspondence established in earlier work Ref.~\cite{qaoa_spin_boson_mapping} to a broader variety of problems. We further introduce a variant of the spin-boson observables computation procedure, allowing a factor $p$ reduction of the memory footprint of the classical simulation.

\subsubsection{The spin-glass QAOA iteration}
\label{si:sec:spin_glass_qaoa_g_iteration}

An analytic formula for the energy density of QAOA on a vast family of dense and sparse optimization problems was derived in Ref.~\cite{qaoa_sparse_hypergraphs_spin_glass_models}. The evaluation procedure has time complexity $4^{pq_{\mathrm{max}}}$ for a constraint satisfaction problem with up to $q_{\mathrm{max}}$-body interactions and QAOA depth $p$. In this paragraph, we give an account of this method and that for an arbitrary mixed spin glass, the formula for the infinite-size energy desnity simplifies to a $\mathcal{O}\left(p^24^p\right)$ complexity procedure generalizing the procedure for pure spin glasses reported in this earlier work. By universality of QAOA established in the same study, this simplifies formula for the energy density also applies to sparse constraint satisfaction problems in the large degree limit.

We now introduce the setting and notation from study Ref.~\cite{qaoa_sparse_hypergraphs_spin_glass_models}, which slightly differ from the present work. The authors consider a binary constraint satisfaction problem given by random cost function:
\begin{align}
    C_{\bm{J}}\left(\bm{\sigma}\right) & := \sum_{1 \leq q \leq q_{\mathrm{max}}}c_q\sum_{1 \leq j_1, \ldots, j_q \leq q_{\mathrm{max}}}J^{(q, n)}_{j_1, \ldots, j_q}\sigma_{j_1} \ldots \sigma_{j_q},\label{si:eq:random_cost_function_previous_work}
\end{align}
where the combinatorial variable is a spin configuration $\bm{\sigma} \in \{1, -1\}^n$.  In the above, tensors $\bm{J}^{(q)}$, $q \geq 1$ have i.i.d sampled entries, with each entry having characteristic function
\begin{align}
    \varphi_{q, n}\left(\lambda\right) & := \mathbb{E}\left[e^{i\lambda J^{(q, n)}_{j_1, \ldots, j_q}}\right].
\end{align}
$\bm{J} = \left(\bm{J}^{(q, n)}\right)$ refers to the collection of tensors $\bm{J}^{(q, n)}$ for all $1 \leq q \leq q_{\mathrm{max}}$. The following technical assumptions are made on characteristic functions:
\begin{itemize}
    \item For all $\lambda \in \mathbb{R}$, $\varphi_{q, n}\left(\lambda\right)$ is real positive for sufficiently large $n$. This allows to define $g_{q, n}\left(\lambda\right) := n^{q - 1}\log\varphi_{q, n}\left(\lambda\right)$ for sufficiently large $n$. In the following, we assume $n$ sufficiently large.
    \item For all $\lambda \in \mathbb{R}$, $\lim_{n \to \infty}g_{q, n}\left(\lambda\right)/n = 0$.
    \item $g_q\left(\lambda\right) := \lim_{n \to \infty}g_{q, n}\left(\lambda\right)$ exists and is differentiable.
    \item $\lim_{n \to \infty}g'_{q, n}\left(\lambda\right) = g_q'\left(\lambda\right)$.
\end{itemize}
We now describe the procedure to evaluate the disorder-averaged $p$-layer QAOA energy density:
\begin{align}
    \mathbb{E}_{\bm{J}}\left[\bra{\bm{\gamma}, \bm{\beta}}C_{\bm{J}}\ket{\bm{\gamma}, \bm{\beta}}\right]
\end{align}
on such a problem. Define the set of $2p$-bits bitstrings
\begin{align}
    A & := \left\{(b_1, b_2, \ldots, b_{p - 1}, b_p, b_{-p}, b_{-(p - 1)}, \ldots, b_{-2}, b_{-1})\,:\,b_{\pm t} \in \{\pm 1\} \quad \forall t \in [p]\right\}.
\end{align}
Define the following bitstring-indexed tensors:
\begin{align}
    Q_{\bm{b}} & := \prod_{1 \leq r \leq p}\left(\cos\beta_r\right)^{1 + \left(b_r + b_{-r}\right)/2}\left(\sin\beta_r\right)^{1 - \left(b_r + b_{-r}\right)/2}\left(i\right)^{\left(b_{-r} - b_r\right)/2},\\
    \Phi_{\bm{b}} & = \sum_{1 \leq r \leq p}\gamma_r\left(b_rb_{r + 1} \ldots b_{p - 1}b_p - b_{-p}b_{-(p - 1)} \ldots b_{-2}b_{-1}\right).
\end{align}
For bitstrings $\bm{b}, \bm{c} \in A$, denote by $\bm{b}\bm{c}$ their bitwise product. Let $\left(W_{\bm{b}}\right)_{\bm{b} \in A}$ be defined as the unique solution of self-consistent equation:
\begin{align}
    W_{\bm{b}} & = Q_{\bm{b}}\exp\left(\sum_{1 \leq q \leq q_{\mathrm{max}}}q\sum_{\bm{b}_1, \ldots, \bm{b}_{q - 1} \in A}g_q\left(c_q\Phi_{\bm{b}\bm{b}_1 \ldots \bm{b}_{q - 1}}\right)W_{\bm{b}_1} \ldots W_{\bm{b}_{q - 1}}\right).\label{si:eq:self_consistent_equation_previous_work}
\end{align}
Then, the infinite-size energy density output by QAOA is given by:
\begin{align}
    \lim_{n \to \infty}\mathbb{E}\left[\bra{\bm{\gamma}, \bm{\beta}}C_J/n\ket{\bm{\gamma}, \bm{\beta}}\right] & = -\sum_{1 \leq q \leq q_{\mathrm{max}}}ic_q\sum_{\bm{b}_1, \ldots, \bm{b}_q \in A}g_q'\left(c_q\Phi_{\bm{b}_1 \ldots \bm{b}_q}\right)W_{\bm{b}_1} \ldots W_{\bm{b}_q}.\label{si:eq:energy_density_previous_work}
\end{align}
Existence and uniqueness of the solution are non-trivial but established as part of Ref.~\cite{qaoa_sparse_hypergraphs_spin_glass_models}'s derivations. In this paragraph, we will exhibit a concrete candidate for $\left(W_{\bm{b}}\right)_{\bm{b} \in A}$ and use uniqueness to argue it provides the desired solution.

We now construct an explicit solution to Eq.~\ref{si:eq:self_consistent_equation_previous_work} for a mixed dense Gaussian spin glass. The latter corresponds to the following choice of random couplings:
\begin{align}
    \bm{J}^{(q, n)} & := \frac{1}{n^{(q - 1)/2}}\bm{J}^{(q)},\\
    \bm{J}^{{(q)}} & = \left(J^{(q)}_{j_1, \ldots, j_q}\right)_{j_1, \ldots, j_q \in [n]},\\
    J^{(q)}_{j_1, \ldots, j_q} & \overset{\mathrm{i.i.d}}{\sim} \mathcal{N}(0, 1).
\end{align}
The procedure from Ref.~\cite{qaoa_sparse_hypergraphs_spin_glass_models} applies to this problem, since the couplings's characteristic functions are given by
\begin{align}
    \varphi_{q, n}\left(\lambda\right) & = \mathbb{E}_{X \sim \mathcal{N}\left(0, 1\right)}\left[\exp\left(i\lambda X\right)\right]\nonumber\\
    & = \exp\left(-\frac{\lambda^2}{2n^{q - 1}}\right),
\end{align}
hence
\begin{align}
    g_{q, n}\left(\lambda\right) = -\frac{\lambda^2}{2}, && g_q\left(\lambda\right) = -\frac{\lambda^2}{2}, && g_q'\left(\lambda\right) = -\lambda.\label{si:eq:dense_spin_glass_characteristic_functions}
\end{align}
Now, define the following polynomial from coefficients $c_q$:
\begin{align}
    \xi\left(x\right) & := \sum_{1 \leq q \leq q_{\mathrm{max}}}c_q^2x^q.
\end{align}
In spin glass theory, it is known as the mixture polynomial of the spin glass. For convenience, define the set of QAOA layer signed indices (with an additional dummy layer $p + 1$):
\begin{align}
    \mathcal{T}_p & := \left\{1, 2, \ldots, p, p + 1, -p - 1, -p, \ldots, -2, -1\right\}.
\end{align}
This index set will be used to index the entries of a family of matrices, as well as bitstring bits. It will also be convenient to use the same set with $p$ one unit less, i.e.
\begin{align}
    \mathcal{T}_{p - 1} & := \left\{1, 2, \ldots, p - 1, p, -p, -(p - 1), \ldots, -2, -1\right\}.
\end{align}
Define the following bitstring-indexed tensor, depending only on QAOA mixer angles:
\begin{align}
    f\left(\bm{a}\right) & := \frac{\bm{1}\left[a_{p + 1} = a_{-p - 1}\right]}{2}\prod_{1 \leq t \leq p}\bra{a_t}e^{i\beta_tX}\ket{a_{t + 1}}\bra{a_{-t - 1}}e^{-i\beta_tX}\ket{a_{-t}}, \qquad \bm{a} \in \{1, -1\}^{\mathcal{T}_p}.
\end{align}
Also, define signed $\bm{\gamma}$ angles by
\begin{align}
    \bm{\Gamma} & := \left(\Gamma_1, \ldots, \Gamma_p, \Gamma_{-p}, \ldots, \Gamma_{-1}\right)\nonumber\\
    & = \left(\gamma_1, \ldots, \gamma_{p - 1}, \gamma_p, -\gamma_p, \ldots, -\gamma_1\right) \in \mathbb{R}^{\mathcal{T}_{p - 1}}.
\end{align}
Next, construct a family of matrices $\bm{G}^{(0)}, \bm{G}^{(1)}, \ldots, \bm{G}^{(p - 1)}, \bm{G}^{(p)}$ inductively in the superscript $0 \leq m \leq p$:
\begin{align}
    G^{(0)}_{r, s} & := 1,\\
    G^{(m + 1)}_{r, s} & := \sum_{\bm{a} \in \{1, -1\}^{\mathcal{T}_p}}a_ra_sf\left(\bm{a}\right)\exp\left(-\frac{1}{2}\sum_{t, u \in \mathcal{T}_{p - 1}}\xi'\left(G^{(m)}_{t, u}\right)\Gamma_t\Gamma_ua_ta_u\right), \qquad 0 \leq m \leq p - 1.\label{si:eq:g_iteration}
\end{align}
We denote the final iterate by $\bm{G} := \bm{G}^{(p)}$. This iteration generalizes the one developed in Ref.~\cite{basso_et_al:LIPIcs.TQC.2022.7} for Max-$q$-XOR, or equivalently (as established in Ref.~\cite{qaoa_sparse_hypergraphs_spin_glass_models}) the pure-$q$ spin glass. By repeating verbatim the arguments from Ref.~\cite{basso_et_al:LIPIcs.TQC.2022.7}, $\bm{G}$ can be shown to satisfy a certain number of symmetries, summarized in the following proposition:

\begin{proposition}[Symmetries of $\bm{G}$ matrix]
\label{si:prop:g_symmetries}
The $\bm{G}^{(m)}$ matrices computed by iteration Eq.~\ref{si:eq:g_iteration} are symmetric:
\begin{align}
    G^{(m)}_{r, s} & = G^{(m)}_{s, r} \qquad \forall r, s \in \mathcal{T}_p.
\end{align}
It has unit $2 \times 2$ diagonal blocks (indices of same absolute value):
\begin{align}
    G^{(m)}_{r, r} = G^{(m)}_{-r, -r} = G^{(m)}_{r, -r} = G^{(m)}_{-r, r} = 1, \qquad \forall 1 \leq r \leq p + 1.
\end{align}
Finally, it satisfies the following index-flip symmetries:
\begin{align}
    G^{(m)}_{-r, s} = \overline{G^{(m)}_{r, s}}, && G^{(m)}_{r, -s} = G^{(m)}_{r, s}, && G^{(m)}_{-r, -s} = \overline{G^{(m)}_{r, s}}, \qquad \forall 1 \leq r \leq s \leq p + 1.
\end{align}
Recursion Eq.~\ref{si:eq:g_iteration} is stationary from level $m$ for indices of absolute values bounded by $(m + 1)$:
\begin{align}
    G^{(m)}_{r, s} = G^{(m + 1)}_{r, s} = \ldots = G^{(p)}_{r, s} \qquad |r|, |s| \leq m + 1.\label{si:eq:g_iteration_stationarity}
\end{align}
In particular, the final iterate $\bm{G} := \bm{G}^{(p)}$ is a fixed-point of the iterative mapping:
\begin{align}
    G_{r, s} & = \sum_{\bm{a} \in \{1, -1\}^{\mathcal{T}_p}}a_ra_sf\left(\bm{a}\right)\exp\left(-\frac{1}{2}\sum_{t, u \in \mathcal{T}_{p - 1}}\xi'\left(G_{t, u}\right)\Gamma_t\Gamma_ua_ta_u\right).\label{si:eq:g_fixed_point_equation}
\end{align}
Finally, when evaluating a matrix entry $\left(r, s\right)$ using summation-over-bitstrings Eq.~\ref{si:eq:g_iteration}, $p$ may be decreased to $q := \max\left\{|r| - 1, |s| - 1\right\}$ in the right-hand side bitstring summation without affecting its value:
\begin{align}
    G^{(m + 1)}_{r, s} & = \sum_{\bm{a} \in \{1, -1\}^{\mathcal{T}_q}}a_ra_sf\left(\bm{a}\right)\exp\left(-\frac{1}{2}\sum_{t, u \in \mathcal{T}_{q - 1}}\xi'\left(G^{(m)}_{t, u}\right)\Gamma_t\Gamma_ua_ta_u\right), \qquad q + 1 := \max\left\{|r|, |s|\right\}.\label{si:eq:g_iteration_restricted}
\end{align}
\end{proposition}

As a consequence of the stationarity property Eq.~\ref{si:eq:g_iteration_stationarity} of the $\bm{G}$ iteration, the $\bm{G}$ iteration can be evaluated in time $\mathcal{O}\left(p^24^p\right)$ instead of the naive $\mathcal{O}\left(p^34^p\right)$ suggested by Eq.~\ref{si:eq:g_iteration}. We are now ready to state a solution for self-consistent equation Eq.~\ref{si:eq:self_consistent_equation_previous_work} in terms of the $\bm{G}$ matrix:

\begin{theorem}[Thermodynamic limit spin-glass QAOA energy density from $\bm{G}$ matrix]
\label{si:th:spin_glass_qaoa_energy_from_g_matrix}
Consider operation $*$, mapping a $(2p)$-bit bitstring $\bm{b}$ to a $(2p + 2)$-bit bitstring as follows:
\begin{align}
    \bm{b} & := \left(b_1, \ldots, b_p, b_{-p}, \ldots, b_{-1}\right) \in \{1, -1\}^{\mathcal{T}_{p - 1}},\\
    \bm{b}^* & := \left(b^*_1, \ldots, b^*_p, b^*_{p + 1}, b^*_{-p - 1}, b^*_{-p}, \ldots, b^*_{-1}\right) \in \{1, -1\}^{\mathcal{T}_p},\\
    b^*_{\pm r} & := b_{\pm r}b_{\pm (r + 1)} \ldots b_{\pm (p - 1)}b_{\pm p}, \qquad 1 \leq r \leq p,\\
    b^*_{\pm (p + 1)} & := 1.
\end{align}
Then,
\begin{align}
    W_{\bm{b}} & := 2f\left(\bm{b}^*\right)H\left(\bm{b}^*\right),\\
    H\left(\bm{a}\right) & := \exp\left(-\frac{1}{2}\sum_{t, u \in \mathcal{T}_{p - 1}}\xi'\left(G_{t, u}\right)\Gamma_t\Gamma_ua_ta_u\right), \qquad \bm{a} \in \{1, -1\}^{\mathcal{T}_p}
\end{align}
solves self-consistent equation Eq.~\ref{si:eq:self_consistent_equation_previous_work}, and infinite-size energy density is given by:
\begin{align}
    \lim_{n \to \infty}\mathbb{E}\left[\bra{\bm{\gamma}, \bm{\beta}}C_J/n\ket{\bm{\gamma}, \bm{\beta}}\right] & = \sum_{t \in \mathcal{T}_{p - 1}}i\Gamma_t\xi\left(G_{p + 1, t}\right).
\end{align}
\begin{proof}
We first note that
\begin{align}
    Q_{\bm{b}} & = \prod_{1 \leq r \leq p}\left(\cos\beta_r\right)^{1 + \left(b_r + b_{-r}\right)/2}\left(\sin\beta_r\right)^{1 - \left(b_r + b_{-r}\right)/2}\left(i\right)^{\left(b_{-r} - b_r\right)/2}\nonumber\\
    & = \prod_{1 \leq r \leq p}\bra{1}e^{i\beta_rX}\ket{b_r}\bra{b_{-r}}e^{-i\beta_rX}\ket{1}\nonumber\\
    & = \prod_{1 \leq r \leq p}\bra{1}e^{i\beta_rX}\ket{b^*_rb^*_{r + 1}}\bra{b^*_{-r}b^*_{-r - 1}}e^{-i\beta_rX}\ket{1}\nonumber\\
    & = \prod_{1 \leq r \leq p}\bra{b^*_r}e^{i\beta_rX}\ket{b^*_{r + 1}}\bra{b^*_{-r - 1}}e^{-i\beta_rX}\ket{b^*_{-r}}\nonumber\\
    & = 2f\left(\bm{b}^*\right),
\end{align}
where the last equality uses $b^*_{\pm (p + 1)} = 1$, so the indicator function in the definition of $f$ indeed equals $1$. Besides,
\begin{align}
    \Phi_{\bm{b}} & := \sum_{1 \leq r \leq p}\gamma_r\left(b_r \ldots b_p - b_{-p} \ldots b_{-r}\right)\nonumber\\
    & = \sum_{1 \leq r \leq p}\gamma_r\left(b^*_r - b^*_{-r}\right)\nonumber\\
    & = \sum_{t \in \mathcal{T}_{p - 1}}\Gamma_tb^*_t.
\end{align}
Hence, the right-hand side of the self-consistent equation Eq.~\ref{si:eq:self_consistent_equation_previous_work} can be expressed:
\begin{align}
    & Q_{\bm{b}}\exp\left(\sum_{1 \leq q \leq q_{\mathrm{max}}}q\sum_{\bm{b}_1, \ldots, \bm{b}_{q - 1} \in A}g_q\left(c_q\Phi_{\bm{b}\bm{b}_1 \ldots \bm{b}_{q - 1}}\right)W_{\bm{b}_1}\ldots W_{\bm{b}_{q - 1}}\right)\nonumber\\
    & = 2f\left(\bm{b}^*\right)\exp\left(-\frac{1}{2}\sum_{1 \leq q \leq q_{\mathrm{max}}}q\sum_{\bm{b}_1, \ldots, \bm{b}_{q - 1} \in A}c_q^2\Phi_{\bm{b}\bm{b}_1 \ldots \bm{b}_{q - 1}}^2W_{\bm{b}_1} \ldots W_{\bm{b}_{q - 1}}\right)\nonumber\\
    & = 2f\left(\bm{b}^*\right)\exp\left(-\frac{1}{2}\sum_{1 \leq q \leq q_{\mathrm{max}}}q\sum_{\bm{b}_1, \ldots, \bm{b}_{q - 1} \in A}c_q^2\sum_{t, u \in \mathcal{T}_{p - 1}}\Gamma_t\Gamma_ub^*_{t}b^*_{1, t} \ldots b^*_{q - 1, t}b^*_{u}b^*_{1, u} \ldots b^*_{q - 1, u}W_{\bm{b}_1} \ldots W_{\bm{b}_{q - 1}}\right)
\end{align}
The sums over $\bm{b}_1, \ldots, \bm{b}_{q - 1}$ inside the exponential can be independently computed as:
\begin{align}
    \sum_{\bm{b}_r \in A}b^*_{r, t}b^*_{r, u}W_{\bm{b}_r} & = \sum_{\bm{b}_r \in A}b^*_{r, t}b^*_{r, u}2f\left(\bm{b}_r^*\right)H\left(\bm{b}_r^*\right)\nonumber\\
    & = \frac{1}{2}\sum_{\bm{a} \in \{1, -1\}^{\mathcal{T}_p}}a_ta_u2f\left(\bm{a}\right)H\left(\bm{a}\right)\nonumber\\
    & = G_{t, u},
\end{align}
where in the first equality, we used length-$(2p + 2)$ bitstrings $\bm{a} \in \{1, -1\}^{\mathcal{T}_p}$ satisfying $a_{p + 1} = a_{-p - 1}$ (required for $f\left(\bm{a}\right) \neq 0$) may be enumerated as the disjoint union of bitstrings $\bm{b}^*$ ($\bm{b} \in \{1, -1\}^{\mathcal{T}_{p - 1}}$), for which $b^*_{p + 1} = b^*_{-p - 1} = 1$ by definition, and the set of bitstrings $-\bm{b}^*$, having bits $\pm (p + 1)$ set to $-1$. The sums over the two disjoint sets of bitstrings are identical, given $f\left(\bm{a}\right) = f\left(-\bm{a}\right)$, $H\left(-\bm{a}\right) = H\left(\bm{a}\right)$. In the third equality, we used the fixed-point equation Eq.~\ref{si:eq:g_fixed_point_equation}. Hence,
\begin{align}
    & Q_{\bm{b}}\exp\left(\sum_{1 \leq q \leq q_{\mathrm{max}}}q\sum_{\bm{b}_1, \ldots, \bm{b}_{q - 1} \in A}g_q\left(c_q\Phi_{\bm{b}\bm{b}_1 \ldots \bm{b}_{q - 1}}\right)W_{\bm{b}_1}\ldots W_{\bm{b}_{q - 1}}\right)\nonumber\\
    & = 2f\left(\bm{b}^*\right)\exp\left(-\frac{1}{2}\sum_{1 \leq q \leq q_{\mathrm{max}}}qc_q^2\sum_{t, u \in \mathcal{T}_{p - 1}}\Gamma_t\Gamma_u\left(G_{t, u}\right)^{q - 1}b_t^*b_u^*\right)\nonumber\\
    & = 2f\left(\bm{b}^*\right)\exp\left(-\frac{1}{2}\sum_{t, u \in \mathcal{T}_{p - 1}}\xi'\left(G_{t, u}\right)\Gamma_t\Gamma_ub_t^*b_u^*\right)\nonumber\\
    & = 2f\left(\bm{b}^*\right)H\left(\bm{b}^*\right)\nonumber\\
    & = W_{\bm{b}}.
\end{align}
Hence, the candidate $\left(W_{\bm{b}}\right)_{\bm{b} \in A}$ satisfies the self-consistent equation, and by uniqueness of the solution, it coincides with the solution $\left(W_{\bm{b}}\right)_{\bm{b} \in A}$ of the self-consistent equation introduced in Ref.~\cite{qaoa_sparse_hypergraphs_spin_glass_models}. For the energy density, starting from Eq.~\ref{si:eq:energy_density_previous_work} established in Ref.~\cite{qaoa_sparse_hypergraphs_spin_glass_models},
\begin{align}
    \lim_{n \to \infty}\mathbb{E}\left[\bra{\bm{\gamma}, \bm{\beta}}C_J/n\ket{\bm{\gamma}, \bm{\beta}}\right] & = -i\sum_{1 \leq q \leq q_{\mathrm{max}}}c_q\sum_{\bm{b}_1, \ldots, \bm{b}_q \in A}g_q'\left(c_q\Phi_{\bm{b}_1 \ldots \bm{b}_q}\right)W_{\bm{b}_1} \ldots W_{\bm{b}_q}\nonumber\\
    & = i\sum_{1 \leq q \leq q_{\mathrm{max}}}c_q^2\sum_{\bm{b}_1, \ldots, \bm{b}_q \in A}\Phi_{\bm{b}_1 \ldots \bm{b}_q}W_{\bm{b}_1} \ldots W_{\bm{b}_q}\nonumber\\
    & = i\sum_{1 \leq q \leq q_{\mathrm{max}}}c_q^2\sum_{\bm{b}_1, \ldots, \bm{b}_q \in A}\sum_{t \in \mathcal{T}_{p - 1}}\Gamma_tb^*_{1, t} \ldots b^*_{q, t}W_{\bm{b}_1} \ldots W_{\bm{b}_q}.
\end{align}
As above, sums over $\bm{b}_1, \ldots, \bm{b}_q$ can be taken independently:
\begin{align}
    \sum_{\bm{b} \in A}b^*_tW_{\bm{b}} & = \sum_{\bm{b} \in A}b^*_t2f\left(\bm{b}^*\right)H\left(\bm{b}^*\right)\nonumber\\
    & = \sum_{\bm{b} \in A}b^*_tb^*_{p + 1}2f\left(\bm{b}^*\right)H\left(\bm{b}^*\right)\nonumber\\
    & = \sum_{\bm{a} \in \{1, -1\}^{\mathcal{T}_p}}a_ta_{p + 1}f\left(\bm{a}\right)H\left(\bm{a}\right)\nonumber\\
    & = G_{p + 1, t}.
\end{align}
where in the second line, we recalled $b^*_{p + 1} = 1$ for all $\bm{b} \in A$ (by definition of the $*$ operation) and in the third line, we used evenness of the summand in the bitstring. All in all, the infinite-size energy density simplifies to:
\begin{align}
    \lim_{n \to \infty}\mathbb{E}\left[\bra{\bm{\gamma}, \bm{\beta}}C_J/n\ket{\bm{\gamma}, \bm{\beta}}\right] & = i\sum_{1 \leq q \leq q_{\mathrm{max}}}c_q^2\sum_{t \in \mathcal{T}_{p - 1}}\Gamma_tG_{p + 1, t}^q\nonumber\\
    & = \sum_{t \in \mathcal{T}_{p - 1}}i\Gamma_t\xi\left(G_{p + 1, t}\right).
\end{align}
\end{proof}
\end{theorem}

\subsubsection{Quantum circuit spin path integrals and spin-boson systems}
\label{si:sec:unitary_circuit_path_integrals_and_spin_boson_systems}

We start with the general definition and notation for a \textit{layered quantum circuit} spin path integral, applicable beyond QAOA.

\begin{definition}[Layered quantum circuit spin path integral measure]
\label{si:def:layered_quantum_circuit_path_integral_measure}
A layered quantum circuit with $p$ layers acting on qubits indexed by $S$ is defined by an initial state $\ket{\psi}$ and $p$ unitary layers $U_1, \ldots, U_p \in \mathbb{C}^{2^S \times 2^S}$. The spin path integral measure of this layered quantum circuit is then defined as:
\begin{align}
    \mu^{[p]}\left(\bm{z}\right) & := \braket{\psi|\bm{z}^{[1]}}\braket{\bm{z}^{[-1]}|\psi}\mathbf{1}\left[\bm{z}^{[p + 1]} = \bm{z}^{[-p - 1]}\right]\prod_{1 \leq t \leq p}\bra{\bm{z}^{[t]}}U_t^{\dagger}\ket{\bm{z}^{[t + 1]}}\bra{\bm{z}^{[-t - 1]}}U_t\ket{\bm{z}^{[-t]}},\label{si:eq:layered_quantum_circuit_path_integral_measure}
\end{align}
where
\begin{align}
    \bm{z} & := \left(z_j^{[t]}\right)_{j \in S,\,t \in \mathcal{T}_p} \in \{1, -1\}^{S \times \mathcal{T}_p}
\end{align}
is a matrix of computational basis states, with rows indexed by spins $j \in S$, and columns indexed by layer indices $t \in \mathcal{T}_p$, where
\begin{align}
    \mathcal{T}_p & := \left\{1, \ldots, p + 1, -p - 1, \ldots, -1\right\}.
\end{align}
Given such a bit matrix, we denote by
\begin{align}
    \bm{z}^{[t]} & := \left(z^{[t]}_j\right)_{j \in S} \in \{1, -1\}^S
\end{align}
the computational basis states of all qubits at layer $t$, and by
\begin{align}
    \bm{z}_j & := \left(z_j^{[t]}\right)_{t \in \mathcal{T}_p}
\end{align}
the computational basis trajectory of spin $j$. Consistent with these conventions, Eq.~\ref{si:eq:layered_quantum_circuit_path_integral_measure} denotes
\begin{align}
    \ket{\bm{z}^{[t]}} & := \bigotimes_{t \in S}\ket{z_j^{[t]}},\\
    \bra{\bm{z}^{[t]}} & := \bigotimes_{t \in S}\bra{z_j^{[t]}},\\
    \ket{\bm{z}^{[t]}}\bra{\bm{z}^{[t]}} & := \bigotimes_{t \in S}\ket{z_j^{[t]}}\bra{z_j^{[t]}}.
\end{align}
\end{definition}

The quantum circuit path integral measure introduced in Eq.~\ref{si:eq:layered_quantum_circuit_path_integral_measure} is obtained by inserting computational basis projectors between all circuit layers. For instance, for a $2$-layer circuit acting on qubit set $S = [n]$, Eq.~\ref{si:eq:layered_quantum_circuit_path_integral_measure} reads:
\begin{align}
    \mu^{[2]}\left(\bm{z}\right) & = \braket{\psi|\bm{z}^{[1]}}\bra{\bm{z}^{[1]}}U_1^{\dagger}\ket{\bm{z}^{[2]}}\bra{\bm{z}^{[2]}}U_2^{\dagger}\ket{\bm{z}^{[3]}}\mathbf{1}\left[\bm{z}^{[3]} = \bm{z}^{[-3]}\right]\bra{\bm{z}^{[-3]}}U_2\ket{\bm{z}^{[-2]}}\bra{\bm{z}^{[-2]}}U_1\ket{\bm{z}^{[-1]}}\braket{\bm{z}^{[-1]}|\psi}.
\end{align}
Negative layer indices relate to the state vector produced by the circuit, while positive indices relate to the transpose conjugate of this state. Summing the above over bit matrices $\bm{z} = \begin{pmatrix}
    \bm{z}_1 & \bm{z}_2 & \bm{z}_3 & \bm{z}_{-3} & \bm{z}_{-2} & \bm{z}_{-1}
\end{pmatrix} \in \{1, -1\}^{S \times \mathcal{T}_2}$, using completeness relation
\begin{align}
    \sum_{\bm{z}^{[t]} \in \{1, -1\}^S}\ket{\bm{z}^{[t]}}\bra{\bm{z}^{[t]}} & = \bm{I}_{\mathbb{C}^{2^S}}
\end{align}
for all layer $t$, we obtain:
\begin{align}
    \sum_{\bm{z} \in \{1, -1\}^{S \times \mathcal{T}_p}}\mu^{[2]}\left(\bm{z}\right) & = \bra{\psi}U_1^{\dagger}U_2^{\dagger}U_2U_1\ket{\psi}\nonumber\\
    & = 1.
\end{align}
Hence, the path integral measure can be interpreted as a quasiprobability distribution. The path integral measure of a quantum circuit allows to express computational basis observables (measured after the evolution) but also time autocorrelations (operators inserted through the evolution) under the layered quantum circuit. An example of time autocorrelation for the previous $2$-layer circuit is:
\begin{align}
    \bra{\psi}\textcolor{blue}{Z_5}U_1^{\dagger}U_2^{\dagger}\textcolor{blue}{Z_6}U_2\textcolor{blue}{Z_8}U_1\ket{\psi} & = \sum_{\bm{z} \in \{1, -1\}^{S \times \mathcal{T}_2}}\mu^{[2]}\left(\bm{z}\right)z_5^{[1]}z_6^{[3]}z_8^{[-2]}.
\end{align}
Hence, a time autocorrelation of $Z$ operators can be expressed as a ``standard'' expectation under the path integral measures, with spin variables carrying layer indices corresponding to their insertion positions in the circuit. Note that layer indices $\pm\left(p + 1\right)$ are equivalent by construction; for instance, the above time autocorrelation can also be expressed:
\begin{align}
    \bra{\psi}\textcolor{blue}{Z_5}U_1^{\dagger}U_2^{\dagger}\textcolor{blue}{Z_6}U_2\textcolor{blue}{Z_8}U_1\ket{\psi} & = \sum_{\bm{z} \in \{1, -1\}^{S \times \mathcal{T}_2}}\mu^{[2]}\left(\bm{z}\right)z_5^{[1]}z_6^{[-3]}z_8^{[-2]}.
\end{align}
As another example, the expectation of any Hamiltonian diagonal in the computational basis, given by a real-valued function $C\left(\bm{x}\right)$, $\bm{x} \in \{1, -1\}^S$, can be expressed from the quantum circuit path integral measure
\begin{align}
    \bra{\psi}U_1^{\dagger}U_2^{\dagger}\textcolor{blue}{C}U_2U_1\ket{\psi} & = \sum_{\bm{z} \in \{1, -1\}^{S \times \mathcal{T}_p}}\mu^{[2]}\left(\bm{z}\right)C\left(\bm{z}^{[3]}\right)\\
    & = \sum_{\bm{z} \in \{1, -1\}^{S \times \mathcal{T}_p}}\mu^{[2]}\left(\bm{z}\right)C\left(\bm{z}^{[-3]}\right).
\end{align}
One can finally consider a hybrid example involving a final layer observable and $Z$ operator insertions at intermediate layers:
\begin{align}
    \bra{\psi}\textcolor{blue}{Z_5}U_1^{\dagger}U_2^{\dagger}\textcolor{blue}{C}\textcolor{blue}{Z_6}U_2\textcolor{blue}{Z_8}U_1\ket{\psi} & = \sum_{\bm{z} \in \{1, -1\}^{S \times \mathcal{T}_p}}\mu^{[2]}\left(\bm{z}\right)z_5^{[1]}z_6^{[3]}z_8^{[-2]}C\left(\bm{z}^{[3]}\right).
\end{align}
By expanding cost Hamiltonian as a Boolean polynomial:
\begin{align}
    C\left(\bm{x}\right) & = \sum_{T \subset S}\hat{C}_Tx_T,\\
    x_T & := \prod_{j \in T}x_j,\\
    \hat{C}_T & := \mathbb{E}_{\bm{x} \sim \mathrm{Unif}\left(\{1, -1\}^S\right)}\left[x_TC\left(\bm{x}\right)\right],
\end{align}
we see all the above expectations reduce to time autocorrelations of $Z$ operators pertaining to arbitrary qubits and inserted at arbitrary layers. The following result summarizes the path integral representation of such quantities:

\begin{proposition}[Path integral representation of $Z$ operators autocorrelations]
\label{si:prop:z_autocorrelation_path_integral_representation}
Consider a layered quantum circuit and its associated path integral measure as introduced in Definition~\ref{si:def:layered_quantum_circuit_path_integral_measure}. Let $I$ a multiset of pairs $\left(j, t\right)$ of spin $j$ and insertion layer $t$. Then, the autocorrelation of $Z$ operators defined by qubit indices and insertion positions $I$ can be expressed as the following ``standard expectation'' under the path integral measure:
\begin{align}
    \left\langle \prod_{(j, t) \in I}z_j^{[t]} \right\rangle_{\mu^{[p]}\left(\bm{z}\right)} & := \sum_{\bm{z} \in \{1, -1\}^{S \times \mathcal{T}_p}}\mu^{[p]}\left(\bm{z}\right)\prod_{(j, t) \in I}z_j^{[t]}.
\end{align}
\end{proposition}

By reversing the interpretation of Proposition~\ref{si:prop:z_autocorrelation_path_integral_representation} (which expresses an autocorrelation as a path integral expectation), we obtain the following important boundedness property for a generic path integral measure:

\begin{corollary}[Boundedness of monomial expectations under path integral measure]
\label{si:cor:path_integral_measure_boundedness_monomials}
Consider a layered quantum circuit and its associated path integral measure as introduced in Definition~\ref{si:def:layered_quantum_circuit_path_integral_measure}. For all multiset $I$ of pairs $\left(j, t\right)$ of spin $j$ and insertion layer $t$,
\begin{align}
    \left|\left\langle \prod_{(j, t) \in I}z_j^{[t]} \right\rangle_{\mu^{[p]}\left(\bm{z}\right)}\right| & \leq 1.
\end{align}
\begin{proof}
This follows from reinterpreting the expectation under the path integral measure as a time autocorrelations of $Z$ operators, according to Proposition~\ref{si:prop:z_autocorrelation_path_integral_representation}. The boundedness then follows from normalization of state $\ket{\psi}$, unitarity of operators $U_t$ and $\left\lVert Z \right\rVert_{\infty} \leq 1$.
\end{proof}
\end{corollary}

The following Proposition provides a path integral measure translation of unitary cancellation $U_t^{\dagger}U_t = I$ in the circuit:

\begin{proposition}[Unitary cancellation in path integral measure]
\label{si:prop:path_integral_measure_unitary_cancellation}
Consider a $p$-layer quantum circuit and its associated spin path integral measure $\mu^{[p]}$ as specified in Definition~\ref{si:def:layered_quantum_circuit_path_integral_measure}. Besides, for any $0 \leq q \leq p$, consider the layered quantum circuit defined by the restriction of the original circuit to the first $q$ layers (for $q = 0$, the restricted circuit merely prepares state $\ket{\psi})$, and denote by $\mu^{[q]}$ the associated path integral measure. Then, for all $1 \leq q \leq p$, summing over final spin configuration indices marginalizes $\mu^{[q]}$ to $\mu^{[q - 1]}$:
\begin{align}
    \sum_{\bm{z}^{[q + 1]}, \bm{z}^{[-q - 1]} \in \{1, -1\}^{S \times \mathcal{T}_q}}\mu^{[q]}\left(\bm{z}^{\left[\mathcal{T}_{q - 1}\right]}, \bm{z}^{[q + 1]}, \bm{z}^{[-q - 1]}\right) & = \mu^{[q - 1]}\left(\bm{z}^{\left[\mathcal{T}_{q - 1}\right]}\right),\label{si:eq:path_integral_measure_unitary_cancellation}
\end{align}
for all $(q - 1)$-layer bit matrix
\begin{align}
    \bm{z}^{\left[\mathcal{T}_{q - 1}\right]} & := \left(z_j^{[t]}\right)_{j \in S,\,t \in \mathcal{T}_{q - 1}}.
\end{align}
In particular, consider a multiset $I$ of pairs $(j, t)$ of spin $j$ and $Z$ insertion layer $t$, where insertion layer indices have absolute value at most $(q + 1)$. Then, for all number of layers $r \in [q, p]$,
\begin{align}
    \left\langle \prod_{(j, t) \in I}z_j^{[t]} \right\rangle_{\mu^{[r]}\left(\bm{z}\right)} & = \left\langle \prod_{(j, t) \in I}z_j^{[t]} \right\rangle_{\mu^{[p]}\left(\bm{z}\right)}.
\end{align}
\begin{proof}
Layer indices $\pm(q + 1)$ occur in $\mu^{[q]}$ only through the boundary indicator $\mathbf{1}\left[\bm{z}^{[q + 1]} = \bm{z}^{[-q - 1]}\right]$ and the layer-$q$ factor $\bra{\bm{z}^{[q]}}U_q^{\dagger}\ket{\bm{z}^{[q + 1]}}\bra{\bm{z}^{[-q - 1]}}U_q\ket{\bm{z}^{[-q]}}$. Summing over these two indices, the indicator identifies $\bm{z}^{[q + 1]} = \bm{z}^{[-q - 1]} =: \bm{x}$, and the completeness relation gives
\begin{align}
    \sum_{\bm{x} \in \{1, -1\}^S}\bra{\bm{z}^{[q]}}U_q^{\dagger}\ket{\bm{x}}\bra{\bm{x}}U_q\ket{\bm{z}^{[-q]}} & = \bra{\bm{z}^{[q]}}U_q^{\dagger}U_q\ket{\bm{z}^{[-q]}} = \braket{\bm{z}^{[q]}|\bm{z}^{[-q]}} = \mathbf{1}\left[\bm{z}^{[q]} = \bm{z}^{[-q]}\right],
\end{align}
which is precisely the boundary indicator of $\mu^{[q - 1]}$; the surviving factors ($1 \leq t \leq q - 1$) are already those of $\mu^{[q - 1]}$, establishing the marginalization identity. For the autocorrelation identity, a monomial with insertion layers of absolute value at most $(q + 1)$ is independent of the spin indices at layers $|t| > q + 1$; iterating the marginalization from $\mu^{[r]}$ down to $\mu^{[q]}$ therefore leaves its expectation unchanged for every $q \leq r \leq p$.
\end{proof}
\end{proposition}

This concludes the generic exposition of layered unitary quantum circuit path integrals. We now turn our attention to spin-boson systems defined by a discrete sequence of unitary evolutions. In such systems, the spin dynamics can be described by a path integral measure taking a specific form. Technically, this path integral measure does not fit the setting of Definition~\ref{si:def:layered_quantum_circuit_path_integral_measure} as the channel evolving the spin is non-unitary due to tracing out bosonic modes. However, it will be convenient to port most of the notation. Besides, results such as the boundedness Corollary~\ref{si:cor:path_integral_measure_boundedness_monomials} and the unitary cancellation Proposition~\ref{si:prop:path_integral_measure_unitary_cancellation} will continue to apply. We start by introducing the definition of a discrete spin-boson evolution relevant to this work, with notation closely mirroring that of the QAOA circuit for convenience.

\begin{definition}[Spin-boson system evolution]
\label{si:def:spin_boson_system}
Let $p \geq 1$ a number of layers, $p' \geq 1$ a number of bosonic modes and $\bm{L} \in \mathbb{C}^{p' \times p}$ an arbitrary matrix with columsn normalized to $1$. Let be given for each layer $t \in [p]$ spin-boson interaction strengths $\bm{\gamma} = \left(\gamma_1, \ldots, \gamma_p\right)$ and spin mixing strengths $\bm{\beta} = \left(\beta_1, \ldots, \beta_p\right)$. To these parameters, we associate discrete unitary spin-boson evolution defined by initial state
\begin{align}
    \ket{\psi} & := \ket{+}_{\mathrm{spin}} \otimes \ket{\overline{0}}_{\mathrm{boson}}^{\otimes p'},
\end{align}
and unitaries:
\begin{align}
    U_t & := \exp\left(-i\beta_tX\right)D\left(-i\gamma_tZ\bm{L}_{:,\,t}\right), \qquad 1 \leq t \leq p.
\end{align}
In the initial state, we denoted by $\ket{\overline{0}}_{\mathrm{boson}}$ the bosonic vacuum. Each layer's unitary includes a displacement operator displacing the $p'$ modes with amplitude $-i\gamma_t\sigma\bm{L}_{:,\,t}$, with $\sigma$ the spin $Z$ projection before layer $t$ and $\bm{L}_{:,\,t} \in \mathbb{C}^{p'}$ column $t$ of $\bm{L}$. Explicitly:
\begin{align}
    D\left(-i\gamma_tZ\bm{L}_{:,\,t}\right) & := \sum_{\sigma \in \{1, -1\}}\ket{\sigma}\bra{\sigma}_{\mathrm{spin}} \otimes \bigotimes_{\mathrm{modes}\,l \in \left[p'\right]} D_l\left(-i\gamma_t\sigma L_{l, t}\right),
\end{align}
with displacement operator $D_l$ acting on mode $l$.
\end{definition}

The spin dynamics in such a system can be described by a path integral measure similar to that of unitary circuits introduced in Definition~\ref{si:def:layered_quantum_circuit_path_integral_measure}. Note, however, that a path integral measure for the spin only cannot fit into this general framework, given the evolution of the spin is non-unitary after tracing out bosonic modes.

\begin{definition}[Path integral measure of spin-boson system]
\label{si:def:spin_boson_path_integral_measure}
Consider a discrete unitary spin-boson evolution as specified in Definition~\ref{si:def:spin_boson_system}. The path integral measure of such a system is defined as:
\begin{align}
    \mu_{SB}\left(\bm{a}\right) & := \bra{+}_{\mathrm{spin}}\bra{\overline{0}}_{\mathrm{bosons}}\ket{a_1}_{\mathrm{spin}}\left(\overrightarrow{\prod_{t = 1}^p}\bra{a_t}_{\mathrm{spin}}U_t^{\dagger}\ket{a_{t + 1}}_{\mathrm{spin}}\right)\left(\overleftarrow{\prod_{t = 1}^p}\bra{a_{-t - 1}}_{\mathrm{spin}}U_t\ket{a_{-t}}_{\mathrm{spin}}\right)\bra{a_1}_{\mathrm{spin}}\ket{+}_{\mathrm{spin}}\ket{\overline{0}}_{\mathrm{bosons}},\label{si:eq:spin_boson_path_integral_measure_circuit_insertion_definition}
\end{align}
where similar to a layered unitary circuit's path integral measure (Definition~\ref{si:def:layered_quantum_circuit_path_integral_measure}), bitstring $\bm{a} \in \{1, -1\}^{\mathcal{T}_p}$ represents the spin's computational basis trajectory. We naturally extend the definition of the path integral measure to a collection of independent spin-boson systems (with each spin coupled to its independent collection of modes). Given a finite spin index set $S$, we let
\begin{align}
    \mu_{SB}\left(\bm{z};\,S\right) & := \prod_{j \in S}\mu_{SB}\left(\bm{z}_j\right).
\end{align}
\end{definition}

Even if the spin-boson path integral measure introduced in Definition~\ref{si:def:spin_boson_path_integral_measure} does not fit the unitary circuit framework from Definition~\ref{si:def:layered_quantum_circuit_path_integral_measure}, all general properties of path integral measures discussed in the current paragraph stand by using the circuit partial basis insertion picture from Eq.~\ref{si:eq:spin_boson_path_integral_measure_circuit_insertion_definition}. In particular, for any pair of layer indices $r, s \in \mathcal{T}_p$,
\begin{align}
    \left\langle a_ra_s \right\rangle_{\mu_{SB}\left(\bm{a}\right)} & := \sum_{\bm{a} \in \{1, -1\}^{\mathcal{T}_p}}\mu_{SB}\left(\bm{a}\right)a_ra_s
\end{align}
is the autocorrelation between $Z$ operators inserted at layers $r, s$ (generalization of Proposition~\ref{si:prop:z_autocorrelation_path_integral_representation}). Hence, the argument from Corollary~\ref{si:cor:path_integral_measure_boundedness_monomials} carries over, and any expectation monomials in bits $\bm{a}$ under the spin-boson path integral measure is bounded by $1$. Finally, unitary cancellation is also reflected in the spin-boson path integral measure similar to Proposition~\ref{si:prop:path_integral_measure_unitary_cancellation} in the unitary case: for any multiset $I$ of layer indices bounded by $(q + 1)$ in absolute value,
\begin{align}
    \left\langle \prod_{t \in I}a_t \right\rangle_{\mu^{[p]}_{SB}\left(\bm{a}\right)} & = \left\langle \prod_{t \in I}a_t \right\rangle_{\mu_{SB}^{[r]}}, \qquad \forall q \leq r \leq p, \qquad q + 1 := \max_{t \in I}|t|.
\end{align}

Remarkably, for any matrix $\bm{L} \in \mathbb{C}^{p' \times p}$ parametrizing spin-boson coupling, the spin-boson path integral measure can be expressed as a quadratic action; this representation can be regarded as a discrete form of the Feynman-Vernon influence functional. A proof is provided in Ref.~\cite[Supplementary Material Proposition 1]{qaoa_spin_boson_mapping}; since the result was phrased somehow differently and to be self-contained, we reproduce a proof below:

\begin{proposition}[Expressing path integral measure of spin-boson system {\cite{qaoa_spin_boson_mapping}}]
\label{si:prop:spin_boson_path_integral_measure_expression}
Consider a spin-boson system subject to a discrete unitary evolution as specified in Definition~\ref{si:def:spin_boson_system}, and consider its spin path integral measure $\mu_{SB}\left(\bm{a}\right)$ introduced in Definition~\ref{si:def:spin_boson_path_integral_measure}. Then, $\mu_{SB}\left(\bm{a}\right)$ admits the following expression from a quadratic action:
\begin{align}
    \mu_{SB}\left(\bm{a}\right) & = f\left(\bm{a}\right)\exp\left(-\frac{1}{2}\sum_{t, u \in \mathcal{T}_{p - 1}}\Gamma_t\Gamma_uK_{t, u}a_ta_u\right), \qquad \bm{a} \in \{1, -1\}^{\mathcal{T}_p},
\end{align}
of the computational basis spin trajectory across $p$ layers. In the above formula,
\begin{align}
    f\left(\bm{a}\right) & := \frac{\bm{1}\left[a_{p + 1} = a_{-p - 1}\right]}{2}\prod_{1 \leq t \leq p}\bra{a_t}e^{i\beta_tX}\ket{a_{t + 1}}\bra{a_{-t - 1}}e^{-i\beta_tX}\ket{a_{-t}}
\end{align}
represents the multiplicative contribution of $X$ rotations to the path integral weight. Vector
\begin{align}
    \bm{\Gamma} & := \left(\Gamma_1, \ldots, \Gamma_p, \Gamma_{-p}, \ldots, \Gamma_{-1}\right)\\
    & := \left(\gamma_1, \ldots, \gamma_p, -\gamma_p, \ldots, -\gamma_1\right) \in \mathbb{R}^{\mathcal{T}_{p - 1}}
\end{align}
collects the spin-boson interactions strengths applied in the direct layers $U_t$ and their inverses $U_t^{\dagger}$. Finally, the quadratic action
\begin{align}
    \exp\left(-\frac{1}{2}\sum_{t, u \in \mathcal{T}_{p - 1}}\Gamma_t\Gamma_uK_{t, u}a_ta_u\right)
\end{align}
accounts for effect of the displacement operators after integrating out the bosonic modes. The action's quadratic form is defined from spin-boson coupling matrix $\bm{L}$ as follows. First, define a non-negative Hermitian matrix $\bm{K}^{\mathrm{herm}}$ as the Gram matrix of $\bm{L}$ columns:
\begin{align}
    \bm{K}^{\mathrm{herm}} & := \bm{L}^{\dagger}\bm{L} \in \mathbb{C}^{p \times p}.
\end{align}
From this Hermitian matrix, define a symmetric matrix indexed by $\mathcal{T}_{p - 1} = \{1, \ldots, p\} \sqcup \{-p \ldots, -1\}$ by the following equalities for $1 \leq t, u \leq p$:
\begin{align}
    K_{-t, -u} & := K^{\mathrm{herm}}_{\max\{t, u\}, \min\{t, u\}},\\
    K_{t, -u} & := K^{\mathrm{herm}}_{t, u},\\
    K_{-t, u} & := K^{\mathrm{herm}}_{u, t},\\
    K_{t, u} & := K^{\mathrm{herm}}_{\min\{t, u\}, \max\{t, u\}}.
\end{align}
We call $\bm{K}$ the influence functional of the spin-boson system.
\begin{proof}
We need to evaluate the ``partial matrix elements" (more accurately, the partial traces) in the path integral measure expression given in Definition~\ref{si:def:spin_boson_path_integral_measure}:
\begin{align}
    \mu_{SB}\left(\bm{a}\right) & := \bra{+}_{\mathrm{spin}}\bra{\overline{0}}_{\mathrm{bosons}}\ket{a_1}_{\mathrm{spin}}\left(\overrightarrow{\prod_{t = 1}^p}\bra{a_t}_{\mathrm{spin}}U_t^{\dagger}\ket{a_{t + 1}}_{\mathrm{spin}}\right)\left(\overleftarrow{\prod_{t = 1}^p}\bra{a_{-t - 1}}_{\mathrm{spin}}U_t\ket{a_{-t}}_{\mathrm{spin}}\right)\bra{a_1}_{\mathrm{spin}}\ket{+}_{\mathrm{spin}}\ket{\overline{0}}_{\mathrm{bosons}},\label{si:eq:spin_boson_path_integral_measure_circuit_insertion_definition_repeated}
\end{align}
For the overlaps with the $\ket{+}_{\mathrm{spin}}$ state:
\begin{align}
    \bra{+}_{\mathrm{spin}}\bra{\overline{0}}_{\mathrm{bosons}}\ket{a_1}_{\mathrm{spins}} & = \frac{1}{\sqrt{2}}\bra{\overline{0}}_{\mathrm{bosons}},\\
    \bra{a_{-1}}_{\mathrm{spin}}\ket{+}_{\mathrm{spin}}\ket{\overline{0}}_{\mathrm{bosons}} & = \frac{1}{\sqrt{2}}\ket{\overline{0}}_{\mathrm{bosons}}.
\end{align}
For the spin-boson unitary matrix elements:
\begin{align}
    \bra{a_{-t - 1}}_{\mathrm{spin}}U_t\ket{a_{-t}}_{\mathrm{spin}} & = \bra{a_{-t - 1}}_{\mathrm{spin}}\exp\left(-i\beta_tX\right)D\left(-i\gamma_tZ\bm{L}_{:,\,t}\right)\ket{a_{-t}}_{\mathrm{spin}}\nonumber\\
    & = \bra{a_{-t - 1}}_{\mathrm{spin}}\exp\left(-i\beta_tX\right)\ket{a_{-t}}_{\mathrm{spin}}D\left(-i\gamma_ta_{-t}\bm{L}_{:,\,t}\right)\nonumber\\
    & = \bra{a_{-t - 1}}\exp\left(-i\beta_tX\right)\ket{a_{-t}}D\left(-i\gamma_ta_{-t}\bm{L}_{:,\,t}\right),
\end{align}
and similarly
\begin{align}
    \bra{a_t}_{\mathrm{spin}}U_t^{\dagger}\ket{a_{t + 1}}_{\mathrm{spin}} & = \bra{a_t}\exp\left(i\beta_tX\right)\ket{a_{t + 1}}D\left(i\gamma_ta_t\bm{L}_{:,\,t}\right).
\end{align}
Collecting all matrix elements,
\begin{align}
    \mu_{SB}\left(\bm{a}\right) & = f\left(\bm{a}\right)\bra{\overline{0}}_{\mathrm{bosons}}\overrightarrow{\prod_{t = 1}^p}D\left(i\gamma_ta_t\bm{L}_{:,\,t}\right)\overleftarrow{\prod_{t = 1}^p}D\left(-i\gamma_ta_{-t}\bm{L}_{:,\,t}\right)\ket{\overline{0}}_{\mathrm{bosons}}.\label{si:eq:spin_boson_path_integral_measure_expression_mixer_displacement_decomposition}
\end{align}
We now express the ket and bra from a single displacement operator using formula for products of (multimode) displacement operators:
\begin{align}
    \overleftarrow{\prod_{j = 1}^n}D\left(\bm{\alpha}_j\right) & = \exp\left(\sum_{1 \leq j < k \leq n}i\Im\left\langle \bm{\alpha}_j, \bm{\alpha}_k \right\rangle\right)D\left(\sum_{1 \leq j \leq n}\bm{\alpha}_j\right).
\end{align}
The above formula follows from iterative application of the $n = 2$ multimode formula:
\begin{align}
    D\left(\bm{\alpha}_2\right)D\left(\bm{\alpha}_1\right) & = \exp\left(i\Im\left\langle \bm{\alpha}_1, \bm{\alpha}_2 \right\rangle\right)D\left(\bm{\alpha}_1 + \bm{\alpha}_2\right),
\end{align}
following in turn from the well-known single-mode multiplication formula:
\begin{align}
    D\left(\alpha_2\right)D\left(\alpha_1\right) & = \exp\left(i\Im\left(\overline{\alpha_1}\alpha_2\right)\right)D\left(\alpha_1 + \alpha_2\right).
\end{align}
For the ket:
\begin{align}
    \overleftarrow{\prod_{t = 1}^p}D\left(-i\gamma_ta_{-t}\bm{L}_{:,\,t}\right)\ket{\overline{0}}_{\mathrm{bosons}} & = D\left(\sum_{1 \leq t < u \leq p}i\gamma_t\gamma_ua_{-t}a_{-u}\Im\left\langle \bm{L}_{:,\,t}, \bm{L}_{:,\,u} \right\rangle\right)\ket{\overline{0}}_{\mathrm{bosons}}\nonumber\\
    & = \exp\left(\sum_{1 \leq t < u \leq p}i\gamma_t\gamma_ua_{-t}a_{-u}\Im K^{\mathrm{herm}}_{t, u}\right)D\left(-i\sum_{1 \leq t \leq p}\gamma_ta_{-t}\bm{L}_{:,\,t}\right)\ket{\overline{0}}_{\mathrm{bosons}}\nonumber\\
    & = \exp\left(-\sum_{1 \leq t < u \leq p}i\gamma_t\gamma_ua_{-t}a_{-u}\Im K_{-t, -u}\right)D\left(-i\sum_{1 \leq t \leq p}\gamma_ta_{-t}\bm{L}_{:,\,t}\right)\ket{\overline{0}}_{\mathrm{bosons}}\nonumber\\
    & = \exp\left(-\frac{1}{2}\sum_{1 \leq t, u \leq p}\gamma_t\gamma_ua_{-t}a_{-u}i\Im K_{-t, -u}\right)\left|-i\sum_{1 \leq t \leq p}\gamma_ta_{-t}\bm{L}_{:,\,t}\right\rangle_{\mathrm{bosons}},
\end{align}
where in the final line, we recalled $K_{-t, -t} = \overline{K^{\mathrm{herm}}_{t, t}} \in \mathbb{R}$ to harmlessly include $t = u$ in the sum. Similarly, for the bra:
\begin{align}
    \bra{\overline{0}}_{\mathrm{bosons}}\overrightarrow{\prod_{t = 1}^p}D\left(i\gamma_ta_t\bm{L}_{:,\,t}\right) & = \exp\left(-\frac{1}{2}\sum_{1 \leq t, u \leq p}\gamma_t\gamma_ua_ta_u\Im K_{t, u}\right)\left\langle -i\sum_{1 \leq t \leq p}\gamma_ta_t\bm{L}_{:,\,t} \right|_{\mathrm{bosons}}.
\end{align}
Taking the bra-ket product:
\begin{align}
    & \bra{\overline{0}}_{\mathrm{bosons}}\overrightarrow{\prod_{t = 1}^p}D\left(i\gamma_ta_t\bm{L}_{:,\,t}\right)\overleftarrow{\prod_{t = 1}^p}D\left(-i\gamma_ta_{-t}\bm{L}_{:,\,t}\right)\ket{\overline{0}}_{\mathrm{bosons}}\nonumber\\
    & = \exp\left(-\frac{1}{2}\sum_{1 \leq t, u \leq p}\gamma_t\gamma_ua_{-t}a_{-u}i\Im K_{-t, -u} - \frac{1}{2}\sum_{1 \leq t, u \leq p}\gamma_t\gamma_ua_ta_ui\Im K_{t, u}\right)\left\langle -i\sum_{1 \leq t \leq p}\gamma_ta_t\bm{L}_{:,\,t} \Bigg| -i\sum_{1 \leq t \leq p}\gamma_ta_{-t}\bm{L}_{:,\,t} \right\rangle.\label{si:eq:spin_boson_path_integral_measure_expression_vacuum_expectation_step_1}
\end{align}
The dot product of coherent states evaluates to:
\begin{align}
    & \left\langle -i\sum_{1 \leq t \leq p}\gamma_ta_t\bm{L}_{:,\,t} \Bigg| -i\sum_{1 \leq t \leq p}\gamma_ta_{-t}\bm{L}_{:,\,t} \right\rangle\nonumber\\
    & = \exp\left(-\frac{1}{2}\left\lVert -i\sum_{1 \leq t \leq p}\gamma_ta_t\bm{L}_{:,\,t} \right\rVert_2^2 - \frac{1}{2}\left\lVert - i\sum_{1 \leq t \leq p}\gamma_ta_{-t}\bm{L}_{:,\,t} \right\rVert_2^2 + \left\langle - i\sum_{1 \leq t \leq p}\gamma_ta_t\bm{L}_{:,\,t}, -i\sum_{1 \leq t \leq p}\gamma_ta_{-t}\bm{L}_{:,\,t} \right\rangle\right)\nonumber\\
    & = \exp\left(-\frac{1}{2}\sum_{1 \leq t, u \leq p}\gamma_t\gamma_ua_ta_u\left\langle \bm{L}_{:,\,t}, \bm{L}_{:,\,u} \right\rangle - \frac{1}{2}\sum_{1 \leq t, u \leq p}\gamma_t\gamma_ua_{-t}a_{-u}\left\langle \bm{L}_{:,\,t}, \bm{L}_{:,\,u} \right\rangle + \sum_{1 \leq t, u \leq p}\gamma_t\gamma_ua_ta_{-u}\left\langle \bm{L}_{:,\,t}, \bm{L}_{:,\,u} \right\rangle\right)\nonumber\\
    & = \exp\left(-\frac{1}{2}\sum_{1 \leq t, u \leq p}\gamma_t\gamma_ua_ta_u  K^{\mathrm{herm}}_{t, u} - \frac{1}{2}\sum_{1 \leq t, u \leq p}\gamma_t\gamma_ua_{-t}a_{-u}K^{\mathrm{herm}}_{t, u} + \sum_{1 \leq t, u \leq p}\gamma_t\gamma_ua_ta_{-u}K^{\mathrm{herm}}_{t, u}\right)
\end{align}
The first quadratic form of $\bm{a}$ in the above exponential can be simplified by noting
\begin{align}
    -\frac{1}{2}\sum_{1 \leq t, u \leq p}\gamma_t\gamma_ua_ta_u K^{\mathrm{herm}}_{t, u} & = -\frac{1}{2}\sum_{1 \leq t, u \leq p}\gamma_t\gamma_ua_ta_u\frac{K^{\mathrm{herm}}_{t, u} + K^{\mathrm{herm}}_{u, t}}{2}\nonumber\\
    & = -\frac{1}{2}\sum_{1 \leq t, u \leq p}\gamma_t\gamma_ua_ta_u\Re K^{\mathrm{herm}}_{t, u}\nonumber\\
    & =  -\frac{1}{2}\sum_{1 \leq t, u \leq p}\gamma_t\gamma_ua_ta_u\Re K_{-t, -u}.
\end{align}
The second quadratic form of $\bm{a}$ similarly simplifies:
\begin{align}
    -\frac{1}{2}\sum_{1 \leq t, u \leq p}\gamma_t\gamma_ua_{-t}a_{-u}K^{\mathrm{herm}}_{t, u} & = -\frac{1}{2}\sum_{1 \leq t, u \leq p}\gamma_t\gamma_ua_{-t}a_{-u}\Re K_{t, u}.
\end{align}
The third quadratic form can be expressed as follows in terms of $\bm{K}$:
\begin{align}
    \sum_{1 \leq t, u \leq p}\gamma_t\gamma_ua_ta_{-u}K^{\mathrm{herm}}_{t, u} & = \sum_{1 \leq t, u \leq p}\gamma_t\gamma_ua_ta_{-u}K_{t, -u}.
\end{align}
All in all, the dot product of coherent states can be expressed as:
\begin{align}
    & \left\langle -i\sum_{1 \leq t \leq p}\gamma_ta_t\bm{L}_{:,\,t} \Bigg| -i\sum_{1 \leq t \leq p}\gamma_ta_{-t}\bm{L}_{:,\,t}\right\rangle\nonumber\\
    & = \exp\left(-\frac{1}{2}\sum_{1 \leq t, u \leq p}\Re K_{t, u}\gamma_t\gamma_ua_ta_u - \frac{1}{2}\sum_{1 \leq t, u \leq p}\Re K_{-t, -u}\gamma_t\gamma_ua_{-t}a_{-u} + \sum_{1 \leq t, u \leq p}K_{t, -u}\gamma_t\gamma_ua_ta_{-u}\right).
\end{align}
Plugging this into Eq.~\ref{si:eq:spin_boson_path_integral_measure_expression_vacuum_expectation_step_1}, combining real and imaginary parts of $\bm{K}$ yields:
\begin{align}
    & \bra{\overline{0}}_{\mathrm{bosons}}\overrightarrow{\prod_{t = 1}^p}D\left(i\gamma_ta_t\bm{L}_{:,\,t}\right)\overleftarrow{\prod_{t = 1}^p}D\left(-i\gamma_ta_{-t}\bm{L}_{:,\,t}\right)\ket{\overline{0}}_{\mathrm{bosons}}\nonumber\\
    & = \exp\left(-\frac{1}{2}\sum_{1 \leq t, u \leq p}K_{t, u}\gamma_t\gamma_ua_ta_u - \frac{1}{2}\sum_{1 \leq t, u \leq p}K_{-t, -u}\gamma_t\gamma_ua_{-t}a_{-u} + \sum_{1 \leq t, u \leq p}K_{t, -u}\gamma_t\gamma_ua_ta_{-u}\right)\nonumber\\
    & = \exp\left(-\frac{1}{2}\sum_{t, u \in \mathcal{T}_{p - 1}}K_{t, u}\Gamma_t\Gamma_ua_ta_u\right).
\end{align}
Plugging this vacuum expectation into Eq.~\ref{si:eq:spin_boson_path_integral_measure_expression_mixer_displacement_decomposition} gives the claimed expression for the path integral measure:
\begin{align}
    \mu_{SB}\left(\bm{a}\right) & = f\left(\bm{a}\right)\exp\left(-\frac{1}{2}\sum_{t, u \in \mathcal{T}_{p - 1}}K_{t, u}\Gamma_t\Gamma_ua_ta_u\right).
\end{align}
\end{proof}
\end{proposition}

\subsubsection{Spin-boson mapping of the spin-glass QAOA $\mathbf{G}$ iteration}
\label{si:sec:spin_boson_mapping_qaoa_iteration}

We now explain how the $\bm{G}$ iteration Eq.~\ref{si:eq:g_iteration} can be evaluated from computing observables over spin-boson states. More specifically, we discuss two approaches of extracting $\bm{G}$ matrix entries from spin-boson states. The first method is a simple generalization Ref.~\cite{qaoa_spin_boson_mapping}, using $p$ spin-boson states. The second novel method uses a single spin-boson states but measures $p$ (fixed-time) observables.

The spin-boson evaluation of the spin-glass QAOA iteration Eq.~\ref{si:eq:g_iteration_restricted}:
\begin{align}
    G^{(m + 1)}_{r, s} & := \sum_{\bm{a} \in \{1, -1\}^{\mathcal{T}_q}}a_ra_sf\left(\bm{a}\right)\exp\left(-\frac{1}{2}\sum_{t, u \in \mathcal{T}_{q - 1}}\xi'\left(G^{(m)}_{t, u}\right)\Gamma_t\Gamma_ua_ta_u\right), \qquad q + 1 := \max\left\{|r|, |s|\right\}.
\end{align}
recognizes the path integral weight of a spin-boson system in the $f$-$\exp$ product. Specifically, we claim that
\begin{align}
    \bm{K} & := \left(K_{t, u}\right)_{t, u \in \mathcal{T}_{q - 1}} := \left(\xi'\left(G_{t, u}\right)\right)_{t, u \in \mathcal{T}_{q - 1}}
\end{align}
is a valid influence functional as specified in Proposition~\ref{si:prop:spin_boson_path_integral_measure_expression}. The required symmetries $\bm{K}$ are satisfied given the $\bm{G}$ symmetries (Proposition~\ref{si:prop:g_symmetries}) and since $\xi$ is a polynomial with real coefficients. The Hermitian corner:
\begin{align}
    \bm{K}^{\mathrm{herm}}_{t, u} & := K_{u, -t}
\end{align}
is also non-negative, which follows from non-negativity of $\bm{G}^{\mathrm{herm}}$,
\begin{align}
    G^{\mathrm{herm}}_{t, u} & := G_{t, -u},
\end{align}
and the following Lemma:

\begin{lemma}[Non-negative matrices are stable under elementwise application of non-negative coefficients polynomials]
Consider $\bm{M} \in \mathbb{C}^{n \times n}$ a non-negative Hermitian matrix. Then, for polynomial $P$ with non-negative coefficients, the matrix
\begin{align}
     \bm{P}^{\odot}\left(\bm{M}\right) & := \left(P(M_{j, k})\right)_{j, k \in [n]},
\end{align}
defined by elementwise application of $\bm{P}$ to $\bm{M}$, is non-negative:
\begin{align}
     \bm{P}^{\odot}\left(\bm{M}\right) & \succeq 0.
\end{align}
\begin{proof}
Since non-negative matrices are stable under non-negative scalar multiplication and sums, it suffices to prove the result for each monomial $P\left(x\right) = x^d$, $d \geq 0$. $\bm{M}$ can be decomposed as a linear combination of rank $1$ projectors (spectral theorem) with non-negative coefficients (non-negativity assumption), i.e.:
\begin{align}
    \bm{M} & = \sum_{\alpha \in [n]}\lambda_{\alpha}\bm{v}^{\left(\alpha\right)}\bm{v}^{\left(\alpha\right)\dagger}, \qquad \lambda_{\alpha} \geq 0, \bm{v}_{\alpha} \in \mathbb{C}^n \quad \forall \alpha \in [n].
\end{align}
In coordinates notation, the above decomposition reads:
\begin{align}
    M_{j, k} & = \sum_{\alpha \in [n]}\lambda_{\alpha}v^{\left(\alpha\right)}_{j}\overline{v^{\left(\alpha\right)}_k}.
\end{align}
Applying $P(x) = x^d$ element-wise then gives:
\begin{align}
    P\left(M_{j, k}\right) & = M_{j, k}^d\nonumber\\
    & = \sum_{\alpha_1, \ldots, \alpha_d \in [n]}\lambda_{\alpha_1} \ldots \lambda_{\alpha_d}v^{\left(\alpha_1\right)}_j \ldots v^{\left(\alpha_d\right)}_j \overline{v^{\left(\alpha_1\right)}_k} \ldots \overline{v^{\left(\alpha_d\right)}_k}\nonumber\\
    & = \sum_{\bm{\alpha} \in [n]^d}\mu_{\bm{\alpha}}w^{\left(\bm{\alpha}\right)}_j\overline{w^{\left(\bm{\alpha}\right)}_k},
\end{align}
where we defined a family of vectors and non-negative reals indexed by $d$-tuples of dimension indices $\bm{\alpha} = \left(\alpha_1, \ldots, \alpha_d\right) \in [n]^d$:
\begin{align}
    \mu_{\bm{\alpha}} & := \lambda_{\alpha_1} \ldots \lambda_{\alpha_d},\\
    w^{\left(\bm{\alpha}\right)}_j & := v^{\left(\alpha_1\right)}_j \ldots v^{\left(\alpha_d\right)}_j.
\end{align}
The above writing of $\bm{P}^{\odot}\left(\bm{M}\right)$ is manifestly a linear combination of rank-$1$ projectors with non-negative coefficients. Hence $\bm{P}^{\odot}\left(\bm{M}\right)$ is Hermitian non-negative.
\end{proof}
\end{lemma}

All in all, a unitary spin-boson system can realize influence functional $\bm{K} = \left(\xi'\left(G_{t, u}\right)\right)_{t, u \in \mathcal{T}_{q - 1}}$. The construction of evolution operators requires a choice of factorization $\bm{L}$ of the Hermitian corner of $\bm{K}$:
\begin{align}
    \bm{K}^{\mathrm{herm}} & = \bm{L}^{\dagger}\bm{L}.
\end{align}
Similar to earlier work Ref.~\cite{qaoa_spin_boson_mapping}, we choose an upper-triangular Choleskly factorization to enable an incremental construction of states as $\bm{G}$ entries of increasing layer index are computed. The following Proposition recalls the procedure used in this earlier study to extract $\bm{G}$ entries from dot products between $p$ states:

\begin{proposition}[$\mathbf{G}$ entries from Gram matrix of $p$ states {\cite[Proposition 1]{qaoa_spin_boson_mapping}}]
\label{si:prop:g_iteration_spin_boson_multi_states}
Let $1 \leq q \leq p$ an intermediate QAOA depth index. Assume entries of absolute values up to $q$ of $\bm{G}^{(q)}$ have already been computed; by the stationarity property of the $\bm{G}$ iteration (Proposition~\ref{si:prop:g_symmetries}), these are the same as the entries of absolute index at most $q$ of the final iterate $\bm{G} = \bm{G}^{(p)}$. Let $\bm{\widetilde{L}}^{(q)} \in \mathbb{C}^{q \times q}$ factorize the $q \times q$ hermitian corner of $\xi'^{\odot}\left(\bm{G}^{(q)}\right)$, i.e.:
\begin{align}
    \bm{\widetilde{L}}^{(q)\dagger}\bm{\widetilde{L}}^{(q)} & = \xi'^{\odot}\left(\bm{G}^{\mathrm{herm}}_{1:q,\,1:q}\right),\\
    G^{\mathrm{herm}}_{t, u} & := G_{t, -u}.
\end{align}
Consider the $q$-layer spin-boson evolution given by unitaries
\begin{align}
    U^{(q)}_t & := \exp\left(-i\beta_tX\right)D\left(-i\gamma_tZ\bm{\widetilde{L}}^{(q)}_{:,\,t}\right),
\end{align}
preparing state
\begin{align}
    \ket{\Psi^{(q)}} & := \overleftarrow{\prod_{t = 1}^q}U^{(q)}_t\ket{+}_{\mathrm{spin}}\ket{\overline{0}}_{\mathrm{bosons}}
\end{align}
Finally, consider the family of $(q + 1)$ states $\ket{\Psi^{(q)}_r}$, $1 \leq r \leq q + 1$, with a single $Z$ insertion inside this dynamics:
\begin{align}
    \ket{\Psi^{(q)}_r} & := \left(\overleftarrow{\prod_{t = r}^q}U^{(q)}_t\right)Z\left(\overleftarrow{\prod_{t = 1}^{r - 1}}U^{(q)}_t\right)\ket{+}_{\mathrm{spin}}\ket{\overline{0}}_{\mathrm{bosons}}.
\end{align}
Then, any $\bm{G}$ Hermitian corner entry of index at most $(q + 1)$ is given by dot product:
\begin{align}
    G^{\mathrm{herm}}_{r, s} & = \sum_{\bm{a} \in \{1, -1\}^{\mathcal{T}_q}}a_ra_{-s}f\left(\bm{a}\right)\exp\left(-\frac{1}{2}\sum_{t, u \in \mathcal{T}_{q - 1}}\xi'\left(G_{t, u}\right)\Gamma_t\Gamma_ua_ta_u\right)\nonumber\\
    & = \braket{\Psi^{(q)}_r|\Psi^{(q)}_s}.
\end{align}
\end{proposition}
The construction of states given in Proposition~\ref{si:prop:g_iteration_spin_boson_multi_states} illustrates the interest of the upper-triangular factorization of $\bm{K}^{\mathrm{herm}} := \xi'^{\odot}\left(\bm{G}^{\mathrm{herm}}\right)$. Namely, for this choice of factorization, restriction to the top-left corner commutes with factorization, i.e. the factorization of the top-left corner of $\bm{K}^{\mathrm{herm}}$ is the top-left corner of the factorization of $\bm{K}^{\mathrm{herm}}$:
\begin{align}
    \bm{\widetilde{L}}^{(q + 1)}_{1:q,\,1:q} & = \bm{\widetilde{L}}^{(q)}.
\end{align}
This property in turn allows an iterative construction of the states defined in the Proposition:
\begin{align}
    \ket{\Psi^{(q + 1)}} & = U^{(q + 1)}_{q + 1}\ket{\Psi^{(q)}},\label{si:eq:ut_factorization_iterative_state_update_1}\\
    \ket{\Psi^{(q + 1)}_r} & = U^{(q + 1)}_{q + 1}\ket{\Psi_r^{(q)}}, \qquad 1 \leq r \leq q,\label{si:eq:ut_factorization_iterative_state_update_2}\\
    \ket{\Psi^{(q + 1)}_{q + 1}} & = ZU^{(q + 1)}_{q + 1}\ket{\Psi^{(q)}}.\label{si:eq:ut_factorization_iterative_state_update_3}
\end{align}
We now introduce a novel method for evaluating $\bm{G}$ matrix entries, requiring manipulating of a single rather than $p$ states. The idea is to exploit $p$ fixed-time observables involving the spin and bosons, instead of $p$ time autocorrelations involving the spin only. The procedure is explained in the following Proposition:

\begin{proposition}[$\mathbf{G}$ matrix entries from fixed-time observables of a single state]
\label{si:prop:g_iteration_spin_boson_single_state}
Let $1 \leq q \leq p$ an intermediate QAOA depth index. Assume entries of absolute values up to $q$ of $\bm{G}^{(q)}$ have already been computed; by the stationarity property of the $\bm{G}$ iteration (Proposition~\ref{si:prop:g_symmetries}), these are the same as the entries of absolute index at most $q$ of the final iterate $\bm{G} = \bm{G}^{(p)}$. Let $\bm{\widetilde{L}}^{(q)} \in \mathbb{C}^{q \times q}$ factorize the top $q \times q$ hermitian corner of $\xi'^{\odot}\left(\bm{G}^{(q)}\right)$, i.e.:
\begin{align}
    \bm{\widetilde{L}}^{(q)\dagger}\bm{\widetilde{L}}^{(q)} & = \xi'^{\odot}\left(\bm{G}^{\mathrm{herm}}_{1:q,\,1:q}\right),\\
    G^{\mathrm{herm}}_{t, u} & := G_{t, -u}.
\end{align}
Consider the $q$-layer spin-boson evolution given by unitaries
\begin{align}
    U^{(q)}_t & := \exp\left(-i\beta_tX\right)D\left(-i\gamma_tZ\bm{\widetilde{L}}^{(q)}_{:,\,t}\right),
\end{align}
preparing state
\begin{align}
    \ket{\Psi^{(q)}} & := \overleftarrow{\prod_{t = 1}^q}U^{(q)}_t\ket{+}_{\mathrm{spin}}\ket{\overline{0}}_{\mathrm{bosons}}.
\end{align}
Then, the entries of $\bm{G}^{\mathrm{herm}}$ of row index $q + 1$ and column indices $[q]$ are given by:
\begin{align}
    \bm{G}^{\mathrm{herm}}_{q + 1,\,1:q} & = \left(-i\bm{\widetilde{L}}^{(q)}\mathrm{diag}\left(\bm{\gamma}_{1:q}\right)\right)^{-1}\bra{\Psi^{(q)}}Z\bm{\hat{c}}_{1:q}\ket{\Psi^{(q)}}.
\end{align}
In the above equation, $\mathrm{diag}\left(\bm{\gamma}_{1:q}\right)$ is the diagonal matrix of the first $q$ dephasing angles and $\bra{\Psi^{(q)}}Z\bm{\hat{c}}_{1:q}\ket{\Psi^{(q)}}$ is the $q$-dimensional vector of expectations $\bra{\Psi^{(q)}}Z\hat{c}_l\ket{\Psi^{(q)}}$, $l \in [q]$, where $\hat{c}_l$ is the annihilation operator for mode $l$. Matrix invertibility is assumed.
\begin{proof}
This results from an explicit computation of the spin-boson observables as a path integral. By repeating the calculation of the path integral measure in the proof of Proposition~\ref{si:prop:spin_boson_path_integral_measure_expression}, but with an additional $\hat{c}_l$ inserted at the end of the evolution, and recalling coherent states are eigenstates of the annihilation operator: %
\begin{align}
    \hat{c}\ket{\alpha} & = \alpha\ket{\alpha},
\end{align}
we obtain:
\begin{align}
    \bra{\Psi^{(q)}}Z\hat{c}_l\ket{\Psi^{(q)}} & = \sum_{\bm{a} \in \{1, -1\}^{\mathcal{T}_q}}a_{q + 1}\left(-i\sum_{1 \leq t \leq q}\widetilde{L}^{(q)}_{l, t}\gamma_ta_{-t}\right)f\left(\bm{a}\right)\exp\left(-\frac{1}{2}\sum_{t, u \in \mathcal{T}_{q - 1}}\xi'\left(G_{t, u}\right)\Gamma_t\Gamma_ua_ta_u\right)\nonumber\\
    & = -i\sum_{1 \leq t \leq q}\widetilde{L}^{(q)}_{l, t}\gamma_t\sum_{\bm{a} \in \{1, -1\}^{\mathcal{T}_q}}a_{q + 1}a_{-t}f\left(\bm{a}\right)\exp\left(-\frac{1}{2}\sum_{t, u \in \mathcal{T}_{q - 1}}\xi'\left(G_{t, u}\right)\Gamma_t\Gamma_ua_ta_u\right)\nonumber\\
    & = -i\sum_{1 \leq t \leq q}\widetilde{L}^{(q)}_{l, t}\gamma_tG^{\mathrm{herm}}_{q + 1, t}\nonumber\\
    & = \left[\left(-i\bm{\widetilde{L}}^{(q)}\mathrm{diag}\left(\bm{\gamma}_{1:q}\right)\right)\bm{G}^{\mathrm{herm}}_{q + 1,\,1:q}\right]_l.
\end{align}
Inverting this linear relation between spin-boson observables and $\bm{G}$ matrix entries gives the result.
\end{proof}
\end{proposition}
Compared to Proposition~\ref{si:prop:g_iteration_spin_boson_multi_states} requiring $(q + 1)$ states for computing entries of $\bm{G}$ of maximum absolute index $(q + 1)$, Proposition~\ref{si:prop:g_iteration_spin_boson_single_state} requires a single state over which $q$ measurements are performed. Using an upper triangular factorization for the Hermitian corner of the influence functional, states can again be prepared iteratively.

\subsection{Analysis of QAOA via the cavity method: basic results and heuristics}
\label{si:sec:qaoa_cavity_method_analysis_heuristic}

In this Section, we provide a heuristic sketch of the cavity calculation analysis of spin-glass QAOA thoroughly developed in Section~\ref{si:sec:qaoa_cavity_method_analysis}. Section~\ref{si:sec:path_integral_representation_spin_glass_qaoa_expectations} opens by expressing QAOA observables as quantum circuit path integrals, as introduced in Section~\ref{si:sec:unitary_circuit_path_integrals_and_spin_boson_systems}. The results are elementary and mere reframings of previous results derived in Refs.~\cite{Farhi2022quantumapproximate,qaoa_sparse_hypergraphs_spin_glass_models}, but rederived in full to ensure compatibility with the current study's conventions. Based on these initial results, Section~\ref{si:sec:heuristic_cavity_argument} provides a heuristic account of the cavity argument enabling an alternative derivation of the $\bm{G}$ iteration introduced in Section~\ref{si:sec:spin_glass_qaoa_g_iteration}. Besides recovering the iteration, the heuristic argument provides insight into the many-body structure of the spin-glass QAOA state: namely, any constant-size set of spins approximately behaves as a set of independent spins, each coupled to its own $p$ bosonic modes. More accurately, the heuristic account shows self-consistency of this approximate independence assumption. Section~\ref{si:sec:heuristic_cavity_argument_p1} specializes the argument to the $p = 1$ case, where approximate independence need not be assumed self-consistently but can be derived from first principles. The analysis of the $p = 1$ setting suggests converting the self-consistent analysis from Section~\ref{si:sec:heuristic_cavity_argument} into an induction argument, ultimately unfolded in the rigorous version of the cavity calculation detailed in Section~\ref{si:sec:qaoa_cavity_method_analysis}.

\subsubsection{The path integral representation of spin-glass QAOA expectations}
\label{si:sec:path_integral_representation_spin_glass_qaoa_expectations}

In this Section, we introduce notation used in this work for the representation of QAOA observables as spin path integrals. The results and notation are largely identical to earlier works Refs.~\cite{Farhi2022quantumapproximate,basso_et_al:LIPIcs.TQC.2022.7,qaoa_sparse_hypergraphs_spin_glass_models}. An important difference is that unlike Refs.~\cite{Farhi2022quantumapproximate,qaoa_sparse_hypergraphs_spin_glass_models}, we do not group the exponential number of spin configurations into a polynomial number of configuration basis numbers. Instead, we express all quantities in terms of overlaps between spin configuration, which will suggest a different analysis of the infinite-size limit.

We now derive a path integral representation of the energy density of spin-glass QAOA at arbitrary finite size. This expression was already derived in a different form in prior work Ref.~\cite{qaoa_sparse_hypergraphs_spin_glass_models}. In this study, following original work Ref.~\cite{Farhi2022quantumapproximate}, the author introduced a \textit{configuration basis} to take advantage of the permutation symmetry (after disorder-averaging) of the spin glasses, allowing them to reduce the summation over $2^{n(p + 1)}$ bits ($\bm{z} \in \{1, -1\}^{n \times \mathcal{T}_p}$) to one over $2^{2p + 2}$ integer variables ranging in $[0, n]$. In this alternative analysis, we remain with sums over bit matrices (or equivalently, expectations under path integral measures) and express all relevant quantities in terms of \textit{normalized overlaps} between spin configurations. The overlap between spin configuration is a central object in classical spin glass theory, with definition recalled below:

\begin{definition}[Overlap between spin glass configurations]
\label{si:def:spin_configurations_overlap}
Given two configurations $\bm{\sigma} = \left(\sigma_j\right)_{j \in S} \in \{1, -1\}^{S}$ and $\bm{\tau} = \left(\tau_j\right)_{j \in S} \in \{1, -1\}^S$ of spins indexed by a finite set $S$, the normalized overlap between spin configurations $\bm{\sigma}, \bm{\tau}$ is defined by:
\begin{align}
    R\left(\bm{\sigma}, \bm{\tau}\right) & := \frac{1}{n}\left\langle \bm{\sigma}, \bm{\tau} \right\rangle\\
    & = \frac{1}{n}\sum_{j \in S}\sigma_j\tau_j,
\end{align}
where in this context, the bracket will commonly be used to denote the Euclidean dot product.
\end{definition}

With Definition~\ref{si:def:spin_configurations_overlap} of spin configurations overlap in mind, we are ready to express the disorder-averaged energy density of spin-glass QAOA in terms of normalized overlaps. The spin-glass QAOA cost function, depending on Gaussian disorder $\bm{J} = \left(\bm{J}^{(q)}\right)_{q \geq 1}$, $\bm{J}^{(q)} = \left(J^{(q)}_{j_1, \ldots, j_q}\right)_{j_1, \ldots, j_q \in S}$, $J^{(q)}_{j_1, \ldots, j_q} \sim \mathcal{N}(0, 1)$, is defined by:
\begin{align}
    C & := \sum_{q \geq 1}c_q\sum_{j_1, \ldots, j_q \in S}|S|^{-(q - 1)/2}J^{(q)}_{j_1, \ldots, j_q}Z_{j_1} \ldots Z_{j_q},\label{si:eq:spin_glass_qaoa_cost_function}
\end{align}
where we labelled the set of spins by $S$. The mixer Hamiltonian is the standard transverse field mixer:
\begin{align}
    B & = \sum_{j \in S}X_j.
\end{align}
The corresponding $p$-layer QAOA circuit, defined by unitary layers
\begin{align}
    U_t & := \exp\left(-i\beta_tB\right)\exp\left(-i\gamma_tC\right),
\end{align}
and initial state
\begin{align}
    \ket{\psi} & = \ket{+}^{\otimes n},
\end{align}
has a random path integral measure $\mu^{[p]}\left(\bm{z};\,\xi, S, \bm{J}\right)$. In the latter notation:
\begin{align}
    \xi\left(x\right) & := \sum_{q \geq 1}c_q^2x^q,\label{si:eq:spin_glass_mixture_polynomial}
\end{align}
known as the \textit{spin-glass mixture polynomial}, encodes the coefficients of the mixed spin glass. By Proposition~\ref{si:prop:z_autocorrelation_path_integral_representation}, the energy density of a specific random instance of spin-glass QAOA can be expressed as:
\begin{align}
    \bra{\bm{\gamma}, \bm{\beta}}C\ket{\bm{\gamma}, \bm{\beta}} & = \left\langle C\left(\bm{z}^{[p + 1]}\right) \right\rangle_{\mu^{[p]}\left(\bm{z};\,\xi, S, \bm{J}\right)}\nonumber\\
    & = \sum_{\bm{z} \in \{1, -1\}^{S \times \mathcal{T}_p}}\mu^{[p]}\left(\bm{z};\,\xi, S, \bm{J}\right)C\left(\bm{z}^{[p + 1]}\right).
\end{align}
However, we are ultimately interested in the disorder-averaged energy density
\begin{align}
    \mathbb{E}\bra{\bm{\gamma}, \bm{\beta}}C\ket{\bm{\gamma}, \bm{\beta}} & = \sum_{\bm{z} \in \{1, -1\}^{S \times \mathcal{T}_p}}\mathbb{E}\left[\mu^{[p]}\left(\bm{z};\,\xi, S, \bm{J}\right)C\left(\bm{z}^{[p + 1]}\right)\right],
\end{align}
where in the right-hand side, the disorder average also encompasses $C$ due to the $\bm{J}$ dependence of the random cost. We start by writing an explicit form of the instance-wise, disorder-dependent path integral measure:

\begin{proposition}[Spin-glass QAOA instance-wise path integral measure]
\label{si:prop:path_integral_measure}
Let $S$ a set of qubit indices, and consider the $p$-layers QAOA applied to this qubit set, with cost function:
\begin{align}
    C & := \sum_{q \geq 1}c_q\sum_{j_1, \ldots, j_q \in S}\left|S\right|^{-(q - 1)/2}J^{(q)}_{j_1, \ldots, j_q}Z_{j_1} \ldots Z_{j_q},\\
    J^{(q)}_{j_1, \ldots, j_q} & \overset{\mathrm{i.i.d}}{\sim} \mathcal{N}(0, 1).
\end{align}
The quantum circuit path integral measure (averaged over random Gaussian disorder) is given by:
\begin{align}
    \mu^{[p]}\left(\bm{z}_S;\,\xi, S, \bm{J}\right) & = \mu^{[p]}_C\left(\bm{z}_S;\,\xi, S, \bm{J}\right)\mu^{[p]}_B\left(\bm{z}_S;\,S\right),\label{si:eq:path_integral_measure}
\end{align}
where
\begin{align}
    \mu^{[p]}_C\left(\bm{z}_S;\,\xi, S, \bm{J}\right) & := \exp\left(\sum_{t \in \mathcal{T}_{p - 1}}i\Gamma_tC\left(\bm{z}^{[t]}\right)\right)\\
    & = \exp\left(\sum_{t \in \mathcal{T}_{p - 1}}i\Gamma_t\sum_{q \geq 1}c_q\sum_{j_1, \ldots, j_q \in S}|S|^{-(q - 1)/2}J^{(q)}_{j_1, \ldots, j_q}z^{[t]}_{j_1} \ldots z^{[t]}_{j_q}\right),\label{si:eq:path_integral_measure_cost_contribution}\\
    \left(\Gamma_1, \ldots, \Gamma_p, \Gamma_{-p}, \ldots, \Gamma_{-1}\right) & := \left(\gamma_1, \ldots, \gamma_p, -\gamma_p, \ldots, -\gamma_1\right)
\end{align}
reflects the contribution of cost unitaries, and
\begin{align}
    \mu^{[p]}_B\left(\bm{z}_S;\,S\right) & := \prod_{j \in S}f\left(\bm{z}_j\right),\label{si:eq:path_integral_measure_mixer_contribution}\\
    f\left(\bm{a}\right) & := \frac{1}{2}\mathbf{1}\left[a_{p + 1} = a_{-p - 1}\right]\prod_{1 \leq t \leq p}\bra{a_t}e^{i\beta_tX}\ket{a_{t + 1}}\bra{a_{-t - 1}}e^{-i\beta_tX}\ket{a_{-t}}
\end{align}
reflects the contribution of mixer unitaries. %
\begin{proof}
By Definition~\ref{si:def:layered_quantum_circuit_path_integral_measure} of the path integral measure for a layered quantum circuit defined by unitaries $U_1, \ldots, U_p$, the quantum circuit path integral measure for a given realization of the disorder $\bm{J} = \left(\bm{J}^{(q)}\right)_{q \geq 1}$ is given by:
\begin{align}
    \mu^{[p]}\left(\bm{z};\,\xi, S, \bm{J}\right) & = \bra{+}^{\otimes n}\ket{\bm{z}_1}\prod_{t = 1}^p\bra{\bm{z}^{[t]}}U_t^{\dagger}\ket{\bm{z}^{[t + 1]}}\prod_{t = 1}^p\bra{\bm{z}^{[-t - 1]}}U_t\ket{\bm{z}^{[-t]}}\bra{\bm{z}^{[-1]}}\ket{+}^{\otimes n},\label{si:eq:qaoa_instance_wise_path_integral_measure}\\
    U_t & := \exp\left(-i\beta_tB\right)\exp\left(-i\gamma_tC\right),
\end{align}
where the cost Hamiltonian depends on disorder $\bm{J}$ and $\bm{z} = = \left(z^{[t]}_j\right)_{j \in S,\,t \in S} = \left(\bm{z}^{[t]}\right)_{t \in \mathcal{T}_p} = \left(\bm{z}_k\right)_{j \in S} \in \{1, -1\}^{S \times \mathcal{T}_p}$ is a bit matrix.Reproducing Refs.~\cite{basso_et_al:LIPIcs.TQC.2022.7,Farhi2022quantumapproximate,qaoa_sparse_hypergraphs_spin_glass_models} and borrowing their notation, we evaluate the matrix elements from Eq.~\ref{si:eq:qaoa_instance_wise_path_integral_measure}:
\begin{align}
    \bra{+}^{\otimes n}\ket{\bm{z}^{[1]}} & = \frac{1}{2^n},\\
    \bra{\bm{z}^{[-1]}}\ket{+}^{\otimes n} & = \frac{1}{2^n},\\
    \bra{\bm{z}^{[t]}}U_t^{\dagger}\ket{\bm{z}^{[t + 1]}} & = \bra{\bm{z}^{[t]}}U_t^{\dagger}\ket{\bm{z}^{[t + 1]}}\nonumber\\
    & = \bra{\bm{z}^{[t]}}\exp\left(i\gamma_t C\right)\exp\left(i\beta_t B\right)\ket{\bm{z}^{[t + 1]}}\nonumber\\
    & = \exp\left(i\gamma_tC\left(\bm{z}^{[t]}\right)\right)\bra{\bm{z}^{[t]}}\exp\left(i\beta_tB\right)\ket{\bm{z}^{[t + 1]}}\nonumber\\
    & = \exp\left(i\gamma_t C\left(\bm{z}^{[t]}\right)\right)\prod_{j \in S}\bra{z_j^{[t]}}\exp\left(i\beta_tX\right)\ket{z_j^{[t + 1]}},\\
    \bra{\bm{z}^{[-t - 1]}}U_t\ket{\bm{z}^{[-t]}} & = \bra{\bm{z}^{[-t - 1]}}\exp\left(-i\beta_tB\right)\exp\left(-i\gamma_tC\right)\ket{\bm{z}^{[-t]}}\nonumber\\
    & = \exp\left(-i\gamma_tC\left(\bm{z}^{[-t]}\right)\right)\prod_{j \in S}\bra{z_j^{[-t - 1]}}\exp\left(-i\beta_tX\right)\ket{z_j^{[-t]}}.
\end{align}
Putting all matrix elements together,
\begin{align}
    \mu^{[p]}\left(\bm{z};\,\xi, S, \bm{J}\right) & = \mu_C^{[p]}\left(\bm{z};\,\xi, S, \bm{J}\right)\mu^{[p]}_B\left(\bm{z};\,S\right),
\end{align}
where we collected the multiplicative contribution from the cost function into $\mu_C^{[p]}\left(\bm{z};\,\xi, S, \bm{J}\right)$, and the multiplicative contribution from the mixers into $\mu_B^{[p]}\left(\bm{z};\,\xi, S, \bm{J}\right)$. The multiplicative contribution from the cost function is explicitly given by
\begin{align}
    \mu_C^{[p]}\left(\bm{z};\,\xi, S, \bm{J}\right) & := \exp\left(\sum_{t \in \mathcal{T}_{p - 1}}i\Gamma_tC\left(\bm{z}^{[t]}\right)\right)\\
    & = \exp\left(\sum_{q \geq 1}c_q\sum_{j_1, \ldots, j_q \in S}\left|S\right|^{-(q - 1)/2}J^{(q)}_{j_1, \ldots, j_q}\sum_{t \in \mathcal{T}_{p - 1}}i\Gamma_tz_{j_1}^{[t]} \ldots z_{j_q}^{[t]}\right),\label{si:eq:qaoa_instance_wise_path_integral_measure_cost_contribution}
\end{align}
where we introduced signed $\bm{\gamma}$ angles
\begin{align}
    \bm{\Gamma} & := \left(\Gamma_t\right)_{t \in \mathcal{T}_{p - 1}}\nonumber\\
    & := \left(\Gamma_1, \ldots, \Gamma_p, \Gamma_{-p}, \ldots, \Gamma_{-1}\right)\nonumber\\
    & := \left(\gamma_1, \ldots, \gamma_p, -\gamma_p, \ldots, -\gamma_1\right).
\end{align}
The multiplicative contribution from the mixers is given by:
\begin{align}
    \mu_B^{[p]}\left(\bm{z};\,S\right) & = \prod_{j \in S}f\left(\bm{z}_j\right),
\end{align}
where $f\left(\bm{a}\right)$ is defined for a single bitstring $\bm{a} \in \{1, -1\}^{\mathcal{T}_p}$ by:
\begin{align}
    f\left(\bm{a}\right) & := \frac{1}{2}\bm{1}\left[a_{p + 1} = a_{-p - 1}\right]\bra{a_t}\exp\left(i\beta_tX\right)\ket{a_{t + 1}}\bra{a_{-t - 1}}\exp\left(-i\beta_tX\right)\ket{a_{-t}}.
\end{align}
\end{proof}
\end{proposition}

We derive an expression for the disorder-averaged energy density in two steps: first (Proposition~\ref{si:prop:energy_density_from_disorder_averaged_path_integral_measure}), we express this quantity in terms of the disorder-averaged measure $\mu^{[p]}\left(\bm{z};\,\xi, S\right) := \mathbb{E}\left[\mu^{[p]}\left(\bm{z};\,\xi, S, \bm{J}\right)\right]$; second (Proposition~\ref{si:prop:disorder_averaged_path_integral_measure}), we provide an explicit expression in terms of configuration overlaps for the disorder-averaged path integral measure.

\begin{proposition}[Spin-glass QAOA energy from disorder-averaged spin-glass QAOA path integral measure]
\label{si:prop:energy_density_from_disorder_averaged_path_integral_measure}
The disorder-averaged energy density of spin-glass QAOA defined by cost function Eq.~\ref{si:eq:spin_glass_qaoa_cost_function} is given by:
\begin{align}
    \mathbb{E}\bra{\bm{\gamma}, \bm{\beta}}C/|S|\ket{\bm{\gamma}, \bm{\beta}} & = \left\langle \sum_{t \in \mathcal{T}_{p - 1}}i\Gamma_t\xi\left(\frac{1}{|S|}\left\langle \bm{z}^{[p + 1]}, \bm{z}^{[t]} \right\rangle\right) \right\rangle_{\mu^{[p]}\left(\bm{z};\,\xi, S\right)}\\
    & = \sum_{\bm{z} \in \{1, -1\}^{S \times \mathcal{T}_{p}}}\mu^{[p]}\left(\bm{z};\,\xi, S\right)\sum_{t \in \mathcal{T}_{p - 1}}i\Gamma_t\xi\left(\frac{1}{|S|}\left\langle \bm{z}^{[p + 1]}, \bm{z}^{[t]} \right\rangle\right),
\end{align}
where expectations are taken under the disorder-averaged spin-glass QAOA path integral measure $\mu^{[p]}\left(\bm{z};\,\xi, S\right) := \mathbb{E}\mu^{[p]}\left(\bm{z};\,\xi, S, \bm{J}\right)$.
\begin{proof}
This results from a simple application of the Gaussian integration by parts Lemma~\ref{si:lemma:gaussian_ibp}. Indeed, expliciting the cost function,
\begin{align}
    \mathbb{E}\bra{\bm{\gamma}, \bm{\beta}}C/|S|\ket{\bm{\gamma}, \bm{\beta}} & = \mathbb{E}\sum_{\bm{z} \in \{1, -1\}^{S \times \mathcal{T}_p}}\mu^{[p]}\left(\bm{z};\,\xi, S, \bm{J}\right)\frac{C\left(\bm{z}^{[p + 1]}\right)}{|S|}\nonumber\\
    & = \mathbb{E}\sum_{\bm{z} \in \{1, -1\}^{S \times \mathcal{T}_p}}\mu^{[p]}\left(\bm{z};\,\xi, S, \bm{J}\right)\frac{1}{|S|}\sum_{q \geq 1}c_q\sum_{j_1, \ldots, j_q \in S}|S|^{-(q - 1)/2}J^{(q)}_{j_1, \ldots, j_q}z^{[p + 1]}_{j_1} \ldots z^{[p + 1]}_{j_q}\nonumber\\
    & = \sum_{q \geq 1}c_q|S|^{-(q + 1)/2}\sum_{j_1, \ldots, j_q \in S}z^{[p + 1]}_{j_1} \ldots z^{[p + 1]}_{j_q}\sum_{\bm{z} \in \{1, -1\}^{S \times \mathcal{T}_p}}\mathbb{E}\left[\mu^{[p]}\left(\bm{z};\,\xi, S\right)J^{(q)}_{j_1, \ldots, j_q}\right].
\end{align}
We then compute the inner disorder average from Gaussian integration by parts Lemma~\ref{si:lemma:gaussian_ibp}, noting that $\bm{J} \longmapsto \mu^{[p]}\left(\bm{z};\,\xi, S, \bm{J}\right)$ is differentiable in $\bm{J}$ of exponential growth:
\begin{align}
    & \mathbb{E}\left[J^{(q)}_{j_1, \ldots, j_q}\mu^{[p]}\left(\bm{z};\,\xi, S, \bm{J}\right)\right]\nonumber\\
    & = \mathbb{E}\left[\mathbb{E}\left[\left(J^{(q)}_{j_1, \ldots, j_q}\right)^2\right]\frac{\partial \mu^{[p]}\left(\bm{z};\,\xi, S, \bm{J}\right)}{\partial J^{(q)}_{j_1, \ldots, j_q}}\right]\nonumber\\
    & = \mathbb{E}\left[\mu_B^{[p]}\left(\bm{z};\,S\right)\frac{\partial \mu_C^{[p]}\left(\bm{z};\,\xi, S, \bm{J}\right)}{\partial J^{(q)}_{j_1, \ldots, j_q}}\right]\nonumber\\
    & = \mathbb{E}\left[\mu_B^{[p]}\left(\bm{z};\,S\right)\frac{\partial}{\partial J^{(q)}_{j_1, \ldots, j_q}}\exp\left(\sum_{t \in \mathcal{T}_{p - 1}}i\Gamma_tC\left(\bm{z}^{[t]}\right)\right)\right]\nonumber\\
    & = \mathbb{E}\left[\mu_B^{[p]}\left(\bm{z};\,S\right)\frac{\partial}{\partial J^{(q)}_{j_1, \ldots, j_q}}\exp\left(\sum_{t \in \mathcal{T}_{p - 1}}i\Gamma_t\sum_{r \geq 1}c_r\sum_{k_1, \ldots, k_r \in S}|S|^{-(r - 1)/2}J^{(r)}_{k_1, \ldots, k_r}z^{[t]}_{k_1} \ldots z^{[t]}_{k_r}\right)\right]\nonumber\\
    & = \mathbb{E}\left[\mu_B^{[p]}\left(\bm{z};\,S\right)\left(\sum_{t \in \mathcal{T}_{p - 1}}i\Gamma_tc_q|S|^{-(q - 1)/2}z^{[t]}_{j_1} \ldots z^{[t]}_{j_q}\right)\mu_C^{[p]}\left(\bm{z};\,\xi, S, \bm{J}\right)\right]\nonumber\\
    & = \mu_B^{[p]}\left(\bm{z};\,S\right)\left(\sum_{t \in \mathcal{T}_{p - 1}}i\Gamma_tc_q|S|^{-(q - 1)/2}z^{[t]}_{j_1} \ldots z^{[t]}_{j_q}\right)\mathbb{E}\left[\mu_C^{[p]}\left(\bm{z};\,\xi, S, \bm{J}\right)\right]\\
    & = \mu_B^{[p]}\left(\bm{z};\,S\right)\left(\sum_{t \in \mathcal{T}_{p - 1}}i\Gamma_tc_q|S|^{-(q - 1)/2}z^{[t]}_{j_1} \ldots z^{[t]}_{j_q}\right)\mu_C^{[p]}\left(\bm{z};\,\xi, S\right)\nonumber\\
    & = \mu^{[p]}\left(\bm{z};\,\xi, S\right)\left(\sum_{t \in \mathcal{T}_{p - 1}}i\Gamma_tc_q|S|^{-(q - 1)/2}z^{[t]}_{j_1} \ldots z^{[t]}_{j_q}\right),
\end{align}
where after Gaussian integration by parts, the disorder average now only hits the path integral measure. Plugging this into the above expression for the disorder-averaged energy density,
\begin{align}
    & \mathbb{E}\bra{\bm{\gamma}, \bm{\beta}}C/|S|\ket{\bm{\gamma}, \bm{\beta}}\nonumber\\
    & = \sum_{q \geq 1}c_q|S|^{-(q + 1)/2}\sum_{j_1, \ldots, j_q \in S}z^{[p + 1]}_{j_1} \ldots z^{[p + 1]}_{j_q}\sum_{\bm{z} \in \{1, -1\}^{S \times \mathcal{T}_p}}\mu^{[p]}\left(\bm{z};\,\xi, S\right)\sum_{t \in \mathcal{T}_{p - 1}}i\Gamma_tc_q|S|^{-(q - 1)/2}z^{[t]}_{j_1} \ldots z^{[t]}_{j_q}\nonumber\\
    & = \sum_{\bm{z} \in \{1, -1\}^{S \times \mathcal{T}_p}}\mu^{[p]}\left(\bm{z};\,\xi, S\right)\sum_{t \in \mathcal{T}_{p - 1}}i\Gamma_t\sum_{q \geq 1}c_q^2|S|^{-q}\sum_{j_1, \ldots, j_q \in S}z^{[p + 1]}_{j_1}z^{[t]}_{j_1} \ldots z^{[p + 1]}_{j_q}z^{[t]}_{j_q}\nonumber\\
    & = \sum_{t \in \mathcal{T}_{p - 1}}i\Gamma_t\sum_{\bm{z} \in \{1, -1\}^{S \times \mathcal{T}_p}}\mu^{[p]}\left(\bm{z};\,\xi, S\right)\sum_{q \geq 1}c_q^2|S|^{-q}\left\langle \bm{z}^{[p + 1]}, \bm{z}^{[t]} \right\rangle\nonumber\\
    & = \sum_{t \in \mathcal{T}_{p - 1}}i\Gamma_t\sum_{\bm{z} \in \{1, -1\}^{S \times \mathcal{T}_p}}\mu^{[p]}\left(\bm{z};\,\xi, S\right)\xi\left(\frac{1}{|S|}\left\langle \bm{z}^{[p + 1]}, \bm{z}^{[t]} \right\rangle\right).
\end{align}
\end{proof}
\end{proposition}

\begin{proposition}[Disorder-averaged spin-glass QAOA path integral measure]
\label{si:prop:disorder_averaged_path_integral_measure}
Let $S$ a set of qubit indices, and consider the $p$-layers QAOA applied to this qubit set, with cost function:
\begin{align}
    C & := \sum_{q \geq 1}c_q\sum_{j_1, \ldots, j_q \in S}\left|S\right|^{-(q - 1)/2}J^{(q)}_{j_1, \ldots, j_q}Z_{j_1} \ldots Z_{j_q},\\
    J^{(q)}_{j_1, \ldots, j_q} & \overset{\mathrm{i.i.d}}{\sim} \mathcal{N}(0, 1).
\end{align}
The quantum circuit path integral measure (averaged over random Gaussian disorder) is given by:
\begin{align}
    \mu^{[p]}\left(\bm{z}_S;\,\xi, S\right) & = \mu^{[p]}_C\left(\bm{z}_S;\,\xi, S\right)\mu^{[p]}_B\left(\bm{z}_S;\,S\right),\label{si:eq:disorder_averaged_path_integral_measure}
\end{align}
where
\begin{align}
    \mu^{[p]}_C\left(\bm{z}_S;\,\xi, S\right) & := \exp\left(-\frac{|S|}{2}\sum_{t, u \in \mathcal{T}_{p - 1}}\Gamma_t\Gamma_u\xi\left(\frac{1}{|S|}\left\langle \bm{z}_S^{[t]}, \bm{z}_S^{[u]} \right\rangle\right)\right),\label{si:eq:disorder_averaged_path_integral_measure_cost_contribution}\\
    \left(\Gamma_1, \ldots, \Gamma_p, \Gamma_{-p}, \ldots, \Gamma_{-1}\right) & := \left(\gamma_1, \ldots, \gamma_p, -\gamma_p, \ldots, -\gamma_1\right)
\end{align}
reflects the contribution of cost unitaries, and $\mu_B\left(\bm{z};\,\xi, S\right)$, unchanged from Proposition~\ref{si:prop:path_integral_measure}, reflects the contribution of mixer unitaries. %
\begin{proof}
We compute the disorder-average of the instance-wise path integral measure $\mu^{[p]}\left(\bm{z};\,\xi, S, \bm{J}\right)$, noting the expectation only hits the cost contribution:
\begin{align}
    \mu^{[p]}\left(\bm{z};\,\xi, S\right) & := \mathbb{E}\left[\mu_C^{[p]}\left(\bm{z};\,\xi, S, \bm{J}\right)\mu_B\left(\bm{z};\,S\right)\right]\nonumber\\
    & = \mathbb{E}\left[\mu_C^{[p]}\left(\bm{z};\,\xi, S, \bm{J}\right)\right]\mu_B\left(\bm{z};\,S\right).
\end{align}
The disorder-average of the cost contribution can be computed the characteristic function of a normal variable: $\mathbb{E}_{X \sim \mathcal{N}(0, 1)}\left[e^{zX}\right] = e^{z^2/2}$, $\forall z \in \mathbb{C}$, and using independence of all $J^{(q)}_{j_1, \ldots, j_q}$:
\begin{align}
    \mathbb{E}\left[\mu_C^{[p]}\left(\bm{z};\,\xi, S, \bm{J}\right)\right] & = \mathbb{E}\left[\exp\left(\sum_{q \geq 1}\sum_{j_1, \ldots, j_q \in S}J_{j_1, \ldots, j_q}c_q\left|S\right|^{-(q - 1)/2}\sum_{t \in \mathcal{T}_{p - 1}}i\Gamma_tz^{[t]}_{j_1} \ldots z^{[t]}_{j_q}\right)\right]\nonumber\\
    & = \exp\left(\sum_{q \geq 1}\sum_{j_1, \ldots, j_q \in S}\frac{1}{2}\left(c_q\left|S\right|^{-(q - 1)/2}\sum_{t \in \mathcal{T}_{p - 1}}i\Gamma_tz^{[t]}_{j_1} \ldots z^{[t]}_{j_q}\right)^2\right)\nonumber\\
    & = \exp\left(-\frac{1}{2}\sum_{q \geq 1}c_q^2\left|S\right|^{-(q - 1)}\sum_{j_1, \ldots, j_q \in S}\sum_{t, u \in \mathcal{T}_{p - 1}}\Gamma_t\Gamma_uz^{[t]}_{j_1} \ldots z^{[t]}_{j_q}z^{[u]}_{j_1} \ldots z^{[u]}_{j_q}\right)\nonumber\\
    & = \exp\left(-\frac{1}{2}\sum_{q \geq 1}c_q^2\left|S\right|^{-(q - 1)}\sum_{t, u \in \mathcal{T}_{p - 1}}\Gamma_t\Gamma_u\sum_{j_1, \ldots, j_q \in S}z^{[t]}_{j_1} \ldots z^{[t]}_{j_q}z^{[u]}_{j_1} \ldots z^{[u]}_{j_q}\right)\nonumber\\
    & = \exp\left(-\frac{1}{2}\sum_{q \geq 1}c_q^2\left|S\right|^{-(q - 1)}\sum_{t, u \in \mathcal{T}_{p - 1}}\Gamma_t\Gamma_u\sum_{j_1, \ldots, j_q \in S}z^{[t]}_{j_1} \ldots z^{[t]}_{j_q}z^{[u]}_{j_1} \ldots z^{[u]}_{j_q}\right)\nonumber\\
    & = \exp\left(-\frac{1}{2}\sum_{q \geq 1}c_q^2\left|S\right|^{-(q - 1)}\sum_{t, u \in \mathcal{T}_{p - 1}}\Gamma_t\Gamma_u\left\langle \bm{z}^{[t]}, \bm{z}^{[u]} \right\rangle^q\right)\nonumber\\
    & = \exp\left(-\frac{\left|S\right|}{2}\sum_{t, u \in \mathcal{T}_{p - 1}}\Gamma_t\Gamma_u\xi\left(\frac{1}{|S|}\left\langle \bm{z}^{[t]}, \bm{z}^{[u]} \right\rangle \right)\right).
\end{align}
This establishes Eqns.~\ref{si:eq:disorder_averaged_path_integral_measure}, \ref{si:eq:disorder_averaged_path_integral_measure_cost_contribution}, \ref{si:eq:path_integral_measure_mixer_contribution} for the disorder-averaged path integral measure of the full QAOA circuit.
\end{proof}
\end{proposition}

\subsubsection{Cavity derivation of the $\mathbf{G}$ iteration: heuristic argument}
\label{si:sec:heuristic_cavity_argument}

In this Section, we give a heuristic account of the derivation of the $\mathbf{G}$ iteration for the spin-glass QAOA energy through a cavity argument. The central ingredient of this argument is the following decomposition of the disorder-averaged path integral measure of spin-glass QAOA derived in Proposition~\ref{si:prop:disorder_averaged_path_integral_measure}:

\begin{lemma}[Cavity decomposition of spin-glass QAOA path integral measure]
\label{si:lemma:cavity_decomposition_path_integral_measure}
Consider the disorder-averaged spin-glass QAOA path integral measure
\begin{align}
    \mu^{[p]}\left(\bm{z};\,\xi, S\right) & = \mu_C^{[p]}\left(\bm{z};\,\xi, S\right)\mu_B^{[p]}\left(\bm{z};\,S\right)
\end{align}
computed in Proposition~\ref{si:prop:disorder_averaged_path_integral_measure}.  Let $S = S' \sqcup S''$ an arbitrary partition of qubit indices and denote $s := |S|, s' := \left|S'\right|, s'' := \left|S''\right|$ for brevity. Then, the cost part of the quantum circuit path integral measure decomposes as follows according to this partition:
\begin{align}
    \mu^{[p]}_C\left(\bm{z}_S;\,\xi, S\right) & = \mu^{[p]}_C\left(\bm{z}_{S'};\,\xi_{s'/s}, S'\right)\mu^{[p]}_{C, 1}\left(\bm{z}_{S'}, \bm{z}_{S''};\,\xi_{s'/s}, S', S''\right)\mu^{[p]}_{C, 2}\left(\bm{z}_{S'}, \bm{z}_{S''};\,\xi, S', S''\right),
\end{align}
where $\mu^{[p]}_C\left(\bm{z}_{S'};\,\xi_{s'/s}, S'\right)$ is the cost part of the QAOA path integral measure for circuit over smaller set of qubits $S'$ (replacing $S \to S'$ in the above definitions, including in cost function scalings $\left|S\right|^{-(q - 1)/2}$), but with transformed spin glass mixture polynomial
\begin{align}
    \xi_{s'/s}\left(x\right) & := \frac{s}{s'}\xi\left(\frac{s'}{s}x\right).
\end{align}
The other multiplicative contributions $\mu_{C, 1}$ and $\mu_{C, 2}$ are not necessarily path integral measures and are defined as:
\begin{align}
    \mu^{[p]}_{C, 1}\left(\bm{z}_{S'}, \bm{z}_{S''};\,\xi_{s'/s}, S', S''\right) & := \exp\left(-\frac{1}{2}\sum_{t, u \in \mathcal{T}_{p - 1}}\Gamma_t\Gamma_u\xi'_{s'/s}\left(\frac{1}{\left|S'\right|}\left\langle \bm{z}^{[t]}_{S'}, \bm{z}^{[u]}_{S'} \right\rangle\right)\left\langle \bm{z}_{S''}^{[t]}, \bm{z}_{S''}^{[u]} \right\rangle\right),\label{si:eq:cavity_decomposition_path_integral_contribution_1}\\
    \mu^{[p]}_{C, 2}\left(\bm{z}_{S'}, \bm{z}_{S''};\,\xi, S', S''\right) & := \exp\left(-\frac{1}{2}\sum_{t, u \in \mathcal{T}_{p - 1}}\Gamma_t\Gamma_u\sum_{m \geq 2}\frac{1}{m!}\xi^{(m)}\left(\frac{1}{|S|}\left\langle \bm{z}_{S'}^{[t]}, \bm{z}_{S'}^{[u]} \right\rangle\right)\frac{1}{|S|^{m - 1}}\left\langle \bm{z}_{S''}^{[t]}, \bm{z}_{S''}^{[u]} \right\rangle^m\right).\label{si:ddd}
\end{align}
\end{lemma}

The decomposition from Lemma~\ref{si:lemma:cavity_decomposition_path_integral_measure} relies on an arbitrary partition $S = S' \sqcup S''$ of the qubit set; intuitively, $S'$ should be considered a large set, and $S''$ a small set. For the purpose of the heuristic argument, one may assume $S''$ of constant size, though the full rigorous calculation in principle allows $S''$ to scale with size.

We now outline the argument. Throughout the argument, it will be very insightful to think of quasiprobability measure $\mu^{[p]}\left(\bm{z};\,\xi, S\right)$ as an ordinary path integral measure. Consider a bipartition of spins $S = S' \sqcup S''$ as in Lemma~\ref{si:lemma:cavity_decomposition_path_integral_measure}, and consider evaluating a moment of the form:
\begin{align}
    \left\langle \prod_{j \in S''}z_j^{\left[\alpha_j\right]} \right\rangle_{\mu^{[p]}\left(\bm{z};\,\xi, S\right)}.
\end{align}
In the above, for all $j \in S''$, $\alpha_j \subset \mathcal{T}_{p + 1}$ is a set of signed layer indices defining a monomial
\begin{align}
    z_j^{\left[\alpha_j\right]} & := \prod_{t \in \alpha_j}z_j^{[t]}.
\end{align}
in spin $j$ trajectory $\bm{z}_j$. The idea is to estimate this pseudo-moment over variables $\bm{z} = \begin{pmatrix}
    \bm{z}_{S'}\\
    \bm{z}_{S''}
\end{pmatrix}$ by separating summation between spin sets $S'$ and $S''$. From Lemma~\ref{si:lemma:cavity_decomposition_path_integral_measure}, the full path integral measure (including cost and mixer contributions) can be decomposed as:
\begin{align}
    \mu^{[p]}\left(\bm{z};\,\xi, S\right) & = \mu^{[p]}\left(\bm{z};\,\xi, S'\right)\mu^{[p]}_{C, 1}\left(\bm{z}_{S'}, \bm{z}_{S''};\,\xi_{s'/s}, S', S''\right)\mu_B\left(\bm{z}_{S''};\,S''\right)\mu^{[p]}_{C, 2}\left(\bm{z}_{S'}, \bm{z}_{S''};\,\xi, S', S''\right).
\end{align}
From this decomposition, the pseudo-moment can be expressed as:
\begin{align}
    & \left\langle \prod_{j \in S''}z_j^{\left[\alpha_j\right]} \right\rangle_{\mu^{[p]}\left(\bm{z};\,\xi, S\right)}\nonumber\\
    & = \left\langle \sum_{\bm{z}_{S''} \in \{1, -1\}^{S'' \times \mathcal{T}_p}}\mu^{[p]}_{C, 1}\left(\bm{z}_{S'}, \bm{z}_{S''};\,\xi_{s'/s}, S', S''\right)\mu^{[p]}_B\left(\bm{z}_{S''};\,S''\right)\mu^{[p]}_{C, 2}\left(\bm{z}_{S'}, \bm{z}_{S''};\,\xi, S',S''\right)\prod_{j \in S''}z_j^{\left[\alpha_j\right]}  \right\rangle_{\mu^{[p]}\left(\bm{z}_{S'};\,\xi_{s'/s}, S'\right)}
\end{align}
We start by simplifying the above expression by dropping $\mu^{[p]}_{C, 2}$; indeed, due to restriction $m \geq 2$ in the sum inside the exponential defining $\mu^{[p]}_{C, 2}$ :
\begin{align}
    \mu^{[p]}_{C, 2}\left(\bm{z}_{S'}, \bm{z}_{S''};\,\xi, S', S''\right) & := \exp\left(-\frac{1}{2}\sum_{t, u \in \mathcal{T}_{p - 1}}\Gamma_t\Gamma_u\sum_{m \geq 2}\frac{1}{m!}\xi^{(m)}\left(\frac{1}{|S|}\left\langle \bm{z}^{[t]}_{S'}, \bm{z}^{[u]}_{S'} \right\rangle\right)\frac{1}{|S|^{m - 1}}\left\langle \bm{z}^{[t]}_{S''}, \bm{z}^{[u]}_{S''} \right\rangle^m\right),
\end{align}
the exponential's argument is at most of order $\mathcal{O}\left(1/n\right)$; this assumes $|S''| = \mathcal{O}(1)$ and uses that $m$ is bounded given $\xi$ is polynomial. Hence, one may write
\begin{align}
    \mu^{[p]}_{C, 2}\left(\bm{z}_{S'}, \bm{z}_{S''};\,\xi, S', S''\right) & \approx 1
\end{align}
inside the path integral expectation. Note this reasoning of bounding and approximating terms inside expectations is only rigorously valid for true probability distributions. However, we will show in the rigorous argument the reasoning ultimately also applies to quantum circuit path integral measures, as a consequence of fundamental boundedness property Corollary~\ref{si:cor:path_integral_measure_boundedness_monomials}. Hence, we presently guess
\begin{align}
    \left\langle \prod_{j \in S''}z_j^{\left[\alpha_j\right]} \right\rangle_{\mu^{[p]}\left(\bm{z};\,\xi, S\right)} & \approx \left\langle \sum_{\bm{z}_{S''} \in \{1, -1\}^{S'' \times \mathcal{T}_p}}\mu^{[p]}_{C, 1}\left(\bm{z}_{S'}, \bm{z}_{S''};\,\xi_{s'/s}, S', S''\right)\mu^{[p]}_B\left(\bm{z}_{S''};\,S''\right)\prod_{j \in S''}z_j^{\left[\alpha_j\right]} \right\rangle_{\mu^{[p]}\left(\bm{z}_{S'};\,\xi_{s'/s}, S'\right)}.\label{si:eq:heuristic_cavity_argument_step_2}\\
    & = \sum_{\bm{z}_{S''} \in \{1, -1\}^{S'' \times \mathcal{T}_p}}\left\langle \mu^{[p]}_{C, 1}\left(\bm{z}_{S'}, \bm{z}_{S''};\,\xi_{s'/s}, S', S''\right)\right\rangle_{\mu^{[p]}\left(\bm{z}_{S'};\,\xi_{s'/s}, S'\right)}\mu^{[p]}_B\left(\bm{z}_{S''};\,S''\right)\prod_{j \in S''}z_j^{\left[\alpha_j\right]}\nonumber\\
    & = \sum_{\bm{z}_{S''} \in \{1, -1\}^{S'' \times \mathcal{T}_p}}\left\langle \mu^{[p]}_{C, 1}\left(\bm{z}_{S'}, \bm{z}_{S''};\,\xi_{s'/s}, S', S''\right)\right\rangle_{\mu^{[p - 1]}\left(\bm{z}_{S'};\,\xi_{s'/s}, S'\right)}\mu^{[p]}_B\left(\bm{z}_{S''};\,S''\right)\prod_{j \in S''}z_j^{\left[\alpha_j\right]}.\label{si:eq:heuristic_cavity_decomposition_approximation_step_1}
\end{align}
In the final line, we replaced measure $\mu^{[p]}\left(\bm{z}_{S'};\,\xi_{s'/s}, S'\right)$ by $\mu^{[p - 1]}\left(\bm{z};\,\xi_{s'/s}, S'\right)$ in the expectation of $\mu^{[p]}_{C, 1}$ using unitary cancellation in path integral measures (Proposition~\ref{si:prop:path_integral_measure_unitary_cancellation}), given $\mu^{[p]}_{C, 1}\left(\bm{z}_{S'}, \bm{z}_{S''};\,\xi_{s'/s}, S', S''\right)$, seen as a function of spin trajectories $\bm{z}_{S'}$:
\begin{align}
    \mu^{[p]}_{C, 1}\left(\bm{z}_{S'}, \bm{z}_{S''};\,\xi_{s'/s}, S', S''\right) & = \exp\left(-\frac{1}{2}\sum_{t, u \in \mathcal{T}_{p - 1}}\Gamma_t\Gamma_u\xi'_{s'/s}\left(\frac{1}{\left|S'\right|}\left\langle \bm{z}_{S'}^{[t]}, \bm{z}_{S'}^{[u]} \right\rangle\right)\left\langle \bm{z}^{[t]}_{S''}, \bm{z}^{[u]}_{S''} \right\rangle\right)
\end{align}
only depends on layer indices $\mathcal{T}_{p - 1}$ of maximum absolute value $p$. We now consider the summation over spins $S''$ bit matrix $\bm{z}_{S''}$ for a given realization of spins $S'$ bit matrix $\bm{z}_{S'}$. This sum can be interpreted as an expectation over spins $S''$, conditioned in the trajectories of spins $S'$. We observe that conditioned on any realization $\bm{z}_{S'}$ of $S'$ trajectories, spins $S''$ are independent since:
\begin{align}
    & \mu^{[p]}_{C, 1}\left(\bm{z}_{S'}, \bm{z}_{S''};\,\xi, S', S''\right)\mu^{[p]}_B\left(\bm{z}_{S''};\,S''\right)\nonumber\\
    & = \exp\left(-\frac{1}{2}\sum_{t, u \in \mathcal{T}_{p - 1}}\Gamma_t\Gamma_u\xi'_{s'/s}\left(\frac{1}{\left|S'\right|}\left\langle \bm{z}_{S'}^{[t]}, \bm{z}_{S'}^{[u]} \right\rangle\right)\left\langle \bm{z}^{[t]}_{S''}, \bm{z}^{[u]}_{S''} \right\rangle\right)\mu^{[p]}_B\left(\bm{z}_{S''};\,S''\right)\nonumber\\
    & = \prod_{j \in S''}\exp\left(-\frac{1}{2}\sum_{t, u \in \mathcal{T}_{p - 1}}\Gamma_t\Gamma_u\xi'_{s'/s}\left(\frac{1}{\left|S'\right|}\left\langle \bm{z}_{S'}^{[t]}, \bm{z}_{S'}^{[u]} \right\rangle\right)z_j^{[t]}z_j^{[u]}\right)f\left(\bm{z}_j\right).\label{si:eq:heuristic_cavity_argument_step_2_cavity_expectand}
\end{align}
We would like to say that spins $S''$ are in fact (approximately) independent, which is weaker than conditional independence given $S'$. We then argue that approximate independence of any constant-size set of spins is at least self-consistent. We first sketch a proof that under this assumption, the path integral measure concentrates normalized overlaps of the large system $\left\langle \bm{z}_{S}^{[t]}, \bm{z}_{S}^{[u]} \right\rangle/|S|$ around the path integral measure's ``second order moments"
\begin{align}
    G^{\left|S\right|}_{t, u} & := \left\langle z_j^{[t]}z_j^{[u]} \right\rangle_{\mu^{[p]}\left(\bm{z};\,\xi, S\right)}, \qquad j \in S.
\end{align}
By permutation invariance of the disorder-averaged path integral measure, the above quantity only depends on spin set size $|S|$ and is independent of spin $j \in S$. Besides, by Corollary~\ref{si:cor:path_integral_measure_boundedness_monomials}, $\left|G^{|S|}_{t, u}\right| \leq 1$. The qualitative reason for overlap concentration is that the overlap
\begin{align}
    \frac{1}{|S|}\left\langle \bm{z}^{[t]}_S, \bm{z}^{[u]}_S \right\rangle & = \frac{1}{|S|}\sum_{j \in S}z_j^{[t]}z_j^{[u]}
\end{align}
can be seen as the average of many approximately independent variables; hence, we expect a law of large numbers to apply. More specifically, one can estimate the lowest-order moments using linearity of expectation, permutation invariance of the disorder-averaged measure, and the approximate independence assumption for constant-size spin sets. For the first order moment.
\begin{align}
    \left\langle \frac{1}{|S|}\left\langle \bm{z}_S^{[t]}, \bm{z}_S^{[u]} \right\rangle\right\rangle_{\mu^{[p]}\left(\bm{z};\,\xi, S\right)} & = \frac{1}{|S|}\sum_{j \in S}\left\langle z_j^{[t]}z_j^{[u]} \right\rangle_{\mu^{[p]}\left(\bm{z};\,\xi, S\right)}\nonumber\\
    & \approx \frac{1}{|S|}\sum_{j \in S}G^{|S|}_{t, u}\nonumber\\
    & \approx G^{|S|}_{t, u}
\end{align}
For the second order moment,
\begin{align}
    \left\langle \left(\frac{1}{|S|}\left\langle \bm{z}_S^{[t]}, \bm{z}_S^{[u]} \right\rangle\right)^2\right\rangle_{\mu^{[p]}\left(\bm{z};\,\xi, S\right)} & = \frac{1}{|S|^2}\sum_{j, k \in S}\left\langle z_j^{[t]}z_j^{[u]}z_k^{[t]}z_k^{[u]} \right\rangle_{\mu^{[p]}\left(\bm{z};\,\xi, S\right)}\nonumber\\
    & = \frac{1}{|S|^2}\sum_{j \in S}\left\langle z_j^{[t]}z_j^{[u]}z_j^{[t]}z_j^{[u]} \right\rangle_{\mu^{[p]}\left(\bm{z};\,\xi, S\right)} + \frac{1}{|S|^2}\sum_{\substack{j, k \in S\\j \neq k}}\left\langle z_j^{[t]}z_j^{[u]}z_k^{[t]}z_k^{[u]} \right\rangle_{\mu^{[p]}\left(\bm{z};\,\xi, S\right)}\nonumber\\
    & = \frac{1}{|S|} + \frac{1}{|S|^2}\sum_{\substack{j, k \in S\\j \neq k}}\left\langle z_j^{[t]}z_j^{[u]}z_k^{[t]}z_k^{[u]} \right\rangle_{\mu^{[p]}\left(\bm{z};\,\xi, S\right)}\nonumber\\
    & \approx \frac{1}{|S|} + \frac{1}{|S|^2}\sum_{\substack{j, k \in S\\j \neq k}}\left\langle z_j^{[t]}z_j^{[u]} \right\rangle_{\mu^{[p]}\left(\bm{z};\,\xi, S\right)}\left\langle z_k^{[t]}z_k^{[u]} \right\rangle_{\mu^{[p]}\left(\bm{z};\,\xi, S\right)}\nonumber\\
    & = \frac{1}{|S|} + \frac{1}{|S|^2}\sum_{\substack{j, k \in S\\j \neq k}}\left(G^{|S|}_{t, u}\right)^2\nonumber\\
    & = \frac{1}{|S|} + \frac{|S|\left(|S| - 1\right)}{|S|^2}\left(G^{|S|}_{t, u}\right)^2\nonumber\\
    & = \left(G^{|S|}_{t, u}\right)^2 + \mathcal{O}\left(\frac{1}{|S|}\right).
\end{align}
Hence, the order 2 moment is the square of the order 1 moment up to a vanishing finite-size correction. This heuristically establishes concentration of the normalized overlap; note the argument only required approximate independence (to a suitable degree of accuracy) of a constant number of spins, rather than the much stronger pairwise independence of all $|S|$ spins. Due to concentration, one may argue overlaps
\begin{align}
    \frac{1}{\left|S'\right|}\left\langle \bm{z}^{[t]}_{S'}, \bm{z}^{[u]}_{S'} \right\rangle
\end{align}
approximately behave as deterministic random variables ---with values $G^{\left|S\right|}_{t, u}$--- under path integral measure $\mu^{[p]}\left(\bm{z};\,\xi, S\right)$. Simplifying Eq.~\ref{si:eq:heuristic_cavity_argument_step_2_cavity_expectand} accordingly, and plugging into Eq.~\ref{si:eq:heuristic_cavity_argument_step_2} then yields further approximation:
\begin{align}
    \left\langle \prod_{j \in S''}z_j^{[\alpha_j]} \right\rangle_{\mu^{[p]}\left(\bm{z};\,\xi, S\right)} & \approx \left\langle \prod_{j \in S''}\sum_{\bm{z}_j \in \{1, -1\}^{\mathcal{T}_p}}\exp\left(-\frac{1}{2}\sum_{t, u \in \mathcal{T}_{p - 1}}\Gamma_t\Gamma_u\xi'_{s'/s}\left(G^{\left|S'\right|}_{t, u}\right)z_j^{[t]}z_j^{[u]}\right)f\left(\bm{z}_j\right)z_j^{\left[\alpha_j\right]} \right\rangle_{\mu^{[p]}\left(\bm{z};\,\xi, S'\right)}\nonumber\\
    & = \prod_{j \in S''}\sum_{\bm{z}_j \in \{1, -1\}^{\mathcal{T}_p}}\exp\left(-\frac{1}{2}\sum_{t, u \in \mathcal{T}_{p - 1}}\Gamma_t\Gamma_u\xi'_{s'/s}\left(G^{\left|S'\right|}_{t, u}\right)z_j^{[t]}z_j^{[u]}\right)f\left(\bm{z}_j\right)z_j^{\left[\alpha_j\right]}.\label{si:eq:heuristic_cavity_argument_step_3}
\end{align}
Thanks to the approximate determinism of normalized overlaps, the expectation under the $S'$ spins has been removed, and spins from $S''$ were shown approximately independent. In essence, approximate independence of finite-size distributions and concentration of overlaps are equivalent.

We now make the assumption that $G_{t, u}^{s}$ converges to a well-defined limit $G_{t, u}$ as the set size $s \to \infty$ (note this can always be ensured over a subsequence of set sizes as all $G$ entries are bounded by $1$). Also recalling the constant size assumption over $S''$, so that $s'/s \approx 1$, one can further write $\xi'_{s'/s} \approx \xi$, yielding ultimate approximation:
\begin{align}
    \left\langle \prod_{j \in S''}z_j^{\left[\alpha_j\right]} \right\rangle_{\mu^{[p]}\left(\bm{z};\,\xi, S\right)} & \approx \left\langle \prod_{j \in S''}\sum_{\bm{z}_j \in \{1, -1\}^{\mathcal{T}_p}}\exp\left(-\frac{1}{2}\sum_{t, u \in \mathcal{T}_{p - 1}}\Gamma_t\Gamma_u\xi'\left(G_{t, u}\right)z_j^{[t]}z_j^{[u]}\right)f\left(\bm{z}_j\right)z_j^{\left[\alpha_j\right]} \right\rangle_{\mu^{[p]}\left(\bm{z};\,\xi, S\right)}.
\end{align}
Applying the above approximate identity to $S'' := \{j\}$ and $\alpha_j = \{t, u\}$, with $j \in S$ an arbitrary spin index, yields:
\begin{align}
    \left\langle z_j^{[t]}z_j^{[u]} \right\rangle_{\mu^{[p]}\left(\bm{z};\,\xi, S\right)} & \approx \sum_{\bm{z}_j \in \{1, -1\}^{\mathcal{T}_p}}z_j^{[t]}z_j^{[u]}f\left(\bm{z}_j\right)\exp\left(-\frac{1}{2}\sum_{t, u \in \mathcal{T}_{p - 1}}\Gamma_t\Gamma_u\xi'\left(G_{t, u}\right)z_j^{[t]}z_j^{[u]}\right).
\end{align}
Recalling $\left\langle z_j^{[t]}z_j^{[u]} \right\rangle_{\mu^{[p]}\left(\bm{z};\,\xi, S\right)} = G^{|S|}_{t, u}$, and taking the infinite-size limit $|S| \to \infty$, yields
\begin{align}
    G_{t, u} & = \sum_{\bm{a} \in \{1, -1\}^{\mathcal{T}_p}}a_ta_uf\left(\bm{a}\right)\exp\left(-\frac{1}{2}\sum_{t, u \in \mathcal{T}_{p - 1}}\Gamma_t\Gamma_u\xi'\left(G_{t, u}\right)a_ta_u\right),
\end{align}
which we recognize as the fixed-point equation satisfied by the spin-glass QAOA $\bm{G}$ matrix.

\begin{remark}[Correctness of substituting overlap by deterministic value]
In Eq.~\ref{si:eq:heuristic_cavity_argument_step_3}, using approximate determinism of normalized overlaps:
\begin{align}
    \frac{1}{\left|S'\right|}\left\langle \bm{z}_{S'}^{[t]}, \bm{z}_{S'}^{[u]} \right\rangle & \approx G^{\left|S'\right|}_{t, u},
\end{align}
we substituted the normalized overlap in $S'$ by $G_{t, u}$ inside the exponential:
\begin{align}
    \exp\left(-\frac{1}{2}\sum_{t, u \in \mathcal{T}_{p - 1}}\Gamma_t\Gamma_u\xi'\left(\frac{1}{\left|S'\right|}\left\langle \bm{z}_{S'}^{[t]}, \bm{z}_{S'}^{[u]} \right\rangle\right)z_j^{[t]}z_j^{[u]}\right) & \approx \exp\left(-\frac{1}{2}\sum_{t, u \in \mathcal{T}_{p - 1}}\Gamma_t\Gamma_u\xi'\left(G^{\left|S'\right|}_{t, u}\right)z_j^{[t]}z_j^{[u]}\right).
\end{align}
This approximation hides an important technicality: we only assumed normalized overlaps approached deterministic random variables in the infinite-size limit in the sense of their constant-order moments. That is, for any fixed moment order, the overlap moments of this order converge to that of a deterministic variable as size is sent to infinity. This notion of convergence proves sufficient to perform the substitution inside the exponential, as will be further detailed in the rigorous version of the argument. Namely, for any fixed moment order, we will perform a Taylor expansion of the exponential to approximate it as a polynomial function of normalized overlaps of the chosen maximum moment order. We will prove smallness of the Taylor remainder similar to a classical measure case by using boundedness property Corollary~\ref{si:cor:path_integral_measure_boundedness_monomials} of path integral measures. The main term of the Taylor expansion will now only involve overlap moments of bounded order, properly covered by weak convergence. The key criterion enabling application of this method is that for any constant target approximation accuracy, the Taylor expansion may be truncated to constant order. This is in turn possible since the function of normalized overlaps is size-independent (or can be in a certain sense bounded by such a function), and the overlap is normalized by size, hence bounded by $1$.
\end{remark}

\subsubsection{Cavity derivation of the $\mathbf{G}$ iteration: the heuristic argument at $p = 1$}
\label{si:sec:heuristic_cavity_argument_p1}

In the previous paragraph, we sketched the cavity calculation derivation of the $\bm{G}$ matrix fixed-point equation at arbitrary $p$. The argument relied on a self-consistent approximate independence in spin index of the path integral measure's finite-size distribution. Note we did not sketch an ab-initio argument for this approximate independence, but rather only introduced it as a natural hypothesis given the structure of the path integral measure. In this Section, we provide a more logically sound account fo the argument at $p = 1$. This special case suggest converting the ill-founded self-consistent argument from the previous section into one by induction over the number of QAOA layers.

Our starting point is the approximation of finite-dimensional expectation Eq.~\ref{si:eq:heuristic_cavity_argument_step_2}:
\begin{align}
    \left\langle \prod_{j \in S''}z_j^{\left[\alpha_j\right]} \right\rangle_{\mu^{[1]}\left(\bm{z};\,\xi, S\right)} & \approx \sum_{\bm{z}_{S''} \in \{1, -1\}^{S'' \times \mathcal{T}_1}}\left\langle\mu^{[1]}_{C, 1}\left(\bm{z}_{S'}, \bm{z}_{S''};\,\xi_{s'/s}, S', S''\right)\right\rangle_{\mu^{[0]}\left(\bm{z}_{S'};\,\xi_{s'/s}, S'\right)}\prod_{j \in S''}f\left(\bm{z}_j\right)z_j^{\left[\alpha_j\right]}\label{si:eq:heuristic_cavity_decomposition_approximation_step_1_p1}
\end{align}
We recall this approximation was obtained by letting
\begin{align}
    \mu^{[1]}_{C, 2}\left(\bm{z}_{S'}, \bm{z}_{S''};\,\xi, S', S''\right) & \approx 1.
\end{align}
This relied on the heuristic argument of $\left(\mu^{[1]}_{C, 2} - 1\right)$ being pointwise suppressed inversely in size $|S|$ and did not rely on the circular approximate independence argument; the argument could easily be made rigorous by invoking Corollary~\ref{si:cor:path_integral_measure_boundedness_monomials}. We now express the approximate measure densities inside the expectation specificially for $p = 1$. We recall the expression for general $p$ from Eq.~\ref{si:eq:heuristic_cavity_argument_step_2_cavity_expectand}, and specialize it to $p = 1$:
\begin{align}
    \mu^{[1]}_{C, 1}\left(\bm{z}_{S'}, \bm{z}_{S''};\,\xi_{s'/s}, S', S''\right) & = \exp\left(-\frac{1}{2}\sum_{t, u \in \mathcal{T}_{0}}\Gamma_t\Gamma_u\xi'_{s'/s}\left(\frac{1}{\left|S'\right|}\left\langle \bm{z}^{[t]}_{S'}, \bm{z}^{[u]}_{S'} \right\rangle\right)\left\langle \bm{z}^{[t]}_{S''}, \bm{z}^{[u]}_{S''} \right\rangle\right)\label{si:eq:mu_c_1_p1}\\
    & = \exp\left(-\frac{\gamma^2}{2}\xi'\left(\frac{1}{\left|S'\right|}\left\langle \bm{z}_{S'}^{[1]}, \bm{z}_{S'}^{[1]} \right\rangle\right)\left\langle \bm{z}^{[1]}_{S''}, \bm{z}^{[1]}_{S''} \right\rangle\right.\nonumber\\
    & \left. \hspace*{40px} - \frac{\gamma^2}{2}\xi'\left(\frac{1}{\left|S'\right|}\left\langle \bm{z}_{S'}^{[-1]}, \bm{z}_{S'}^{[-1]} \right\rangle\right)\left\langle \bm{z}^{[-1]}_{S''}, \bm{z}^{[-1]}_{S''} \right\rangle\right.\nonumber\\
    & \left. \hspace*{40px} + \gamma^2\xi'\left(\frac{1}{\left|S'\right|}\left\langle \bm{z}_{S'}^{[1]}, \bm{z}_{S'}^{[-1]} \right\rangle\right)\left\langle \bm{z}^{[t]}_{S''}, \bm{z}^{[u]}_{S''} \right\rangle\right)\nonumber\\
    & = \exp\left(-\gamma^2\xi'(1)\left|S''\right| + \gamma^2\xi'\left(\frac{1}{\left|S'\right|}\left\langle \bm{z}_{S'}^{[1]}, \bm{z}_{S'}^{[-1]} \right\rangle\right)\left\langle \bm{z}^{[1]}_{S''}, \bm{z}^{[-1]}_{S''} \right\rangle\right),
\end{align}
where in the final line, we simplified configuration overlaps with identical layer indices to $1$ (note this simplification applied to $p > 1$ as well). We now argue that under the $S'$ path integral measure $\mu^{[p]}\left(\bm{z}_{S'};\,\xi_{s'/s}, S'\right)$, normalized overlaps between layers $1$ and $-1$ can also be replaced by $1$, i.e. $\left\langle \bm{z}_{S''}^{[1]}, \bm{z}_{S''}^{[-1]} \right\rangle/\left|S''\right| \longrightarrow 1$. This follows from the following elementary Lemma:

\begin{lemma}[Monomials of maximum layer index $1$ have expectation $1$]
\label{si:lemma:maximum_layer_index_1_monomial_expectation}
Consider any $p$-layer unitary circuit as specified in Definition~\ref{si:def:layered_quantum_circuit_path_integral_measure}, given by initial state $\ket{\psi}$ and unitaries $U_t$ ($1 \leq t \leq p$), together with its path integral measure $\mu^{[p]}$. Also, consider the circuits obtained from truncating the initial circuit to $q \in [1, p]$ layers, and denote by $\mu^{[q]}$ the corresponding path integral measure. Consider a multiset $I$ of pairs $(j, t)$ of spin indices $j \in S$ and layer indices $t \in \mathcal{T}_1$ (i.e. all time indices are $\pm 1$). For all $j \in S$, denote by $I_j$ the multiset with an occurrence of $t$ for each $(j, t) \in I$. Finally, assume the circuit's initial state is $\mathbb{Z}_2$-symmetric, i.e. $X^{\otimes S}\ket{\psi} = \ket{\psi}$. Then, the following holds for the expectation of $\bm{z}$ monomials defined by $I$:
\begin{align}
    \left\langle \prod_{(j, t) \in I}z_j^{[t]} \right\rangle_{\mu^{[p]}\left(\bm{z}\right)} & = \prod_{j \in S}\bm{1}\left[|I_j|\,\mathrm{even}\right].
\end{align}
\begin{proof}
Since by assumption all layer indices are bounded by $1$ in absolute value, Proposition~\ref{si:prop:path_integral_measure_unitary_cancellation} applied to rephrase the expectation as one under $\mu^{[0]}$:
\begin{align}
    \left\langle \prod_{(j, t) \in I}z_j^{[t]} \right\rangle_{\mu^{[p]}\left(\bm{z}\right)} & = \left\langle \prod_{(j, t) \in I}z_j^{[t]} \right\rangle_{\mu^{[0]}\left(\bm{z}\right)}.
\end{align}
From the time autocorrelation interpretation of $\bm{z}$ monomials expectation under the path integral measure (Proposition~\ref{si:prop:z_autocorrelation_path_integral_representation}), the above can be interpreted as the expectation of a product of $Z$ operators in the $0$-layer restricted circuit, i.e. state $\ket{\psi}$. The spin indices and insertion positions are defined by $I$, and since the circuit reduces to initial state $\ket{\psi}$ and $Z$ operators commute, insertion positions $\pm 1$ are equivalent. If any spin $j \in S$, has an odd number of $Z$ insertions, the $Z$ operators pertaining to this spin simplify to $Z_j$, and the expectation is $0$ by $\mathbb{Z}_2$ symmetry of the state ($X^{\otimes S}Z_j = -Z_jX^{\otimes S}$). If all spins $j$ have an even number of $Z$ operators, these simplify to the identity and the expectation reduces to $\braket{\psi|\psi} = 1$. This proves the claim.
\end{proof}
\end{lemma}

Lemma~\ref{si:lemma:maximum_layer_index_1_monomial_expectation} allows to exactly average Eq.~\ref{si:eq:mu_c_1_p1} under the path integral measure over spins $S'$:
\begin{align}
    & \left\langle \exp\left(-\frac{1}{2}\sum_{t, u \in \mathcal{T}_0}\Gamma_t\Gamma_u\xi'_{s'/s}\left(\frac{1}{\left|S'\right|}\left\langle \bm{z}^{[t]}_{S'}, \bm{z}^{[u]}_{S'} \right\rangle\right)\left\langle \bm{z}_{S''}^{[t]}, \bm{z}_{S''}^{[u]} \right\rangle\right) \right\rangle_{\mu^{[p]}\left(\bm{z}_{S'};\,\xi_{s'/s}, S'\right)}\nonumber\\
    & = \exp\left(-\frac{1}{2}\sum_{t, u \in \mathcal{T}_0}\Gamma_t\Gamma_u\xi'_{s'/s}\left(1\right)\left\langle \bm{z}_{S''}^{[t]}, \bm{z}_{S''}^{[u]} \right\rangle\right).
\end{align}
The evenness condition from the Lemma is automatically satisfied since the normalized overlap:
\begin{align}
    \frac{1}{\left|S'\right|}\left\langle \bm{z}_{S'}^{[t]}, \bm{z}_{S'}^{[u]} \right\rangle & = \frac{1}{\left|S'\right|}\sum_{k \in S'}z_k^{[t]}z_k^{[u]}
\end{align}
is a polynomial with even occurrences of index $k$ in each monomial. Hence, the Boolean Fourier expansion of the exponential in variables $\bm{z}_{S'}$ is a sum of monomials satisfying the same property. As a result, the Lemma transforms all these monomials to $1$ after averaging, and everything goes as if setting $\bm{z}_{S'} \longleftarrow \bm{1}_{S' \times \mathcal{T}_1}$ in the original exponential. Plugging this path integral average into Eq.~\ref{si:eq:heuristic_cavity_decomposition_approximation_step_1_p1}
\begin{align}
    \left\langle \prod_{j \in S''}z_j^{\left[\alpha_j\right]} \right\rangle_{\mu^{[1]}\left(\bm{z};\,\xi, S\right)} & \approx \sum_{\bm{z}_{S''} \in \{1, -1\}^{S'' \times \mathcal{T}_p}}\prod_{j \in S''}f\left(\bm{z}_j\right)\exp\left(-\frac{1}{2}\sum_{t, u \in \mathcal{T}_0}\Gamma_t\Gamma_u\xi'_{s'/s}(1)\left\langle \bm{z}_{S''}^{[t]}, \bm{z}_{S''}^{[u]} \right\rangle\right)z_j^{\left[\alpha_j\right]},
\end{align}
showing approximate independence of spins $S''$ under the path integral measure. We reiterate the only approximation in this calculation was to drop $\mu^{[1]}_{C, 2} - 1$ from the path integral expectation given its suppression in the inverse size $1/|S|$ ---an argument that could easily be made rigorous from the boundedness property of path integral measures Corollary~\ref{si:cor:path_integral_measure_boundedness_monomials}. Now using assumption $s'' = \mathcal{O}(1)$, the previous approximation can be further loosened up to an error in inverse size:
\begin{align}
    \left\langle \prod_{j \in S''}z_j^{\left[\alpha_j\right]} \right\rangle_{\mu^{[1]}\left(\bm{z};\,\xi, S\right)} & \approx \sum_{\bm{z}_{S''} \in \{1, -1\}^{S'' \times \mathcal{T}_p}}\prod_{j \in S''}f\left(\bm{z}_j\right)\exp\left(-\frac{1}{2}\sum_{t, u \in \mathcal{T}_0}\Gamma_t\Gamma_u\xi'(1)\left\langle \bm{z}^{[t]}_{S''}, \bm{z}^{[u]}_{S''} \right\rangle\right)z_j^{\left[\alpha_j\right]}.\label{si:eq:heuristic_cavity_decomposition_approximation_step_3_p1}
\end{align}
From Proposition~\ref{si:prop:spin_boson_path_integral_measure_expression}, we see the independent quasiprobability distribution of each spin:
\begin{align}
    f\left(\bm{a}\right)\exp\left(-\frac{1}{2}\sum_{t, u \in \mathcal{T}_0}\Gamma_t\Gamma_u\xi'(1)a_ta_u\right) & = f\left(\bm{a}\right)\exp\left(-\gamma^2\xi'(1)\left(1 - a_1a_{-1}\right)\right)
\end{align}
is the path integral measure of a spin inside a single-layer spin-boson system (Definition~\ref{si:def:spin_boson_system}) of influence functional $\begin{pmatrix}
    \xi'(1) & \xi'(1)\\
    \xi'(1) & \xi'(1)
\end{pmatrix} \in \mathbb{R}^{\mathcal{T}_0 \times \mathcal{T}_0}$. From approximate independence of finite-dimensional spin-distributions at $p = 1$ in the infinite-size limit, one may deduce the thermodynamic-limit of second-order moments. The case of a moment with all layer indices bounded by $1$ is trivial and covered by Lemma~\ref{si:lemma:maximum_layer_index_1_monomial_expectation}, so let us consider a moment of maximum layer index $2$:
\begin{align}
    \left\langle z_j^{[1]}z_j^{[2]} \right\rangle_{\mu^{[p]}\left(\bm{z};\,\xi, S\right)} & \approx \sum_{\bm{z}_j \in \{1, -1\}^{\mathcal{T}_1}}z_j^{[1]}z_j^{[2]}f\left(\bm{z}_j\right)\exp\left(-\frac{1}{2}\sum_{t, u \in \mathcal{T}_0}\Gamma_t\Gamma_u\xi'(1)z_j^{[t]}z_j^{[u]}\right)\nonumber\\
    & = G_{1, 2},
\end{align}
for the $\bm{G}$ matrix associated to the spin glass (see Section~\ref{si:sec:spin_glass_qaoa_g_iteration} and specifically Eq.~\ref{si:eq:g_iteration}). Hence, the (heuristic) cavity argument establishes that second-order moments up to layer index $2$ converge to $\bm{G}$ matrix entries in the infinite-size limit.

By approximate independence in spin index of finite-dimensional spin distribution up to layer index $2$, replaying the argument from Section~\ref{si:sec:heuristic_cavity_argument} gives that normalized overlaps:
\begin{align}
    \frac{1}{\left|S\right|}\left\langle \bm{z}_S^{[t]}, \bm{z}_S^{[u]} \right\rangle, \qquad |t|, |u| \leq 2.
\end{align}
concentrate around entries $G_{t, u}$ under the path integral measure. From this observation, the cavity decomposition at $p = 2$:
\begin{align}
    \left\langle \prod_{j \in S''}z_j^{\left[\alpha_j\right]} \right\rangle_{\mu^{[2]}\left(\bm{z};\,\xi, S\right)} & \approx \sum_{\bm{z}_{S''} \in \{1, -1\}^{S'' \times \mathcal{T}_2}}\left\langle \mu^{[2]}_{C, 1}\left(\bm{z}_{S'}, \bm{z}_{S''};\,\xi_{s'/s}, S', S''\right) \right\rangle_{\mu^{[1]}\left(\bm{z}_{S'};\,\xi, S\right)}\prod_{j \in S''}f\left(\bm{z}_j\right)z_j^{\left[\alpha_j\right]},\label{si:eq:heuristic_cavity_decomposition_approximation_step_1_p2}\\
    \mu^{[2]}_{C, 1}\left(\bm{z}_{S'}, \bm{z}_{S''};\,\xi_{s'/s}, S', S''\right) & = \exp\left(-\frac{1}{2}\sum_{t, u \in \,\mathcal{T}_1}\Gamma_t\Gamma_u\xi_{s'/s}'\left(\frac{1}{\left|S'\right|}\left\langle \bm{z}^{[t]}_{S'}, \bm{z}^{[u]}_{S'} \right\rangle\right)\left\langle \bm{z}^{[t]}_{S''}, \bm{z}^{[u]}_{S''} \right\rangle\right)
\end{align}
can be simplified by approximating:
\begin{align}
    \left\langle \mu^{[2]}_{C, 1}\left(\bm{z}_{S'}, \bm{z}_{S''};\,\xi_{s'/s}, S', S''\right)\right\rangle_{\mu^{[1]}\left(\bm{z}_{S'};\,\xi_{s'/s}, S'\right)} & \approx \exp\left(-\frac{1}{2}\sum_{t, u \in \,\mathcal{T}_1}\Gamma_t\Gamma_u\xi_{s'/s}'\left(G_{t, u}\right)\left\langle \bm{z}^{[t]}_{S''}, \bm{z}^{[u]}_{S''} \right\rangle\right)\\
    & \approx \exp\left(-\frac{1}{2}\sum_{t, u \in \,\mathcal{T}_1}\Gamma_t\Gamma_u\xi'\left(G_{t, u}\right)\left\langle \bm{z}^{[t]}_{S''}, \bm{z}^{[u]}_{S''} \right\rangle\right)
\end{align}
Plugging this approximation into approximate cavity expansion Eq.~\ref{si:eq:heuristic_cavity_decomposition_approximation_step_1_p2} yields:
\begin{align}
    \left\langle \prod_{j \in S''}z_j^{\left[\alpha_j\right]} \right\rangle_{\mu^{[2]}\left(\bm{z};\,\xi, S\right)} & \approx \sum_{\bm{z}_{S''} \in \{1, -1\}^{S'' \times \mathcal{T}_2}}\prod_{j \in S''}f\left(\bm{z}_j\right)\exp\left(-\frac{1}{2}\sum_{t, u \in \mathcal{T}_2}\Gamma_t\Gamma_u\xi'\left(G_{t, u}\right)z_j^{[t]}z_j^{[u]}\right)z_j^{\left[\alpha_j\right]},\label{si:eq:heuristic_cavity_decomposition_approximation_step_3_p2}
\end{align}
showing approximate independence of spins $S''$ under the path integral measure, now with layer indices up to absolute values $3$ (i.e., layer indices in $\mathcal{T}_2$). Summarizing, we started by observing that for layer indices of absolute value $1$, concentration of overlap under the path integral measure held trivially (as a consequence of Lemma~\ref{si:lemma:maximum_layer_index_1_monomial_expectation}). Using concentration of overlap at level $1$, we established approximate independence of spins for absolute layer index up to $2$ (Eq.~\ref{si:eq:heuristic_cavity_decomposition_approximation_step_3_p1}). By the low-order moment calculation argument from Section~\ref{si:sec:heuristic_cavity_argument}, we deduced from there concentration of overlaps for layer index bounded by $2$. Finally, from the latest cavity calculation, resulting in final estimate Eq.~\ref{si:eq:heuristic_cavity_decomposition_approximation_step_3_p2}, we deduced from overlap concentration at level $2$ approximate independence of spins of layer index up to $3$. This sketch of reasoning suggests that overlap concentration and approximate independence in spin index can be simultaneously established by an induction argument, instead of the circular self-consistent argument outlined in Section~\ref{si:sec:heuristic_cavity_argument}. More specifically, concentration of overlaps in layer indices $\mathcal{T}_{q - 1}$ yields approximate independence of spins with time layer indices in $\mathcal{T}_q$.

\section{Analysis of spin-glass QAOA via the cavity method: the rigorous argument}
\label{si:sec:qaoa_cavity_method_analysis}

In this Section, we develop a fully rigorous version of the cavity analysis of spin-glass QAOA expectation, heuristically skeched in Section~\ref{si:sec:qaoa_cavity_method_analysis_heuristic}. Section~\ref{si:sec:rigorous_argument_notation_outline} starts with a general outline of the derivation and relevant additional notation. The fundamental tool enabling the analysis is the cavity decomposition of the spin-glass QAOA path integral measure stated in Lemma~\ref{si:lemma:cavity_decomposition_path_integral_measure}. The analysis of this cavity expansion requires expanding cavity terms to a fixed Taylor order in overlaps. After abstracting the principle of this expansion in Section~\ref{si:sec:rigorous_argument_notation_outline}, Section~\ref{si:sec:non_colliding_taylor_expansion} derives explicit error bounds for this expansion. A crucial difference compared to the heuristic argument is that Taylor errors are expressed in terms of \textit{majorizing functions} as opposed to pointwise bounds. This stronger notion of functional error translates to scalar error estimates when passing to path integrals, even though path integrals are not genuine probability measures. Finally, Section~\ref{si:sec:noncolliding_taylor_expansion_application_spin_glass_qaoa_cavity_decomposition} applies the abstract Taylor expansion results from the previous section to the cavity expansion of QAOA, leading to a fully rigorous form of the heuristic argument outlined in Section~\ref{si:sec:heuristic_cavity_argument}.  The final argument is not only rigorous but quantitative, leading for instance to the following explicit error bound between the finite-size energy density and its thermodynamic limit:

\begin{corollary}[Upper-bounding finite size for target error on energy density]
\label{si:cor:finite_size_energy_density_error_bound}
Let $x > 0$ bound the derivatives of the spin glass mixture polynomial:
\begin{align}
    \xi(2), \xi'(1), \xi''(33) & \leq x
\end{align}
and further satisfy
\begin{align}
    \gamma^2p^2x & \geq 1
\end{align}
(which can always be satisfied by enlarging it if necessary). Then, for all $\varepsilon > 0$ satisfying technical assumption
\begin{align}
    \varepsilon & \leq \min\left\{e^{-1/2}, 2^{-(8p^2 + 3)/100}\right\}4\gamma p x,
\end{align}
and all instance size
\begin{align}
    n & \geq \max\left\{2^{8p^2 + 3}b(x)\left(\frac{2^{\mathrm{deg}(\xi) + 4}\gamma^2p^2x^2}{\varepsilon^2}\right)^{\left(\frac{2e(x)}{\log 2}\right)^{2p}}\left(2\log_2\left(\frac{4\gamma p x}{\varepsilon}\right) + \mathrm{deg}(\xi)\right)^2, \frac{16\gamma px}{\varepsilon}\right\},
\end{align}
the finite-size error on the energy density is at most $\varepsilon$: $\left|\nu_n - \nu_{\infty}\right| \leq \varepsilon$. In the above formula, $b(x) = 256p^4\left(2e\gamma^2p^2x\right)^p$ and $e(x) = 70\gamma^2p^2x + p\log 2$ are functions of upper bounds on $\xi$ derivatives.
\end{corollary}

\subsection{Notation and outline}
\label{si:sec:rigorous_argument_notation_outline}

In all this section, we consider a fixed spin-glass problem over a varying set $S$, characterized by random cost function:
\begin{align}
    C\left(\bm{\sigma}\right) & = \sum_{q \geq 1}c_q\left|S\right|^{-(q - 1)/2}\sum_{j_1, \ldots, j_q \in S}J^{(q)}_{j_1, \ldots, j_q}\sigma_{j_1} \ldots \sigma_{j_q}.
\end{align}
This spin glass problem is associated to mixture polynomial (independent of spin set $S$)
\begin{align}
    \xi\left(x\right) & := \sum_{q \geq 1}c_q^2x^q.
\end{align}
We study QAOA on this spin glass problem for a fixed number of layers $p$ and a fixed set of angles $\bm{\gamma} = \left(\gamma_1, \ldots, \gamma_p\right) \in \mathbb{R}^p$, $\bm{\beta} = \left(\beta_1, \ldots, \beta_p\right) \in \mathbb{R}^p$. For convenience, we introduce an upper bound $\gamma$ on all phase separator angles:
\begin{align}
    \left|\gamma_t\right| & \leq \gamma, \qquad \forall t \in [p].
\end{align}
Our goal is to make rigorous the heuristic cavity argument sketched in Section~\ref{si:sec:heuristic_cavity_argument} to arrive at the spin-glass QAOA $\bm{G}$ iteration. The iteration was already rigorously proven in Section~\ref{si:sec:spin_glass_qaoa_g_iteration} from a uniqueness result based on earlier work Ref.~\cite{qaoa_sparse_hypergraphs_spin_glass_models}. However, as hinted in the non-rigorous discussion, the cavity calculation further provides insight into the many-body structure of spin-glass QAOA, showing each constant-size set of qubits behaves as independent spin-boson system. Besides, we show the cavity calculation can provide concrete bounds between finite-size and infinite-size QAOA energy densities. While the bounds we obtain in final result Corollary~\ref{si:cor:finite_size_energy_density_error_bound} are daunting and of little practical use, we also suspect they are quite loose and call for improvement of our techniques.

The guiding thread of the proof consists in formalizing the induction argument sketched in Section~\ref{si:sec:heuristic_cavity_argument_p1}. Recalling approximate cavity decomposition Eq.~\ref{si:eq:heuristic_cavity_decomposition_approximation_step_1},
\begin{align}
    \left\langle \prod_{j \in S}z_j^{\left[\alpha_j\right]} \right\rangle_{\mu^{[p]}\left(\bm{z};\,\xi, S\right)} & \approx \sum_{\bm{z}_{S''} \in \{1, -1\}^{S'' \times \mathcal{T}_p}}\left\langle \mu^{[p]}_{C, 1}\left(\bm{z}_{S'}, \bm{z}_{S''};\,\xi_{s'/s}, S', S''\right) \right\rangle_{\mu^{[p - 1]}\left(\bm{z}_{S'};\,\xi_{s'/s}, S'\right)}\mu^{[p]}_B\left(\bm{z}_{S''}\,S''\right)\prod_{j \in S''}z_j^{\left[\alpha_j\right]},
\end{align}
the induction argument consisted in replacing the normalized $S'$ overlaps in
\begin{align}
    \mu^{[p]}_{C, 1}\left(\bm{z}_{S'}, \bm{z}_{S''};\,\xi_{s'/s}, S', S''\right) & = \exp\left(-\frac{1}{2}\sum_{t, u \in \mathcal{T}_{p - 1}}\Gamma_t\Gamma_u\xi'_{s'/s}\left(\frac{1}{\left|S'\right|}\left\langle \bm{z}^{[t]}_{S'}, \bm{z}^{[u]}_{S'} \right\rangle\right)\left\langle \bm{z}^{[t]}_{S''}, \bm{z}^{[u]}_{S''} \right\rangle\right)
\end{align}
by $\bm{G}$ matrix entries under subsystem $S'$ path integral measure:
\begin{align}
    \frac{1}{\left|S'\right|}\left\langle \bm{z}_{S'}^{[t]}, \bm{z}^{[u]}_{S'} \right\rangle & \approx G_{t, u}
\end{align}
The key observation is that spins in overlaps have layer indices in $\mathcal{T}_{p - 1}$ (bounded in absolute value by $p$), while the current induction step considers expectations of spins with layer indices in $\mathcal{T}_p$. Hence, we invoked the induction hypothesis to infer the distribution of the normalized overlap under the large subsystem's path integral measure. Informally, the induction hypothesis gives approximate independence in spin index of a constant-size spin set; from there, the law of large numbers implies the normalized overlaps behave as a deterministic random variable in the infinite-size limit ($\left|S'\right| \to \infty$, equivalent to $\left|S\right| \to \infty$). The delicate point is the constant size condition for approximate independence in the induction hypothesis, so one cannot rely on pairwise independence of all $S'$ spins. We address this difficulty by approximating $\mu^{[p]}_{C, 1}\left(\bm{z}_{S'}, \bm{z}_{S''};\,\xi_{s'/s}, S', S''\right)$ as a constant-degree polynomial in centered spin products $z_j^{[t]}z_j^{[u]}$, with $j \in S'$, $t, u \in \mathcal{T}_{p - 1}$. We then invoke the induction hypothesis to obtain that each monomial without spin index collision term, i.e. where each $j$ appears at most once, approximately converges to product of $\bm{G}$ entries under the path integral measure.

At a technical level, the polynomial approximation is obtained by Taylor expansion. Indeed, the magnitude of terms inside the exponential defining $\mu^{[p]}_{C, 1}$ is bounded by a constant independent of the large spin set size $\left|S'\right|$:
\begin{align}
    \left|-\frac{1}{2}\sum_{t, u \in \mathcal{T}_{p - 1}}\Gamma_t\Gamma_u\xi'_{s'/s}\left(\frac{1}{\left|S'\right|}\left\langle \bm{z}_{S'}^{[t]}, \bm{z}_{S'}^{[u]} \right\rangle\right)\left\langle \bm{z}_{S''}^{[t]}, \bm{z}_{S''}^{[u]} \right\rangle\right| & \leq \frac{\gamma^2}{2}\left|\mathcal{T}_{p - 1}\right|^2\xi'_{s'/s}\left(1\right)s''\nonumber\\
    & = 2\gamma^2p^2\xi'(1)s'',
\end{align}
where we introduced a bound $\gamma$ on the magnitude of phase separator angles. While the bound is independent of the large set size $\left|S'\right|$, it does depend on the small set size $\left|S''\right|$; since the latter is assumed of constant order, the same holds of the bound. Therefore, a Taylor expansion of constant order may achieve arbitrarily small error, uniformly in the choice of large spin set and small spin set (assuming a small spin set of bounded size). The previous reasoning on the magnitude of the Taylor expansion holds pointwise, while the estimate must be used under the path integral measure. Fortunately, we will see this step is fully justified by the generic boundedness property of path integral measures. Finally, since spin products converge to the $\bm{G}$ entries in the infinite-size limit, it is more convenient to work with centered spin products or overlaps, i.e.
\begin{align}
    z^{[t]}_jz^{[u]}_j - G_{t, u}, && \frac{1}{\left|S'\right|}\left\langle \bm{z}_{S'}^{[t]}, \bm{z}_{S'}^{[u]} \right\rangle.
\end{align}
The previous reasoning, bounding the Taylor expansion order to constant, continues to hold by translation change of variable inside polynomial $\xi'_{s'/s}$:
\begin{align}
    \mu^{[p]}_{C, 1}\left(\bm{z}_{S'}, \bm{z}_{S''};\,\xi, S', S''\right) & = \exp\left(-\frac{1}{2}\sum_{t, u \in \mathcal{T}_{p - 1}}\Gamma_t\Gamma_u\xi'_{s'/s}\left(\frac{1}{\left|S'\right|}\left\langle \bm{z}_{S'}^{[t]}, \bm{z}_{S'}^{[u]} \right\rangle\right)\left\langle \bm{z}^{[t]}_{S''}, \bm{z}^{[u]}_{S''} \right\rangle\right)\nonumber\\
    & = \exp\left(-\frac{1}{2}\sum_{t, u \in \mathcal{T}_{p - 1}}\Gamma_t\Gamma_u\sum_{m \geq 0}\xi_{s'/s}^{(1 + m)}\left(G_{t, u}\right)\left(\frac{1}{\left|S'\right|}\left\langle \bm{z}_{S'}^{[t]}, \bm{z}_{S'}^{[u]} \right\rangle - G_{t, u}\right)^m\left\langle \bm{z}_{S''}^{[t]}, \bm{z}_{S''}^{[u]} \right\rangle\right).
\end{align}
Since $\xi'_{s'/s}$ is a polynomial, the $m \geq 0$ summation is finite.

We abstract the desired Taylor expansion procedure as the problem of expanding a function of the form:
\begin{align}
    \psi: \left\{
    \begin{array}{ccc}
        \mathbb{C}^D & \longrightarrow & \mathbb{C}\\
        \bm{w} & \longmapsto & \varphi\left(\bm{S}\left(\bm{w}\right)\right)
    \end{array}
    \right.,
\end{align}
where $\bm{S} := \left(S_l\left(\bm{w}\right)\right)_{l \in L}$ is a collection of sums:
\begin{align}
    S_l\left(\bm{w}\right) & := \sum_{j \in J}f_{j, l}\left(\bm{w}\right),
\end{align}
with $f_{j, w}: \mathbb{C}^D \longrightarrow \mathbb{C}$. $D, J, L$ are finite index sets. The application to $\mu^{[p]}_{C, 1}\left(\bm{z}_{S'}, \bm{z}_{S''};\,\xi_{s'/s}, S', S'\right)$ is given by promoting the first bit matrix variable $\bm{z}_{S'} \in \{1, -1\}^{S \times \mathcal{T}_{p - 1}}$ to a complex matrix $\bm{w}_{S'} \in \mathbb{C}^{S' \times \mathcal{T}_{p - 1}}$, and letting:
\begin{align}
    D & := S' \times \mathcal{T}_{p - 1},\\
    J & := S',\\
    L & := \mathcal{T}_{p - 1} \times \mathcal{T}_{p - 1},\\
    f_{j,\,(t, u)}\left(\bm{w}\right) & := \left(w_j^{[t]}w_j^{[u]} - G_{t, u}\right)/\left|S'\right|,\\
    \varphi\left(\bm{x}\right) & := \exp\left(-\frac{1}{2}\sum_{t, u \in \mathcal{T}_{p - 1}}\Gamma_t\Gamma_u\sum_{m \geq 0}\xi^{\left(1 + m\right)}_{s'/s}\left(G_{t, u}\right)x^m_{t, u}\left\langle \bm{z}_{S''}^{[t]}, \bm{z}_{S''}^{[u]} \right\rangle\right).
\end{align}
Hence, sums $\bm{S} = \left(S_{t, u}\right)_{t, u \in \mathcal{T}_{p - 1}}$ are given by:
\begin{align}
    S_{t, u}\left(\bm{w}\right) & := \frac{1}{\left|S'\right|}\sum_{j \in S'}\left(w^{[t]}_jw^{[u]}_j - G_{t, u}\right)\\
    & = \frac{1}{\left|S'\right|}\left\langle \bm{w}_{S'}^{[t]}, \bm{w}_{S'}^{[u]} \right\rangle - G_{t, u}.
\end{align}
From these definitions, one can write
\begin{align}
    \mu^{[p]}_{C, 1}\left(\bm{w}_{S'}, \bm{z}_{S''};\,\xi_{s'/s}, S', S''\right) & = \varphi\left(\bm{S}\left(\bm{w}\right)\right).
\end{align}
Section~\ref{si:sec:non_colliding_taylor_expansion} develops Taylor approximations of $\psi\left(\bm{w}\right) = \varphi\left(\bm{S}\left(\bm{w}\right)\right)$ as a polynomial in indeterminates $f_{j, l}\left(\bm{w}\right)$. We further require the polynomial to have no collision term in index $j$ i.e. each $j$ can appear at most once in each monomial in indeterminates $f_{j, l}$. Coming back to the $\mu^{[p]}_{C, 1}\left(\bm{w}_{S'}, \bm{z}_{S''};\,\xi_{s'/s}, S', S''\right)$ expansion application, this means avoiding monomials of centered spin products with a repeated spin index $j \in S'$, for instance
\begin{align}
    \left(w_j^{[r]}w_j^{[s]} - G_{r, s}\right)\left(w_j^{[t]}w_j^{[u]} - G_{t, u}\right).
\end{align}
As the Taylor approximation will ultimately be used under a path integral measure, special care must be taken in bounding the remainder. In particular, a pointwise bound (namely, on all bit matrices $\bm{w}_{S'} \in \{1, -1\}^{S' \times \mathcal{T}_{p - 1}}$) is insufficient given the path integral measure is not a genuine probability measure. Our solution is to find a simple \textit{majorant} for discarded terms. A majorant, or majorizing function for a $d$-variate complex analytic function admitting a power series around $\bm{0}$:
\begin{align}
    f\left(\bm{w}\right) & = \sum_{\bm{m} \in \mathbb{N}^d}a_{\bm{m}}\prod_{j \in [d]}w_j^{m_j},
\end{align}
is a function with non-negative power series coefficients upper-bounding those of $f$:
\begin{align}
    g\left(\bm{w}\right) = \sum_{\bm{m} \in \mathbb{N}^d}b_{\bm{m}}\prod_{j \in [d]}w_j^{m_j}, && b_{\bm{m}} & \geq 0, && \left|a_{\bm{m}}\right| \leq b_{\bm{m}}.
\end{align}
We denote $f \preceq g$ to indicate $g$ majorizes $f$. In this work, we will frequently denote $\overline{f}$ (not to be confused with complex conjugation) for a majorant of $f$. Note majorants are not unique ---for instance, $f \preceq g \implies f \preceq 2g$. The interest of majorant is their compatibility with path integral measure expectations, as justified in the following Lemma:

\begin{lemma}[Bounding functions under path integral measures from majorant]
\label{si:lemma:path_integral_expectation_bound_from_majorant}
Let $\mu^{[p]}\left(\bm{z};\,S\right)$ a $p$-layer unitary quantum circuit path integral measure over qubit set $S$ as specified in Definition~\ref{si:def:layered_quantum_circuit_path_integral_measure}. Let $f\left(\bm{w}\right)$, $\bm{w} \in \mathbb{C}^{S \times \mathcal{T}_p}$, a multivariate analytic function of a matrix that is analytic over an open set containing closed polydisk $\overline{B}(0, 1)^{S' \times \mathcal{T}_p}$, and let us assume $f$ admits a majorant $\overline{f}$ satisfying the same analyticity condition. Then, the path integral expectation of $f\left(\bm{z}\right)$ over bit matrices $\bm{z} \in \{1, -1\}^{S \times \mathcal{T}_p}$ is upper-bounded by the majorant evaluated at the all-$1$ matrix:
\begin{align}
    \left|\left\langle f\left(\bm{z}\right) \right\rangle_{\mu^{[p]}\left(\bm{z};\,S\right)}\right| & \leq \overline{f}\left(\bm{1}_{S \times \mathcal{T}_p}\right).
\end{align}
\begin{proof}
We express the path integral expectation by linearity using the power series expansion of $f$:
\begin{align}
    \left\langle f\left(\bm{z}\right) \right\rangle_{\mu^{[p]}\left(\bm{z};\,S\right)} & =  \left\langle \sum_{\bm{m} \in \mathbb{N}^{S \times \mathcal{T}_p}}a_{\bm{m}}\prod_{(j, t) \in S \times \mathcal{T}_p}\left(z_j^{[t]}\right)^{m_j^{[t]}} \right\rangle_{\mu^{[p]}\left(\bm{z};\,S\right)}\nonumber\\
    & = \sum_{\bm{m} \in \mathbb{N}^{S \times \mathcal{T}_p}}a_{\bm{m}}\left\langle \prod_{(j, t) \in S \times \mathcal{T}_p}\left(z_j^{[t]}\right)^{m_j^{[t]}} \right\rangle_{\mu^{[p]}\left(\bm{z};\,S\right)},
\end{align}
where the inversion between path integral expectation and series is legitimate, given the path integral expectation is a finite sum over bit configurations, whose interchange with the convergent power series preserves absolute and uniform convergence. Each path integral in the above sum reduces to a that of a monomial of spin variables; by Corollary~\ref{si:cor:path_integral_measure_boundedness_monomials}, these are bounded by $1$:
\begin{align}
    \left|\left\langle \prod_{(j, t) \in S \times \mathcal{T}_p}\left(z_j^{[t]}\right)^{m_j^{[t]}} \right\rangle\right| & \leq 1.
\end{align}
Then, for any majorant
\begin{align}
    \overline{f}\left(\bm{w}\right) & := \sum_{\bm{m} \in \mathbb{N}^{S \times \mathcal{T}_p}}b_{\bm{m}}\prod_{(j, t) \in S \times \mathcal{T}_p}\left(w_j^{[t]}\right)^{m_j^{[t]}},
\end{align}
one can write:
\begin{align}
    \left|\left\langle f\left(\bm{z}\right) \right\rangle_{\mu^{[p]}\left(\bm{z};\,S\right)}\right| & \leq \sum_{\bm{m} \in \mathbb{N}^{S \times \mathcal{T}_p}}\left|a_{\bm{m}}\right|\nonumber\\
    & \leq \sum_{\bm{m} \in \mathbb{N}^{S \times \mathcal{T}_p}}b_{\bm{m}}\nonumber\\
    & = \sum_{\bm{m} \in \mathbb{N}^{S \times \mathcal{T}_p}}b_{\bm{m}}\prod_{(j, t) \in S \times \mathcal{T}_p}1^{m_j^{[t]}}\nonumber\\
    & = \overline{f}(\bm{1}_{S \times \mathcal{T}_p}).
\end{align}
\end{proof}
\end{lemma}
Lemma~\ref{si:lemma:path_integral_expectation_bound_from_majorant} provides a rigorous way of approximating a path integral expectation given approximations of the underlying functions. Namely, given two ``close'' functions of a bit matrix $f\left(\bm{z}\right)$, $g\left(\bm{z}\right)$, with difference majorized as $f\left(\bm{w}\right) - g\left(\bm{w}\right) \preceq h\left(\bm{w}\right)$, one may write
\begin{align}
    \left|\left\langle f\left(\bm{w}\right) \right\rangle_{\mu^{[p]}\left(\bm{z};\,S\right)} - \left\langle g\left(\bm{w}\right) \right\rangle_{\mu^{[p]}\left(\bm{z};\,S\right)}\right| & \leq h\left(\bm{1}_{S \times \mathcal{T}_p}\right)
\end{align}

\subsection{Non-colliding Taylor expansion under quantum circuit path integral measure}
\label{si:sec:non_colliding_taylor_expansion}

This Section develops the main technical ingredient of the rigorous version of the cavity argument. Namely, we consider a function of the form
\begin{align}
    \psi:\left\{\begin{array}{ccc}
        \mathbb{C}^D & \longrightarrow & \mathbb{C}\\
        \bm{w} & \longmapsto & \varphi\left(\bm{S}\left(\bm{w}\right)\right)
    \end{array}\right.,
\end{align}
where $\varphi: \mathbb{C}^L \longrightarrow \mathbb{C}$,
\begin{align}
    \bm{S}: & \left\{\begin{array}{ccc}
        \mathbb{C}^D & \longrightarrow & \mathbb{C}^L\\
        \bm{w} & \longmapsto & \left(S_l\left(\bm{w}\right)\right)_{l \in L}
    \end{array}\right.,\\
    S_l\left(\bm{w}\right) & := \sum_{j \in J}f_{j, l}\left(\bm{w}\right),
\end{align}
$f_{j, l}: \mathbb{C}^D \longrightarrow \mathbb{C}$, and index sets $D, J, L$ are finite. The application of this functional form to the cavity calculation of QAOA is explained in Section~\ref{si:sec:rigorous_argument_notation_outline}. We seek to approximate $\psi\left(\bm{w}\right) = \varphi\left(\bm{S}\left(\bm{w}\right)\right)$ as a low-degree polynomial in indeterminates $f_{j, l}\left(\bm{w}\right)$ defining sums $\bm{S}\left(\bm{w}\right)$. We further require this low-degree polynomial to be collision-free in index $j$, i.e. each index $j$ to appear at most once among the $f_{j, l}\left(\bm{w}\right)$ making each monomial.

The main idea to derive such an expansion is to define an interpolating function:
\begin{align}
    \psi_t\left(\bm{w}\right) & := \varphi\left(t\bm{S}\left(\bm{w}\right)\right), \qquad t \in [0, 1],
\end{align}
such that $\psi_0\left(\bm{w}\right) = \varphi\left(\bm{0}\right)$ and $\psi_1\left(\bm{w}\right) = \varphi\left(\bm{S}\left(\bm{w}\right)\right)$. We then relate $\psi_1\left(\bm{w}\right)$ to $\psi_0\left(\bm{w}\right)$ by repeated differentiation over $t$ and integration, similar to a standard proof technique for the ordinary Taylor expansion formula with integral remainder. An important difference with the usual Taylor formula is that we seek an approximation as a polynomial in indeterminates $f_{j, l}\left(\bm{w}\right)$ rather than variable $\bm{w}$. Another important difference is that we will remove collision terms in index $j \in J$ while gradually increasing the order of expansion. Illustrating the lowest-order step,
\begin{align}
    \psi\left(\bm{w}\right) & = \psi_1\left(\bm{w}\right)\nonumber\\
    & = \psi_0\left(\bm{w}\right) + \int_0^1\!\mathrm{d}t\,\frac{\mathrm{d}\psi_t\left(\bm{w}\right)}{\mathrm{d}t}\nonumber\\
    & = \varphi\left(\bm{0}\right) + \int_0^1\!\mathrm{d}t\,\sum_{l \in L}S_l\left(\bm{w}\right)\partial_l\varphi\left(t\bm{S}\left(\bm{w}\right)\right)\nonumber\\
    & = \varphi\left(\bm{0}\right) + \int_0^1\!\mathrm{d}t\,\sum_{\substack{j \in J\\l \in L}}f_{j, l}\left(\bm{w}\right)\partial_l\varphi\left(t\bm{S}\left(\bm{w}\right)\right).
\end{align}
Approximating the $\varphi$ derivative by its $0$ value would give an affine approximation of $\psi\left(\bm{w}\right)$ in indeterminates $f_{j, l}\left(\bm{w}\right)$:
\begin{align}
    \psi\left(\bm{w}\right) & \approx \varphi\left(\bm{0}\right) + \sum_{\substack{j \in J\\l \in L}}f_{j, l}\left(\bm{w}\right).
\end{align}
For our application, we will only be able to discard higher order derivatives for a satisfying approximation error, and we therefore need to iterate the above expansion. Namely, going to the next order:
\begin{align}
    \psi\left(\bm{w}\right) & = \varphi\left(\bm{0}\right) + \int_0^1\!\mathrm{d}t_1\,\sum_{\substack{j_1 \in J\\l_1 \in L}}f_{j_1, l_1}\left(\bm{w}\right)\partial_{l_1}\varphi\left(t_1\bm{S}\left(\bm{w}\right)\right)\nonumber\\
    & =  \varphi\left(\bm{0}\right) + \int_0^1\!\mathrm{d}t_1\,\sum_{\substack{j_1 \in J\\l_1 \in L}}f_{j_1, l_1}\left(\bm{w}\right)\left(\partial_{l_1}\varphi\left(\bm{0}\right) + \int_0^1\!\mathrm{d}t_2\,\sum_{\substack{j_2 \in J\\l_2 \in L}}f_{j_2, l_2}\left(\bm{w}\right)t_1\partial_{l_1l_2}\varphi\left(t_1t_2\bm{S}\left(\bm{w}\right)\right)\right)\nonumber\\
    & = \varphi\left(\bm{0}\right) + \sum_{\substack{j_1 \in J\\l_1 \in L}}\partial_{l_1}\varphi\left(\bm{0}\right)f_{j_1, l_1}\left(\bm{w}\right) + \int_0^1\!\mathrm{d}t_1\int_0^1\!\mathrm{d}t_2\,\sum_{\substack{j_1, j_2 \in J\\l_1, l_2 \in L}}\partial_{l_1l_2}\varphi\left(t_1t_2\bm{S}\left(\bm{w}\right)\right)f_{j_1, l_1}\left(\bm{w}\right)f_{j_2, l_2}\left(\bm{w}\right).
\end{align}
We see the quadratic term in indeterminates $f_{j, l}\left(\bm{w}\right)$ includes collision terms $j_1 = j_2$. These were pulled down from $\partial_{l_1}\varphi\left(t_1t_2\bm{S}\left(\bm{w}\right)\right)$ when taking the $t_2$ derivative. To avoid these, we will remove all $f$ with first index $j_1$ from inside $\varphi$ before iterating the derivative. The following two Lemmas quantify this approximation error in terms of majorizing function:

\begin{lemma}[Majorizing error of removing sum term inside non-linear function]
\label{si:lemma:remove_single_sum_term_nonlinear_function}
Let $d \geq 1$ an integer, $D, J, L$ finite index sets and a family of analytic functions $\bm{f} = \left(f_{j, l}\right)_{j \in J, l \in L}$, $f_{j, l}: \mathbb{C}^D \longrightarrow \mathbb{C}$, indexed by $J \times L$. Let further be given additional such function $g: \mathbb{C}^D \longrightarrow \mathbb{C}$, and finally an analytic function $\varphi: \mathbb{C}^L \longrightarrow \mathbb{C}$. From these analytic functions, define new analytic functions $\psi_0, \psi_1: \mathbb{C}^D \longrightarrow \mathbb{C}$ by:
\begin{align}
    \psi_0\left(\bm{z}\right) & := \varphi\left(\left(\sum_{j \in J}f_{j, l}\left(
    \bm{z}\right)\right)_{l \in L}\right),\\
    \psi_1\left(\bm{z}\right) & := \varphi\left(\left(\sum_{j \in J}f_{j, l}\left(
    \bm{z}\right) + \delta_{l, l_0}g\left(\bm{z}\right)\right)_{l \in L}\right),
\end{align}
for some $l_0 \in L$. Assume analytic functions $f_{j, l}, g, \partial_l\varphi$ ($j \in J, l \in L$) admit majorants, denoted as follows:
\begin{align}
    f_{j, l} & \preceq \overline{f_{j, l}}, && g \preceq \overline{g}, && \partial_l\varphi \preceq \overline{\partial_l\varphi} \quad \forall l \in L.
\end{align}
Then, the following majorization bounds holds between $\psi_0$ and $\psi_1$:
\begin{align}
    \psi_1\left(\bm{z}\right) - \psi_0\left(\bm{z}\right) & \preceq \overline{g}\left(\bm{z}\right)
    \overline{\partial_{l_0}\varphi}\left(\left(\sum_{j \in J}\overline{f_{j, l}}\left(\bm{z}\right) + \delta_{l, l_0}\overline{g}\left(\bm{z}\right)\right)_{l \in L}\right).
\end{align}
\begin{proof}
We introduce the following vectorized notations for families of functions $\bm{f} = \left(f_{j, l}\right)_{j \in J, l \in L}$, $f_{j, l}: \mathbb{C}^D \longrightarrow \mathbb{C}$ indexed by $J \times L$. Let:
\begin{align}
    S_{\bm{f}, l}\left(\bm{z}\right) & := \sum_{j \in J}f_{j, l}\left(\bm{z}\right),\\
    S_{\bm{f}}\left(\bm{z}\right) & := \left(S_{\bm{f}, l}\left(\bm{z}\right)\right)_{l \in L}.
\end{align}
Hence, $S_{\bm{f}}: \mathbb{C}^D \longrightarrow \mathbb{C}^L$. We define:
\begin{align}
    \psi_t\left(\bm{z}\right) & := \varphi\left(S_{\bm{h}^t}\right),
\end{align}
where for all $t \in [0, 1]$, $(J \times L)$-indexed family of functions $\bm{h}^t = \left(h^t_{j, l}\right)_{j \in J, l \in L}$ is defined by:
\begin{align}
    h^t_{j, l}\left(\bm{z}\right) & := f_{j, l}\left(\bm{z}\right) + \frac{1}{|J|}\delta_{l, l_0}tg\left(\bm{z}\right).
\end{align}
The Lemma's statement then amounts to majorize difference
\begin{align}
    \psi_1\left(\bm{z}\right) - \psi_0\left(\bm{z}\right),
\end{align}
seen as a function of complex variables $\bm{z} \in \mathbb{C}^D$. We express this difference by computing the time derivative of $\psi_t$. By the chain rule, it holds:
\begin{align}
    \frac{\mathrm{d}\psi_t\left(\bm{z}\right)}{\mathrm{d}t} & = \frac{\mathrm{d} S_{\bm{h}^t}\left(\bm{z}\right)}{\mathrm{d}t}\partial_{l_0}\varphi\left(S_{\bm{h}^t}\left(\bm{z}\right)\right).
\end{align}
The $t$ derivative is simply:
\begin{align}
    \frac{\mathrm{d}S_{\bm{h}^t}\left(\bm{z}\right)}{\mathrm{d}t}\left(\bm{z}\right) & = \frac{\mathrm{d}}{\mathrm{d}t}\left(\sum_{j \in J}f_{j, l_0}\left(\bm{z}\right) + tg\left(\bm{z}\right)\right)\\
    & = g\left(\bm{z}\right).
\end{align}
Hence, we obtained the following final expression for the time derivative of $\psi_t$:
\begin{align}
    \frac{\mathrm{d}\psi_t\left(\bm{z}\right)}{\mathrm{d}t} & = g\left(\bm{z}\right)\partial_{l_0}\varphi\left(S_{\bm{h}^t}\left(\bm{z}\right)\right).
\end{align}
Integrating this equality:
\begin{align}
    \psi_1\left(\bm{z}\right) - \psi_0\left(\bm{z}\right) & = g\left(\bm{z}\right)\int_0^1\!\mathrm{d}t\,\partial_{l_0}\varphi\left(S_{\bm{h}^t}\left(\bm{z}\right)\right).
\end{align}
Noting elementary majorization bound:
\begin{align}
    S_{\bm{h}^t}\left(\bm{z}\right) & \preceq \sum_{j \in J}\overline{f_{j, l}}\left(\bm{z}\right) + \delta_{l, l_0}t\overline{g}\left(\bm{z}\right)\nonumber\\
    & \preceq \sum_{j \in J}\overline{f_{j, l}}\left(\bm{z}\right) + \delta_{l, l_0}\overline{g}\left(\bm{z}\right)\nonumber\\
    & =: \overline{S_{\bm{h}^1}}\left(\bm{z}\right),
\end{align}
and using composition of majorization bounds, the above integral representation yields the majorization bound:
\begin{align}
    \psi_1\left(\bm{z}\right) - \psi_0\left(\bm{z}\right) & \preceq \overline{g}\left(\bm{z}\right)\overline{\partial_{l_0}\varphi}\left(\overline{S_{\bm{h}^t}}\left(\bm{z}\right)\right),
\end{align}
The convergence hypothesis of Proposition~\ref{si:prop:majorization_composition} is met here even though the outer majorant $\overline{\partial_{l_0}\varphi}$ is not entire (it has poles from the Cauchy-estimate factors of Proposition~\ref{si:prop:majorization_entire_function}): the composition is ultimately evaluated at the all-ones matrix, where $\overline{S_{\bm{h}^t}}\left(\bm{1}\right)$ lies within the polydisk of convergence of $\overline{\partial_{l_0}\varphi}$.
where
\begin{align}
    \overline{S_{\bm{h}^t}}\left(\bm{z}\right) & := \left(\overline{S_{\bm{h}^t, l}}\left(\bm{z}\right)\right)_{l \in L}
\end{align}
collects the element-wise absolute bounds over $S_{\bm{h}^t} = \left(S_{\bm{h}^t, l}\right)_{l \in L}$.
\end{proof}
\end{lemma}

\begin{lemma}[Non-colliding Taylor expansion of function of sums of functions, first-order]
\label{si:lemma:remove_collision_terms_nonlinear_function}
Let be given a family $\left(f_{j, l}\right)_{j \in J,\,l \in L}$ of analytic functions $\mathbb{C}^d \longrightarrow \mathbb{C}$, indexed by finite sets $J$ and $L$, and with corresponding majorants $\left(\overline{f_{j, l}}\right)_{j \in J,\,l \in L}$. Let $\varphi$ an analytic function $\mathbb{C}^L \longrightarrow \mathbb{C}$, with second derivatives $\left(\partial_{lm}\varphi\right)_{l, m \in L}$ majorized by $\left(\overline{\partial_{lm}\varphi}\right)_{l, m \in L}$. Let us define shorthand notations:
\begin{align}
    S_l\left(\bm{z}\right) & := \sum_{j \in J}f_{j, l}\left(\bm{z}\right),\\
    \hat{S}_{j, l}\left(\bm{z}\right) & := \sum_{k \in J - \{j\}}f_{k, l}\left(\bm{z}\right), \qquad j \in J.
\end{align}
By the majorization assumption on functions $f_{j, l}$, $S_l$ admits the following majorant:
\begin{align}
    \overline{S}_l & := \sum_{j \in J}\overline{f_{j, l}}.
\end{align}
For convenience, we also define the following vectorized notations over $l \in L$:
\begin{align}
    \bm{S}\left(\bm{z}\right) & := \left(S_l\left(\bm{z}\right)\right)_{l \in L},\\
    \bm{\hat{S}}_j\left(\bm{z}\right) & := \left(\hat{S}_{j, l}\left(\bm{z}\right)\right)_{l \in L},\\
    \bm{\overline{S}}\left(\bm{z}\right) & := \left(\overline{S}_l\left(\bm{z}\right)\right)_{l \in L}.\label{si:eq:s_majorant_definition}
\end{align}
Then, the following majorization estimate holds for $\varphi\left(\bm{S}\left(\bm{z}\right)\right)$:
\begin{align}
    \varphi\left(\bm{S}\left(\bm{z}\right)\right) - \varphi\left(\bm{0}_L\right) - \int_0^1\!\mathrm{d}t\,\sum_{\substack{j \in J\\l \in L}}f_{j, l}\left(\bm{z}\right)\partial_l\varphi\left(t\bm{\hat{S}}_j\left(\bm{z}\right)\right) & \preceq \sum_{\substack{j \in J\\l, r \in L}}\overline{f_{j, l}}\left(\bm{z}\right)\overline{f_{j, r}}\left(\bm{z}\right)\overline{\partial_{lr}\varphi}\left(\bm{\overline{S}}\left(\bm{z}\right)\right).\label{si:eq:remove_collision_terms_nonlinear_function}
\end{align}
    Importantly, the integral term on the left-hand side has no ``collision term", i.e. for all $j \in J$, $f_{j, l}$ does not occur in $\bm{\hat{S}}_j$. Assuming majorants $\left(\overline{\partial_l\varphi}\right)_{l \in L}$ for the first derivatives of $\varphi$, the previous majorization bound implies the following weaker one:
\begin{align}
    \varphi\left(S\left(\bm{z}\right)\right) - \varphi\left(S\left(\bm{0}\right)\right) & \preceq \sum_{\substack{j \in J\\l \in L}}\overline{f_{j, l}}\left(\bm{z}\right)\overline{\partial_l\varphi}\left(\bm{\overline{S}}\left(\bm{z}\right)\right).
\end{align}
\begin{proof}
Let:
\begin{align}
    \psi_t\left(\bm{z}\right) & := \varphi\left(t\left(\sum_{j \in J}f_{j, l}\left(\bm{z}\right)\right)_{l \in L}\right).
\end{align}
We compute:
\begin{align}
    \chi_t\left(\bm{z}\right) & := \frac{\mathrm{d}\psi_t}{\mathrm{d}t}\left(\bm{z}\right)\nonumber\\
    & = \frac{\mathrm{d}}{\mathrm{d}t}\varphi\left(t\bm{S}\left(\bm{z}\right)\right)\nonumber\\
    & = \sum_{l \in L}S_l\left(\bm{z}\right)\partial_l\varphi\left(t\bm{S}\left(\bm{z}\right)\right)\nonumber\\
    & = \sum_{\substack{j \in J\\l \in L}}f_{j, l}\left(\bm{z}\right)\partial_l\varphi\left(t\bm{S}\left(\bm{z}\right)\right).\label{si:eq:remove_collision_terms_nonlinear_function_first_derivative_expression}
\end{align}
for all $j \in J, l \in L$, we now aim to replace $\bm{S}\left(\bm{z}\right)$ by $\bm{\hat{S}}_j\left(\bm{z}\right)$ inside $\partial_l\varphi$. That is, we aim to replace $S_l$ by $\hat{S}_{j, l}$. We perform this sequentially in elements of $L$, denoted without loss of generality $L := \left\{1, 2, \ldots, |L| - 1, |L|\right\}$. Fixing arbitrary $j \in J, l \in L$ in the outer sum, we will prove sequentially on the $L$ element index $r$ that:
\begin{align}
    \partial_l\varphi\left(t\bm{\hat{S}}_{j,\,1:r}\left(\bm{z}\right), t\bm{S}_{r + 1:|L|}\left(\bm{z}\right)\right) - \partial_l\varphi\left(t\bm{\hat{S}}_{j,\,1:r - 1}\left(\bm{z}\right), t\bm{S}_{r:|L|}\left(\bm{z}\right)\right) & \preceq t\overline{\partial_{lr}\varphi}\left(t\bm{\overline{S}}\left(\bm{z}\right)\right),\label{si:eq:remove_collision_terms_nonlinear_function_single_argument_removal_bound}
\end{align}
where we introduced for all $1 \leq r \leq s \leq |L|$:
\begin{align}
    \bm{\hat{S}}_{j,\,r:s}\left(\bm{z}\right) & := \left(\hat{S}_{j, r}\left(\bm{z}\right), \hat{S}_{j, r + 1}\left(\bm{z}\right), \ldots, \hat{S}_{j, s - 1}\left(\bm{z}\right) + \hat{S}_{j, s}\left(\bm{z}\right)\right),\\
    \bm{S}_{r:s}\left(\bm{z}\right) & := \left(S_r\left(\bm{z}\right), S_{r + 1}\left(\bm{z}\right), \ldots, S_{s - 1}\left(\bm{z}\right), S_s\left(\bm{z}\right)\right),
\end{align}
and
\begin{align}
    \bm{\overline{S}}\left(\bm{z}\right) & := \left(S_l\left(\bm{z}\right)\right)_{l \in L},\\
    \overline{S}_l\left(\bm{z}\right) & := \sum_{j \in J}\overline{f_{j, l}}\left(\bm{z}\right).
\end{align}

Let us produce an absolute bound of:
\begin{align}
    & \partial_l\varphi\left(t\bm{\hat{S}}_{j,\,1:r}\left(\bm{z}\right), t\bm{S}_{r + 1:|L|}\left(\bm{z}\right)\right) - \partial_l\varphi\left(t\bm{\hat{S}}_{j,\,1:r - 1}\left(\bm{z}\right), t\bm{S}_{r:|L|}\left(\bm{z}\right)\right)\nonumber\\
    & = \partial_l\varphi\left(t\bm{\hat{S}}_{j,\,1:r - 1}\left(\bm{z}\right), t\hat{S}_{j,\,r}\left(\bm{z}\right), t\bm{S}_{r + 1:|L|}\left(\bm{z}\right)\right) - \partial_l\varphi\left(t\bm{\hat{S}}_{j,\,1:r - 1}, tS_r\left(\bm{z}\right), t\bm{S}_{r + 1:|L|}\left(\bm{z}\right)\right)
\end{align}
for an arbitrary $L$ element index $r$. Noting that
\begin{align}
    tS_r\left(\bm{z}\right) & = t\hat{S}_{j, r}\left(\bm{z}\right) + tf_{j, r}\left(\bm{z}\right),
\end{align}
we apply Lemma~\ref{si:lemma:remove_single_sum_term_nonlinear_function} with parameters:
\begin{align}
    J & \longleftarrow J,\\
    L & \longleftarrow L = \left\{1, 2, \ldots, |L| - 1, |L|\right\},\\
    \varphi & \longleftarrow \partial_l\varphi,\\
    f_{k, l} & \longleftarrow tf_{k, l} && \textrm{ for } k \in J - \{j\}, l \in L,\\
    f_{j, l} & \longleftarrow 0 && \textrm{ for } 1 \leq l \leq r,\\
    f_{j, l} & \longleftarrow tf_{j, l} && \textrm{ for } r + 1 \leq l \leq |L|,\\
    g & \longleftarrow tf_{j, r},\\
    l_0 & \longleftarrow r.
\end{align}
This produces absolute bound:
\begin{align}
    & \partial_l\varphi\left(t\bm{\hat{S}}_{j,\,1:r - 1}\left(\bm{z}\right), t\hat{S}_{j,\,r}\left(\bm{z}\right), t\bm{S}_{r + 1:|L|}\left(\bm{z}\right)\right) - \partial_l\varphi\left(t\bm{\hat{S}}_{j,\,1:r - 1}, tS_r\left(\bm{z}\right), t\bm{S}_{r + 1:|L|}\left(\bm{z}\right)\right)\nonumber\\
    & \preceq t\overline{f_{j, r}}\left(\bm{z}\right)\overline{\partial_{lr}\varphi}\left(t\bm{\overline{\hat{S}}}_{j,\,1:r - 1}\left(\bm{z}\right), t\overline{\hat{S}_{j,\,r}}\left(\bm{z}\right) + t\overline{f_{j, r}}\left(\bm{z}\right), t\bm{\overline{S}}_{r + 1:|L|}\right)\nonumber\\
    & \preceq t\overline{f_{j, r}}\left(\bm{z}\right)\overline{\partial_{lr}\varphi}\left(t\bm{\overline{\hat{S}}}_{j,\,1:r - 1}\left(\bm{z}\right), t\overline{S_r}\left(\bm{z}\right), t\bm{\overline{S}}_{r + 1:|L|}\right)\nonumber\\
    & \preceq t\overline{f_{j, r}}\left(\bm{z}\right)\overline{\partial_{lr}\varphi}\left(t\bm{\overline{S}}_{1:r - 1}\left(\bm{z}\right), t\overline{S_r}\left(\bm{z}\right), t\overline{S}_{r + 1:|L|}\left(\bm{z}\right)\right)\nonumber\\
    & = t\overline{f_{j, r}}\left(\bm{z}\right)\overline{\partial_{lr}\varphi}\left(t\overline{\bm{S}}\left(\bm{z}\right)\right).
\end{align}
Telescopically summing absolute bound Eq.~\ref{si:eq:remove_collision_terms_nonlinear_function_single_argument_removal_bound} over $r = 1, \ldots, |L|$ gives final bound:
\begin{align}
    \partial_l\varphi\left(t\bm{\hat{S}}_j\left(\bm{z}\right)\right) - \partial_l\varphi\left(t\bm{S}\left(\bm{z}\right)\right) & \preceq \sum_{r \in L}t\overline{f_{j, r}}\left(\bm{z}\right)\overline{\partial_{lr}\varphi}\left(t\bm{\overline{S}}\left(\bm{z}\right)\right).
\end{align}
Plugging this absolute bound estimate into the $t$ derivative $\chi_t$ of $\psi_t$ in Eq.~\ref{si:eq:remove_collision_terms_nonlinear_function_first_derivative_expression} yields the following majorization bound on this derivative:
\begin{align}
    \chi_t\left(\bm{z}\right) - \sum_{\substack{j \in J\\l \in L}}f_{j, l}\left(\bm{z}\right)\partial_l\varphi\left(t\bm{\hat{S}}_j\left(\bm{z}\right)\right) & = \sum_{\substack{j \in J\\l \in L}}f_{j, l}\left(\bm{z}\right)\left(\partial_l\varphi\left(t\bm{S}\left(\bm{z}\right)\right) - \partial_l\varphi\left(t\bm{\hat{S}}_j\left(\bm{z}\right)\right)\right)\nonumber\\
    & \preceq \sum_{\substack{j \in J\\l, r \in L}}t\overline{f_{j, l}}\left(\bm{z}\right)\overline{f_{j, r}}\left(\bm{z}\right)\overline{\partial_{lr}\varphi}\left(t\overline{\bm{S}}\left(\bm{z}\right)\right).
\end{align}
Integrating this over $t \in [0, 1]$,
\begin{align}
    \psi_1\left(\bm{z}\right) - \psi_0\left(\bm{z}\right) - \int_0^1\!\mathrm{d}t\,\sum_{\substack{j \in J\\l \in L}}f_{j, l}\left(\bm{z}\right)\partial_l\varphi\left(t\bm{\hat{S}}_j\left(\bm{z}\right)\right) & \preceq \int_0^1\!\mathrm{d}t\,\sum_{\substack{j \in J\\l, r \in L}}t\overline{f_{j, l}}\left(\bm{z}\right)\overline{f_{j, r}}\left(\bm{z}\right)\overline{\partial_{lr}\varphi}\left(t\overline{\bm{S}}\left(\bm{z}\right)\right)\nonumber\\
    & \preceq \sum_{\substack{j \in J\\l, r \in L}}\overline{f_{j, l}}\left(\bm{z}\right)\overline{f_{j, r}}\left(\bm{z}\right)\overline{\partial_{lr}\varphi}\left(\bm{\overline{S}}\left(\bm{z}\right)\right).
\end{align}
\end{proof}
\end{lemma}

Lemma~\ref{si:lemma:remove_collision_terms_nonlinear_function} performed a first-order Taylor expansion of $\varphi\left(S\left(\bm{z}\right)\right)$ in indeterminates $f_{j, l}\left(\bm{z}\right)$, where for each linear term $f_{j, l}\left(\bm{z}\right)$, $f_{j, l}\left(\bm{z}\right)$ is removed from inside the derivative of $\varphi$ multiplying this term:
\begin{align}
    f_{j, l}\left(\bm{z}\right)\partial_l\varphi\left(t\bm{S}\left(\bm{z}\right)\right) & \longrightarrow f_{j, l}\left(\bm{z}\right)\partial_l\varphi\left(t\bm{\hat{S}}_j\left(\bm{z}\right)\right).
\end{align}
As discussed in the Section's introduction, this ensures that no collision terms in $j$ will be pulled out of $\varphi$ when iterating the first-order expansion. Recalling the interpretation of $j$ as the spin index (see outline from Section~\ref{si:sec:rigorous_argument_notation_outline}), the absence of collision means the Taylor polynomial will present as a product of degree $1$ indeterminates $f_{j, (r, s)}\left(z^{[r]}_jz^{[s]}_j - G_{r, s}\right)$ pertaining to different spins $j$; by the heuristic approximate independence argument (Sections~\ref{si:sec:heuristic_cavity_argument}, \ref{si:sec:heuristic_cavity_argument_p1}), the path integral expectation of such a monomial is expected to vanish in the infinite-size limit. Note the error from removal of collision terms in Lemma~\ref{si:lemma:remove_collision_terms_nonlinear_function} was bounded in terms of a majorant rather than pointwise, which will make it suitable for path integration. The following Lemma iterates the single-step expansion and trimming of collision terms to obtain an approximate expansion of arbitrary order:

\begin{lemma}[Non-colliding Taylor expansion of function of sums of functions, higher-order]
\label{si:lemma:remove_collision_terms_nonlinear_function_iterative}
Let be given a family $\left(f_{j, l}\right)_{j \in J,\,l \in L}$ of analytic functions $\mathbb{C}^D \longrightarrow \mathbb{C}$, indexed by finite sets $J$ and $L$, and with corresponding majorants $\left(\overline{f_{j, l}}\right)_{j \in J,\,l \in L}$. Let $\varphi: \mathbb{C}^L \longrightarrow \mathbb{C}$ an analytic function, with derivatives of order $d$ majorized by $\left(\overline{\partial_{l_1, \ldots, l_d}\varphi}\right)_{l_1, \ldots, l_d \in L}$. Let us define shorthand notations:
\begin{align}
    S_l\left(\bm{w}\right) & := \sum_{j \in J}f_{j, l}\left(\bm{w}\right), && l \in L,\\
    S_{K, l}\left(\bm{w}\right) & := \sum_{j \in J - K}f_{j, l}\left(\bm{w}\right), && K \subset J, l \in L.
\end{align}
For convenience, we also define the following vectorized notations over $l \in L$:
\begin{align}
    \bm{S}\left(\bm{w}\right) & := \left(S_l\left(\bm{w}\right)\right)_{l \in L},\\
    \bm{\hat{S}}_K\left(\bm{w}\right) & := \left(\hat{S}_{K, l}\left(\bm{w}\right)\right)_{l \in L},\\
    \bm{\overline{S}}\left(\bm{w}\right) & := \left(\overline{S}_l\left(\bm{w}\right)\right)_{l \in L}.
\end{align}
Then, for all integer $d \geq 1$, the following estimate holds for $\varphi\left(S\left(\bm{w}\right)\right)$:
\begin{align}
    & \varphi\left(\bm{S}\left(\bm{w}\right)\right) - \sum_{\substack{J' \subset J,\,\left|J'\right| < d\\\bm{l} \in L^{J'}}}\partial_{\bm{l}}\varphi\left(\bm{0}_L\right)\prod_{k \in J'}f_{k, l_k}\left(\bm{w}\right) - \sum_{\substack{J' \subset J,\,\left|J'\right| = d\\\bm{l} \in L^{J'}}}\int_0^1\!\mathrm{d}s\,d\left(1 - s\right)^{d - 1}\partial_{\bm{l}}\varphi\left(s\bm{\hat{S}}_{J'}\left(\bm{w}\right)\right)\prod_{k \in J'}f_{k, l_k}\left(\bm{w}\right)\nonumber\\
    & \preceq \sum_{0 \leq m < d}\sum_{\substack{J' \subset J,\,\left|J'\right| = m\\\bm{l} \in L^{J'}\\j \in J - J'\\l, l' \in L}}\overline{\partial_{\bm{l}, l, l'}\varphi}\left(\bm{\overline{S}}\left(\bm{w}\right)\right)\overline{f_{j, l}}\left(\bm{w}\right)\overline{f_{j, l'}}\left(\bm{w}\right)\prod_{k \in J'}\overline{f_{k, l_k}}\left(\bm{w}\right).\label{si:eq:remove_collision_terms_nonlinear_function_iterative}
\end{align}
When writing $\partial_{\bm{l}}\varphi$ in the above formula, $\bm{l}$ is interpreted as a tuple (the indexing by $J'$ is irrelevant since derivatives according to different variables commute); besides, $\partial_{\bm{l}, l, l'}\varphi := \partial_{l, l'}\partial_{\bm{l}}\varphi$. For $d = 1$, the integral over zero-dimensional vector being interpreted as the identity, the above estimate recovers Eq.~\ref{si:eq:remove_collision_terms_nonlinear_function} from Lemma~\ref{si:eq:remove_collision_terms_nonlinear_function}.
\begin{proof}
For convenience, introduce for all integer $m \geq 0$ notations:
\begin{align}
    I_m^{(0)}\left(\bm{w}\right) & := \sum_{\substack{J' \subset J,\,\left|J'\right| = m\\\bm{l} \in L^{J'}}}\partial_{\bm{l}}\varphi\left(\bm{0}_L\right)\prod_{k \in J'}f_{k, l_k}\left(\bm{w}\right),\\
    I_m'\left(\bm{w}\right) & := \sum_{\substack{J' \subset J,\,\left|J'\right| = m\\\bm{l} \in L^{J'}}}\int_0^1\!\mathrm{d}s\,m(1 - s)^{m - 1}\partial_{\bm{l}}\varphi\left(s\bm{\hat{S}}_{J'}\left(\bm{w}\right)\right)\prod_{k \in J'}f_{k, l_k}\left(\bm{w}\right).\label{si:eq:remove_collision_terms_nonlinear_function_iterative_integral_term}\\
    \overline{I}_m\left(\bm{w}\right) & := \sum_{\substack{J' \subset J,\,\left|J'\right| = m\\\bm{l} \in L^{J'}\\j \in J - J'\\l, l' \in L}}\overline{\partial_{\bm{l}, l, l'}}\left(\bm{\overline{S}}\left(\bm{w}\right)\right)\overline{f_{j, l}}\left(\bm{w}\right)\overline{f_{j, l'}}\left(\bm{w}\right)\prod_{k \in J'}\overline{f_{k, l_k}}\left(\bm{w}\right).
\end{align}
The statement can then be phrased:
\begin{align}
    \varphi\left(\bm{S}\left(\bm{w}\right)\right) - \sum_{0 \leq m < d}I^{(0)}_m\left(\bm{w}\right) - I'_d\left(\bm{w}\right) & \preceq \sum_{0 \leq m < d}\overline{I}_m\left(\bm{w}\right).\label{si:eq:remove_collision_terms_nonlinear_function_iterative_rephrased}
\end{align}
We prove the result by induction on $d$. For $d = 1$, this is Lemma~\ref{si:lemma:remove_collision_terms_nonlinear_function}. Now, assume the estimate in Eq.~\ref{si:eq:remove_collision_terms_nonlinear_function_iterative} proven up to some degree $d$ and let us deduce the corresponding estimate at degree $d + 1$. We then expand term $I'_d\left(\bm{w}\right)$. For all $J' \subset J, \left|J'\right| = d$ and $\bm{l} \in L^{J'}$ and each value of integration variable $s \in [0, 1]$, let us apply Lemma~\ref{si:lemma:remove_collision_terms_nonlinear_function} with parameters:
\begin{align}
    J & \longrightarrow J - J',\\
    L & \longrightarrow L,\\
    f_{j, l} & \longrightarrow sf_{j, l}, \qquad j \in J - J', l \in L\\
    \varphi & \longrightarrow \partial_{\bm{l}}\varphi.
\end{align}
Eq.~\ref{si:eq:remove_collision_terms_nonlinear_function} from the Lemma then produces estimate:
\begin{align}
    & \partial_{\bm{l}}\varphi\left(s\bm{\hat{S}}_{J'}\left(\bm{w}\right)\right) - \partial_{\bm{l}}\varphi\left(\bm{0}_L\right) - \int_0^1\!\mathrm{d}t\,\sum_{\substack{j \in J - J'\\l \in L}}sf_{j, l}\left(\bm{w}\right)\partial_{\bm{l}, l}\varphi\left(ts\bm{\hat{S}}_{J' \sqcup \{j\}}\left(\bm{w}\right)\right)\nonumber\\
    & \preceq \sum_{\substack{j \in J - J'\\l, l' \in L}}s\overline{f_{j, l}}\left(\bm{z}\right)s\overline{f_{j, l'}}\left(\bm{w}\right)\overline{\partial_{\bm{l}, l, l'}\varphi}\left(s\bm{\overline{S}}\left(\bm{w}\right)\right)\nonumber\\
    & \preceq \sum_{\substack{j \in J\\l, l' \in L}}\overline{f_{j, l}}\left(\bm{z}\right)\overline{f_{j, l'}}\left(\bm{w}\right)\overline{\partial_{\bm{l}, l, l'}\varphi}\left(\bm{\overline{S}}\left(\bm{w}\right)\right).
\end{align}
Plugging this majorization bound into $I'_d\left(\bm{w}\right)$ defined in Eq.~\ref{si:eq:remove_collision_terms_nonlinear_function_iterative_integral_term} implies (by linearity) majorization bound:
\begin{align}
    I'_d\left(\bm{w}\right) - K_d^{(0)}\left(\bm{w}\right) - K_d'\left(\bm{w}\right) & \preceq \overline{K}_d\left(\bm{w}\right),\label{si:eq:remove_collision_terms_nonlinear_function_iterative_integral_term_majorization_estimate}
\end{align}
where
\begin{align}
    K_d^{(0)}\left(\bm{w}\right) & := \sum_{\substack{J' \subset J,\,\left|J'\right| = d\\\bm{l} \in L^{J'}}}\int_0^1\!\mathrm{d}s\,d\left(1 - s\right)^{d - 1}\,\partial_{\bm{l}}\varphi\left(\bm{0}_L\right)\prod_{k \in J'}f_{k, l_k}\left(\bm{w}\right),\\
    K_d'\left(\bm{w}\right) & := \sum_{\substack{J' \subset J,\,\left|J'\right| = d\\\bm{l} \in L^{J'}}}\int_0^1\!\mathrm{d}s\,d\left(1 - s\right)^{d - 1}\int_0^1\!\mathrm{d}t\sum_{\substack{j \in J - J'\\l \in L}}sf_{j, l}\left(\bm{w}\right)\partial_{\bm{l}, l}\varphi\left(ts\bm{\hat{S}}_{J' \sqcup \{j\}}\left(\bm{w}\right)\right)\prod_{k \in J'}f_{k, l_k}\left(\bm{w}\right),\\
    \overline{K}_d\left(\bm{w}\right) & := \sum_{\substack{J' \subset J,\,\left|J'\right| = d\\\bm{l} \in L^{J'}}}\int_0^1\!\mathrm{d}s\,d\left(1 - s\right)^{d - 1}\sum_{\substack{j \in J - J'\\l, l' \in L}}\overline{f_{j, l}}\left(\bm{w}\right)\overline{f_{j, l'}}\left(\bm{w}\right)\overline{\partial_{\bm{l}, l, l'}\varphi}\left(\bm{\overline{S}}\left(\bm{w}\right)\right)\prod_{k \in J'}\overline{f_{k, l_k}}\left(\bm{w}\right).
\end{align}
We now transform $K^{(0)}_d\left(\bm{w}\right)$, $K'_d\left(\bm{w}\right)$ and $\overline{K}_d\left(\bm{w}\right)$ appropriately. For $K^{(0)}_d\left(\bm{w}\right)$,
\begin{align}
    K^{(0)}_d\left(\bm{w}\right) & = \sum_{\substack{J' \subset J,\,\left|J'\right| = d\\\bm{l} \in L^{J'}}}\partial_{\bm{l}}\varphi\left(\bm{0}_L\right)\int_0^1\!\mathrm{d}s\,d\left(1 - s\right)^{d - 1}\prod_{k \in J'}f_{k, l_k}\left(\bm{w}\right)\nonumber\\
    & = \sum_{\substack{J' \subset J,\,\left|J'\right| = d\\\bm{l} \in L^{J'}}}\partial_{\bm{l}}\varphi\left(\bm{0}_L\right)\prod_{k \in J'}f_{k, l_k}\left(\bm{w}\right)\nonumber\\
    & = I^{(0)}_d\left(\bm{w}\right)\label{si:eq:remove_collision_terms_nonlinear_function_iterative_integral_term_k0}
\end{align}
For $K'_d\left(\bm{w}\right)$, performing change of variables $\left(u, v\right) := \left(s, st\right)$,
\begin{align}
    K'_d\left(\bm{w}\right) & = \sum_{\substack{J' \subset J,\,\left|J'\right| = d\\\bm{l} \in L^{J'}\\j \in J - J'\\l \in L}}f_{j, l}\left(\bm{w}\right)\int_0^1\!\mathrm{d}s\int_0^1\!\mathrm{d}t\,d\left(1 - s\right)^{d - 1}s\,\partial_{\bm{l}, l}\varphi\left(ts\bm{\hat{S}}_{J' \sqcup \{j\}}\left(\bm{w}\right)\right)\prod_{k \in J'}f_{k, l_k}\left(\bm{w}\right)\nonumber\\
    & = \sum_{\substack{J' \subset J,\,\left|J'\right| = d\\\bm{l} \in L^{J'}\\j \in J - J'\\l \in L}}f_{j, l}\left(\bm{w}\right)\int_0^1\!\mathrm{d}v\int_v^1\!\mathrm{d}u\,d\left(1 - u\right)^{d - 1}\partial_{\bm{l}, l}\varphi\left(v\bm{\hat{S}}_{J' \sqcup \{j\}}\left(\bm{w}\right)\right)\prod_{k \in J'}f_{k, l_k}\left(\bm{w}\right)\nonumber\\
    & = \sum_{\substack{J' \subset J,\,\left|J'\right| = d\\\bm{l} \in L^{J'}\\j \in J - J'\\l \in L}}f_{j, l}\left(\bm{w}\right)\int_0^1\!\mathrm{d}v\,\left(1 - v\right)^d\partial_{\bm{l}, l}\varphi\left(v\bm{\hat{S}}_{J' \sqcup \{j\}}\left(\bm{w}\right)\right)\prod_{k \in J'}f_{k, l_k}\left(\bm{w}\right)\nonumber\\
    & = \sum_{\substack{J' \subset J,\,\left|J'\right| = d\\j \in J - J'}}\sum_{\bm{l} \in L^{J' \sqcup \{j\}}}\int_0^1\!\mathrm{d}v\,\left(1 - v\right)^d\partial_{\bm{l}}\varphi\left(v\bm{\hat{S}}_{J' \sqcup \{j\}}\left(\bm{w}\right)\right)\prod_{k \in J' \sqcup \{j\}}f_{k, l_k}\left(\bm{w}\right).
\end{align}
We now observe that each term of the sum only depends on $(d + 1)$-elements set $J'' := J' \sqcup \{j\}$. Each such set $J''$ occurs for $(d + 1)$ pairs $\left(J', j\right)$ of the outer sum (corresponding to a choice of any element of $J''$ for $j$). Hence,
\begin{align}
    K'_d\left(\bm{w}\right) & = \sum_{\substack{J'' \subset J,\,\left|J''\right| = d + 1\\\bm{l} \in L^{J''}}}\int_0^1\!\mathrm{d}v\,(d + 1)\left(1 - v\right)^d\partial_{\bm{l}}\varphi\left(v\bm{\hat{S}}_{J''}\left(\bm{w}\right)\right)\prod_{k \in J''}f_{k, l_k}\left(\bm{w}\right)\\
    & = I'_{d + 1}\left(\bm{w}\right).\label{si:eq:remove_collision_terms_nonlinear_function_iterative_integral_term_iprime}
\end{align}
Similarly, for $\overline{K}_d\left(\bm{w}\right)$,
\begin{align}
    \overline{K}_d\left(\bm{w}\right) & = \sum_{\substack{J' \subset J,\,\left|J'\right| = d\\\bm{l} \in L^{J'}\\j \in J - J'\\l, l' \in L}}\overline{\partial_{\bm{l}, l, l'}\varphi}\left(\bm{\overline{S}}\left(\bm{w}\right)\right)\overline{f_{j, l}}\left(\bm{w}\right)\overline{f_{j, l'}}\left(\bm{w}\right)\prod_{k \in J'}\overline{f_{k, l_k}}\left(\bm{w}\right)\nonumber\\
    & = \overline{I}_d\left(\bm{w}\right).\label{si:eq:remove_collision_terms_nonlinear_function_iterative_integral_term_ibar}
\end{align}
From these computations of $K^{(0)}_d\left(\bm{w}\right)$, $K'_d\left(\bm{w}\right)$, $\overline{K}_d\left(\bm{w}\right)$, estimate Eq.~\ref{si:eq:remove_collision_terms_nonlinear_function_iterative_integral_term_majorization_estimate} of $I'_d\left(\bm{w}\right)$ can be reexpressed:
\begin{align}
    I'_d\left(\bm{w}\right) - I^{(0)}_d\left(\bm{w}\right) - I'_{d + 1}\left(\bm{w}\right) & \preceq \overline{I}_d\left(\bm{w}\right).
\end{align}
Plugging this estimate of $I'_d\left(\bm{w}\right)$ into the induction hypothesis Eq.~\ref{si:eq:remove_collision_terms_nonlinear_function_iterative_rephrased} yields
\begin{align}
    \varphi\left(\bm{S}\left(\bm{w}\right)\right) - \sum_{\substack{J' \subset J,\,\left|J'\right| < d\\\bm{l} \in L^{J'}}}I^{(0)}_m\left(\bm{w}\right) - I_d^{(0)}\left(\bm{w}\right) - I_{d + 1}'\left(\bm{w}\right) & \preceq \sum_{0 \leq m < d}\overline{I_m}\left(\bm{w}\right) + \overline{I_d}\left(\bm{w}\right),
\end{align}
that is
\begin{align}
    \varphi\left(\bm{S}\left(\bm{w}\right)\right) - \sum_{0 \leq m < d + 1}I^{(0)}_m\left(\bm{w}\right) - I'_{d + 1}\left(\bm{w}\right) & \preceq \sum_{0 \leq m < d + 1}\overline{I}_m\left(\bm{w}\right),
\end{align}
which is the induction statement (as formulated in Eq.~\ref{si:eq:remove_collision_terms_nonlinear_function_iterative_rephrased}).
\end{proof}
\end{lemma}

Lemma \ref{si:lemma:remove_collision_terms_nonlinear_function_iterative} provides an approximate expansion of $\varphi\left(\bm{S}\left(\bm{z}\right)\right)$ as a polynomial in indeterminates $f_{j, l}\left(\bm{z}\right)$, with error bounded in terms of absolute majorants of $\varphi$ and its derivatives. To simplify this error term in Eq.~\ref{si:eq:remove_collision_terms_nonlinear_function_iterative}, we will rely on generic bounds on derivatives of multivariate analytic functions. Essentially, these bounds state the derivatives of analytic functions decrease faster than any geometric sequence, with constant factors related by the function's maximum over a polydisk. Substituting this simple ansatz for derivative bounds will in turn allow us to sum error term Eq.~\ref{si:eq:remove_collision_terms_nonlinear_function_iterative}. The following Lemma states the precise estimate on derivatives, resulting from Cauchy's formula for multivariate analytic functions on polydisks:

\begin{lemma}[Cauchy estimate on polydisks]
\label{si:lemma:cauchy_estimate_polydisks}
Let $L$ be a finite index set and $\varphi: \mathbb{C}^L \longrightarrow \mathbb{C}$ be entire, i.e. analytic over $\mathbb{C}^L$. For all $R > 0$, let
\begin{align}
    M_R & := \sup_{\bm{w} \in B(0, R)^L}\left|\varphi\left(\bm{w}\right)\right|
\end{align}
the maximum of $\varphi$ over all coordinates individually bounded by $R$. Then, for all $\bm{w} \in B(0, R)^L$, and all differentiation orders $\bm{m} \in \mathbb{N}^L$, the following bound on the $\varphi$ derivative evaluated at $\bm{w}$ holds:
\begin{align}
    \left|\partial^{\bm{m}}\varphi\left(\bm{w}\right)\right| & \leq M_{2R}\prod_{l \in L}\frac{m_l!}{R^{m_l}}.
\end{align}
\begin{proof}
According to Cauchy's formula on polydisks,
\begin{align}
    \partial^{\bm{m}}\varphi\left(\bm{w}\right) & = \left(\prod_{l \in L}\frac{m_l!}{2\pi}\oint\limits_{\mathcal{C}_+\left(w_l, R\right)}\mathrm{d}\omega_l\right)\frac{\varphi\left(\bm{\omega}\right)}{\prod_{l \in L}\left(\omega_l - w_l\right)^{1 + m_l}},
\end{align}
where $\mathcal{C}_+\left(w_l, R\right)$ denotes a positively oriented circular contour of radius $R$ centered at $w_l$. Applying the triangular inequality, and using the length of each such contour is $2\pi R$:
\begin{align}
    \oint_{\mathcal{C}_+\left(w_l, R\right)}\left|\mathrm{d}\omega_l\right| & = 2\pi R,
\end{align}
yields:
\begin{align}
    \left|\partial^{\bm{m}}\varphi\left(\bm{w}\right)\right| & \leq \left(\prod_{l \in L}\frac{m_l!}{2\pi}\oint\limits_{\mathcal{C}_+\left(w_l, R\right)}\,\left|\mathrm{d}\omega_l\right|\right)\frac{\left|\varphi\left(\bm{\omega}\right)\right|}{\prod_{l \in L}\left|\omega_l - w_l\right|^{1 + m_l}}\nonumber\\
    & \leq \left(\prod_{l \in L}\frac{m_l!}{2\pi}\oint\limits_{\mathcal{C}_+\left(w_l, R\right)}\,\left|\mathrm{d}\omega_l\right|\right)\frac{M_{2R}}{R^{ |L| + \left|\bm{m}\right|}}\nonumber\\
    & = \prod_{l \in L}\frac{m_l!}{2\pi}2\pi R \times \frac{M_{2R}}{R^{|L| + |\bm{m}|}}\nonumber\\
    & = M_{2R}\prod_{l \in L}\frac{m_l!}{R^{m_l}},
\end{align}
where we denoted by $|\bm{m}| := \sum_{l \in L}m_l$ the total differentiation order.
\end{proof}
\end{lemma}

Cauchy's estimate Lemma~\ref{si:lemma:cauchy_estimate_polydisks} states a pointwise bound on derivatives, which is a priori too weak to take path integral averages. Fortunately, the following Proposition lifts the bound to a majorization one:

\begin{proposition}[Majorization bounds for entire function]
\label{si:prop:majorization_entire_function}
Consider a multivariate entire function $\varphi: \mathbb{C}^L \longrightarrow \mathbb{C}$. Following notations from Lemma~\ref{si:lemma:cauchy_estimate_polydisks}, let for all $R > 0$
\begin{align}
    M_R & := \sup_{\bm{w} \in B(0, R)^L}\left|\varphi\left(\bm{w}\right)\right|
\end{align}
the maximum of $\varphi$ over coordinates individually bounded by $R$. Then, for any differentiation orders $\bm{m} \in \mathbb{N}^L$ and all radius $R > 0$ the following majorization bound holds for the derivative of $\varphi$ over polydisk $B\left(0, R/2\right)^L$:
\begin{align}
    \partial^{\bm{m}}\varphi\left(\bm{w}\right) & \preceq M_{2R}\prod_{l \in L}\frac{m_l!}{R^{m_l}\left(1 - w_l/R\right)^{1 + m_l}}.
\end{align}
\begin{proof}
Since $\varphi$, hence also $\partial^{\bm{m}}\varphi$ is entire, the derivative is the sum of its absolutely convergent power series at every $\bm{w} \in \mathbb{C}^L$:
\begin{align}
    \partial^{\bm{m}}\varphi\left(\bm{w}\right) & = \sum_{\bm{n} \in \mathbb{N}^L}\partial^{\bm{m} + \bm{n}}\varphi\left(\bm{0}\right)\prod_{l \in L}\frac{w_l^{n_l}}{n_l!}.
\end{align}
From Lemma~\ref{si:lemma:cauchy_estimate_polydisks}, the coefficients of the power series can be upper-bounded as:
\begin{align}
    \left|\partial^{\bm{m} + \bm{n}}\varphi\left(\bm{0}\right)\right| & \leq M_{2R}\prod_{l \in L}\frac{\left(m_l + n_l\right)!}{R^{m_l + n_l}}.
\end{align}
Substituting this upper bound in the power series yields a majorant
\begin{align}
    \psi\left(\bm{w}\right) & = \sum_{\bm{n} \in \mathbb{N}^L}M_{2R}\prod_{l \in L}\frac{\left(n_l + m_l\right)!}{n_l!}\frac{w_l^{n_l}}{R^{n_l + m_l}}\nonumber\\
    & = M_{2R}\prod_{l \in L}\sum_{n_l \geq 0}\frac{\left(n_l + m_l\right)!}{n_l!}\frac{w_l^{n_l}}{R^{n_l + m_l}}\nonumber\\
    & = M_{2R}\prod_{l \in L}\frac{m_l!}{R^{m_l}\left(1 - w_l/R\right)^{1 + m_l}},
\end{align}
which is analytic over polydisk $B(0, R/2)^L$.
\end{proof}
\end{proposition}

Combining non-colliding Taylor expansion Lemma~\ref{si:lemma:remove_collision_terms_nonlinear_function_iterative} and the majorization estimate of entire functions Proposition~\ref{si:prop:majorization_entire_function} yields a tractable error estimate when approximating $\varphi\left(\bm{S}\left(\bm{w}\right)\right)$ by the zero-derivatives term of the expansion.

\begin{proposition}[Approximating $\varphi\left(\bm{S}\left(\bm{w}\right)\right)$ as a non-colliding polynomial in $f_{j, l}\left(\bm{w}\right)$]
\label{si:prop:polynomial_approximation_entire_function_path_integral_measure}
Let entire functions $\varphi$ and $f_{j, l}$ be as in Lemma~\ref{si:lemma:cauchy_estimate_polydisks}. Recalling notation from Proposition~\ref{si:prop:majorization_entire_function}, let be given the maximum of $\varphi$ over any polydisk of radii $R$:
\begin{align}
    M_R & := \sup_{\forall l \in L,\,\left|x_l\right| \leq R}\left|\varphi\left(\bm{x}\right)\right|.
\end{align}
Besides, assume existence of a constant $C$ such that:
\begin{align}
    \overline{f_{j, l}}\left(\bm{1}_d\right) & \leq \frac{C}{\left|J\right|}.
\end{align}
Then, for all integer parameter $d \geq 1$ and any positive real parameter $R > 2C$, there exists a majorant $\overline{\psi}\left(\bm{w}\right)$ for:
\begin{align}
    \psi\left(\bm{w}\right) & := \varphi\left(\bm{S}\left(\bm{w}\right)\right) - \sum_{\substack{J' \subset J,\,\left|J'\right| < d\\\bm{l} \in L^{J'}}}\partial_{\bm{l}}\varphi\left(\bm{0}_L\right)\prod_{k \in J'}f_{k, l_k}\left(\bm{w}\right),
\end{align}
bounded as follows at the all-$1$ variables:
\begin{align}
    \overline{\psi}\left(\bm{1}_{D}\right) & \leq 2^{|L|}M_{2R}\left(\left(\frac{2C}{R}\right)^d\binom{d + |L| - 1}{|L| - 1} + \frac{1}{|J|}|L|\left(|L| + 1\right)\frac{\left(2C/R\right)^2}{\left(1 - 2C/R\right)^{|L| + 2}}\right).
\end{align}
Note the bound only depends on the polydisk bounds $M_R$ of $\varphi$ as well as constant $C$ bounding $f_{j, l}$, and remains valid by replacing these constants with larger ones. The bound is therefore uniform over all entire functions $\varphi$ characterized by the appropriate constants.
\begin{proof}
We seek a majorant of function
\begin{align}
    \psi\left(\bm{w}\right) & := \varphi\left(\bm{S}\left(\bm{w}\right)\right) - \sum_{\substack{J' \subset J,\,\left|J'\right| < d\\\bm{l} \in L^{J'}}}\partial_{\bm{l}}\varphi\left(\bm{0}_L\right)\prod_{k \in J'}f_{k, l_k}\left(\bm{w}\right).
\end{align}
Lemma~\ref{si:lemma:remove_collision_terms_nonlinear_function_iterative} implies the following majorization bound for $\psi$:
\begin{align}
    \psi\left(\bm{w}\right) & \preceq \sum_{\substack{J' \subset J,\,\left|J'\right| = d\\\bm{l} \in L^{J'}}}\int_0^1\!\mathrm{d}s\,d\left(1 - s\right)^{d - 1}\overline{\partial_{\bm{l}}\varphi}\left(s\bm{\overline{S}}\left(\bm{w}\right)\right)\prod_{k \in J'}\overline{f_{k, l_k}}\left(\bm{w}\right)\nonumber\\
    & \hspace*{20px} + \sum_{0 \leq m < d}\sum_{\substack{J' \subset J,\,\left|J'\right| = m\\\bm{l} \in L^{J'}\\j \in J - J'\\l, l' \in L}}\overline{\partial_{\bm{l}, l, l'}\varphi}\left(\bm{\overline{S}}\left(\bm{w}\right)\right)\overline{f_{j, l}}\left(\bm{w}\right)\overline{f_{j, l'}}\left(\bm{w}\right)\prod_{k \in J'}\overline{f_{k, l_k}}\left(\bm{w}\right)\nonumber\\
    & \preceq \sum_{\substack{J' \subset J,\,\left|J'\right| = d\\\bm{l} \in L^{J'}}}\int_0^1\!\mathrm{d}s\,d\left(1 - s\right)^{d - 1}\overline{\partial_{\bm{l}}\varphi}\left(\bm{\overline{S}}\left(\bm{w}\right)\right)\prod_{k \in J'}\overline{f_{k, l_k}}\left(\bm{w}\right)\nonumber\\
    & \hspace*{20px} + \sum_{0 \leq m < d}\sum_{\substack{J' \subset J,\,\left|J'\right| = m\\\bm{l} \in L^{J'}\\j \in J - J'\\l, l' \in L}}\overline{\partial_{\bm{l}, l, l'}\varphi}\left(\bm{\overline{S}}\left(\bm{w}\right)\right)\overline{f_{j, l}}\left(\bm{w}\right)\overline{f_{j, l'}}\left(\bm{w}\right)\prod_{k \in J'}\overline{f_{k, l_k}}\left(\bm{w}\right)\nonumber\\
    & = \sum_{\substack{J' \subset J,\,\left|J'\right| = d\\\bm{l} \in L^{J'}}}\overline{\partial_{\bm{l}}\varphi}\left(\bm{\overline{S}}\left(\bm{w}\right)\right)\prod_{k \in J'}\overline{f_{k, l_k}}\left(\bm{w}\right) + \sum_{0 \leq m < d}\sum_{\substack{J' \subset J,\,\left|J'\right| = m\\\bm{l} \in L^{J'}\\j \in J - J'\\l, l' \in L}}\overline{\partial_{\bm{l}, l, l'}\varphi}\left(\bm{\overline{S}}\left(\bm{w}\right)\right)\overline{f_{j, l}}\left(\bm{w}\right)\overline{f_{j, l'}}\left(\bm{w}\right)\prod_{k \in J'}\overline{f_{k, l_k}}\left(\bm{w}\right)\nonumber\\
    & =: \overline{\psi}\left(\bm{w}\right).\label{si:eq:entire_function_low_degree_polynomial_approximation_error_majorant}
\end{align}
where $\bm{\overline{S}}$ is the majorant of $\bm{S}$ defined from $\overline{f_{j, l}}$ according to Eq.~\ref{si:eq:s_majorant_definition}, and for all tuple of $L$ elements $\bm{l}$, $\overline{\partial_{\bm{l}}\varphi}$ is assumed to be a majorant of $\partial_{\bm{l}}\varphi$ analytic on an open polydisk containing the image $\bm{\overline{S}}\left(\overline{B}(0, 1)^{D}\right)$ of the unit polydisk under $\bm{\overline{S}}$. We now explicitly specify majorant $\overline{\partial_{\bm{l}}\varphi}$ of $\partial_{\bm{l}}\varphi$ for all tuple $\bm{l}$ of $L$ elements. First, note the following bound for the range of $\bm{\overline{S}} = \left(\overline{S}_l\right)_{l \in L}$:
\begin{align}
    \forall \bm{w} \in \overline{B}\left(0, 1\right)^{D}, \quad \left|\overline{S}_l\left(\bm{w}\right)\right| & = \left|\sum_{j \in J}\overline{f_{j, l}}\left(\bm{w}\right)\right|\nonumber\\
    & \leq \sum_{j \in J}\left|\overline{f_{j, l}}\left(\bm{w}\right)\right|\nonumber\\
    & \leq \sum_{j \in J}\frac{C}{|J|}\nonumber\\
    & \leq C,
\end{align}
i.e. $\bm{S}\left(\overline{B}\left(0, 1\right)^{D}\right) \subset \overline{B}(0, C)^L$. Let us now consider a majorant for $\partial_{\bm{l}}\varphi$, with $\bm{l}$ an arbitrary $L$-tuple of differentiation indices. Let us instead characterize this tuple by the multiplicities $\bm{m} = \left(m_l\right)_{l \in L}$ of each element $l \in L$. By Proposition~\ref{si:prop:majorization_entire_function}, for any $R > 0$, it holds
\begin{align}
    \partial_{\bm{l}}\varphi\left(\bm{x}\right) & \preceq \overline{\partial_l\varphi}\left(\bm{x}\right) := M_{2R}\prod_{l \in L}\frac{m_l!}{R^{m_l}\left(1 - x_l/R\right)^{1 + m_l}},
\end{align}
which is analytic over polydisk $B\left(0, R\right)^{L}$. By composition of majorants, for all $R > 2C$,
\begin{align}
    \partial_{\bm{l}}\varphi\left(\bm{S}\left(\bm{w}\right)\right) & \preceq \overline{\partial_{\bm{l}}\varphi}\left(\bm{\overline{S}}\left(\bm{w}\right)\right)\nonumber\\
    & = M_{2R}\prod_{l \in L}\frac{m_l!}{R^{m_l}\left(1 - \overline{S_l}\left(\bm{w}\right)/R\right)^{1 + m_l}}.
\end{align}
The majorant evaluated at all-$1$ variables is upper-bounded as:
\begin{align}
    \overline{\partial_{\bm{l}}\varphi}\left(\bm{\overline{S}}\left(\bm{1}_{D}\right)\right) & \leq M_{2R}\prod_{l \in L}\frac{m_l!}{R^{m_l}\left(1 - C/R\right)^{1 + m_l}}\nonumber\\
    & \leq M_{2R}\prod_{l \in L}\frac{2^{1 + m_l}m_l!}{R^{m_l}}\nonumber\\
    & \leq 2^{|L|}M_{2R}\left(\frac{2}{R}\right)^{|\bm{m}|}\prod_{l \in L}m_l!,
\end{align}
where $|\bm{m}| := \sum_{l \in L}m_l$ is the total differentiation order.

Based on the latest bound on $\overline{\partial_{\bm{l}}\varphi}\left(\bm{\overline{S}}\left(\bm{1}_{D}\right)\right)$, we bound the two terms of $\overline{\psi}\left(\bm{w}\right)$ from Eq.~\ref{si:eq:entire_function_low_degree_polynomial_approximation_error_majorant} separately. For the first term,
\begin{align}
    \sum_{\substack{J' \subset J,\,\left|J'\right| = d\\\bm{l} \in L^{J'}}}\overline{\partial_{\bm{l}}\varphi}\left(\bm{\overline{S}}\left(\bm{1}_{D}\right)\right)\prod_{k \in J'}\overline{f_{k, l_k}}\left(\bm{1}_{D}\right) & \leq \sum_{\substack{J' \subset J,\,\left|J'\right| = d\\\bm{l} \in L^{J'}}}\left(\frac{C}{|J|}\right)^{\left|J'\right|}\overline{\partial_{\bm{l}}\varphi}\left(\bm{\overline{S}}\left(\bm{1}_{D}\right)\right)\nonumber\\
    & \leq \sum_{\substack{J' \subset J,\,\left|J'\right| = d\\\bm{m} \in \mathbb{N}^{L},\,\left|\bm{m}\right| = d}}\left(\frac{C}{|J|}\right)^{\left|J'\right|}\binom{d}{\bm{m}}2^{|L|}M_{2R}\left(\frac{2}{R}\right)^{|\bm{m}|}\prod_{l \in L}m_l!\nonumber\\
    & = \binom{|J|}{d}\sum_{\bm{m} \in \mathbb{N}^L,\,|\bm{m}| = d}\left(\frac{C}{|J|}\right)^{d}d!2^{|L|}M_{2R}\left(\frac{2}{R}\right)^d\nonumber\\
    & = \binom{|J|}{d}\binom{d + |L| - 1}{|L| - 1}\left(\frac{C}{|J|}\right)^dd!2^{|L|}M_{2R}\left(\frac{2}{R}\right)^d\nonumber\\
    & \leq 2^{|L|}M_{2R}\left(\frac{2C}{R}\right)^d\binom{d + |L| - 1}{|L| - 1}.\label{si:eq:entire_function_low_degree_polynomial_approximation_taylor_error_estimate}
\end{align}
In the second line, we reindexed the sum over ordered $|J'|$-tuples $\bm{l} \in L^{J'}$ as a sum over multiplicities, counting as $\binom{d}{\bm{m}} = \frac{d!}{\prod_{l \in L}m_l!}$ (where $|\bm{m}| = \sum_lm_l = d$) the number of ordered tuples with $m_l$ occurrences of $l$.

For the second term of Eq.~\ref{si:eq:entire_function_low_degree_polynomial_approximation_error_majorant}, let us first consider a single value of $m$ in the outer sum. Similar to the previous bounds, it holds:
\begin{align}
    & \sum_{\substack{J' \subset J,\,\left|J'\right| = m\\\bm{l} \in L^{J'}\\j \in J - J'\\l, l' \in L}}\overline{\partial_{\bm{l}, l, l'}\varphi}\left(\bm{\overline{S}}\left(\bm{1}_{D}\right)\right)\overline{f_{j, l}}\left(\bm{1}_{D}\right)\overline{f_{j, l'}}\left(\bm{1}_{D}\right)\prod_{k \in J'}\overline{f_{k, l_k}}\left(\bm{1}_{D}\right)\nonumber\\
    & \leq \sum_{\substack{J' \subset J,\,\left|J'\right| = m\\\bm{l} \in L^{J'}\\j \in J - J'\\l, l' \in L}}\left(\frac{C}{|J|}\right)^{m + 2}\overline{\partial_{\bm{l}, l, l'}\varphi}\left(\bm{\overline{S}}\left(\bm{1}_{D}\right)\right)\nonumber\\
    & = \sum_{\substack{J' \subset J,\,\left|J'\right| = m\\j \in J - J'\\\bm{m} \in \mathbb{N}^{L},\,\left|\bm{m}\right| = m + 2}}\left(\frac{C}{|J|}\right)^{m + 2}\binom{m + 2}{\bm{m}}2^{|L|}M_{2R}\left(\frac{2}{R}\right)^{m + 2}\prod_{l \in L}m_l!\nonumber\\
    & = \binom{|J|}{m}\left(|J| - m\right)2^{|L|}M_{2R}\left(\frac{C}{|J|}\right)^{m + 2}\left(\frac{2}{R}\right)^{m + 2}\sum_{\bm{m} \in \mathbb{N}^{L},\,\left|\bm{m}\right| = m + 2}\binom{m + 2}{\bm{m}}\prod_{l \in L}m_l!\nonumber\\
    & = \binom{|J|}{m}\left(|J| - m\right)2^{|L|}M_{2R}\left(\frac{2C}{|J|R}\right)^{m + 2}\binom{m + 1 + |L|}{|L| - 1}(m + 2)!\nonumber\\
    & \leq \frac{1}{|J|}2^{|L|}M_{2R}\left(\frac{2C}{R}\right)^{m + 2}\binom{m + 1 + |L|}{|L| - 1}(m + 2)(m + 1)\nonumber\\
    & = \frac{1}{|J|}2^{|L|}|L|\left(|L| + 1\right)M_{2R}\left(\frac{2C}{R}\right)^{m + 2}\binom{m + 1 + |L|}{|L| + 1}
\end{align}
We finally sum this over $0 \leq m < d$, weakened to $m \geq 0$, to bound the second term of Eq.~\ref{si:eq:entire_function_low_degree_polynomial_approximation_error_majorant}:
\begin{align}
    & \sum_{0 \leq m < d}\sum_{\substack{J' \subset J,\,\left|J'\right| = m\\\bm{l} \in L^{J'}\\j \in J - J'\\l, l' \in L}}\overline{\partial_{\bm{l}, l, l'}\varphi}\left(\bm{\overline{S}}\left(\bm{1}_{D}\right)\right)\overline{f_{j, l}}\left(\bm{1}_{D}\right)\overline{f_{j, l'}}\left(\bm{1}_{D}\right)\prod_{k \in J'}\overline{f_{k, l_k}}\left(\bm{1}_{D}\right)\nonumber\\
    & \leq \frac{1}{|J|}2^{|L|}|L|\left(|L| + 1\right)M_{2R}\sum_{0 \leq m < d}\binom{m + 1 + |L|}{|L| + 1}\left(\frac{2C}{R}\right)^{m + 2}\nonumber\\
    & \leq \frac{1}{|J|}2^{|L|}|L|\left(|L| + 1\right)M_{2R}\sum_{m \geq 0}\binom{m + 1 + |L|}{|L| + 1}\left(\frac{2C}{R}\right)^{m + 2}\nonumber\\
    & \leq \frac{1}{|J|}2^{|L|}|L|\left(|L| + 1\right)M_{2R}\frac{\left(2C/R\right)^2}{\left(1 - 2C/R\right)^{|L| + 2}}.\label{si:eq:entire_function_low_degree_polynomial_approximation_collision_error_estimate}
\end{align}
Note the $1/|J|$ suppression factor compared to the estimate Eq.~\ref{si:eq:entire_function_low_degree_polynomial_approximation_taylor_error_estimate} for the first term of Eq.~\ref{si:eq:entire_function_low_degree_polynomial_approximation_error_majorant}. Adding Eqns.~\ref{si:eq:entire_function_low_degree_polynomial_approximation_taylor_error_estimate} and \ref{si:eq:entire_function_low_degree_polynomial_approximation_collision_error_estimate} yields
\begin{align}
    \overline{\psi}\left(\bm{1}_{D}\right) & \leq 2^{|L|}M_{2R}\left(\frac{2C}{R}\right)^d\binom{d + |L| - 1}{|L| - 1} + \frac{1}{|J|}2^{|L|}|L|\left(|L| + 1\right)M_{2R}\frac{\left(2C/R\right)^2}{\left(1 - 2C/R\right)^{|L| + 2}}\nonumber\\
    & = 2^{|L|}M_{2R}\left(\left(\frac{2C}{R}\right)^d\binom{d + |L| - 1}{|L| - 1} + \frac{1}{|J|}|L|\left(|L| + 1\right)\frac{\left(2C/R\right)^2}{\left(1 - 2C/R\right)^{|L| + 2}}\right),
\end{align}
which is the desired result.
\end{proof}
\end{proposition}

Proposition~\ref{si:prop:polynomial_approximation_entire_function_path_integral_measure} bounds the error (in the majorization sense) of approximating $\varphi\left(\bm{S}\left(\bm{w}\right)\right)$ by a constant-degree polynomial in indeterminates $f_{j, l}\left(\bm{w}\right)$. The polynomial is further ``non-colliding" in the sense each $j \in J$ occurs at most once in each monomial. If $f_{j, l}\left(\bm{w}\right)$ is itself polynomial in $\bm{w}$, the result bounds the error of approximation $\varphi\left(\bm{S}\left(\bm{w}\right)\right)$ by a constant-degree polynomial in $\bm{w}$. The bound on the majorant value at the all-$1$ variables provides a bound on the path integral expectation as established in Lemma~\ref{si:lemma:path_integral_expectation_bound_from_majorant}.

\subsection{Application to spin-glass QAOA cavity decomposition}
\label{si:sec:noncolliding_taylor_expansion_application_spin_glass_qaoa_cavity_decomposition}

Proposition~\ref{si:prop:polynomial_approximation_entire_function_path_integral_measure} derived in Section~\ref{si:sec:non_colliding_taylor_expansion} provides an approximation of a function of the form $\varphi\left(\bm{S}\left(\bm{w}\right)\right)$, where
\begin{align}
    \varphi & : \mathbb{C}^L \longrightarrow \mathbb{C},\\
    \bm{S}\left(\bm{w}\right) & := \left(S_l\left(\bm{w}\right)\right)_{l \in L},\\
    S_l\left(\bm{w}\right) & := \sum_{j \in J}f_{j, l}\left(\bm{w}\right),\\
    f_{j, l} & : \mathbb{C}^D \longrightarrow \mathbb{C},
\end{align}
as a polynomial in indeterminates $f_{j, l}\left(\bm{w}\right)$. As discussed in Section~\ref{si:sec:rigorous_argument_notation_outline}, in the context of spin-glass QAOA, functions $f_{j, (t, u)}\left(\bm{w}\right)$ is the centered product of spin $j$ at layers $t, u$: $f_{j, (t, u)}\left(\bm{z}\right) := z^{[t]}_jz^{[u]}_j - G_{t, u}$, and the goal of the expansion is to obtain an approximation of cavity factor $\mu_{C, 1}^{[p]}\left(\bm{z}_{S'}, \bm{z}_{S''};\,\xi, S', S''\right)$ as a constant-degree polynomial in centered spin products.

The goal of this Section is to formalize the heuristic argument from Section~\ref{si:sec:heuristic_cavity_argument}, suggesting approximate independence of small spin sets and concentration of normalized overlaps for large spin sets $\left\langle \bm{z}^{[t]}_S, \bm{z}^{[u]}_S \right\rangle/|S|$ around $\bm{G}$ matrix entries $G_{t, u}$. The general proof strategy for the result follows the induction idea sketched in the $p = 1$ specialization of the heuristic argument (Section~\ref{si:sec:heuristic_cavity_argument_p1}). Namely, we phrase approximate independence of constant-size sets of spins by introducing the maximum path integral expectation of a product of centered spin products, given a large set size $s$ and a small set size $s''$: %
\begin{align}
    \varepsilon^{[q]}\left(\xi, s, s''\right) & := \max_{\substack{\alpha_1, \ldots, \alpha_{s''} \subset \mathcal{T}_q\\\left|\alpha_j\right|\,\mathrm{even}}}\left|\left\langle \prod_{j \in \left[s''\right]}\left(z^{[\alpha_j]}_j - G^{\left(\left|\alpha_j\right|\right)}_{\alpha_j}\right) \right\rangle_{\mu^{[q]}\left(\bm{z};\,\xi, [s]\right)}\right|, \qquad s \geq s''\label{si:eq:centered_moments_maximum_informal}
\end{align}
The above quantity depends on a spin glass polynomial $\xi$ and intermediate layer index $q$. This explicit dependence is needed as both will vary in the course of the argument. $\alpha_1, \ldots, \alpha_{s''} \subset \mathcal{T}_q$ are layer indices of absolute value bounded by $(q + 1)$.
\begin{align}
    z^{\left[\alpha_j\right]}_j & := \prod_{t \in \alpha_j}z^{[t]}_j
\end{align}
denotes the product of computational basis states of spin $j$ at layers $\alpha_j$, and
\begin{align}
    G^{\left(\left|\alpha_j\right|\right)}_{\alpha_j} & := \left\langle \prod_{t \in \alpha_j}a_t \right\rangle_{\mu^{[q]}_{SB}\left(\bm{a};\,\xi\right)}
\end{align}
is a multipoint function (of degree $|\alpha_j|$) of the spin-boson system defined by spin-glass polynomial $\xi$, and associated self-consistent influence functional $\bm{G}$. By construction of the $\bm{G}$ iteration (Section~\ref{si:sec:spin_glass_qaoa_g_iteration}) and related spin-boson systems (Sections~\ref{si:sec:unitary_circuit_path_integrals_and_spin_boson_systems}, \ref{si:sec:spin_boson_mapping_qaoa_iteration}),
\begin{align}
    \bm{G}^{(2)} & = \bm{G}.
\end{align}
In the following, we may omit degree superscript on $\bm{G}^{(d)}$ tensors given it is specified by the number of indices, and consistency with the $\bm{G}$ matrix expressed in the above equation.

The maximum from Eq.~\ref{si:eq:centered_moments_maximum_informal} is over a finite set, hence is always finite (provided $s \geq s''$, otherwise it is vacuously $-\infty$). By permutation invariance of the disorder-averaged spin-glass QAOA path integral measure, it holds
\begin{align}
    \max_{\substack{\alpha_j \subset \mathcal{T}_p, |\alpha_j|\,\mathrm{even}} \,\, \forall j \in S''}\left|\left\langle \prod_{j \in S''}\left(z_j^{\left[\alpha_j\right]} - G^{\left(|\alpha_j|\right)}_{\alpha_j}\right) \right\rangle_{\mu^{[q]}\left(\bm{z};\,\xi, S\right)}\right| & = \varepsilon^{[p]}\left(\xi, \left|S\right|, \left|S''\right|\right)
\end{align}
for all set $S$ and all subset $S'' \subset S$. These quantities allow to relate moments of spin variables to $\bm{G}$ tensor entries; namely:
\begin{align}
    \left|\left\langle \prod_{j \in S''}z_j^{\left[\alpha_j\right]} \right\rangle_{\mu^{[q]}\left(\bm{z};\,\xi, S\right)} - \prod_{j \in S''}G_{\alpha_j}\right| & = \left|\left\langle \prod_{j \in S''}\left(z_j^{\left[\alpha_j\right]} - G_{\alpha_j} + G_{\alpha_j}\right) \right\rangle_{\mu^{[q]}\left(\bm{z};\,\xi, S\right)} - \prod_{j \in S''}G_{\alpha_j}\right|\nonumber\\
    & = \left|\left\langle \sum_{T'' \subset S''}\prod_{j \in T''}\left(z^{[\alpha_j]}_j - G_{\alpha_j}\right)\prod_{j \in S'' - T''}G_{\alpha_j} \right\rangle_{\mu^{[q]}\left(\bm{z};\,\xi, S\right)} - \prod_{j \in S''}G_{\alpha_j}\right|\nonumber\\
    & = \left|\sum_{\substack{T'' \subset S''\\T'' \neq \varnothing}}\left\langle \prod_{j \in T''}\left(z^{[\alpha_j]}_j - G_{\alpha_j}\right) \right\rangle_{\mu^{[q]}\left(\bm{z};\,\xi, S\right)}\prod_{j \in S'' - T''}G_{\alpha_j}\right|\nonumber\\
    & \leq \sum_{\substack{T'' \subset S''\\T'' \neq \varnothing}}\left|\left\langle \prod_{j \in T''}\left(z^{[\alpha_j]}_j - G_{\alpha_j}\right) \right\rangle_{\mu^{[q]}\left(\bm{z};\,\xi, S\right)}\right|\prod_{j \in S'' - T''}\left|G_{\alpha_j}\right|\nonumber\\
    & \leq \sum_{\substack{T'' \subset S''\\T'' \neq \varnothing}}\varepsilon^{[q]}\left(\xi, \left|S\right|, \left|T''\right|\right)\nonumber\\
    & = \sum_{1 \leq t \leq t''}\binom{s''}{t''}\varepsilon^{[q]}\left(\xi, |S|, t''\right)\nonumber\\
    & \leq 2^{s''}\max_{1 \leq t'' \leq s''}\varepsilon^{[q]}\left(\xi, |S|, t''\right).
\end{align}
Hence, showing vanishing of the maximum centered moments in the thermodynamic limit $|S| \to \infty$ proves
\begin{align}
    \left\langle \prod_{j \in S''}z^{[\alpha_j]}_j \right\rangle_{\mu^{[q]}\left(\bm{z};\,\xi, S''\right)} & \underset{|S| \to \infty}{\sim} \prod_{j \in S''}G_{\alpha_j},
\end{align}
hence in particular (applying the above identity to the singleton sets given by elements of $S''$)
\begin{align}
    \left\langle z^{[\alpha_j]}_j \right\rangle_{\mu^{[q]}\left(\bm{z};\,\xi, S\right)} & \underset{|S| \to \infty}{\sim} G_{\alpha_j},
\end{align}
establishing approximate independence:
\begin{align}
    \left\langle \prod_{j \in S''}z^{[\alpha_j]}_j \right\rangle_{\mu^{[q]}\left(\bm{z};\,\xi, S\right)} & \underset{|S| \to \infty}{\sim} \prod_{j \in S''}\left\langle z_j^{[\alpha_j]} \right\rangle_{\mu^{[q]}\left(\bm{z};\,\xi, S\right)}.
\end{align}
Our strategy to bound $\varepsilon^{[q]}\left(\xi, s, s''\right)$ will be to derive a recursive inequality in $q$ on this quantity, leading to final outcome of this Section Corollary~\ref{si:cor:centered_moments_maximum_recursion_simplified_special_case}. The derivation of the recursion proceeds by a cavity calculation similar to the one outlined in the heuristic argument Section~\ref{si:sec:heuristic_cavity_argument}. Namely, by definition $\varepsilon^{[q]}\left(\xi, s, s''\right)$ is an upper bound for any expectation of the form:
\begin{align}
    \left\langle \prod_{j \in S''}\left(z^{[\alpha_j]}_j - G_{\alpha_j}\right) \right\rangle_{\mu^{[q]}\left(\bm{z};\,\xi, S\right)}.
\end{align}
The difference between this quantity and the one expressed in the heuristic argument Section~\ref{si:sec:heuristic_cavity_argument} is the centering around $\bm{G}$.  According to the cavity decomposition of the path integral measure Lemma~\ref{si:lemma:cavity_decomposition_path_integral_measure}, the above can be recast:
\begin{align}
    & \left\langle \prod_{j \in S''}\left(z^{[\alpha_j]}_j - G_{\alpha_j}\right) \right\rangle_{\mu^{[q]}\left(\bm{z};\,\xi, S\right)}\nonumber\\
    & = \sum_{\bm{z}_{S''} \in \{1, -1\}^{S'' \times \mathcal{T}_q}}\left\langle \mu^{[q]}_{C, 1}\left(\bm{z}_{S'}, \bm{z}_{S''};\,\xi_{s'/s}, S', S'\right)\mu^{[q]}_{C, 2}\left(\bm{z}_{S'}, \bm{z}_{S''};\,\xi, S', S''\right) \right\rangle_{\mu^{[q - 1]}\left(\bm{z}_{S'};\,\xi_{s'/s}, S'\right)}\mu_B\left(\bm{z}_{S''};\,S\right)\nonumber\\
    & \hspace*{90px} \times \prod_{j \in S''}\left(z_j^{[\alpha_j]} - G_{\alpha_j}\right)
\end{align}
The above calculation is exact. Next, we will see the $\mu^{[q]}_{C, 2}$ factor can be dropped at the cost of a small error. This follows the intuition from the heuristic argument as this quantity is of order $1/|S|$ for a constant-size small set $S''$. However, the current Section will make this argument rigorous by expressing smallness in terms of a majorant rather than pointwise. We are then left with bounding
\begin{align}
    \left\langle \prod_{j \in S''}\left(z^{[\alpha_j]}_j - G_{\alpha_j}\right) \right\rangle_{\mu^{[q]}\left(\bm{z};\,\xi, S\right)} & \approx \sum_{\bm{z}_{S''} \in \{1, -1\}^{S'' \times \mathcal{T}_q}}\left\langle \mu^{[q]}_{C, 1}\left(\bm{z}_{S'}, \bm{z}_{S''};\,\xi_{s'/s}, S', S''\right) \right\rangle_{\mu^{[q - 1]}\left(\bm{z}_{S'};\,\xi_{s'/s}, S'\right)}\mu_B\left(\bm{z}_{S''};\,S\right)\nonumber\\
    & \hspace*{85px} \times \prod_{j \in S''}\left(z^{[\alpha_j]}_j - G_{\alpha_j}\right),\label{si:eq:centered_spin_products_cavity_expansion_step_1}
\end{align}
where we recall
\begin{align}
    \mu^{[q]}_{C, 1}\left(\bm{z}_{S'}, \bm{z}_{S''};\,\xi_{s'/s}, S', S''\right) & = \exp\left(-\frac{1}{2}\sum_{t, u \in \mathcal{T}_{q - 1}}\Gamma_t\Gamma_u\xi'_{s'/s}\left(\frac{1}{\left|S'\right|}\left\langle \bm{z}^{[t]}_{S'}, \bm{z}^{[u]}_{S'} \right\rangle\right)\left\langle \bm{z}^{[t]}_{S''}, \bm{z}^{[u]}_{S''} \right\rangle\right).
\end{align}
The above can be rewritten in terms of centered overlaps or centered spin products by Taylor-expanding $\xi_{s'/s}'$ around $\bm{G}$:
\begin{align}
    \mu^{[q]}_{C, 1}\left(\bm{z}_{S'}, \bm{z}_{S''};\,\xi_{s'/s}, S', S''\right) & = \exp\left(-\frac{1}{2}\sum_{m \geq 0}\frac{1}{m!}\sum_{t, u \in \mathcal{T}_{p - 1}}\Gamma_t\Gamma_u\xi_{s'/s}^{(1 + m)}\left(G_{t, u}\right)\left(\frac{1}{\left|S'\right|}\left\langle \bm{z}^{[t]}_{S'}, \bm{z}^{[u]}_{S'} \right\rangle - G_{t, u}\right)^m\left\langle \bm{z}^{[t]}_{S''}, \bm{z}^{[u]}_{S''} \right\rangle\right)\nonumber\\
    & = \exp\left(-\frac{1}{2}\sum_{m \geq 0}\frac{1}{m!}\sum_{t, u \in \mathcal{T}_{p - 1}}\Gamma_t\Gamma_u\xi^{(1 + m)}_{s'/s}\left(G_{t, u}\right)\left(\sum_{j \in S'}\frac{z^{[t]}_jz^{[u]}_j - G_{t, u}}{\left|S'\right|}\right)^m\left\langle \bm{z}^{[t]}_{S''}, \bm{z}^{[u]}_{S''} \right\rangle\right).
\end{align}
We then invoke the framework of Section~\ref{si:sec:non_colliding_taylor_expansion}, in particular Proposition~\ref{si:prop:polynomial_approximation_entire_function_path_integral_measure}, to approximate the above as a polynomial in centered spin products $z^{[t]}_jz^{[u]}_j - G_{t, u}$ with $j \in S'$ and $t, u \in \mathcal{T}_{q - 1}$. By linearity, this polynomial approximation reduces the estimation of cavity expectation $\left\langle \mu^{[q]}_{C, 1}\left(\bm{z}_{S'}, \bm{z}_{S''};\,\xi_{s'/s}, S', S''\right) \right\rangle_{\mu^{[q]}\left(\bm{z}_{S'};\,\xi, S', S''\right)}$ to estimating expectations of the form
\begin{align}
    \left\langle \prod_{j \in T''}\left(z_j^{[t_j]}z_j^{[u_j]} - G_{t_j, u_j}\right) \right\rangle_{\mu^{[q - 1]}\left(\bm{z}_{S'};\,\xi_{s'/s}, S'\right)},
\end{align}
where $T'' \subset S'$ will be a set of constant size (indeed, we will be able to truncate the expansion to constant order), and $\bm{t}, \bm{u} \in \mathcal{T}_{q - 1}^{T''}$. Given layer indices in the above lie in $\mathcal{T}_{q - 1}$, the moment can be bounded from $\varepsilon^{[q - 1]}\left(\xi_{s'/s}, \left|S'\right|, \left|S''\right|\right)$. Summing over all polynomial terms then ultimately bounds
\begin{align}
    \left\langle \prod_{j \in S''}\left(z^{[\alpha_j]} - G_{\alpha_j}\right) \right\rangle_{\mu^{[q]}\left(\bm{z}_{S};\,\xi, S\right)},
\end{align}
hence $\varepsilon^{[q]}\left(\xi, |S|, \left|S''\right|\right)$ in terms of $\varepsilon^{[q - 1]}$ and other contributions vanishing in the thermodynamic limit. Note this informal overview of the main argument overlooked a few technicalities; for instance, shifted mixture polynomial $\xi_{s'/s}$ does not have the same $\bm{G}$ matrix as original mixture polynomial $\xi$, so that
\begin{align}
    \left\langle \prod_{j \in T''}\left(z_j^{[t_j]}z_j^{[u_j]} - G_{t_j, u_j}\right) \right\rangle_{\mu^{[q - 1]}\left(\bm{z}_{S'};\,\xi_{s'/s}, S'\right)}
\end{align}
cannot immediately be bounded in terms of $\varepsilon^{[q - 1]}$. Besides, when expanding $\mu_{C, 1}^{[q]}\left(\bm{z}_{S'}, \bm{z}_{S''};\,\xi_{s'/s}, S', S''\right)$ as a low-degree polynomial in centered spin products $z_j^{[t]}z_j^{[u]} - G_{t, u}$, we did not explain how to handle the degree zero term in the expansion. This term, corresponding to empty $T'' = \varnothing$, indeed does not yield a vanishing cavity expectation, since
\begin{align}
    \left\langle \prod_{j \in T''}\left(z^{[t_j]}_jz^{[u_j]}_j - G_{t_j, u_j}\right) \right\rangle_{\mu^{[q - 1]}\left(\bm{z}_{S'};\,\xi_{s'/s}, S', S''\right)} & = \left\langle 1 \right\rangle_{\mu^{[q - 1]}\left(\bm{z}_{S'};\,\xi_{s'/s}, S', S''\right)}\nonumber\\
    & = 1.
\end{align}
We defer the treatment of these gaps and others to the detailed proof.

\subsubsection{Proof of main technical result}
\label{si:sec:centered_moments_maximum_recursion_proof}

We now turn our attention to the proof of the main technical result. As sketched in the introduction to Section~\ref{si:sec:noncolliding_taylor_expansion_application_spin_glass_qaoa_cavity_decomposition}, this result, expressed in Proposition~\ref{si:prop:centered_moments_maximum_recursion_simplified} and Corollary~\ref{si:cor:centered_moments_maximum_recursion_simplified_special_case}, is a recursive inequality in intermediate layer index $q$ for the centered moments function %
\begin{align}
    \varepsilon^{[q]}\left(\xi, s, s''\right) & := \max_{\substack{\alpha_1, \ldots, \alpha_{s''} \subset \mathcal{T}_q\\|\alpha_j|\,\mathrm{even}}}\left|\left\langle \prod_{j \in \left[s''\right]}\left(z^{[\alpha_j]} - G_{\alpha_j}\right) \right\rangle_{\mu^{[q]}\left(\bm{z};\,\xi, S\right)}\right|.
\end{align}
Following the idea outlined in the introduction, the recursive inequality will be derived from the cavity decomposition of the QAOA disorder-average path integral measure, stated in Lemma~\ref{si:lemma:cavity_decomposition_path_integral_measure} which we now prove:

\begin{replemma}{si:lemma:cavity_decomposition_path_integral_measure}
Consider the disorder-averaged spin-glass QAOA path integral measure
\begin{align}
    \mu^{[p]}\left(\bm{z};\,\xi, S\right) & = \mu_C^{[p]}\left(\bm{z};\,\xi, S\right)\mu_B^{[p]}\left(\bm{z};\,S\right)
\end{align}
computed in Proposition~\ref{si:prop:disorder_averaged_path_integral_measure}.  Let $S = S' \sqcup S''$ an arbitrary partition of qubit indices and denote $s := |S|, s' := \left|S'\right|, s'' := \left|S''\right|$ for brevity. Then, the cost part of the quantum circuit path integral measure decomposes as follows according to this partition:
\begin{align}
    \mu^{[p]}_C\left(\bm{z}_S;\,\xi, S\right) & = \mu^{[p]}_C\left(\bm{z}_{S'};\,\xi_{s'/s}, S'\right)\mu^{[p]}_{C, 1}\left(\bm{z}_{S'}, \bm{z}_{S''};\,\xi_{s'/s}, S', S''\right)\mu^{[p]}_{C, 2}\left(\bm{z}_{S'}, \bm{z}_{S''};\,\xi, S', S''\right),
\end{align}
where $\mu^{[p]}_C\left(\bm{z}_{S'};\,\xi_{s'/s}, S'\right)$ is the cost part of the QAOA path integral measure for circuit over smaller set of qubits $S'$ (replacing $S \to S'$ in the above definitions, including in cost function scalings $\left|S\right|^{-(q - 1)/2}$), but with transformed spin glass mixture polynomial
\begin{align}
    \xi_{s'/s}\left(x\right) & := \frac{s}{s'}\xi\left(\frac{s'}{s}x\right).
\end{align}
The other multiplicative contributions $\mu_{C, 1}$ and $\mu_{C, 2}$ are not necessarily path integral measures and are defined as:
\begin{align}
    \mu^{[p]}_{C, 1}\left(\bm{z}_{S'}, \bm{z}_{S''};\,\xi_{s'/s}, S', S''\right) & := \exp\left(-\frac{1}{2}\sum_{t, u \in \mathcal{T}_{p - 1}}\Gamma_t\Gamma_u\xi'_{s'/s}\left(\frac{1}{\left|S'\right|}\left\langle \bm{z}^{[t]}_{S'}, \bm{z}^{[u]}_{S'} \right\rangle\right)\left\langle \bm{z}_{S''}^{[t]}, \bm{z}_{S''}^{[u]} \right\rangle\right),\\
    \mu^{[p]}_{C, 2}\left(\bm{z}_{S'}, \bm{z}_{S''};\,\xi, S', S''\right) & := \exp\left(-\frac{1}{2}\sum_{t, u \in \mathcal{T}_{p - 1}}\Gamma_t\Gamma_u\sum_{m \geq 2}\frac{1}{m!}\xi^{(m)}\left(\frac{1}{|S|}\left\langle \bm{z}_{S'}^{[t]}, \bm{z}_{S'}^{[u]} \right\rangle\right)\frac{1}{|S|^{m - 1}}\left\langle \bm{z}_{S''}^{[t]}, \bm{z}_{S''}^{[u]} \right\rangle^m\right).
\end{align}
\end{replemma}
\begin{proof}
We start with the expression of the cost contribution to the path integral measure established in Proposition~\ref{si:prop:disorder_averaged_path_integral_measure}, Eq.~\ref{si:eq:disorder_averaged_path_integral_measure_cost_contribution}:
\begin{align}
    \mu^{[p]}_C\left(\bm{z}_S;\,\xi, S\right) & = \exp\left(-\frac{|S|}{2}\sum_{t, u \in \mathcal{T}_{p - 1}}\Gamma_t\Gamma_u\xi\left(\frac{1}{|S|}\left\langle \bm{z}_S^{[t]}, \bm{z}_S^{[u]} \right\rangle\right)\right).\label{si:eq:path_integral_measure_cost_contribution_repeated}
\end{align}
The main idea of the cavity decomposition is too linearly decompose the overlap into contribution from $S'$ and another from $S''$:
\begin{align}
    \frac{1}{|S|}\left\langle \bm{z}^{[t]}_S, \bm{z}^{[u]}_S \right\rangle & = \frac{1}{|S|}\left\langle \bm{z}^{[t]}_{S'}, \bm{z}^{[u]}_{S'} \right\rangle + \frac{1}{|S|}\left\langle \bm{z}^{[t]}_{S''}, \bm{z}^{[u]}_{S''} \right\rangle\nonumber\\
    & = \frac{\left|S'\right|}{|S|}\frac{1}{\left|S'\right|}\left\langle \bm{z}^{[t]}_{S'}, \bm{z}^{[u]}_{S'} \right\rangle + \frac{1}{|S|}\left\langle \bm{z}^{[t]}_{S''}, \bm{z}^{[u]}_{S''} \right\rangle,
\end{align}
and to expand the (non-linear) polynomial $\xi$ of the $S$ overlap in terms the $S'$ overlap. This is done by Taylor expansion (which only has a finite number of terms as $\xi$ is polynomial):
\begin{align}
    \xi\left(\frac{1}{|S|}\left\langle \bm{z}^{[t]}_S, \bm{z}^{[u]}_S \right\rangle\right) & = \xi\left(\frac{s'}{s}\frac{1}{\left|S'\right|}\left\langle \bm{z}^{[t]}_{S'}, \bm{z}^{[u]}_{S'} \right\rangle + \frac{1}{|S|}\left\langle \bm{z}^{[t]}_{S''}, \bm{z}^{[u]}_{S''} \right\rangle\right)\nonumber\\
    & = \xi\left(\frac{s'}{s}\frac{1}{\left|S'\right|}\left\langle \bm{z}^{[t]}_{S'}, \bm{z}^{[u]}_{S'} \right\rangle\right) + \xi'\left(\frac{s'}{s}\frac{1}{\left|S'\right|}\left\langle \bm{z}^{[t]}_{S'}, \bm{z}^{[u]}_{S'} \right\rangle\right)\frac{1}{|S|}\left\langle \bm{z}^{[t]}_{S''}, \bm{z}^{[u]}_{S''} \right\rangle\nonumber\\
    & \hspace*{15px} + \sum_{m \geq 2}\frac{1}{m!}\xi^{(m)}\left(\frac{1}{|S|}\left\langle \bm{z}^{[t]}_{S'}, \bm{z}^{[u]}_{S'} \right\rangle\right)\left(\frac{1}{|S|}\left\langle \bm{z}^{[t]}_{S''}, \bm{z}^{[u]}_{S''} \right\rangle\right)^m,
\end{align}
where we deliberately singled out the order $0$ and $1$ terms.
Plugging this into the cost contribution expression Eq.~\ref{si:eq:path_integral_measure_cost_contribution_repeated},
\begin{align}
    & \exp\left(-\frac{|S|}{2}\sum_{t, u \in \mathcal{T}_{p - 1}}\Gamma_t\Gamma_u\xi\left(\frac{1}{|S|}\left\langle \bm{z}^{[t]}_S, \bm{z}^{[u]}_S \right\rangle\right)\right)\nonumber\\
    & = \exp\left(-\frac{|S|}{2}\sum_{t, u \in \mathcal{T}_{p - 1}}\Gamma_t\Gamma_u\xi\left(\frac{s'}{s}\frac{1}{\left|S'\right|}\left\langle \bm{z}^{[t]}_{S'}, \bm{z}^{[u]}_{S'} \right\rangle\right)\right)\nonumber\\
    & \hspace*{20px} \times \exp\left(-\frac{|S|}{2}\sum_{t, u \in \mathcal{T}_{p - 1}}\Gamma_t\Gamma_u\xi'\left(\frac{s'}{s}\frac{1}{\left|S'\right|}\left\langle \bm{z}^{[t]}_{S'}, \bm{z}^{[u]}_{S'} \right\rangle\right)\frac{1}{|S|}\left\langle \bm{z}^{[t]}_{S''}, \bm{z}^{[u]}_{S''} \right\rangle\right)\nonumber\\
    & \hspace*{20px} \times \exp\left(-\frac{|S|}{2}\sum_{t, u \in \mathcal{T}_{p - 1}}\Gamma_t\Gamma_u\sum_{m \geq 2}\frac{1}{m!}\xi^{(m)}\left(\frac{1}{|S|}\left\langle \bm{z}_{S'}^{[t]}, \bm{z}_{S'}^{[u]} \right\rangle\right)\left(\frac{1}{|S|}\left\langle \bm{z}_{S''}^{[t]}, \bm{z}_{S''}^{[u]} \right\rangle\right)^{m}\right)\nonumber\\
    & =  \exp\left(-\frac{\left|S'\right|}{2}\sum_{t, u \in \mathcal{T}_{p - 1}}\Gamma_t\Gamma_u\frac{s}{s'}\xi\left(\frac{s'}{s}\frac{1}{\left|S'\right|}\left\langle \bm{z}^{[t]}_{S'}, \bm{z}^{[u]}_{S'} \right\rangle\right)\right)\nonumber\\
    & \hspace*{20px} \times \exp\left(-\frac{1}{2}\sum_{t, u \in \mathcal{T}_{p - 1}}\Gamma_t\Gamma_u\xi'\left(\frac{s'}{s}\frac{1}{\left|S'\right|}\left\langle \bm{z}^{[t]}_{S'}, \bm{z}^{[u]}_{S'} \right\rangle\right)\left\langle \bm{z}^{[t]}_{S''}, \bm{z}^{[u]}_{S''} \right\rangle\right)\nonumber\\
    & \hspace*{20px} \times \exp\left(-\frac{1}{2}\sum_{t, u \in \mathcal{T}_{p - 1}}\Gamma_t\Gamma_u\sum_{m \geq 2}\frac{1}{m!}\xi^{(m)}\left(\frac{1}{|S|}\left\langle \bm{z}_{S'}^{[t]}, \bm{z}_{S'}^{[u]} \right\rangle\right)\frac{1}{|S|^{m - 1}}\left\langle \bm{z}_{S''}^{[t]}, \bm{z}_{S''}^{[u]} \right\rangle^{m}\right).
\end{align}
Defining
\begin{align}
    \xi_{s'/s}\left(x\right) & := \frac{s}{s'}\xi\left(\frac{s'}{s}x\right),
\end{align}
so that $\xi_{s'/s}'\left(x\right) = \xi'\left(\frac{s'}{s}x\right)$, the above provides the desired decomposition.
\end{proof}

Lemma~\ref{si:lemma:cavity_decomposition_path_integral_measure} is the key ingredient allowing a recursive reasoning in the bounding of maxmimum centered moments. This is because the decomposition of path integral measure $\mu^{[p]}\left(\bm{z};\,\xi, S\right)$, over a bit matrix indexed by layers $\mathcal{T}_p$, involves $\mu^{[p]}_{C, 1}\left(\bm{z}_{S'}, \bm{z}_{S''};\,\xi_{s'/s}, S', S''\right)$ and $\mu^{[p]}_{C, 2}\left(\bm{z}_{S'}, \bm{z}_{S''};\,\xi, S', S''\right)$, over layer indices $\mathcal{T}_{p - 1}$ only. Before reasoning about maximum centered moments, we state their formal definition hereafter:

\begin{definition}[Maximal value of centered moments given system size and moment degree]
\label{si:def:centered_moments_maximum}
Given a system size $n$ and centered moment degree $s'' \leq n$, we define:
\begin{align}
    \varepsilon^{[p]}\left(\xi, n, s''\right) & := \max_{\substack{\alpha_1, \ldots, \alpha_{s''} \subset \mathcal{T}_p\\\left|\alpha_j\right|\,\mathrm{even}}}\left|\left\langle \prod_{j \in [s'']}\left(z_j^{[\alpha_j]} - G_{\alpha_j}\right) \right\rangle_{\mu^{[p]}\left(\bm{z};\,\xi,[n]\right)}\right|.
\end{align}
Note this is well-defined for all system size $n$ and centered moment degree $s''$ as the maximum of a finite set. Also, by invariance of the disorder-averaged path integral measure $\mu^{[p]}\left(\,\cdot\,;\,\xi, S\right)$ under permutation of spins, for all qubit labels set $S$ and all subset $S'' \subset S$,
\begin{align}
    \max_{\substack{\alpha_{j} \subset \mathcal{T}_p\,\forall j \in S''\\\left|\alpha_j\right|\,\mathrm{even}}}\left|\left\langle \prod_{j \in S''}\left(z_j^{[\alpha_j]} - G_{\alpha_j}\right) \right\rangle_{\mu\left(\bm{z};\,\xi, S\right)}\right|
\end{align}
only depends on $\left|S''\right|$. Likewise, for all set of qubit labels $S$ and all $s'' \leq |S|$,
\begin{align}
    \max_{S'' \subset S\,:\,\left|S''\right| = s''}\max_{\substack{\alpha_j \subset \mathcal{T}_p\,\forall j \in S''\\\left|\alpha_j\right|\,\mathrm{even}}}\left|\left\langle \prod_{j \in S''}\left(z_j^{[\alpha_j]} - G_{\alpha_j}\right) \right\rangle_{\mu\left(\bm{z};\,\xi, S\right)}\right| & = \varepsilon^{[p]}\left(\xi, \left|S\right|, s''\right).
\end{align}
From $\varepsilon^{[p]}$, we also define the maximum over upper-bounded degree:
\begin{align}
    \varepsilon_{\mathrm{max}}^{[q]}\left(\xi, n, s''\right) & := \sup_{1 \leq d' \leq s''}\varepsilon^{[p]}\left(\xi, n, d'\right).
\end{align}
\end{definition}

The following Lemma is the core technical result bounding the maximum centered moments at level $q$ recursively in level $q$. The proof is based on two approximations of the cavity decomposition of the path integral measure derived in Lemma~\ref{si:lemma:cavity_decomposition_path_integral_measure}:
\begin{itemize}
    \item Approximating $\mu^{[q + 1]}_{C, 2}\left(\bm{z}_{S'}, \bm{z}_{S''};\,\xi, S', S''\right) \approx 1$ (in the majorization sense), effectively dropping this factor from the cavity decomposition. This step is elementary and based on the boundedness property of path integral measures (Corollary \ref{si:cor:path_integral_measure_boundedness_monomials}).
    \item For each fixed small subsystem trajectory $\bm{z}_{S''} \in \{1, -1\}^{S'' \times \mathcal{T}_{q + 1}}$, approximating $\mu^{[q + 1]}_{C, 1}\left(\bm{z}_{S'}, \bm{z}_{S''};\,\xi_{s'/s}, S', S''\right)$ as a fixed-degree polynomial in centered spin products $z^{[t]}_jz^{[u]}_j - G_{t, u}$ for $j \in S'$ and $t, u \in \mathcal{T}_q$. Bounding these new centered moments over a different subsystem brings in the maximum centered moments at previous order $\varepsilon^{[q]}$.
\end{itemize}
While the bounds derived in this Lemma are cumbersome, they will be iteratively simplified in subsequent Proposition~\ref{si:prop:centered_moments_maximum_recursion_simplified} and Corollary~\ref{si:cor:centered_moments_maximum_recursion_simplified_special_case} at the cost of weakening upper bounds on $\bm{\gamma}$ angles.

\begin{lemma}[Recursion relation for maximum value of centered moment]
\label{si:lemma:centered_moments_maximum_recursion}
Consider the maximum value $\varepsilon^{[q]}_{\mathrm{max}}\left(\xi, n, s''\right)$ of centered moments of maximum degree $s''$, as introduced in Definition~\ref{si:def:centered_moments_maximum}. This quantity satisfies the following recursive inequality in $q$ for all $0 \leq q \leq p$, for qubit set size $s = s' + s''$, partitioned into a ``small qubit set" size $s''$ and a ``large qubit set" size $s'$:
\begin{align}
    \varepsilon_{\mathrm{max}}^{[q + 1]}\left(\xi, s, s''\right) & \leq \overline{\varepsilon_1}\left(\xi, s', s''\right) + \overline{\varepsilon_2}\left(\xi, s', s''\right) + \overline{\varepsilon_3}\left(\xi, s', s'', R\right) + \overline{\varepsilon_4}\left(\xi, s', s'', R\right)\nonumber\\
    & \hspace*{10px} + c\left(\xi, s'', R, d\right)\varepsilon^{[q]}_{\mathrm{max}}\left(\xi_{s'/s}, s', d\right),\label{si:eq:centered_moments_maximum_recursion}
\end{align}
where
\begin{align}
    \overline{\varepsilon_1}\left(\xi, s', s''\right) & := \gamma^2p^2\xi''(2)\frac{\left(s''\right)^2}{s}\exp\left(\left(2\xi'(1) + \xi''(2)\right)\gamma^2p^2s''\right),\\
    \overline{\varepsilon_2}\left(\xi, s', s''\right) & := 2\gamma^2p^2\xi''(1)\frac{\left(s''\right)^2}{s}4^{s''},\\
    \overline{\varepsilon_3}\left(\xi, s', s'', R, d\right) & := 2^{8p^2 + 1}e\left(2e\gamma^2p^2x''\right)^p\frac{s''}{s}\exp\left(4\gamma^2p^2R\xi''(1 + 2R)s''\right),\\
    \overline{\varepsilon_4}\left(\xi, s', s'', R\right) & := 2^{4p^2}\left(\left(\frac{4}{R}\right)^d\binom{d + 4p^2 - 1}{4p^2 - 1} + \frac{1}{s'}4p^2\left(4p^2 + 1\right)\frac{\left(4/R\right)^2}{\left(1 - 4/R\right)^{4p^2 + 2}}\right)\nonumber\\
    & \hspace*{15px} \times \exp\left(\left(4\gamma^2p^2R\xi''(1 + 2R) + 4\gamma^2p^2\xi'(1) + p \log 2\right)s''\right),\\
    c\left(\xi, s'', R, d\right) & := 2^{8p^2}\exp\left(4\gamma^2p^2R\xi''(1 + 2R)s''\right),
\end{align}
and $R > 4$, $d \geq 0$ are free parameters. The definition of $\overline{\varepsilon_3}$ in the above equations depends on an arbitrary bound
\begin{align}
    \xi''(1) & \leq x''
\end{align}
on the second derivative of the spin glass mixture polynomial. For these bounds to hold, we enforce constraints:
\begin{align}
    2\gamma^2p^2x'' & \geq 2,\\
    \frac{s''}{s} & \leq \kappa(x''),
\end{align}
where we introduce the shorthand
\begin{align}
    \kappa(x) & := \frac{1}{2\left(2e\gamma^2p^2x\right)^{p + 1}}
\end{align}
for the maximal admissible relative set size, as a function of a bound $x$ on the mixture polynomial's second derivative; here it is applied at $x''$.
The first constraint can always be satisfied by choosing a large enough bound on $\gamma$. The second constraint can be satisfied by choosing a sufficiently small (by an amount depending only on $p$, $\gamma$, $x''$ and not on $s$) relative set size $s''/s$.

\begin{proof}
We wish to bound
\begin{align}
    \left\langle \prod_{j \in S''}\left(z^{[\alpha_j]}_j - G_{\alpha_j}\right) \right\rangle_{\mu^{[q + 1]}\left(\bm{z}_S;\,\xi, S\right)},
\end{align}
where for all $j \in S''$, $\alpha_j \in \binom{\mathcal{T}_{q + 1}}{d_j}$ is a set of $d_j$ layer indices from $\mathcal{T}_{q + 1} = \{1, 2, \ldots, q + 1, q + 2, -q - 2, -q - 1, \ldots, -2, -1\}$. $G_{\alpha_j}$ is the $d_j$-point time autocorrelation of $Z$ evaluated at time indices $\alpha_j$, where for conciseness the autocorrelation degree $d_j$ is omitted from the notation. From the cavity expansion of the spin-glass QAOA path integral measure Lemma~\ref{si:lemma:cavity_decomposition_path_integral_measure}, the above pseudo-expectation can be expanded:
\begin{align}
    & \left\langle \prod_{j \in S''}\left(z^{\left[\alpha_j\right]}_j - G_{\alpha_j}\right) \right\rangle_{\mu^{[q + 1]}\left(\bm{z}_S;\,\xi,\,S\right)}\nonumber\\
    & = \sum_{\bm{z}_{S''} \in \{1, -1\}^{S'' \times \mathcal{T}_{q + 1}}}\left\langle \mu^{[q + 1]}_{C, 1}\left(\bm{z}_{S'}, \bm{z}_{S''};\,\xi_{s'/s}, S', S''\right)\mu^{[q + 1]}_{C, 2}\left(\bm{z}_{S'}, \bm{z}_{S''};\,\xi, S', S''\right)\right\rangle_{\mu^{[q + 1]}\left(\bm{z}_{S'};\,\xi_{s'/s},\,S'\right)}\nonumber\\
    & \hspace*{90px} \times \mu^{[q + 1]}_B\left(\bm{z}_{S''};\,S''\right)\prod_{j \in S''}\left(z^{\left[\alpha_j\right]}_j - G_{\alpha_j}\right)\nonumber\\
    & = \sum_{\bm{z}_{S''} \in \{1, -1\}^{S'' \times \mathcal{T}_{q + 1}}}\left\langle \mu^{[q + 1]}_{C, 1}\left(\bm{z}_{S'}, \bm{z}_{S''};\,\xi_{s'/s}, S', S''\right)\mu^{[q + 1]}_{C, 2}\left(\bm{z}_{S'}, \bm{z}_{S''};\,\xi, S', S''\right)\right\rangle_{\mu^{[q]}\left(\bm{z}_{S'};\,\xi_{s'/s},\,S'\right)}\nonumber\\
    & \hspace*{90px} \times \mu^{[q + 1]}_B\left(\bm{z}_{S''};\,S''\right)\prod_{j \in S''}\left(z^{\left[\alpha_j\right]}_j - G_{\alpha_j}\right),\label{si:eq:centered_moment_cavity_expansion_exact}
\end{align}
where the path integral expectation $\left\langle\,\cdot\,\right\rangle_{\mu^{[q]}\left(\bm{z}_{S'};\,\xi_{s'/s}, S'\right)}$ is according to the disorder-averaged spin-glass $q$-layers QAOA (with modified mixture polynomial $\xi_{s'/s}$) acting on spins $S'$ only. The simplification from $(q + 1)$ layers to $q$ layers between lines $2$ and $3$ is possible since $\mu^{[q + 1]}_{C, 1}$ and $\mu^{[q + 1]}_{C, 2}$ only depend on bit matrices restricted to layers $\mathcal{T}_q = \left\{1, \ldots, q + 1, -q - 1, \ldots, -1\right\}$ (Proposition~\ref{si:prop:path_integral_measure_unitary_cancellation}). We first replace $\mu^{[q + 1]}_{C, 2}\left(\bm{z}_{S'}, \bm{z}_{S''};\,\xi, S', S''\right)$ by $1$ inside the path integral expectation with the help of Lemma~\ref{si:lemma:mu_c_2_constant_degree_polynomial_approximation}:
\begin{align}
    & \left\langle \mu^{[q + 1]}_{C, 1}\left(\bm{z}_{S'}, \bm{z}_{S''};\,\xi_{s'/s}, S', S''\right)\mu^{[q + 1]}_{C, 2}\left(\bm{z}_{S'}, \bm{z}_{S''};\,\xi, S', S''\right) \right\rangle_{\mu^{[q]}\left(\bm{z}_{S'};\,\xi_{s'/s}, S'\right)}\nonumber\\
    & = \left\langle \mu^{[q + 1]}_{C, 1}\left(\bm{z}_{S'}, \bm{z}_{S''};\,\xi_{s'/s}, S', S''\right) \right\rangle_{\mu^{[q]}\left(\bm{z}_{S'};\,\xi_{s'/s}, S'\right)}\nonumber\\
    & \hspace*{15px} + \left\langle \mu^{[q + 1]}_{C, 1}\left(\bm{z}_{S'}, \bm{z}_{S''};\,\xi_{s'/s}, S', S''\right)\widetilde{\mu}^{[q + 1]}_{C, 2}\left(\bm{z}_{S'}, \bm{z}_{S''};\,\xi, S', S''\right) \right\rangle_{\mu^{[q]}\left(\bm{z}_{S'};\,\xi_{s'/s}, S'\right)},
\end{align}
where the second term is bounded as:
\begin{align}
    & \left|\left\langle \mu^{[q + 1]}_{C, 1}\left(\bm{z}_{S'}, \bm{z}_{S''};\,\xi_{s'/s}, S', S''\right)\widetilde{\mu}^{[q + 1]}_{C, 2}\left(\bm{z}_{S'}, \bm{z}_{S''};\,\xi, S', S''\right) \right\rangle_{\mu^{[q]}\left(\bm{z}_{S'};\,\xi_{s'/s}, S'\right)}\right|\nonumber\\
    & \leq \overline{\mu^{[q + 1]}_{C, 1}}\left(\bm{1}_{S' \times \mathcal{T}_q};\,\xi_{s'/s}, S', S''\right)\overline{\widetilde{\mu}^{[q + 1]}_{C, 2}}\left(\bm{1}_{S' \times \mathcal{T}_q};\,\xi, S', S''\right)\nonumber\\
    & \leq \exp\left(2(q + 1)^2\gamma^2\xi'(1)\left|S''\right|\right)\frac{\gamma^2(q + 1)^2\left|S''\right|^2}{|S|}\xi''(2)\exp\left(\frac{\gamma^2(q + 1)^2\left|S''\right|^2}{|S|}\xi''(2)\right)\nonumber\\
    & \leq \frac{\gamma^2p^2\left|S''\right|^2}{|S|}\xi''(2)\exp\left(\left(2\xi'(1) + \frac{\left|S''\right|}{|S|}\xi''(2)\right)p^2\gamma^2\left|S''\right|\right)\nonumber\\
    & \leq \frac{\gamma^2p^2\left|S''\right|^2}{|S|}\xi''(2)\exp\left(\left(2\xi'(1) + \xi''(2)\right)p^2\gamma^2\left|S''\right|\right),
\end{align}
where in the second line, we used majorization bounds for $\mu^{[q + 1]}_{C, 1}$ and $\widetilde{\mu}^{[q + 1]}_{C, 2}$ provided by Lemmas~\ref{si:lemma:mu_c_1_majorization_bound}, \ref{si:lemma:mu_c_2_constant_degree_polynomial_approximation} respectively. Therefore, it holds
\begin{align}
    & \left\langle \prod_{j \in S''}\left(z^{[\alpha_j]}_j - G_{\alpha_j}\right) \right\rangle_{\mu^{[q + 1]}\left(\bm{z}_{S'};\,\xi, S\right)}\nonumber\\
    & = \sum_{\bm{z}_{S''} \in \{1, -1\}^{S'' \times \mathcal{T}_{q + 1}}}\left\langle \mu^{[q + 1]}_{C, 1}\left(\bm{z}_{S'}, \bm{z}_{S''};\,\xi_{s'/s}, S', S''\right) \right\rangle_{\mu^{[q]}\left(\bm{z}_{S'};\,\xi_{s'/s}, S'\right)}\mu^{[q + 1]}_B\left(\bm{z}_{S''};\,S''\right)\prod_{j \in S''}\left(z^{[\alpha_j]}_j - G_{\alpha_j}\right)\nonumber\\
    & \hspace*{20px} + \varepsilon_1\left(\xi, S', S''\right),\label{si:eq:centered_moment_cavity_expansion_approximation_1}
\end{align}
with
\begin{align}
    \varepsilon_1\left(\xi, S', S''\right) & \leq \overline{\varepsilon_1}\left(\xi, S', S''\right) := \frac{\gamma^2p^2\left|S''\right|^2}{|S|}\xi''(2)\exp\left(\left(2\xi'(1) + \xi''(2)\right)p^2\gamma^2\left|S''\right|\right).
\end{align}
We now analyze the path integral expectation of $\mu_{C, 1}$ using the additive decomposition of this function of a bit matrix provided in Lemma~\ref{si:lemma:mu_c_1_constant_degree_polynomial_approximation}. By Eq.~\ref{si:eq:mu_c_1_additive_contributions} of the Lemma,
\begin{align}
    & \left\langle \mu^{[q + 1]}_{C, 1}\left(\bm{z}_{S'}, \bm{z}_{S''};\,\xi_{s'/s}, S', S''\right) \right\rangle_{\mu^{[q]}\left(\bm{z}_{S'};\,\xi_{s'/s}, S'\right)}\nonumber\\
    & = \left\langle 1 + \mu^{[q + 1]}_{C, 1', d}\left(\bm{z}_{S'}, \bm{z}_{S''};\,\xi_{s'/s}, S', S''\right) + \mu^{[q + 1]}_{C, 1'', d}\left(\bm{z}_{S'}, \bm{z}_{S''};\,\xi_{s'/s}, S', S''\right) \right\rangle_{\mu^{[q]}\left(\bm{z}_{S'};\,\xi_{s'/s}, S'\right)}\mu^{[q + 1]}_{C,SB}\left(\bm{z}_{S''};\,\xi_{s'/s}, S''\right),
\end{align}
where Taylor expansion order $d$ is left unspecified. Combining the displacement contribution of the spin-boson path integral measure $\mu^{[q + 1]}_{C, SB}\left(\bm{z}_{S''};\,\xi_{s'/s}, S''\right)$ with the mixer contribution $\mu^{[q + 1]}_B\left(\bm{z}_{S''};\,S''\right)$ of the original QAOA circuit in Eq.~\ref{si:eq:centered_moment_cavity_expansion_approximation_1},
\begin{align}
    & \left\langle \mu^{[q + 1]}_{C, 1}\left(\bm{z}_{S'}, \bm{z}_{S''};\,\xi_{s'/s}, S', S''\right) \right\rangle_{\mu^{[q]}\left(\bm{z}_{S'};\,\xi_{s'/s}, S'\right)}\mu_B\left(\bm{z}_{S''};\,S''\right)\nonumber\\
    & = \left\langle 1 + \mu^{[q + 1]}_{C, 1', d}\left(\bm{z}_{S'}, \bm{z}_{S''};\,\xi_{s'/s}, S', S''\right) + \mu^{[q + 1]}_{C, 1'', d}\left(\bm{z}_{S'}, \bm{z}_{S''};\,\xi_{s'/s}, S', S''\right) \right\rangle_{\mu^{[q]}\left(\bm{z}_{S'};\,\xi_{s'/s}, S'\right)}\mu^{[q + 1]}_{SB}\left(\bm{z}_{S''};\,\xi_{s'/s}, S''\right),\label{si:eq:centered_moment_cavity_expansion_cavity_expectation_decomposition}
\end{align}
where $\mu^{[q + 1]}_{SB}\left(\bm{z}_{S''};\,\xi_{s'/s}, S''\right)$ is the full path integral measure of a $(q + 1)$-layer spin-boson circuit with spins labelled $S''$ and single-spin influence functional given by $\left(\xi'_{s'/s}\left(G_{t, u}\right)\right)_{t, u \in \mathcal{T}_q} \in \mathbb{C}^{\mathcal{T}_q \times \mathcal{T}_q}$. By linearity, Eq.~\ref{si:eq:centered_moment_cavity_expansion_cavity_expectation_decomposition} decomposes the cavity expectation under measure $\mu^{[q]}\left(\bm{z}_{S'};\,\xi_{s'/s}, S'\right)$ into three terms:
\begin{align}
    & \left\langle 1 \right\rangle_{\mu^{[q]}\left(\bm{z}_{S'};\,\xi_{s'/s}, S'\right)},\\
    & \left\langle \mu^{[q + 1]}_{C, 1', d}\left(\bm{z}_{S'}, \bm{z}_{S''};\,\xi_{s'/s}, S', S''\right) \right\rangle_{\mu^{[q]}\left(\bm{z}_{S'};\,\xi_{s'/s}, S'\right)},\\
    & \left\langle \mu^{[q + 1]}_{C, 1'', d}\left(\bm{z}_{S'}, \bm{z}_{S''};\,\xi_{s'/s}, S', S''\right) \right\rangle_{\mu^{[q]}\left(\bm{z}_{S'};\,\xi_{s'/s}, S'\right)}
\end{align}
which we treat successively.

\paragraph{The $1$ term}\mbox{}

For this term, the cavity expectation in Eq.~\ref{si:eq:centered_moment_cavity_expansion_cavity_expectation_decomposition} is trivial:
\begin{align}
    \left\langle 1 \right\rangle_{\mu^{[q]}\left(\bm{z}_{S'};\,\xi_{s'/s}, S'\right)} & = 1.
\end{align}
The contribution of this term to the centered moment is then:
\begin{align}
    \left\langle \prod_{j \in S''}\left(z_j^{\left[\alpha_j\right]} - G_{\alpha_j}\right) \right\rangle_{\mu^{[q + 1]}\left(\bm{z}_S;\,\xi, S\right)} & \supset \sum_{\bm{z}_{S''} \in \{1, -1\}^{S'' \times \mathcal{T}_q}}\mu^{[q + 1]}_{SB}\left(\bm{z}_{S''};\,\xi_{s'/s}, S''\right)\prod_{j \in S''}\left(z_j^{\left[\alpha_j\right]} - G_{\alpha_j}\right)\nonumber\\
    & = \left\langle \prod_{j \in S''}\left(z_j^{\left[\alpha_j\right]} - G_{\alpha_j}\right) \right\rangle_{\mu_{SB}^{[q + 1]}\left(\bm{z}_{S''};\,\xi_{s'/s}, S''\right)}.
\end{align}
The intuition is that this term is small, since $\xi_{s'/s} \approx \xi$ and by definition of $\bm{G}$ tensors and independence of spins under spin-boson measure $\mu^{[q + 1]}_{SB}\left(\bm{z}_{S''};\,\xi, S''\right)$, the expectation under this measure vanishes:
\begin{align}
    \left\langle \prod_{j \in S''}\left(z_j^{\left[\alpha_j\right]} - G_{\alpha_j}\right) \right\rangle_{\mu^{[q + 1]}_{SB}\left(\bm{z}_{S''};\,\xi, S''\right)} & = \prod_{j \in S''}\left\langle z_j^{\left[\alpha_j\right]} - G_{\alpha_j} \right\rangle_{\mu_{SB}^{[q + 1]}\left(\bm{z}_{S''};\,\xi, S''\right)}\nonumber\\
    & = 0.
\end{align}
The expectation under the perturbed spin-boson measure $\mu^{[q + 1]}_{SB}\left(\bm{z}_{S''};\,\xi_{s'/s}, S''\right)$ is then small by change of measure:
\begin{align}
    \left\langle \prod_{j \in S''}\left(z_j^{\left[\alpha_j\right]} - G_{\alpha_j}\right) \right\rangle_{\mu^{[q + 1]}_{SB}\left(\bm{z}_{S''};\,\xi_{s'/s}, S''\right)} & = \left\langle \frac{\mu^{[q + 1]}_{SB}\left(\bm{z}_{S''};\,\xi_{s'/s}, S''\right)}{\mu^{[q + 1]}_{SB}\left(\bm{z}_{S''};\,\xi, S''\right)}\prod_{j \in S''}\left(z_j^{\left[\alpha_j\right]} - G_{\alpha_j}\right) \right\rangle_{\mu^{[q + 1]}_{SB}\left(\bm{z}_{S''};\,\xi, S''\right)}\nonumber\\
    & = \left\langle \left(\frac{\mu^{[q + 1]}_{SB}\left(\bm{z}_{S''};\,\xi_{s'/s}, S''\right)}{\mu^{[q + 1]}_{SB}\left(\bm{z}_{S''};\,\xi, S''\right)} - 1\right)\prod_{j \in S''}\left(z_j^{\left[\alpha_j\right]} - G_{\alpha_j}\right) \right\rangle_{\mu^{[q + 1]}_{SB}\left(\bm{z}_{S''};\,\xi, S''\right)},
\end{align}
where in the second line, we subtracted one by virtue of the previous vanishing identity. We then invoke the majorization bound on ratio of path integral spin-boson measures
\begin{align}
    \psi\left(\bm{w}_{S''}\right) & := \frac{\mu^{[q + 1]}_{SB}\left(\bm{w}_{S''};\,\xi_{s'/s}, S''\right)}{\mu^{[q + 1]}_{SB}\left(\bm{w}_{S''};\,\xi, S''\right)} - 1
\end{align}
stated in Lemma~\ref{si:lemma:spin_boson_measures_ratio_majorization_bound}:
\begin{align}
    \psi\left(\bm{w}_{S''}\right) & \preceq \overline{\psi}\left(\bm{w}_{S''}\right),
\end{align}
where
\begin{align}
    \overline{\psi}\left(\bm{w}_{S''}\right) & := \frac{\gamma^2}{2}\sum_{t, u \in \mathcal{T}_q}\left|\xi'_{s'/s}\left(G_{t, u}\right) - \xi'\left(G_{t, u}\right)\right|\left\langle \bm{w}^{[t]}_{S''}, \bm{w}^{[u]}_{S''} \right\rangle\exp\left(\frac{\gamma^2}{2}\sum_{t, u \in \mathcal{T}_q}\left|\xi'_{s'/s}\left(G_{t, u}\right) - \xi'\left(G_{t, u}\right)\right|\left\langle \bm{w}^{[t]}_{S''}, \bm{w}^{[u]}_{S''} \right\rangle\right).
\end{align}
Using trivial majorization bound for the product of centered spins:
\begin{align}
    \prod_{j \in S''}\left(z_j^{\left[\alpha_j\right]} - G_{\alpha_j}\right) & \preceq \prod_{j \in S''}\left(z_j^{\left[\alpha_j\right]} + 1\right),
\end{align}
evaluating to $2^{\left|S''\right|}$ at $\bm{z}_{S''} = \bm{1}_{S'' \times \mathcal{T}_{q + 1}}$, one can then write:
\begin{align}
    \left|\left\langle \prod_{j \in S''}\left(z_j^{[\alpha_j]} - G_{\alpha_j}\right) \right\rangle_{\mu^{[q + 1]}_{SB}\left(\bm{z}_{S''};\,\xi_{s'/s}, S''\right)}\right| & \leq 2^{\left|S''\right|}\overline{\psi}\left(\bm{1}_{S'' \times \mathcal{T}_{q + 1}}\right).
\end{align}
We finally bound
\begin{align}
    \overline{\psi}\left(\bm{1}_{S'' \times \mathcal{T}_{q + 1}}\right) & \leq \frac{\gamma^2}{2}\sum_{t, u \in \mathcal{T}_q}\left|\xi'_{s'/s}\left(G_{t, u}\right) - \xi'\left(G_{t, u}\right)\right|\left|S''\right|\exp\left(\frac{\gamma^2}{2}\sum_{t, u \in \mathcal{T}_q}\left|\xi'_{s'/s}\left(G_{t, u}\right) - \xi'\left(G_{t, u}\right)\right|\left|S''\right|\right)\nonumber\\
    & \leq \frac{\gamma^2}{2}\left|\mathcal{T}_q\right|^2\frac{\left|S''\right|^2}{|S|}\xi''(1)\exp\left(\frac{\gamma^2}{2}\left|\mathcal{T}_q\right|^2\frac{\left|S''\right|^2}{|S|}\xi''(1)\right)\nonumber\\
    & \leq 2\gamma^2p^2\xi''(1)\frac{\left|S''\right|^2}{|S|}\exp\left(2\gamma^2p^2\xi''(1)\frac{\left|S''\right|^2}{|S|}\right).
\end{align}
Finally,
\begin{align}
    \left|\left\langle \prod_{j \in S''}\left(z_j^{[\alpha_j]} - G_{\alpha_j}\right) \right\rangle_{\mu^{[q + 1]}_{SB}\left(\bm{z}_{S''};\,\xi_{s'/s}, S''\right)}\right| & \leq 2^{\left|S''\right|}2\gamma^2p^2\xi''(1)\frac{\left|S''\right|^2}{|S|}\exp\left(2\gamma^2p^2\xi''(1)\frac{\left|S''\right|^2}{|S|}\right)\nonumber\\
    & \leq 2^{\left|S''\right|}2\gamma^2p^2\xi''(1)\frac{\left|S''\right|^2}{|S|}2^{\left|S''\right|}\nonumber\\
    & \leq 2\gamma^2p^2\xi''(1)\frac{\left|S''\right|^2}{|S|}4^{\left|S''\right|}\nonumber\\
    & \leq 2\gamma^2p^2x''\frac{\left|S''\right|^2}{|S|}4^{\left|S''\right|}\nonumber\\
    & =: \overline{\varepsilon_2}\left(\xi, S', S''\right).
\end{align}
where in the final looser bound, we assumed
\begin{align}
    |S| & \geq \frac{2\gamma^2p^2x''}{\log 2}\left|S''\right| \qquad \textrm{(assumption 1)}\label{si:eq:centered_moments_maximum_recursion_assumption_1_proof_statement}.
\end{align}

\paragraph{The $\mu_{C, 1', d}$ term}\mbox{}

Plugging expansion of $\mu^{[p]}_{C, 1', d}$ stated in Eq.~\ref{si:eq:mu_c_1p_d_definition},
\begin{align}
    & \left\langle \mu^{[q + 1]}_{C, 1', d}\left(\bm{z}_{S'}, \bm{z}_{S''};\,\xi_{s'/s}, S', S''\right) \right\rangle_{\mu^{[q]}\left(\bm{z}_{S'};\,\xi_{s'/s}, S'\right)}\nonumber\\
    & = \sum_{\substack{J' \subset S'\\1 \leq \left|J'\right| < d\\\bm{t}, \bm{u} \in \mathcal{T}_q^{J'}}}\mu^{[q + 1]}_{C, 1', \bm{t}, \bm{u}}\left(\bm{z}_{S''};\,\xi_{s'/s}, S', S''\right)\left\langle \prod_{j \in J'}\left(z_j^{[t_j]}z_j^{[u_j]} - G_{t_j, u_j}\right) \right\rangle_{\mu^{[q]}\left(\bm{z}_{S'};\,\xi_{s'/s}, S'\right)}.\label{si:eq:centered_moments_maximum_recursion_1p_term_expansion}
\end{align}
The idea is then to bound all path integral expectations under $\mu^{[q]}\left(\bm{z}_{S'};\,\xi_{s'/s}, S'\right)$ on the right-hand side by a small number ---namely, $\varepsilon^{[q]}_{\mathrm{max}}$ from Definition~\ref{si:def:centered_moments_maximum}--- and apply the triangular inequality. A minor complication is that $\mu^{[q + 1]}\left(\bm{z}_S;\,\xi_{s'/s}, S'\right)$ concentrates overlap around a slightly perturbed $\bm{G}$ matrix given the spin glass mixture polynomial is a mild perturbation $\xi_{s'/s}$ of $\xi$. Let us then temporarily denote $\bm{G}^{s'/s} = \left(G^{s'/s}_{t, u}\right)_{t, u \in \mathcal{T}_q} \in \mathbb{C}^{\mathcal{T}_q \times \mathcal{T}_q}$ the $\bm{G}$ matrix generated by $q$-layer spin-glass QAOA applied to mixture polynomial $\xi_{s'/s}$. We then decompose the expectation between the $\bm{G} - \bm{G}^{s'/s}$ error on the one hand, and the overlap concentration $z_j^{[t]}z_j^{[u]} - G^{s'/s}_{t, u}$ on the other hand:
\begin{align}
    & \left\langle \prod_{j \in J'}\left(z_j^{\left[t_j\right]}z_j^{\left[u_j\right]} - G_{t_j, u_j}\right) \right\rangle_{\mu^{[q]}\left(\bm{z}_{S'};\,\xi_{s'/s}, S'\right)}\nonumber\\
    & = \left\langle \prod_{j \in J'}\left(z_j^{\left[t_j\right]}z_j^{\left[u_j\right]} - G^{s'/s}_{t_j, u_j} + G^{s'/s}_{t_j, u_j} - G_{t_j, u_j}\right) \right\rangle_{\mu^{[q]}\left(\bm{z}_{S'};\,\xi_{s'/s}, S'\right)}\nonumber\\
    & = \left\langle \sum_{K' \subset J'}\prod_{j \in J' - K'}\left(G^{s'/s}_{t_j, u_j} - G_{t_j, u_j}\right)\prod_{j \in K'}\left(z_j^{\left[t_j\right]}z_j^{\left[u_j\right]} - G^{s'/s}_{t_j, u_j}\right) \right\rangle_{\mu^{[q]}\left(\bm{z}_{S'};\,\xi_{s'/s}, S'\right)}\nonumber\\
    & = \sum_{K' \subset J'}\prod_{j \in J' - K'}\left(G^{s'/s}_{t_j, u_j} - G_{t_j, u_j}\right)\left\langle \prod_{j \in K'}\left(z_j^{\left[t_j\right]}z_j^{\left[u_j\right]} - G^{s'/s}_{t_j, u_j}\right) \right\rangle_{\mu^{[q]}\left(\bm{z}_{S'};\,\xi_{s'/s}, S'\right)}\nonumber\\
    & = \prod_{j \in J'}\left(G^{s'/s}_{t_j, u_j} - G_{t_j, u_j}\right) + \sum_{\varnothing \subsetneq K' \subset J'}\prod_{j \in J' - K'}\left(G^{s'/s}_{t_j, u_j} - G_{t_j, u_j}\right)\left\langle \prod_{j \in K'}\left(z_j^{\left[t_j\right]}z_j^{\left[u_j\right]} - G^{s'/s}_{t_j, u_j}\right) \right\rangle_{\mu^{[q]}\left(\bm{z}_{S'};\,\xi_{s'/s}, S'\right)},\label{si:eq:centered_moments_maximum_recursion_1p_term_cavity_expectation_decomposition}
\end{align}
where in the final line, we singled out $K' = \varnothing$ term, without centered moments under path integral measure $\mu^{[q]}\left(\bm{z}_{S'};\,\xi_{s'/s}, S'\right)$. The first term can then be bounded from Corollary~\ref{si:cor:simple_g_matrix_variation_bound_parametrized_xi} as follows:
\begin{align}
    \left|\prod_{j \in J'}\left(G^{s'/s}_{t_j, u_j} - G_{t_j, u_j}\right)\right| & \leq \delta_G^{\left|J'\right|},\label{si:eq:centered_moments_maximum_recursion_1p_term_cavity_expectation_decomposition_term_1_bound}
\end{align}
where
\begin{align}
    \delta_G & := 2e\left(2e\gamma^2p^2\right)^p\left(x''\right)^{p - 1}\delta_{\xi},
\end{align}
with
\begin{align}
    \delta_{\xi} & := \left\lVert \xi'_{s'/s} - \xi' \right\rVert_1\nonumber\\
    & = \sum_{m \geq 1}m\left|\left(\frac{s'}{s}\right)^{m - 1}c_m - c_m\right|\nonumber\\
    & \leq \sum_{m \geq 1}m(m - 1)\left|\frac{s'}{s} - 1\right|c_m\nonumber\\
    & \leq \frac{\left|S''\right|}{\left|S\right|}\sum_{m \geq 1}m(m - 1)c_m\nonumber\\
    & \leq \frac{\left|S''\right|}{\left|S\right|}\xi''(1)\nonumber\\
    & \leq \frac{\left|S''\right|}{\left|S\right|}x''
\end{align}
hence
\begin{align}
    \delta_G & \leq 2e\left(2e\gamma^2p^2x''\right)^p\frac{\left|S''\right|}{|S|}.
\end{align}
To apply Corollary~\ref{si:cor:simple_g_matrix_variation_bound_parametrized_xi}, we need to satisfy constraints
\begin{align}
    2e\gamma^2p^2x'' & \geq 2,\\
    \delta_{\xi} & \leq \min\left\{\frac{1}{2\gamma^2p^3}, \frac{1}{2\left(2e\gamma^2p^2\right)^{p + 1}\left(x''\right)^p}\right\}.
\end{align}
Using upper-bound $\delta_{\xi} \leq \left|S''\right|x''/\left|S\right|$ established above, the second constraint can be weakened to
\begin{align}
    \left|S\right| & \geq \max\left\{2\gamma^2p^3x'', 2\left(2e\gamma^2p^2x''\right)^{p + 1}\right\}\left|S''\right|.
\end{align}
The $\max$ can be reduced to its second term by observing that
\begin{align}
    2\left(2e\gamma^2p^2x''\right)^{p + 1} & \geq 2 \gamma^2p^3x'' \cdot \frac{e}{p} \cdot 2\left(2e\gamma^2p^2x''\right)^p\nonumber\\
    & \geq 2\gamma^2p^3x'' \cdot \frac{e}{p} \cdot (2e)^p\nonumber\\
    & \geq 2\gamma^2p^3x'',
\end{align}
where in the second line, we used the current Lemma's assumption $\gamma^2p^2x'' \geq 1$. Hence, it is sufficient to assume
\begin{align}
    \left|S\right| & \geq 2\left(2e\gamma^2p^2x''\right)^{p + 1}\left|S''\right| \qquad \textrm{(assumption 2)}\label{si:eq:centered_moments_maximum_recursion_assumption_2_proof_statement},
\end{align}
to ensure $\delta_{\xi}$ bounded as required by Corollary~\ref{si:cor:simple_g_matrix_variation_bound_parametrized_xi}. With this constraint satisfied,
\begin{align}
    \delta_G & \leq 2e\left(2e\gamma^2p^2x''\right)^p\frac{\left|S''\right|}{|S|}\nonumber\\
    & \leq \frac{2e\left(2e\gamma^2p^2x''\right)^p}{2\left(2e\gamma^2p^2x''\right)^{p + 1}}\nonumber\\
    & \leq \frac{1}{2\gamma^2p^2x''}\nonumber\\
    & \leq \frac{1}{2}\nonumber\\
    & \leq 1.
\end{align}

The second term of Eq.~\ref{si:eq:centered_moments_maximum_recursion_1p_term_cavity_expectation_decomposition} can be controlled by bounding $\bm{G}$ matrix entries by $1$ and using definition of $\varepsilon^{[q]}$, given $\left|K'\right| \leq \left|J'\right| \leq d$:
\begin{align}
    & \left|\sum_{\varnothing \subsetneq K' \subset J'}\prod_{j \in J' - K'}\left(G^{s'/s}_{t_j, u_j} - G_{t_j, u_j}\right)\left\langle \prod_{j \in K'}\left(z_j^{\left[t_j\right]}z_j^{\left[u_j\right]} - G^{s'/s}_{t_j, u_j}\right) \right\rangle_{\mu^{[q]}\left(\bm{z}_{S'};\,\xi_{s'/s}, S'\right)}\right|\nonumber\\
    & \leq \sum_{\varnothing \subsetneq K' \subset J'}\left|\prod_{j \in J' - K'}\left(G^{s'/s}_{t_j, u_j} - G_{t_j, u_j}\right)\right|\cdot\left|\left\langle \prod_{j \in K'}\left(z_j^{\left[t_j\right]}z_j^{\left[u_j\right]} - G^{s'/s}_{t_j, u_j}\right) \right\rangle_{\mu^{[q]}\left(\bm{z}_{S'};\,\xi_{s'/s}, S'\right)}\right|\nonumber\\
    & \leq \sum_{\varnothing \subsetneq K' \subset J'}2^{\left|J'\right| - \left|K'\right|}\varepsilon^{[q]}_{\mathrm{max}}\left(\xi_{s'/s}, \left|S'\right|, d\right)\nonumber\\
    & \leq 3^{\left|J'\right|}\varepsilon^{[q]}_{\mathrm{max}}\left(\xi_{s'/s}, \left|S'\right|, d\right).\label{si:eq:centered_moments_maximum_recursion_1p_term_cavity_expectation_decomposition_term_2_bound}
\end{align}
Plugging bounds Eqns.~\ref{si:eq:centered_moments_maximum_recursion_1p_term_cavity_expectation_decomposition_term_1_bound}, \ref{si:eq:centered_moments_maximum_recursion_1p_term_cavity_expectation_decomposition_term_2_bound} into Eq.~\ref{si:eq:centered_moments_maximum_recursion_1p_term_cavity_expectation_decomposition} yields:
\begin{align}
    \left|\left\langle \prod_{j \in J'}\left(z_j^{\left[t_j\right]}z_j^{\left[u_j\right]} - G_{t_j, u_j}\right) \right\rangle_{\mu^{[q]}\left(\bm{z}_{S'};\,\xi_{s'/s}, S'\right)}\right| & \leq \delta_G^{\left|J'\right|} + 3^{\left|J'\right|}\varepsilon^{[q]}_{\mathrm{max}}\left(\xi_{s'/s}, \left|S'\right|, d\right)
\end{align}
Plugging this bound into Eq.~\ref{si:eq:centered_moments_maximum_recursion_1p_term_expansion}, and using Lemma~\ref{si:lemma:mu_c_1p_d_coefficients_series_bound} to bound absolute sums of coefficients $\left|\mu^{[q + 1]}_{C, 1', d, \bm{t}, \bm{u}}\right|$, then yields:
\begin{align}
    & \left|\left\langle \mu^{[q + 1]}_{C, 1', d}\left(\bm{z}_{S'}, \bm{z}_{S''};\,\xi_{s'/s}, S', S''\right) \right\rangle_{\mu^{[q]}\left(\bm{z}_{S'};\,\xi_{s'/s}, S'\right)}\right|\nonumber\\
    & = \sum_{\substack{J' \subset S'\\1 \leq \left|J'\right| < d\\\bm{t}, \bm{u} \in \mathcal{T}_q}}\left|\mu^{[q + 1]}_{C, 1', \bm{t}, \bm{u}}\left(\bm{z}_{S''};\,\xi_{s'/s}, S', S''\right)\right|\cdot\left|\left\langle \prod_{j \in J'}\left(z_j^{[t_j]}z_j^{[u_j]} - G_{t_j, u_j}\right) \right\rangle_{\mu^{[q]}\left(\bm{z}_{S'};\,\xi_{s'/s}, S'\right)}\right|\nonumber\\
    & \leq \sum_{\substack{J' \subset S'\\1 \leq \left|J'\right| < d\\\bm{t}, \bm{u} \in \mathcal{T}_q}}\left|\mu^{[q + 1]}_{C, 1', \bm{t}, \bm{u}}\left(\bm{z}_{S''};\,\xi_{s'/s}, S', S''\right)\right|\cdot\left|\left\langle \prod_{j \in J'}\left(z_j^{[t_j]}z_j^{[u_j]} - G_{t_j, u_j}\right) \right\rangle_{\mu^{[q]}\left(\bm{z}_{S'};\,\xi_{s'/s}, S'\right)}\right|\nonumber\\
    & \leq \delta_G\sum_{\substack{J' \subset S'\\1 \leq \left|J'\right| < d\\\bm{t}, \bm{u} \in \mathcal{T}_q}}\left|\mu^{[q + 1]}_{C, 1', \bm{t}, \bm{u}}\left(\bm{z}_{S''};\,\xi_{s'/s}, S', S''\right)\right| + \varepsilon^{[q]}_{\mathrm{max}}\left(\xi_{s'/s}, \left|S'\right|, d\right)\sum_{\substack{J' \subset S'\\1 \leq \left|J'\right| < d\\\bm{t}, \bm{u} \in \mathcal{T}_q}}3^{\left|J'\right|}\left|\mu^{[q + 1]}_{C, 1', \bm{t}, \bm{u}}\left(\bm{z}_{S''};\,\xi_{s'/s}, S', S''\right)\right|\nonumber\\
    & \leq \exp\left(4p^2\gamma^2R\xi''(1 + 2R)\left|S''\right|\right)\left(\frac{\delta_G}{\left(1 - 1/R\right)^{4p^2}} + \frac{\varepsilon_{\mathrm{max}}^{[q]}\left(\xi_{s'/s}, \left|S'\right|, d\right)}{\left(1 - 3/R\right)^{4p^2}}\right)\nonumber\\
    & \leq \frac{\exp\left(4p^2\gamma^2R\xi''(1 + 2R)\left|S''\right|\right)}{\left(1 - 3/R\right)^{4p^2}}\left(\delta_G + \varepsilon_{\mathrm{max}}^{[q]}\left(\xi_{s'/s}, \left|S'\right|, d\right)\right)\nonumber\\
    & \leq 2^{8p^2}\exp\left(4p^2\gamma^2R\xi''(1 + 2R)\left|S''\right|\right)\left(2e\left(2e\gamma^2p^2x''\right)^p\frac{\left|S''\right|}{|S|} + \varepsilon^{[q]}_{\mathrm{max}}\left(\xi_{s'/s}, \left|S'\right|, d\right)\right)\nonumber\\
    & =: \overline{\varepsilon_3}\left(\xi, \left|S'\right|, S'', R\right) + c\left(\xi, \left|S''\right|, R, d\right)\varepsilon^{[q]}_{\mathrm{max}}\left(\xi_{s'/s}, S', d\right).
\end{align}

\paragraph{The $\mu_{C, 1'', d}$ term}\mbox{}

We handle the term involving $\mu_{C, 1'', d}$ in Eq.~\ref{si:eq:centered_moment_cavity_expansion_cavity_expectation_decomposition} thanks to majorization bound Eq.~\ref{si:eq:mu_c_1pp_d_majorization_bound} on this function of bit matrices, extended to matrices of complex numbers for bits labeled by $S'$:
\begin{align}
    \mu^{[q + 1]}_{C, 1'', d}\left(\,\cdot\,, \bm{z}_{S''};\,\xi_{s'/s}, S', S''\right) & \preceq \overline{\mu^{[q + 1]}_{C, 1'', d}}\left(\,\cdot\,, \bm{z}_{S''};\,\xi_{s'/s}, S', S''\right).
\end{align}
This yields:
\begin{align}
    & \left|\left\langle \mu^{[q + 1]}_{C, 1'', d}\left(\bm{z}_{S'}, \bm{z}_{S''};\,\xi_{s'/s}, S', S''\right) \right\rangle_{\mu^{[q]}\left(\bm{z}_{S'};\,\xi_{s'/s}, S'\right)}\right|\nonumber\\
    & \leq \overline{\mu^{[q + 1]}_{C, 1'', d}}\left(\bm{1}_{S' \times \mathcal{T}_q}, \bm{z}_{S''};\,\xi_{s'/s}, S', S''\right)\nonumber\\
    & \leq 2^{4p^2}\left(\left(\frac{4}{R}\right)^d\binom{d + 4p^2 - 1}{4p^2 - 1} + \frac{1}{\left|S'\right|}4p^2\left(4p^2 + 1\right)\frac{\left(4/R\right)^2}{\left(1 - 4/R\right)^{4p^2 + 2}}\right)\exp\left(4p^2\gamma^2R\xi''\left(1 + 2R\right)\left|S''\right|\right).
\end{align}
Summing this over $\bm{z}_{S''}$ against $\mu^{[q + 1]}_{SB}$ in Eq.~\ref{si:eq:centered_moment_cavity_expansion_cavity_expectation_decomposition} (using $\left|\mu^{[q + 1]}_{B}\right| \leq 1$ pointwise) yields the following error contribution to the centered moment;
\begin{align}
    \left|\varepsilon_4\left(\xi, S', S''\right)\right| & \leq 2^{4p^2}\left(\left(\frac{4}{R}\right)^d\binom{d + 4p^2 - 1}{4p^2 - 1} + \frac{1}{\left|S'\right|}4p^2\left(4p^2 + 1\right)\frac{\left(4/R\right)^2}{\left(1 - 4/R\right)^{4p^2 + 2}}\right)\nonumber\\
    & \hspace*{15px} \times \exp\left(\left(4p^2\gamma^2R\xi''\left(1 + 2R\right) + 4p^2\gamma^2\xi'(1) + p \log(2)\right)\left|S''\right|\right)\\
    & =: \overline{\varepsilon_4}\left(\xi, S', S'', R\right).
\end{align}
\end{proof}
\end{lemma}

Our final goal, fulfilled in Section~\ref{si:sec:behaviour_centered_moments_large_size}, will be to use the recursive inequality on maximum centered moments $\varepsilon^{[q]}$ derived in Lemma~\ref{si:lemma:centered_moments_maximum_recursion} to provide an explicit bound on these at arbitrary $q$. As the recursive bounds derived in the Lemma are a bit cumbersome for that purpose, we simplify them in the following Proposition:

\begin{proposition}[Recursion relation for maximum value of centered moment, simplified]
\label{si:prop:centered_moments_maximum_recursion_simplified}
Recall the setting of Lemma~\ref{si:lemma:centered_moments_maximum_recursion}, where the spin set size $s$ is partitioned into a ``large" set size $s'$ and a ``small" set size $s''$: $s = s' + s''$. Assume a bound $x$ on $\xi'(1), \xi''(2)$:
\begin{align}
    \xi'(1), \xi''(2) \leq x.
\end{align}
Note the above automatically implies $\xi''(1) \leq \xi''(2) \leq x$. Besides, assume:
\begin{align}
    \gamma^2p^2x \geq 1, && \frac{s''}{s} \leq \kappa(x).\label{si:eq:centered_moments_maximum_recursion_simplified_constraints}
\end{align}
Note the first constraint can always be satisfied by increasing upper bound $\gamma$ on the phase separator angles if necessary; then, the second bound enforces a sufficiently small (by an amount independent of $s$) relative set size $s''/s$. Then, the following simplified recursion (compare with Eq.~\ref{si:eq:centered_moments_maximum_recursion}) holds for the maximum of centered moments:
\begin{align}
    \varepsilon^{[q + 1]}_{\mathrm{max}}\left(\xi, s, s''\right) & \leq 2^{8p^2}\left(\varepsilon^{[q]}_{\mathrm{max}}\left(\xi_{s'/s}, s', d\right) + b(x)\frac{\left(s''\right)^2}{s} + \left(\frac{8}{R}\right)^d\right)\exp\left(e\left(\xi, x\right)s''\right),
\end{align}
for any integer parameter $d \geq 1$ and positive real parameter $R > 8$, with quantities $b(x), e(\xi, x)$ defined as:
\begin{align}
    b\left(x\right) & := 256p^4\left(2e\gamma^2p^2x\right)^p,\\
    e\left(\xi, x\right) & := 4\gamma^2p^2R\xi''(1 + 2R) + 4\gamma^2p^2x + p \log 2.
\end{align}
Note the same bounds hold replacing $\xi'(1), \xi''(2)$ by upper bounds of the values of these parameters.
\begin{proof}
By assumptions Eq.~\ref{si:eq:centered_moments_maximum_recursion_simplified_constraints}, the recursion inequality from Lemma~\ref{si:lemma:centered_moments_maximum_recursion} holds. The current result will follow from upper-bounding $\overline{\varepsilon_1}, \overline{\varepsilon_2}, \overline{\varepsilon_3}, \overline{\varepsilon_4}, c$ defined in the previous Lemma. For that purpose, we observe these quantities present as sums of terms of the form:
\begin{align}
    c_1\exp\left(c_2s''\right), && \frac{1}{s}c_1\exp\left(c_2s''\right), && \frac{s''}{s}c_1\exp\left(c_2s''\right), && \frac{\left(s''\right)^2}{s}c_1\exp\left(c_2s''\right),
\end{align}
for some constants $c_1, c_2 > 0$. We may then obtain a more concise bound by bounding each $c_1, c_2$ with its maximum value assumed over all terms of this form. For simplicity, we will also bound $1/s, s''/s \leq \left(s''\right)^2/s$

\paragraph{Bounding $\overline{\varepsilon_4}$}

\begin{align}
    \overline{\varepsilon_4}\left(\xi, s', s'', R\right) & = 2^{4p^2}\left(\left(\frac{4}{R}\right)^d\binom{d + 4p^2 - 1}{4p^2 - 1} + \frac{1}{s'}4p^2\left(4p^2 + 1\right)\frac{\left(4/R\right)^2}{\left(1 - 4/R\right)^{4p^2 + 2}}\right)\nonumber\\
    & \hspace*{20px} \times \exp\left(\left(4\gamma^2p^2R\xi''(1 + 2R) + 4\gamma^2p^2\xi'(1) + p \log 2\right)s''\right)\nonumber\\
    & \leq 2^{4p^2}\left(\left(\frac{8}{R}\right)^d2^{4p^2 - 1} + \frac{1}{s'}4p^2\left(8p^2 + 1\right)\frac{\left(4/R\right)^2}{\left(1 - 4/R\right)^{4p^2 + 2}}\right)\nonumber\\
    & \hspace*{20px} \times \exp\left(\left(4\gamma^2p^2R\xi''(1 + 2R) + 4\gamma^2p^2\xi'(1) + p \log 2\right)s''\right)\nonumber\\
    & \leq 2^{4p^2}\left(\left(\frac{8}{R}\right)^d2^{4p^2 - 1} + \frac{1}{s}8p^2\left(4p^2 + 1\right)\frac{\left(4/R\right)^2}{\left(1 - 4/R\right)^{4p^2 + 2}}\right)\nonumber\\
    & \hspace*{20px} \times \exp\left(\left(4\gamma^2p^2R\xi''(1 + 2R) + 4\gamma^2p^2\xi'(1) + p \log 2\right)s''\right)\nonumber\\
    & \leq 2^{4p^2}\left(\left(\frac{8}{R}\right)^d2^{4p^2 - 1} + \frac{1}{s}8p^2\left(4p^2 + 1\right)2^{4p^2}\right)\exp\left(\left(4\gamma^2p^2R\xi''(1 + 2R) + 4\gamma^2p^2\xi'(1) + p \log 2\right)s''\right)\nonumber\\
    & \leq \left(2^{8p^2 - 1}\left(\frac{8}{R}\right)^d + \frac{\left(s''\right)^2}{s}2^{8p^2 + 3}p^2\left(4p^2 + 1\right)\right)\exp\left(\left(4\gamma^2p^2R\xi''(1 + 2R) + 4\gamma^2p^2\xi'(1) + p \log 2\right)s''\right)\nonumber\\
    & \leq \left(2^{8p^2 - 1}\left(\frac{8}{R}\right)^d + \frac{\left(s''\right)^2}{s}2^{8p^2 + 6}p^4\right)\exp\left(\left(4\gamma^2p^2R\xi''(1 + 2R) + 4\gamma^2p^2\xi'(1) + p \log 2\right)s''\right)\nonumber\\
    & \leq \left(2^{8p^2 - 1}\left(\frac{8}{R}\right)^d + \frac{\left(s''\right)^2}{s}2^{8p^2 + 6}p^4\right)\exp\left(\left(4\gamma^2p^2R\xi''(1 + 2R) + 4\gamma^2p^2x + p \log 2\right)s''\right).\label{si:eq:centered_moments_maximum_recursion_simplified_epsilon_4_bound}
\end{align}
In the second line, we used upper bound Lemma~\ref{si:lemma:binomial_coefficient_bound} on binomial coefficients. In the third line, we upper-bounded
\begin{align}
    \frac{1}{s'} & = \frac{1}{s - s''}\nonumber\\
    & \leq \frac{1}{s - \kappa(x)s}\nonumber\\
    & \leq \frac{1}{s - 2^{-1}s}\nonumber\\
    & \leq \frac{2}{s}.
\end{align}
In the fourth line, we plugged assumption $R > 8$. In the fifth line, we loosened the bound by estimating $1/s \leq \left(s''\right)^2/2$. In the sixth line, we simplified the polynomial quantity in $p$ to $p^2\,\cdot\,\left(4p^2 + 1\right) \leq p^2\,\cdot\,8p^2 = 8p^4$.

\paragraph{Bounding $\overline{\varepsilon_1}$}

\begin{align}
    \overline{\varepsilon_1}\left(\xi, s', s''\right) & = \gamma^2p^2\xi''(2)\frac{\left(s''\right)^2}{s}\exp\left(\left(2\xi'(1) + \xi''(2)\right)\gamma^2p^2s''\right)\nonumber\\
    & \leq \gamma^2p^2\xi''(2)\frac{\left(s''\right)^2}{s}\exp\left(\left(4\gamma^2p^2R\xi''(1 + 2R) + 4\gamma^2p^2\xi'(1) + p \log 2\right)s''\right)\nonumber\\
    & \leq \gamma^2p^2x\frac{\left(s''\right)^2}{s}\exp\left(\left(4\gamma^2p^2R\xi''(1 + 2R) + 4\gamma^2p^2x + p \log 2\right)s''\right),\label{si:eq:centered_moments_maximum_recursion_simplified_epsilon_1_bound}
\end{align}
where in the second line, we use sequence of bounds:
\begin{align}
    & 4\gamma^2p^2R\xi''(1 + 2R) + 4\gamma^2p^2\xi'(1) + p \log 2\nonumber\\
    & \geq 4\gamma^2p^2R\xi''(1 + 2R) + 4\gamma^2p^2\xi'(1)\nonumber\\
    & \geq 32\gamma^2p^2\xi''(17) + 4\gamma^2p^2\xi'(1) && \textrm{($R > 8$ and non-negativity of $\xi$ coefficients)}\nonumber\\
    & \geq 32\gamma^2p^2\xi''(2) + 4\gamma^2p^2\xi'(1)\nonumber\\
    & \geq \gamma^2p^2\xi''(2) + 2\gamma^2p^2\xi'(1)
\end{align}
to upper-bound the exponential rate.

\paragraph{Bounding $\overline{\varepsilon_2}$}

\begin{align}
    \overline{\varepsilon_2}\left(\xi, s', s''\right) & = 2\gamma^2p^2\xi''(1)\frac{\left(s''\right)^2}{s}4^{s''}\nonumber\\
    & \leq 2\gamma^2p^2x\frac{\left(s''\right)^2}{s}\exp\left(\left(4\gamma^2p^2R\xi''(1 + 2R) + 4\gamma^2p^2x + p \log 2\right)s''\right),\label{si:eq:centered_moments_maximum_recursion_simplified_epsilon_2_bound}
\end{align}
where we bounded the exponential rate from sequence of inequalities:
\begin{align}
    4\gamma^2p^2R\xi''(1 + 2R) + 4\gamma^2p^2x + p \log 2 & \geq 4\gamma^2p^2x\nonumber\\
    & \geq 4 && \textrm{(assumption Eq.~\ref{si:eq:centered_moments_maximum_recursion_simplified_constraints})}\nonumber\\
    & \geq \log 4.
\end{align}

\paragraph{Bounding $\overline{\varepsilon_3}$}

\begin{align}
    \overline{\varepsilon_3}\left(\xi, s', s'', R, d\right) & = 2^{8p^2 + 1}e\left(2e\gamma^2p^2x\right)^p\frac{s''}{s}\exp\left(4\gamma^2p^2R\xi''(1 + 2R)s''\right)\nonumber\\
    & \le 2^{8p^2 + 1}e\left(2e\gamma^2p^2x\right)^p\frac{\left(s''\right)^2}{s}\exp\left(\left(4\gamma^2p^2R\xi''(1 + 2R) + 4\gamma^2p^2x + p \log 2\right)s''\right),\label{si:eq:centered_moments_maximum_recursion_simplified_epsilon_3_bound}
\end{align}
where we loosened $s'/s$ to $\left(s''\right)^2/s$.

\paragraph{Bounding $c$}

For the $c$ function defined in Lemma~\ref{si:lemma:centered_moments_maximum_recursion}, we use simple bound:
\begin{align}
    c\left(\xi, s'', R, d\right) & = 2^{8p^2}\exp\left(4\gamma^2p^2R\xi''(1 + 2R)s''\right)\nonumber\\
    & \leq 2^{8p^2}\exp\left(\left(4\gamma^2p^2R\xi''(1 + 2R) + 4\gamma^2p^2x + p \log 2\right)s''\right).\label{si:eq:centered_moments_maximum_recursion_simplified_c_bound}
\end{align}

\paragraph{Combining all bounds together}\mbox{}

We now combine bounds Eqns.~\ref{si:eq:centered_moments_maximum_recursion_simplified_epsilon_1_bound}, \ref{si:eq:centered_moments_maximum_recursion_simplified_epsilon_2_bound}, \ref{si:eq:centered_moments_maximum_recursion_simplified_epsilon_3_bound}, \ref{si:eq:centered_moments_maximum_recursion_simplified_epsilon_4_bound}, \ref{si:eq:centered_moments_maximum_recursion_simplified_c_bound} to arrive at the result. Note all these bounds involve multiplicative factor
\begin{align}
    \exp\left(\left(4\gamma^2p^2R\xi''(1 + 2R) + 4\gamma^2p^2x + p \log 2\right)s''\right) & =: \exp\left(e\left(\xi, x\right)s''\right).
\end{align}
Using assumption $\gamma^2p^2x \geq 1$ from Eq.~\ref{si:eq:centered_moments_maximum_recursion_simplified_constraints}, $1 \leq p^4$, and $e^2 \leq 8 = 2^3$, bounds Eqns.~\ref{si:eq:centered_moments_maximum_recursion_simplified_epsilon_1_bound}, \ref{si:eq:centered_moments_maximum_recursion_simplified_epsilon_2_bound}, \ref{si:eq:centered_moments_maximum_recursion_simplified_epsilon_3_bound}, \ref{si:eq:centered_moments_maximum_recursion_simplified_epsilon_4_bound} can be loosened to:
\begin{align}
    \overline{\varepsilon_1}\left(\xi, s', s''\right), \overline{\varepsilon_2}\left(\xi, s', s''\right), \overline{\varepsilon_3}\left(\xi, s', s''\right) & \leq 2^{8p^2 + 6}p^4\left(2e\gamma^2p^2x\right)^p\frac{\left(s''\right)^2}{s}\exp\left(e\left(\xi, x\right)s''\right),\\
    \overline{\varepsilon_4}\left(\xi, s', s''\right) & \leq \left(2^{8p^2}\left(\frac{8}{R}\right)^d + 2^{8p^2 + 6}p^4\left(2e\gamma^2p^2x\right)^p\frac{\left(s''\right)^2}{s}\right)\exp\left(e\left(\xi, x\right)s''\right).
\end{align}
Adding these bounds in Lemma~\ref{si:lemma:centered_moments_maximum_recursion}'s recursion inequality:
\begin{align}
    \varepsilon_{\mathrm{max}}^{[q + 1]}\left(\xi, s, s''\right) & \leq \overline{\varepsilon_1}\left(\xi, s', s''\right) + \overline{\varepsilon_2}\left(\xi, s', s''\right) + \overline{\varepsilon_3}\left(\xi, s', s'', R\right) + \overline{\varepsilon_4}\left(\xi, s', s'', R\right)\nonumber\\
    & \hspace*{15px} + c\left(\xi, s'', R, d\right)\varepsilon^{[q]}_{\mathrm{max}}\left(\xi_{s'/s}, s', d\right),
\end{align}
yields the claimed simplified bound:
\begin{align}
    \varepsilon^{[q + 1]}_{\mathrm{max}}\left(\xi, s, s''\right) & \leq 2^{8p^2}\left(\varepsilon^{[q]}_{\mathrm{max}}\left(\xi_{s'/s}, s', d\right) + b\left(x\right)\frac{\left(s''\right)^2}{s} + \left(\frac{8}{R}\right)^d\right)\exp\left(e\left(\xi, x\right)s''\right),
\end{align}
where
\begin{align}
    b\left(x\right) & := 256p^4\left(2e\gamma^2p^2x\right)^p,\\
    e\left(\xi, x\right) & := 4\gamma^2p^2R\xi''(1 + 2R) + 4\gamma^2p^2x + p \log 2.
\end{align}
\end{proof}
\end{proposition}

In the following Corollary, we instantiate Proposition~\ref{si:prop:centered_moments_maximum_recursion_simplified} for a specific $R$ to obtain even more concrete bounds:

\begin{corollary}[Recursion relation for maximum value of centered moment, special case]
\label{si:cor:centered_moments_maximum_recursion_simplified_special_case}
Recall the setting of Lemma~\ref{si:lemma:centered_moments_maximum_recursion} and Proposition~\ref{si:prop:centered_moments_maximum_recursion_simplified}, where the spin set size $s$ is partitioned into a ``large" set size $s'$ and a ``small" set size $s''$: $s = s' + s''$. Assume a bound $x$ on $\xi'(1), \xi''(33)$:
\begin{align}
    \xi'(1), \xi''(33) \leq x.
\end{align}
Note the above automatically implies $\xi''(2) \leq \xi''(33) \leq x$. Besides, assume
\begin{align}
    \gamma^2p^2x \geq 1, && \frac{s''}{s} \leq \kappa(x).\label{si:eq:centered_moments_maximum_recursion_simplified_special_case_constraints}
\end{align}
Then, the maximum centered moments introduced in Definition~\ref{si:def:centered_moments_maximum} satisfy the following recursive inequality in the number of layers $q$:
\begin{align}
    \varepsilon^{[q + 1]}_{\mathrm{max}}\left(\xi, s, s''\right) & \leq 2^{8p^2}\left(\varepsilon^{[q]}_{\mathrm{max}}\left(\xi_{s'/s}, s', d\right) + b(x)\frac{\left(s''\right)^2}{s} + 2^{-d}\right)\exp\left(e(x)s''\right),
\end{align}
where
\begin{align}
    b\left(x\right) & := 256p^4\left(2e\gamma^2p^2x\right)^p,\\
    e\left(x\right) & := 70\gamma^2p^2x + p \log 2.
\end{align}
\begin{proof}
This follows from letting $R := 16$ in Proposition~\ref{si:prop:centered_moments_maximum_recursion_simplified}. Then, $e(\xi)$ defined in that Proposition can be bounded as
\begin{align}
    e(\xi, x) & = 4\gamma^2p^2R\xi''(1 + 2R) + 4\gamma^2p^2x + p \log 2\nonumber\\
    & \leq 64\gamma^2p^2\xi''(33) + 4\gamma^2p^2x + p \log 2\nonumber\\
    & \leq 70\gamma^2p^2x + p \log 2.
\end{align}
\end{proof}
\end{corollary}
The interest of Corollary~\ref{si:cor:centered_moments_maximum_recursion_simplified_special_case} compared to Proposition~\ref{si:prop:centered_moments_maximum_recursion_simplified} is the manifest uniformity over $\xi$ of functions $b(x), e(x)$ on the right-hand side of the bound: these functions only depend on bound $x$ on the first two derivatives of the mixture polynomial.

\subsubsection{Behaviour of centered moments at large size}
\label{si:sec:behaviour_centered_moments_large_size}

In this Section, we apply main technical result Proposition~\ref{si:prop:centered_moments_maximum_recursion_simplified} proven in Section~\ref{si:sec:centered_moments_maximum_recursion_proof} to study the behaviour of centered moments at large size. We start by proving vanishing of centered moments in the infinite-size limit (Proposition~\ref{si:prop:vanishing_centered_moments_infinite_size}); according to the introductory discussion from Section~\ref{si:sec:noncolliding_taylor_expansion_application_spin_glass_qaoa_cavity_decomposition}, this result entails approximate independence of spins and concentration of normalized overlaps to $\bm{G}$ entries in the infinite-size limit. We then exploit technical Proposition~\ref{si:prop:centered_moments_maximum_recursion_simplified} further to derive concrete finite-size bounds on the maximum centered moments (Proposition~\ref{si:prop:simplified_threshold_function_bound}) and the QAOA energy density (Corollary~\ref{si:cor:finite_size_energy_density_error_bound}). Unlike the infinite-size limit result, these finite-size results require further analysis of the recursion for the maximum centered moments, in particular an explicit approximate solution of this recursion.

The following Proposition directly shows vanishing of centered moments in the infinite-size limit using recursive inequality Proposition~\ref{si:prop:centered_moments_maximum_recursion_simplified} on maximum centered moments $\varepsilon^{[q]}\left(\xi, s, s''\right)$, without relying on an explicit solution to this recursion. Namely, given a target maximum magnitude $\varepsilon$ on some maximum centered moment $\varepsilon^{[q]}\left(\xi, s, s''\right)$, it constructs a threshold size $s^{[q]}$, depending on $\xi$ and $s''$, such that for $s \geq s^{[q]}$, the maximum centered moment is bounded by $\varepsilon$.

\begin{proposition}[Vanishing of centered moments at fixed degree and infinite size]
\label{si:prop:vanishing_centered_moments_infinite_size}
Consider the maximal centered moments of degree $s''$ introduced in Definition~\ref{si:def:centered_moments_maximum}. Consider a family $\mathcal{X}$ of spin glass mixture polynomials where $\xi'(1), \xi''(1), \xi''(2)$ (note $\xi''(1) \leq \xi''(2)$) are uniformly bounded by some constant $x$. Assume technical constraint:
\begin{align}
    \gamma^2p^2x \geq 1,
\end{align}
which can always be satisfied by increasing $\gamma$. Then, for all number of layers $0 \leq q \leq p$, there exists a function $s^{[q]} = s^{[q]}\left(x, \varepsilon, s''\right)$ such that whenever
\begin{align}
    s \geq s^{[q]}\left(x, \varepsilon, s''\right),
\end{align}
then
\begin{align}
    \varepsilon^{[q]}_{\mathrm{max}}\left(\xi, s, s''\right) & \leq \varepsilon \qquad \forall \xi \in \mathcal{X}.
\end{align}
We stress uniformity of the bound in polynomial $\xi \in \mathcal{X}$, with $\mathcal{X}$ only defined by boundedness conditions $\xi'(1), \xi''(2) \leq x$.
\begin{proof}
We construct family of functions $s^{[q]}$ inductively in $q$. For $q = 0$, the maximum centered moment is trivial:
\begin{align}
    \varepsilon^{[0]}_{\mathrm{max}}\left(\xi, s, s''\right) & := \max_{\substack{\alpha_1, \ldots, \alpha_{s''} \subset \mathcal{T}_0\\\left|\alpha_j\right|\,\mathrm{even}}}\left|\left\langle \prod_{j \in \left[s''\right]}\left(z_j^{\left[\alpha_j\right]} - G_{\alpha_j}\right) \right\rangle_{\mu^{[0]}\left(\xi;\,[n]\right)}\right|.
\end{align}
Indeed, in the above $\mathcal{T}_0 = \left\{1, -1\right\}$, the path integral measure is that of the circuit preparing the $\ket{+}$ state, so that for all subset $J \subset [s'']$
\begin{align}
    \left\langle \prod_{j \in J}z_j^{[\alpha]} \right\rangle_{\mu^{[0]}\left(\xi;\,[n]\right)} & = \prod_{j \in J}\left\langle z_j^{\left[\alpha_j\right]} \right\rangle_{\mu^{[0]}\left(\xi;\,[n]\right)}\nonumber\\
    & = \prod_{j \in J}\mathbf{1}\left[|\alpha_j|\,\mathrm{even}\right]
\end{align}
Hence,
\begin{align}
    \varepsilon^{[0]}_{\mathrm{max}}\left(\xi, s, s''\right) & = 0,
\end{align}
and one may set
\begin{align}
    s^{[0]}\left(x, \varepsilon, s''\right) & := s'',
\end{align}
independent of parameters $x, \varepsilon$. This completes the initialization step $q = 0$.

Let us now assume $s^{[q]}$ constructed (for any parameters $x, \varepsilon, s''$) and consider constructing $s^{[q + 1]}$. Fix $\varepsilon > 0$; to find a threshold for $s$ such that $\varepsilon^{[q + 1]}_{\mathrm{max}}\left(\xi, s, s''\right) \leq \varepsilon$, we use recursive inequality from Corollary~\ref{si:cor:centered_moments_maximum_recursion_simplified_special_case}:
\begin{align}
    \varepsilon^{[q + 1]}_{\mathrm{max}}\left(\xi, s, s''\right) & \leq 2^{8p^2}\left(\varepsilon^{[q]}_{\mathrm{max}}\left(\xi_{s'/s}, s', d\right) + b(x)\frac{\left(s''\right)^2}{s} + 2^{-d}\right)\exp\left(e(x)s''\right),
\end{align}
and find parameters such that each additive contribution in the right-hand-side of the inequality is bounded by $\varepsilon/3$.

\paragraph{Satisfying constraints of Corollary~\ref{si:cor:centered_moments_maximum_recursion_simplified_special_case}}\mbox{}

In order to apply the above recursion on $\varepsilon_{\mathrm{max}}$ from Corollary~\ref{si:cor:centered_moments_maximum_recursion_simplified_special_case}, we must ensure validity of the constraints Eq.~\ref{si:eq:centered_moments_maximum_recursion_simplified_constraints}. Hence, we require:
\begin{align}
    s \geq \frac{s''}{\kappa(x)}.
\end{align}

\paragraph{Contribution $2^{-d}$}\mbox{}

For contribution involving term $2^{-d}$, taking
\begin{align}
    d & = \left\lceil \frac{1}{\log 2}\left(e(x)s'' + \log\left(\frac{2^{8p^2 + 2}}{\varepsilon}\right)\right) \right\rceil\nonumber\\
    & =: d^*.
\end{align}
ensures
\begin{align}
    2^{8p^2}2^{-d}\exp\left(e(x)s''\right) & \leq \frac{\varepsilon}{4}.
\end{align}
To ensure $\varepsilon^{[q]}_{\mathrm{max}}\left(\xi_{s'/s}, s', d\right)$ is well-defined for this choice of $d$, we enforce:
\begin{align}
    s & \geq \frac{1}{\log 2}\left(e(x)s'' + \log\left(\frac{2^{8p^2 + 2}}{\varepsilon}\right)\right).
\end{align}

\paragraph{Contribution $b(x)\left(s''\right)^2/s$}\mbox{}

To control this contribution, it is sufficient to choose
\begin{align}
    s & \geq \frac{2^{8p^2 + 2}}{\varepsilon}b(x)\left(s''\right)^2\exp\left(e(x)s''\right),
\end{align}
implying
\begin{align}
    2^{8p^2}b(x)\frac{\left(s''\right)^2}{s}\exp\left(e(x)s''\right) & \leq \frac{\varepsilon}{4}.
\end{align}

\paragraph{Contribution $\varepsilon^{[q]}_{\mathrm{max}}$}\mbox{}

Given the previously fixed $d = d^*$, this contribution can be bounded according to the induction hypothesis by ensuring
\begin{align}
    s' \geq s^{[q]}\left(x_{s'/s}, 2^{-8p^2 - 1}\exp\left(-e(x)s''\right)\varepsilon, d^*\right),
\end{align}
where $x_{s'/s}$ is a constant upper-bounding $\xi_{s'/s}'(1), \xi_{s'/s}''(33)$ uniformly in $\xi \in \mathcal{X}$. Such a constant exists, since for all $y \geq 0$:
\begin{align}
    \xi_{s'/s}'(y) & = \xi'\left(\frac{s'}{s}y\right)\nonumber\\
    & \leq \xi'(y),\\
    \xi''_{s'/s}(y) & = \frac{s'}{s}\xi''\left(\frac{s'}{s}y\right)\nonumber\\
    & \leq \xi''(y).
\end{align}
Hence, one may take $x_{s'/s} = x$. Re-expressing the lower bound on $s'$ in terms of $s$,
\begin{align}
    s & = s'' + s'\nonumber\\
    & \geq s'' + s^{[q]}\left(x, 2^{-8p^2 - 1}\exp\left(-e(x)s''\right)\varepsilon, d^*\right)
\end{align}
ensures that
\begin{align}
    2^{8p^2}\varepsilon^{[q]}_{\mathrm{max}}\left(\xi_{s'/s}, s', d^*\right)\exp\left(e(x)s''\right) & \leq \frac{\varepsilon}{2}
\end{align}

\paragraph{Combining all constraints}\mbox{}

All in all, choosing
\begin{align}
    s & \geq s^{[q + 1]}\left(x, \varepsilon, s''\right)\nonumber\\
    & := \max\left\{s^{[q + 1]}_1\left(x, \varepsilon, s''\right), s^{\log}_2\left(x, \varepsilon, s''\right), s_2\left(x, \varepsilon, s''\right), s_3\left(x, \varepsilon, s''\right)\right\},
\end{align}
where
\begin{align}
    s^{[q + 1]}_1\left(x, \varepsilon, s''\right) & := s'' + s^{[q]}\left(x, 2^{-8p^2 - 1}\exp\left(-e(x)s''\right)\varepsilon, \left\lceil \frac{1}{\log 2}\left(e(x)s'' + \log\left(\frac{2^{8p^2 + 2}}{\varepsilon}\right)\right) \right\rceil \right),\\
    s_2^{\log}\left(x, \varepsilon, s''\right) & := \left\lceil\frac{1}{\log 2}\left(e(x)s'' + \log\left(\frac{2^{8p^2 + 2}}{\varepsilon}\right)\right)\right\rceil,\\
    s_2\left(x, \varepsilon, s''\right) & := \frac{2^{8p^2 + 2}}{\varepsilon}b(x)\left(s''\right)^2\exp\left(e(x)s''\right),\\
    s_3\left(x, \varepsilon, s''\right) & := \frac{s''}{\kappa(x)}.
\end{align}
ensures
\begin{align}
    \varepsilon^{[q + 1]}_{\mathrm{max}}\left(\xi, s, s''\right) & \leq 2^{8p^2}\varepsilon^{[q]}_{\mathrm{max}}\left(\xi_{s'/s}, s', d\right)\exp\left(e(x)s''\right) + 2^{8p^2}b(x)\frac{\left(s''\right)^2}{s}\exp\left(e(x)s''\right) + 2^{8p^2}2^{-d}\exp\left(e(x)s''\right)\nonumber\\
    & \leq \frac{\varepsilon}{2} + \frac{\varepsilon}{4} + \frac{\varepsilon}{4}\nonumber\\
    & = \varepsilon.
\end{align}
\end{proof}
\end{proposition}

Proposition~\ref{si:prop:vanishing_centered_moments_infinite_size} established that for all constant degree $s''$, the maximal centered moments of degree $s''$ (introduced in Definition~\ref{si:def:centered_moments_maximum}) vanish in the infinite-size limit. We now refine this result by deriving finite-size bounds controlling the values of these centered moments at finite size. The analysis is based on the relatively simple threshold size function $s^{[q]}\left(x, \varepsilon, s''\right)$ constructed inductively in $q$ in the proof of Proposition~\ref{si:prop:vanishing_centered_moments_infinite_size}:
\begin{align}
    s^{[0]}\left(x, \varepsilon, s''\right) & := s'',\\
    s^{[q + 1]}\left(x, \varepsilon, s''\right) & := \max\left\{s_1^{[q + 1]}\left(x, \varepsilon, s''\right), s_2^{\log}\left(x, \varepsilon, s''\right), s_2\left(x, \varepsilon, s''\right), s_3\left(x, \varepsilon, s''\right)\right\},
\end{align}
where
\begin{align}
    s_1^{[q + 1]}\left(x, \varepsilon, s''\right) & = s'' + s^{[q]}\left(x, 2^{-8p^2 - 1}\exp\left(-e(x)s''\right)\varepsilon, \left\lceil \frac{1}{\log 2}\left(e(x)s'' + \log\left(\frac{2^{8p^2 + 2}}{\varepsilon}\right)\right) \right\rceil\right),\\
    s^{\log}_2\left(x, \varepsilon, s''\right) & := \left\lceil \frac{1}{\log 2}\left(e(x)s'' + \log\left(\frac{2^{8p^2 + 2}}{\varepsilon}\right)\right) \right\rceil,\\
    s_2\left(x, \varepsilon, s''\right) & := \frac{2^{8p^2 + 2}}{\varepsilon}b(x)\left(s''\right)^2\exp\left(e(x)s''\right),\\
    s_3\left(x, \varepsilon, s''\right) & := \frac{s''}{\kappa(x)}.
\end{align}
The above maximum can be simplified by removing the $s_2^{\log}$ contribution, noting that by the recursive definition of $s^{[q]}$, $s^{[q]}\left(x, \varepsilon, s''\right) \geq s''$, hence
\begin{align}
    s_1^{[q + 1]}\left(x, \varepsilon, s''\right) & = s'' + s^{[q]}\left(x, 2^{-8p^2 - 1}\exp\left(-e(x)s''\right)\varepsilon, \left\lceil \frac{1}{\log 2}\left(e(x)s'' + \log\left(\frac{2^{8p^2 + 2}}{\varepsilon}\right)\right) \right\rceil\right)\nonumber\\
    & \geq s^{[q]}\left(x, 2^{-8p^2 - 1}\exp\left(-e(x)s''\right)\varepsilon, \left\lceil \frac{1}{\log 2}\left(e(x)s'' + \log\left(\frac{2^{8p^2 + 2}}{\varepsilon}\right)\right) \right\rceil\right)\nonumber\\
    & \geq \left\lceil \frac{1}{\log 2}\left(e(x)s'' + \log\left(\frac{2^{8p^2 + 2}}{\varepsilon}\right)\right) \right\rceil\nonumber\\
    & = s_2^{\log}\left(x, \varepsilon, s''\right).
\end{align}
All in all, we can write:
\begin{align}
    s^{[q + 1]}\left(x, \varepsilon, s''\right) & = \max\left\{s_1^{[q + 1]}\left(x, \varepsilon, s''\right), s_2\left(x, \varepsilon, s''\right), s_3\left(x, \varepsilon, s''\right) \right\}.
\end{align}
Our goal is to produce a simple upper bound of $s^{[q]}$ as a function of $x, s'', \varepsilon$ based on this recursive definition. We start by simplifying the above recursion by removing the integral part:

\begin{lemma}[Removing integral part in $s^{[q]}$ recursion]
\label{si:lemma:simplified_size_threshold_function}
For all parameters $x, \varepsilon > 0$ and integer $s'' \geq 0$,
\begin{align}
    s^{[q]}\left(x, \varepsilon, s''\right) & \leq \overline{s}^{[q]}\left(x, \varepsilon, s''\right),
\end{align}
where $\overline{s}^{[q]}$ is a function defined over all real variables $x > 0, \varepsilon \in (0, 1), s'' > 0$ by recursion on $0 \leq q \leq p$:
\begin{align}
    \overline{s}^{[0]}\left(x, \varepsilon, s''\right) & := s'',\\
    \overline{s}^{[q + 1]}\left(x, \varepsilon, s''\right) & := \max\left\{\overline{s}_1^{[q + 1]}\left(x, \varepsilon, s''\right), \overline{s}_2\left(x, \varepsilon, s''\right), \overline{s}_3\left(x, \varepsilon, s''\right)\right\},\label{si:eq:simplified_size_threshold_function_definition}
\end{align}
where
\begin{align}
    \overline{s}_1^{[q + 1]}\left(x, \varepsilon, s''\right) & := s'' + \overline{s}^{[q]}\left(x, 2^{-8p^2 - 1}\exp\left(-e(x)s''\right)\varepsilon, \frac{1}{\log 2}\left(e(x)s'' + \log\left(\frac{2^{8p^2 + 2}}{\varepsilon}\right)\right) + 1\right),\\
    \overline{s}_2\left(x, \varepsilon, s''\right) & := s_2\left(x, \varepsilon, s''\right)\\
    & = \frac{2^{8p^2 + 2}}{\varepsilon}b(x)\left(s''\right)^2\exp\left(e(x)s''\right),\\
    \overline{s}_3\left(x, \varepsilon, s''\right) & := s_3\left(x, \varepsilon, s''\right)\\
    & = \frac{s''}{\kappa(x)}.
\end{align}
Besides, $\overline{s}^{[q]}\left(x, \varepsilon, s''\right)$ is nonincreasing in its second argument $\varepsilon$ and nondecreasing in its third argument $s''$.
\begin{proof}
We prove the inequality by induction of $q$.

At $q = 0$, the result is trivial as $s^{[0]}\left(x, \varepsilon, s''\right) = \overline{s}^{[0]}\left(x, \varepsilon, s''\right) = s''$.

Now, assume the inequality holds up to level $q$. We start by proving the inequality part of the statement, i.e. $s^{[q + 1]} \leq \overline{s}^{[q + 1]}$. By the recursive definition of $s^{[q + 1]}$,
\begin{align}
    s^{[q + 1]}\left(x, \varepsilon, s''\right) & = \max\left\{s_1^{[q + 1]}\left(x, \varepsilon, s''\right), s_2\left(x, \varepsilon, s''\right), s_3\left(x, \varepsilon, s''\right)\right\}\nonumber\\
    & = \max\left\{s_1^{[q + 1]}\left(x, \varepsilon, s''\right), \overline{s}_2\left(x, \varepsilon, s''\right), \overline{s}_3\left(x, \varepsilon, s''\right)\right\}.
\end{align}
Now, by the induction hypothesis,
\begin{align}
    s_1^{[q + 1]}\left(x, \varepsilon, s''\right) & = s'' + s^{[q]}\left(x, 2^{-8p^2 - 1}\exp\left(-e(x)s''\right)\varepsilon, \left\lceil \frac{1}{\log 2}\left(e(x)s'' + \log\left(\frac{2^{8p^2 + 2}}{\varepsilon}\right)\right) \right\rceil\right)\nonumber\\
    & \leq s'' + \overline{s}^{[q]}\left(x, 2^{-8p^2 - 1}\exp\left(-e(x)s''\right)\varepsilon, \left\lceil \frac{1}{\log 2}\left(e(x)s'' + \log\left(\frac{2^{8p^2 + 2}}{\varepsilon}\right)\right) \right\rceil\right)\nonumber\\
    & \leq s'' + \overline{s}^{[q]}\left(x, 2^{-8p^2 - 1}\exp\left(-e(x)s''\right)\varepsilon, \frac{1}{\log 2}\left(e(x)s'' + \log\left(\frac{2^{8p^2 + 2}}{\varepsilon}\right)\right) + 1\right)\nonumber\\
    & = \overline{s}_1^{[q + 1]}\left(x, \varepsilon, s''\right).
\end{align}
In the second line, we used the inequality statement: $s^{[q]} \leq \overline{s}^{[q]}$ from the induction hypothesis; in the third line, we used the monotonicity statement from the induction hypothesis. Hence,
\begin{align}
    s^{[q + 1]}\left(x, \varepsilon, s''\right) & \leq \overline{s}^{[q + 1]}\left(x, \varepsilon, s''\right),
\end{align}
proving the inequality part of the induction statement.

Let us now prove the monotonicity part of the induction statement. This results from each of $\overline{s}^{[q + 1]}_1, \overline{s}_2, \overline{s}_3$ satisfying the required monotonicity properties. The verification is immediate for explicit functions $\overline{s}_2, \overline{s}_3$. Let us verify this more carefully for recursively defined
\begin{align}
    \overline{s}_1^{[q + 1]}\left(x, \varepsilon, s''\right) & := s'' + \overline{s}^{[q]}\left(x, 2^{-8p^2 - 1}\exp\left(-e(x)s''\right)\varepsilon, \frac{1}{\log 2}\left(e(x)s'' + \log\left(\frac{2^{8p^2 + 2}}{\varepsilon}\right)\right) + 1\right).
\end{align}
For instance, increasing $s''$ decreases the second argument and increases the third argument of $\overline{s}^{[q]}$ in the above expression, which by the induction hypothesis cannot decrease $\overline{s}^{[q]}$ evaluated at these points. Likewise, decreasing $\varepsilon$ decreases the second argument and increases the third argument of $\overline{s}^{[q]}$ in the above, which cannot decrease $\overline{s}^{[q]}$ either. All in all, the desired monotonicity properties hold at level $(q + 1)$.
\end{proof}
\end{lemma}

After removing the integral part, the problem then becomes to obtain a simple estimate for the looser threshold function $\overline{s}^{[q]}$. The $\overline{s}^{[q]}$ function is still recursively defined by a maximum, whose first branch involves the previous level function $\overline{s}^{[q - 1]}$, and second branch is an explicit function of $\varepsilon, s''$. One may then track the growth of arguments of $\overline{s}$ while the first maximum branch is taken, and plug these values into the second maximum branch as soon as it exceeds the first. The following Lemma bounds the growth of arguments of $\overline{s}^{[q]}$ while the first maximum branch is taken:

\begin{lemma}[Bounding first branch of $\max$ recursion for simplified threshold function]
\label{si:lemma:simplified_threshold_function_first_maximum_branch_bound}
Let all assumptions be as in Corollary~\ref{si:cor:centered_moments_maximum_recursion_simplified_special_case}. Denote by
\begin{align}
    \varphi\left(\varepsilon, s''\right) & := \left(\varphi_1\left(\varepsilon, s''\right), \varphi_2\left(\varepsilon, s''\right)\right)\\
    & := \left(2^{-8p^2 - 1}\exp\left(-e(x)s''\right)\varepsilon, \frac{1}{\log 2}\left(e(x)s'' + \log\left(\frac{2^{8p^2 + 2}}{\varepsilon}\right)\right) + 1\right)
\end{align}
the flow of parameters $\left(\varepsilon, s''\right)$ when the first branch of the maximum is taken in recursion Eq.~\ref{si:eq:simplified_size_threshold_function_definition}. Let $\left(x, \varepsilon, s''\right)$ be parameters such that the first branch of the maximum is taken $r_0$ times in determining $\overline{s}^{[p]}\left(x, \varepsilon, s''\right)$, i.e.:
\begin{align}
    \overline{s}^{[p - r]}\left(\varphi^{\circ r}\left(\varepsilon, s''\right)\right) & = \left(\varphi^{\circ r}\right)_2\left(\varepsilon, s''\right) + \overline{s}^{[p - r - 1]}\left(\varphi^{\circ (r + 1)}\left(\varepsilon, s''\right)\right) \qquad \forall 0 \leq r < r_0,\label{si:eq:simplified_threshold_function_first_maximum_branch_recursion}
\end{align}
where $\varphi^{\circ r}$ is the composition of $\varphi$ $r$ times with itself, and $\left(\varphi^{\circ r}\right)_2\left(\varepsilon, s''\right)$ is the second coordinate of $\varphi^{\circ r}\left(\varepsilon, s''\right)$, i.e. the ``flowed $s''$ parameter". For conciseness, we omitted parameter $x$ from the notation. Besides, for convenience make technical assumption:
\begin{align}
    \varepsilon & \leq \min\left\{e^{-1/2}, 2^{-(8p^2 + 3)/100}\right\}.
\end{align}
Then, the following bounds hold on $\overline{s}^{[p - r]}\left(\varepsilon, s''\right)$:
\begin{align}
    \overline{s}^{[p]}\left(\varepsilon, s''\right) & \leq \overline{s}^{\left[p - r_0\right]}\left(\varphi^{\circ r_0}\left(\varepsilon, s''\right)\right) +  \frac{\left(2e(x)/\log 2\right)^{r_0}}{200}\left(2\log_2\left(\frac{1}{\varepsilon}\right) + s''\right).
\end{align}
with parameters $\left(\varepsilon, s''\right)$ iterates bounded as:
\begin{align}
    \left(\varphi^{\circ r_0}\right)_1\left(\varepsilon, s''\right) & \geq \exp\left(-\left(\frac{2e(x)}{\log 2}\right)^{r_0}s''\right)\varepsilon^{\frac{2}{\log 2}\left(\frac{2e(x)}{\log 2}\right)^{r_0}},\label{si:eq:simplified_threshold_function_first_maximum_branch_recursion_iterated_epsilon_lower_bound}\\
    \left(\varphi^{\circ r_0}\right)_2\left(\varepsilon, s''\right) & \leq \left(\frac{2e(x)}{\log 2}\right)^{r_0}\left(2\log_2\left(\frac{1}{\varepsilon}\right) + s''\right).\label{si:eq:simplified_threshold_function_first_maximum_branch_recursion_iterated_spp_upper_bound}
\end{align}
\begin{proof}
Iterating recursion Eq.~\ref{si:eq:simplified_threshold_function_first_maximum_branch_recursion},
\begin{align}
    \overline{s}^{[p]}\left(\varepsilon, s''\right) & = \sum_{0 \leq r < r_0}\left(\varphi^{\circ r}\right)_2\left(\varepsilon, s''\right) + \overline{s}^{[p - r_0]}\left(\varphi^{\circ r_0}\left(\varepsilon, s''\right)\right).\label{si:eq:simplified_threshold_function_first_maximum_branch_recursion_solution}
\end{align}
To control the sum in the above expression and obtain bounds on the argument of $\overline{s}^{[p - r_0]}$, we produce a lower bound on $\left(\varphi^{\circ r}\right)_1\left(\varepsilon, s''\right)$ and an upper bound on $\left(\varphi^{\circ r}\right)_2\left(\varepsilon, s''\right)$ for all $0 \leq r \leq r_0$. For that purpose, consider change of variables $\left(y_r, z_r\right) \longrightarrow \left(u_r, v_r\right)$ defined by:
\begin{align}
    u_r & := \log_2\left(\frac{1}{y_r}\right),\\
    v_r & := z_r.
\end{align}
Then, $\left(u_r, v_r\right)$ satisfy recursion:
\begin{align}
    u_{r + 1} & = \log_2\left(\frac{1}{y_{r + 1}}\right)\nonumber\\
    & = \log_2\left(\frac{1}{\varphi_1\left(y_r, z_r\right)}\right)\nonumber\\
    & = \log_2\left(\frac{2^{8p^2 + 1}}{y_r}\exp\left(e(x)z_r\right)\right)\nonumber\\
    & = 8p^2 + 1 + \log_2\left(\frac{1}{y_r}\right) + \frac{1}{\log 2}e(x)z_r\nonumber\\
    & = 8p^2 + 1 + u_r + \frac{1}{\log 2}e(x)v_r,\\
    v_{r + 1} & = z_{r + 1}\nonumber\\
    & = \varphi_2\left(y_r, z_r\right)\nonumber\\
    & = 8p^2 + 3 + \log_2\left(\frac{1}{y_r}\right) + \frac{1}{\log 2}e(x)z_r\nonumber\\
    & = 8p^2 + 3 + u_r + \frac{1}{\log 2}e(x)v_r.
\end{align}
One can write $\left(u_r, v_r\right) \leq \left(\overline{u}_r, \overline{v}_r\right)$, with $\left(\overline{u}_r, \overline{v}_r\right)$ satisfying loosened recursion:
\begin{align}
    \left(\overline{u}_0, \overline{v}_0\right) & := \left(u_0, v_0\right) := \left(\log_2\left(\frac{1}{\varepsilon}\right), s''\right),\\
    \overline{u}_{r + 1} & = 8p^2 + 3 + \frac{e(x)}{\log 2}\left(\overline{u}_r + \overline{v}_r\right),\\
    \overline{v}_{r + 1} & = 8p^2 + 3 + \frac{e(x)}{\log 2}\left(\overline{u}_r + \overline{v}_r\right).
\end{align}
In the above estimate, we used $1 \leq e(x)/\log 2$, which follows from the definition of $e(x)$ and assumption $\gamma^2p^2x \geq 1$ in Corollary~\ref{si:cor:centered_moments_maximum_recursion_simplified_special_case}. All in all, for all $0 \leq r \leq r_0$, it holds
\begin{align}
    u_r, v_r & \leq u_r + v_r\\
    & \leq \overline{u}_r + \overline{v}_r\\
    & \leq \frac{\left(2e(x)/\log 2\right)^r - 1}{2e(x)/\log 2 - 1}(16p^2 + 6) + \left(\frac{2e(x)}{\log 2}\right)^r\left(u_0 + v_0\right)\nonumber\\
    & \leq \left(\frac{2e(x)}{\log 2}\right)^r\left(\frac{16p^2 + 6}{200} + u_0 + v_0\right)\nonumber\\
    & \leq \left(\frac{2e(x)}{\log 2}\right)^r\left(2u_0 + v_0\right),
\end{align}
where in the final line, we used assumption:
\begin{align}
    \varepsilon \leq 2^{-(16p^2 + 6)/200} \implies \frac{16p^2 + 6}{200} & \leq \log_2\left(\frac{1}{\varepsilon}\right) = u_0.
\end{align}
Rephrasing bounds on $u_r, v_r$ in terms of $\left(y_r, z_r\right) = \varphi^{\circ r}\left(\varepsilon, s''\right) = \left(\left(\varphi^{\circ r}\right)_1\left(\varepsilon, s''\right), \left(\varphi^{\circ r}\right)_2\left(\varepsilon, s''\right)\right)$,
\begin{align}
    \left(\varphi^{\circ r}\right)_1\left(\varepsilon, s''\right) & \geq \exp\left(-\left(\frac{2e(x)}{\log 2}\right)^r\left(2\log_2\left(\frac{1}{\varepsilon}\right) + s''\right)\right)\nonumber\\
    & \geq \exp\left(-\left(\frac{2e(x)}{\log 2}\right)^rs''\right)\varepsilon^{\frac{2}{\log 2}\left(\frac{2e(x)}{\log 2}\right)^r},\\
    \left(\varphi^{\circ r}\right)_2\left(\varepsilon, s''\right) & \leq \left(\frac{2e(x)}{\log 2}\right)^r\left(2u_0 + v_0\right)\nonumber\\
    & = \left(\frac{2e(x)}{\log 2}\right)^r\left(2\log_2\left(\frac{1}{\varepsilon}\right) + s''\right).
\end{align}
The estimate on $\left(\varphi^{\circ r}\right)_2\left(\varepsilon, s''\right)$ yields:
\begin{align}
    \sum_{0 \leq r < r_0}\left(\varphi^{\circ r}\right)_2\left(\varepsilon, s''\right) & \leq \frac{\left(2e(x)/\log 2\right)^{r_0} - 1}{2e(x)/\log 2 - 1}\left(2\log_2\left(\frac{1}{\varepsilon}\right) + s''\right)\nonumber\\
    & \leq \frac{\left(2e(x)/\log 2\right)^{r_0}}{200}\left(2\log_2\left(\frac{1}{\varepsilon}\right) + s''\right).
\end{align}
and plugging these into Eq.~\ref{si:eq:simplified_threshold_function_first_maximum_branch_recursion_solution} yields estimate:
\begin{align}
    \overline{s}^{[p]}\left(\varepsilon, s''\right) & \leq \overline{s}^{[p - r_0]}\left(\varphi^{\circ r_0}\left(\varepsilon, s''\right)\right) + \frac{\left(2e(x)/\log 2\right)^{r_0}}{200}\left(2\log_2\left(\frac{1}{\varepsilon}\right) + s''\right).
\end{align}
\end{proof}
\end{lemma}

\begin{proposition}[Bounding simplified threshold function]
\label{si:prop:simplified_threshold_function_bound}
Let $x > 0$ bound derivatives of the spin glass mixture polynomial:
\begin{align}
    \xi'(1), \xi''(33) & \leq x
\end{align}
and further satisfy
\begin{align}
    \gamma^2p^2x & \geq 1
\end{align}
(which can be guaranteed by enlarging it if necessary). Then, the following bound holds on the simplified threshold function:
\begin{align}
    \overline{s}^{[p]}\left(\varepsilon, s''\right) & \leq 2^{8p^2 + 3}b(x)\left(\frac{2^{s''}}{\varepsilon^2}\right)^{\left(\frac{2e(x)}{\log 2}\right)^{2p}}\left(2\log_2\left(\frac{1}{\varepsilon}\right) + s''\right)^2,
\end{align}
with $b(x) = 256p^4\left(2e\gamma^2p^2x\right)^p$ and $e\left(x\right) := 70\gamma^2p^2x + p\log 2$ depending on some bound $x$ on the values of $\xi$: $\xi'(1), \xi''(33) \leq x$ as specified in Corollary~\ref{si:cor:centered_moments_maximum_recursion_simplified_special_case}.
\begin{proof}
Recall the recursion in $q$ satisfied by $\overline{s}^{[q]}$ (Eq.~\ref{si:eq:simplified_size_threshold_function_definition}):
\begin{align}
    \overline{s}^{[q + 1]}\left(\varepsilon, s''\right) & := \max\left\{\overline{s}_1^{[q + 1]}\left(\varepsilon, s''\right), \overline{s}_2\left(\varepsilon, s''\right), \overline{s}_3\left(\varepsilon, s''\right)\right\}.
\end{align}
and consider the smallest integer $r$ ($0 \leq r < p$) such that the second or third maximum branch is taken for computing $\overline{s}^{[p - r]}\left(\varphi^{\circ r}\left(\varepsilon, s''\right)\right)$, i.e. the smallest $r$ such that:
\begin{align}
   \max\left\{\overline{s}_2\left(\varphi^{\circ r}\left(\varepsilon, s''\right)\right), \overline{s}_3\left(\varphi^{\circ r}\left(\varepsilon, s''\right)\right)\right\} & \geq \overline{s}^{[p - r]}_1\left(\varphi^{\circ r}\left(\varepsilon, s''\right)\right).
\end{align}
We first assume such an $r$ exists, i.e. some iteration takes the second branch of the maximum; we leave the case where no such $r$ exists to the end of the proof. By definition of $\overline{s}^{[p - r]}$, the above inequality implies
\begin{align}
    \overline{s}^{[p - r]}\left(\varphi^{\circ r}\left(\varepsilon, s''\right)\right) & = \max\left\{\overline{s}_2\left(\varphi^{\circ r}\left(\varepsilon, s''\right)\right), \overline{s}_3\left(\varphi^{\circ r}\left(\varepsilon, s''\right)\right)\right\}.
\end{align}
We consider equality with each term of the maximum separately.

\paragraph{Case of maximum branch 2 taken}\mbox{}

In this case,
\begin{align}
    \overline{s}^{[p - r]}\left(\varphi^{\circ r}\left(\varepsilon, s''\right)\right) & = \overline{s}_2\left(\varphi^{\circ r}\left(\varepsilon, s''\right)\right)\nonumber\\
    & = \frac{2^{8p^2 + 2}}{\left(\varphi^{\circ r}\right)_1\left(\varepsilon, s''\right)}b(x)\left(\varphi^{\circ r}\right)_2\left(\varepsilon, s''\right)^2\exp\left(e(x)\left(\varphi^{\circ r}\right)_2\left(\varepsilon, s''\right)\right)
\end{align}
Invoking the monotonicity of $\overline{s}^{[p - r]}$ (Lemma~\ref{si:lemma:simplified_size_threshold_function}) and using bounds Eqns.~\ref{si:eq:simplified_threshold_function_first_maximum_branch_recursion_iterated_epsilon_lower_bound}, \ref{si:eq:simplified_threshold_function_first_maximum_branch_recursion_iterated_spp_upper_bound} on iterates $\varphi^{\circ r}\left(\varepsilon, s''\right)$, the above can be bounded by:
\begin{align}
    \overline{s}^{[p - r]}\left(\varphi^{\circ r}\left(\varepsilon, s''\right)\right) & \leq \frac{2^{8p^2 + 2}}{\exp\left(-\left(\frac{2e(x)}{\log 2}\right)^rs''\right)\varepsilon^{\frac{2}{\log 2}\left(\frac{2e(x)}{\log 2}\right)^{r}}}b(x)\left(\left(\frac{2e(x)}{\log 2}\right)^r\left(2\log_2\left(\frac{1}{\varepsilon}\right) + s''\right)\right)^2\nonumber\\
    & \hspace*{20px} \times \exp\left(e(x)\left(\frac{2e(x)}{\log 2}\right)^r\left(2\log_2\left(\frac{1}{\varepsilon}\right) + s''\right)\right)\nonumber\\
    & \leq \frac{2^{8p^2 + 2}}{\varepsilon^{\frac{2}{\log 2}\left(1 + e(x)\right)\left(\frac{2e(x)}{\log 2}\right)^r}}b(x)\left(2\log_2\left(\frac{1}{\varepsilon}\right) + s''\right)^2\exp\left(\left(1 + e(x)\right)\left(\frac{2e(x)}{\log 2}\right)^rs''\right)\nonumber\\
    & \leq \frac{2^{8p^2 + 2}}{\varepsilon^{2\left(\frac{2e(x)}{\log 2}\right)^{r + 1}}}b(x)\left(2\log_2\left(\frac{1}{\varepsilon}\right) + s''\right)^2\exp\left(2e(x)\left(\frac{2e(x)}{\log 2}\right)^rs''\right)
\end{align}

\paragraph{Case of maximum branch 3 taken}\mbox{}

In this case,
\begin{align}
    \overline{s}^{[p - r]}\left(\varphi^{\circ r}\left(\varepsilon, s''\right)\right) & = \overline{s}_3\left(\varphi^{\circ r}\left(\varepsilon, s''\right)\right)\nonumber\\
    & = \frac{\left(\varphi^{\circ r}\right)_2\left(\varepsilon, s''\right)}{\kappa\left(x\right)}\nonumber\\
    & = 2\left(2e\gamma^2p^2x\right)^{p + 1}\left(\varphi^{\circ r}\right)_2\left(\varepsilon, s''\right)\nonumber\\
    & \leq 2\left(2e\gamma^2p^2x\right)^{p + 1}\left(\frac{2e(x)}{\log 2}\right)^r\left(2\log_2\left(\frac{1}{\varepsilon}\right) + s''\right)\nonumber\\
    & \leq 2\left(2e\gamma^2p^2x\right)^{p + 1}\left(\frac{2e(x)}{\log 2}\right)^r\left(\frac{1}{2} \cdot \frac{200}{16p^2 + 6}\right)2\log_2\left(\frac{1}{\varepsilon}\right)\left(2\log_2\left(\frac{1}{\varepsilon}\right) + s''\right)\nonumber\\
    & \leq 2\left(\frac{e \cdot e(x)}{35}\right)^{p + 1}\left(\frac{2e(x)}{\log 2}\right)^r5\left(2\log_2\left(\frac{1}{\varepsilon}\right) + s''\right)^2\nonumber\\
    & \leq 10\left(\frac{2e(x)}{\log 2}\right)^{p + 1 + r}\left(2\log_2\left(\frac{1}{\varepsilon}\right) + s''\right)^2\nonumber\\
    & \leq 10\left(2\log_2\left(\frac{1}{\varepsilon}\right) + s''\right)^2\exp\left(\left(\frac{2e(x)}{\log 2}\right)^{p + 1 + r}\right)\nonumber\\
    & \leq 10\left(2\log_2\left(\frac{1}{\varepsilon}\right) + s''\right)^2\exp\left(\log\left(\frac{1}{\varepsilon^2}\right)\left(\frac{2e(x)}{\log 2}\right)^{p + 1 + r}\right)\nonumber\\
    & = \frac{10}{\varepsilon^{2\left(\frac{2e(x)}{\log 2}\right)^{p + 1 + r}}}\left(2\log_2\left(\frac{1}{\varepsilon}\right) + s''\right)^2
\end{align}

\paragraph{Concluding for maximum branch 2 or 3 taken}\mbox{}

We now merge the bounds on $\overline{s}^{[p - r]}\left(\varphi^{\circ r}\left(\varepsilon, s''\right)\right)$ whether branch 2 or 3 of the maximum is taken after $r$ steps. Accounting for all possible values of $r = 0, \ldots, p - 1$, and $\varepsilon \leq 1$, we obtain looser bound:
\begin{align}
    \overline{s}^{[p - r]}\left(\varphi^{\circ r}\left(\varepsilon, s''\right)\right) & \leq 2^{8p^2 + 2}b(x)\left(\frac{2^{s''}}{\varepsilon^2}\right)^{\left(\frac{2e(x)}{\log 2}\right)^{2p}}\left(2\log_2\left(\frac{1}{\varepsilon}\right) + s''\right)^2.
\end{align}
To deduce this bound from the ``maximum branch 3 taken'' case, we observed $1 \leq b(x) = 256p^4\left(2e\gamma^2p^2x\right)^p$ given $\gamma^2p^2x \geq 1$ and $p \geq 1$. Then, the bound on $\overline{s}^{[p]}\left(\varepsilon, s''\right)$ provided by Lemma~\ref{si:lemma:simplified_threshold_function_first_maximum_branch_bound} can be estimated:
\begin{align}
    \overline{s}^{[p]}\left(\varepsilon, s''\right) & \leq \overline{s}^{[p - r]}\left(\varphi^{\circ r}\left(\varepsilon, s''\right)\right) + \frac{\left(2e(x)/\log 2\right)^{r}}{200}\left(2\log_2\left(\frac{1}{\varepsilon}\right) + s''\right)\nonumber\\
    & \leq 2^{8p^2 + 2}b(x)\left(\frac{2^{s''}}{\varepsilon^2}\right)^{\left(\frac{2e(x)}{\log 2}\right)^{2p}}\left(2\log_2\left(\frac{1}{\varepsilon}\right) + s''\right)^2 + \frac{\left(2e(x)/\log 2\right)^r}{200}\left(2\log_2\left(\frac{1}{\varepsilon}\right) + s''\right).
\end{align}
We now observe the first term dominates from sequence of inequalities:
\begin{align}
    \left(\frac{2e(x)}{\log 2}\right)^r\left(2\log_2\left(\frac{1}{\varepsilon}\right) + s''\right) & \leq \exp\left(\left(\frac{2e(x)}{\log 2}\right)^r\left(2\log_2\left(\frac{1}{\varepsilon}\right) + s''\right)\right)\nonumber\\
    & = \frac{1}{\varepsilon^{\frac{2}{\log 2}\left(\frac{2e(x)}{\log 2}\right)^r}}\exp\left(\left(\frac{2e(x)}{\log 2}\right)^rs''\right)\nonumber\\
    & \leq \frac{1}{\varepsilon^{2\left(\frac{2e(x)}{\log 2}\right)^{r + 1}}}\exp\left(2e(x)\left(\frac{2e(x)}{\log 2}\right)^rs''\right)\nonumber\\
    & \leq \frac{1}{\varepsilon^{2\left(\frac{2e(x)}{\log 2}\right)^{r + 1}}}\left(\frac{1}{2} \cdot \frac{200}{16p^2 + 6}\right)^2\left(2\log_2\left(\frac{1}{\varepsilon}\right)\right)^2\exp\left(2e(x)\left(\frac{2e(x)}{\log 2}\right)^rs''\right)\nonumber\\
    & \leq \frac{1}{\varepsilon^{2\left(\frac{2e(x)}{\log 2}\right)^{r + 1}}}\left(\frac{1}{2} \cdot \frac{200}{16p^2 + 6}\right)^2\left(2\log_2\left(\frac{1}{\varepsilon}\right) + s''\right)^2\exp\left(2e(x)\left(\frac{2e(x)}{\log 2}\right)^rs''\right)\nonumber\\
    & = \left(\frac{1}{\varepsilon^2}\right)^{\left(\frac{2e(x)}{\log 2}\right)^{r + 1}}\left(\frac{1}{2} \cdot \frac{200}{16p^2 + 6}\right)^2\left(2\log_2\left(\frac{1}{\varepsilon}\right) + s''\right)^2\left(2^{s''}\right)^{\left(\frac{2e(x)}{\log 2}\right)^{r + 1}}\nonumber\\
    & \leq \left(\frac{1}{\varepsilon^2}\right)^{\left(\frac{2e(x)}{\log 2}\right)^p}\left(\frac{1}{2} \cdot \frac{200}{16p^2 + 6}\right)^2\left(2\log_2\left(\frac{1}{\varepsilon}\right) + s''\right)^2\left(2^{s''}\right)^{\left(\frac{2e(x)}{\log 2}\right)^{p}}\nonumber\\
    & \leq \left(\frac{2^{s''}}{\varepsilon^2}\right)^{\left(\frac{2e(x)}{\log 2}\right)^{2p}}\left(\frac{1}{2} \cdot \frac{200}{16p^2 + 6}\right)^2\left(2\log_2\left(\frac{1}{\varepsilon}\right) + s''\right)^2.
\end{align}
In the first line, we used bound $y \leq \exp\left(y\right)$, valid for all $y \geq 0$. In the third line, we used $2e(x) \geq 1$, following from the assumptions of Corollary~\ref{si:cor:centered_moments_maximum_recursion_simplified_special_case}. In the fourth line, we used assumption $\varepsilon \leq 2^{-(8p^2 + 3)/100}$. From the above bound, it follows:
\begin{align}
    \frac{\left(2e(x)/\log 2\right)^r}{200}\left(2\log_2\left(\frac{1}{\varepsilon}\right) + s''\right) & \leq \left(\frac{2^{s''}}{\varepsilon^2}\right)^{\left(\frac{2e(x)}{\log 2}\right)^{2p}}\left(2\log_2\left(\frac{1}{\varepsilon}\right) + s''\right)^2.
\end{align}
Finally,
\begin{align}
    \overline{s}^{[p]}\left(\varepsilon, s''\right) & \leq 2^{8p^2 + 3}b(x)\left(\frac{2^{s''}}{\varepsilon^2}\right)^{\left(\frac{2e(x)}{\log 2}\right)^{2p}}\left(2\log_2\left(\frac{1}{\varepsilon}\right) + s''\right)^2.\label{si:eq:simplified_threshold_function_bound_second_max_branch_case}
\end{align}

\paragraph{Case where maximum branch 1 is always taken}\mbox{}

We now discuss the case where the first branch of the maximum is always taken. In this case, Lemma~\ref{si:lemma:simplified_threshold_function_first_maximum_branch_bound} with $r_0 := p$ gives
\begin{align}
    \overline{s}^{[p]}\left(\varepsilon, s''\right) & \leq \overline{s}^{[0]}\left(\varphi^{\circ p}\left(\varepsilon, s''\right)\right) + \frac{\left(2e(x)/\log 2\right)^p}{200}\left(2\log_2\left(\frac{1}{\varepsilon}\right) + s''\right)^2.
\end{align}
This leads to estimate:
\begin{align}
    \overline{s}^{[p]}\left(\varepsilon, s''\right) & \leq \left(\varphi^{\circ p}\right)_2\left(\varepsilon, s''\right) + \frac{\left(2e(x)/\log 2\right)^p}{200}\left(2\log_2\left(\frac{1}{\varepsilon}\right) + s''\right)^2\nonumber\\
    & \leq \left(\frac{2e(x)}{\log 2}\right)^p\left(2\log_2\left(\frac{1}{\varepsilon}\right) + s''\right) + \frac{\left(2e(x)/\log 2\right)^p}{200}\left(2\log_2\left(\frac{1}{\varepsilon}\right) + s''\right)^2\nonumber\\
    & \leq \left(\frac{2e(x)}{\log 2}\right)^p\left(\frac{1}{2}\cdot\frac{200}{16p^2 + 6}\right)\left(2\log_2\left(\frac{1}{\varepsilon}\right)\right)\left(2\log_2\left(\frac{1}{\varepsilon}\right) + s''\right) + \frac{\left(2e(x)/\log 2\right)^p}{200}\left(2\log_2\left(\frac{1}{\varepsilon}\right) + s''\right)^2\nonumber\\
    & \leq 5\left(\frac{2e(x)}{\log 2}\right)^p\left(2\log_2\left(\frac{1}{\varepsilon}\right) + s''\right)^2\nonumber\\
    & \leq 5\exp\left(\left(\frac{2e(x)}{\log 2}\right)^p\right)\left(2\log_2\left(\frac{1}{\varepsilon}\right) + s''\right)^2\nonumber\\
    & \leq 5\exp\left(\log\left(\frac{1}{\varepsilon^2}\right)\left(\frac{2e(x)}{\log 2}\right)^p\right)\left(2\log_2\left(\frac{1}{\varepsilon}\right) + s''\right)^2\nonumber\\
    & \leq \frac{5}{\varepsilon^{2\left(\frac{2e(x)}{\log 2}\right)^p}}\left(2\log_2\left(\frac{1}{\varepsilon}\right) + s''\right)^2\nonumber\\
    & \leq \frac{5}{\varepsilon^{2\left(\frac{2e(x)}{\log 2}\right)^p}}\left(2\log_2\left(\frac{1}{\varepsilon}\right) + s''\right)^2\exp\left(2e(x)\left(\frac{2e(x)}{\log 2}\right)^ps''\right)\nonumber\\
    & = 5\left(\frac{2^{s''}}{\varepsilon^2}\right)^{\left(\frac{2e(x)}{\log 2}\right)^p}\left(2\log_2\left(\frac{1}{\varepsilon}\right) + s''\right)^2.\label{si:eq:simplified_threshold_function_bound_first_max_branch_case}
\end{align}
where in the fifth line, we used assumption $\varepsilon \leq e^{-1/2}$. Unifying Eq.~\ref{si:eq:simplified_threshold_function_bound_first_max_branch_case} and Eq.~\ref{si:eq:simplified_threshold_function_bound_second_max_branch_case}, maximized over all possible termination steps $r = 0, \ldots, p - 1$ of the $\max$ recursion, gives the following general bound for the threshold cost function:
\begin{align}
    \overline{s}^{[p]}\left(\varepsilon, s''\right) & \leq 2^{8p^2 + 3}b(x)\left(\frac{2^{s''}}{\varepsilon^2}\right)^{\left(\frac{2e(x)}{\log 2}\right)^{2p}}\left(2\log_2\left(\frac{1}{\varepsilon}\right) + s''\right)^2.
\end{align}
\end{proof}
\end{proposition}

Proposition~\ref{si:prop:simplified_threshold_function_bound} provides an operational bound on the simplified threshold function, i.e. the minimum size required for centered overlap moments of a given order to be smaller than a given $\varepsilon$. Since the QAOA energy density can be expressed in terms of overlap moments, Proposition~\ref{si:prop:simplified_threshold_function_bound} implies the following concrete finite-size error estimate between the energy density at finite size and its thermodynamic limit:

\begin{repcorollary}{si:cor:finite_size_energy_density_error_bound}
Let $x > 0$ bound the derivatives of the spin glass mixture polynomial:
\begin{align}
    \xi(2), \xi'(1), \xi''(33) & \leq x
\end{align}
and further satisfy
\begin{align}
    \gamma^2p^2x & \geq 1
\end{align}
(which can always be satisfied by enlarging it if necessary). Then, for all $\varepsilon > 0$ satisfying technical assumption
\begin{align}
    \varepsilon & \leq \min\left\{e^{-1/2}, 2^{-(8p^2 + 3)/100}\right\}4\gamma p x,
\end{align}
and all instance size
\begin{align}
    n & \geq \max\left\{2^{8p^2 + 3}b(x)\left(\frac{2^{\mathrm{deg}(\xi) + 4}\gamma^2p^2x^2}{\varepsilon^2}\right)^{\left(\frac{2e(x)}{\log 2}\right)^{2p}}\left(2\log_2\left(\frac{4\gamma p x}{\varepsilon}\right) + \mathrm{deg}(\xi)\right)^2, \frac{16\gamma px}{\varepsilon}\right\},
\end{align}
the finite-size error on the energy density is at most $\varepsilon$: $\left|\nu_n - \nu_{\infty}\right| \leq \varepsilon$. In the above formula, $b(x) = 256p^4\left(2e\gamma^2p^2x\right)^p$ and $e(x) = 70\gamma^2p^2x + p\log 2$ are functions of upper bounds on $\xi$ derivatives introduced in Corollary~\ref{si:cor:centered_moments_maximum_recursion_simplified_special_case}.
\begin{proof}
We start with the exact expression of the disorder-averaged energy density at arbitrary finite size $n$:
\begin{align}
    \nu_n & = \left\langle \sum_{t \in \mathcal{T}_{p - 1}}i\Gamma_t\xi\left(\frac{1}{n}\left\langle \bm{z}^{[p + 1]}, \bm{z}^{[t]} \right\rangle\right) \right\rangle_{\mu^{[p]}\left(\bm{z};\,\xi, [n]\right)}.
\end{align}
Expanding the normalized overlaps $\left\langle \bm{z}^{[p + 1]}, \bm{z}^{[t]} \right\rangle/n$ around $G_{p + 1, t}$,
\begin{align}
    \nu_n & = \left\langle \sum_{t \in \mathcal{T}_{p - 1}}i\Gamma_t\sum_{q \geq 0}\frac{\xi^{(q)}\left(G_{p + 1, t}\right)}{q!}\left(\frac{1}{n}\left\langle \bm{z}^{[p + 1]}, \bm{z}^{[t]} \right\rangle - G_{p + 1, t}\right)^q \right\rangle_{\mu^{[p]}\left(\bm{z};\,\xi, S\right)}\nonumber\\
    & = \sum_{t \in \mathcal{T}_{p - 1}}i\Gamma_t\xi\left(G_{p + 1, t}\right) + \sum_{t \in \mathcal{T}_{p - 1}}i\Gamma_t\sum_{q \geq 1}\frac{\xi^{(q)}\left(G_{p + 1, t}\right)}{q!}\left\langle \left(\frac{1}{n}\left\langle \bm{z}^{[p + 1]}, \bm{z}^{[t]} \right\rangle - G_{p + 1, t}\right)^q \right\rangle_{\mu^{[p]}\left(\bm{z};\,\xi, S\right)}\nonumber\\
    & =: \nu_{\infty} + \sum_{t \in \mathcal{T}_{p - 1}}i\Gamma_t\sum_{q \geq 1}\frac{\xi^{(q)}\left(G_{p + 1, t}\right)}{q!}\left\langle \left(\frac{1}{n}\left\langle \bm{z}^{[p + 1]}, \bm{z}^{[t]} \right\rangle - G_{p + 1, t}\right)^q \right\rangle_{\mu^{[p]}\left(\bm{z};\,\xi, S\right)}.
\end{align}
where $q$ is effectively restricted to $1 \leq q \leq \mathrm{deg}\left(\xi\right)$. We observe that $\nu_n - \nu_{\infty}$ involves normalized overlaps centered about $\bm{G}$. These can in turn be expanded as products of centered monomials $\left(z_j^{[p + 1]}z_j^{[t]} - G_{p + 1, t}\right)$, which under the path integral measure become centered moments (Definition~\ref{si:def:centered_moments_maximum}) of bounded degree $\mathrm{deg}\left(\xi\right)$. Finally, Proposition~\ref{si:prop:simplified_threshold_function_bound} provides a concrete finite size bound to make centered moments arbitrarily small. The only minor obstacle to this direct argument is potential repetition of some $j$ indices when expanding centered overlaps as a product of centered monomials. For instance, for the centered overlap squared,
\begin{align}
    \left(\frac{1}{n}\left\langle \bm{z}^{[p + 1]}, \bm{z}^{[t]} \right\rangle - G_{p + 1, t}\right)^2 & = \frac{1}{n^2}\left(\sum_{j \in [n]}\left(z_j^{[p + 1]}z_j^{[t]} - G_{p + 1, t}\right)\right)^2\nonumber\\
    & = \frac{1}{n^2}\sum_{j, k \in [n]}\left(z_j^{[p + 1]}z_j^{[t]} - G_{p + 1, t}\right)\left(z_k^{[p + 1]}z_k^{[t]} - G_{p + 1, t}\right)\nonumber\\
    & = \frac{1}{n^2}\sum_{j \in [n]}\left(z_j^{[p + 1]}z_j^{[t]} - G_{p + 1, t}\right)^2 + \frac{2}{n^2}\sum_{\{j, k\} \subset [n]}\left(z_j^{[p + 1]}z_j^{[t]} - G_{p + 1, t}\right)\left(z_k^{[p + 1]}z_k^{[t]} - G_{p + 1, t}\right).
\end{align}
The second term becomes a centered moment when averaged under the path integral measure; more specifically, the path integral average can be upper-bounded by $\varepsilon^{[p]}_{\mathrm{max}}\left(\xi, n, 2\right)$. In contrast, the first term collects collision terms $j = k$ and does not reduce to a centered moment when averaged under the path integral. However, extended from bit matrices $\bm{z} \in \{1, -1\}^{n \times \mathcal{T}_p}$ to complex matrices $\mathbb{C}^{n \times \mathcal{T}_p}$, it is majorized by:
\begin{align}
    \frac{1}{n^2}\sum_{j \in [n]}\left(w_j^{[p + 1]}w_j^{[t]} - G_{p + 1, t}\right)^2 & \preceq \frac{1}{n^2}\sum_{j \in [n]}\left(w_j^{[p + 1]}w_j^{[t]} + 1\right)^2,
\end{align}
hence contributes at most
\begin{align}
    \left|\left\langle \frac{1}{n^2}\sum_{j \in [n]}\left(w_j^{[p + 1]}w_j^{[t]} - G_{p + 1, t}\right)^2 \right\rangle_{\mu^{[p]}\left(\bm{z};\,\xi, [n]\right)}\right| & \leq \frac{1}{n^2}\sum_{j \in [n]}\left(1 \cdot 1 + 1\right)^2\nonumber\\
    & = \frac{4}{n},
\end{align}
which vanishes as $n \to \infty$. We now generalize this argument to all monomials of a centered overlap using systematic majorization estimate Lemma~\ref{si:lemma:remove_collision_terms_nonlinear_function_iterative}. We write the finite size error as:
\begin{align}
    \nu_n - \nu_{\infty} & = \sum_{t \in \mathcal{T}_{p - 1}}i\Gamma_t\sum_{1 \leq q \leq \mathrm{deg}(\xi)}\xi^{(q)}\left(G_{p + 1, t}\right)\left\langle \varphi_q\left(S_t\left(\bm{z}\right)\right) \right\rangle_{\mu^{[p]}\left(\bm{z};\,\xi, [n]\right)},\label{si:eq:finite_size_energy_density_error}
\end{align}
with
\begin{align}
    \varphi_q\left(x\right) & := \frac{x^q}{q!},\\
    S_t\left(\bm{w}\right) & := \sum_{j \in [n]}f_{j, t}\left(\bm{w}\right),\\
    f_{j, t}\left(\bm{w}\right) & := \frac{1}{n}\left(w_j^{[p + 1]}w_j^{[t]} - G_{p + 1, t}\right).
\end{align}
To obtain a majorization estimate of $\varphi$ without collision terms, we apply Lemma~\ref{si:lemma:remove_collision_terms_nonlinear_function_iterative} with parameters:
\begin{align}
    D & \longrightarrow [n] \times \mathcal{T}_p,\\
    J & \longrightarrow [n],\\
    L & \longrightarrow \{t\},\\
    f_{j, l} & \longrightarrow f_{j, l} \qquad \forall l \in L = \{t\},\\
    \varphi & \longrightarrow \varphi_q.
\end{align}
We can further choose majorants
\begin{align}
    \overline{f_{j, t}}\left(\bm{w}\right) & := \frac{1}{n}\left(w_j^{[p + 1]}w_j^{[t]} + 1\right),\\
    \overline{\varphi_q^{(m)}} & := \varphi_q^{(m)}
\end{align}
for $f_{j, t}$ and the order $m$ derivative of $\varphi_q$. We wish to only keep the main term of the Lemma's estimate
\begin{align}
    \sum_{\substack{J' \subset J,\,\left|J'\right| < d\\\bm{l} \in L^{J'}}}\partial_{\bm{l}}\varphi\left(\bm{0}_L\right)\prod_{k \in J'}f_{k, l_k}\left(\bm{w}\right),
\end{align}
excluding spin collisions due to the inner product excluding repeated indices $k$. All derivatives of $\varphi = \varphi_q$ at $0$ except the order $q$ one are zero, and we let $d := q + 1$. For this choice of parameters, the main term is computed as
\begin{align}
    \sum_{\substack{J' \subset J,\,\left|J'\right| < d\\\bm{l} \in L^{J'}}}\partial_{\bm{l}}\varphi\left(\bm{0}_L\right)\prod_{k \in J'}f_{k, l_k}\left(\bm{w}\right) & =  \sum_{\substack{J' \subset [n],\,\left|J'\right| = q\\\bm{l} \in L^{J'},\,\bm{l} = \left(t, \ldots, t\right)}}\partial_{\bm{l}}\varphi\left(\bm{0}_L\right)\prod_{k \in J'}f_{k, l_k}\left(\bm{w}\right)\\
    & =  \sum_{\substack{J' \subset [n],\,\left|J'\right| = q}}\prod_{k \in J'}\frac{1}{n}\left(w_k^{[p + 1]}w_k^{[t]} - G_{p + 1, t}\right)\nonumber\\
    & = \frac{1}{n^q}\sum_{\substack{J' \subset [n],\,\left|J'\right| = q}}\prod_{k \in J'}\left(w_k^{[p + 1]}w_k^{[t]} - G_{p + 1, t}\right),\label{si:eq:centered_normalized_overlap_noncolliding_expansion_main_term}
\end{align}
where each term of the sum taken under the path integral measure now gives a centered moment. The integral remainder term vanishes due to the $(q + 1)$ derivative order:
\begin{align}
    R_{\mathrm{int}, q, t}\left(\bm{0}\right) & = \sum_{\substack{J' \subset J,\,\left|J'\right| = q + 1\\\bm{l} \in L^{J'}}}\int_0^1\!\mathrm{d}s\,d\left(1 - s\right)^{d - 1}\partial_{\bm{l}}\varphi\left(s\bm{\hat{S}}_{J'}\left(\bm{w}\right)\right)\prod_{k \in J'}f_{k, l_k}\left(\bm{w}\right)\nonumber\\
    & = 0.\label{si:eq:centered_normalized_overlap_noncolliding_expansion_integral_remainder}
\end{align}
As for the sum remainder term, it specializes to:
\begin{align}
    R_{\mathrm{sum},\,q, t}\left(\bm{w}\right) & = \sum_{0 \leq m < d}\sum_{\substack{J' \subset J,\,\left|J'\right| = m\\\bm{l} \in L^{J'}\\j \in J - J'\\l, l' \in L}}\overline{\partial_{\bm{l}, l, l'}\varphi}\left(\bm{\overline{S}}\left(\bm{w}\right)\right)\overline{f_{j, l}}\left(\bm{w}\right)\overline{f_{j, l'}}\left(\bm{w}\right)\prod_{k \in J'}\overline{f_{k, l_k}}\left(\bm{w}\right)\nonumber\\
    & = \sum_{0 \leq m \leq q}\sum_{\substack{J' \subset [n],\,\left|J'\right| = m\\\bm{l} \in L^{J'},\,\bm{l} = \left(t, \ldots, t\right)\\j \in [n] - J'\\l = l' = t}}\overline{\varphi_q^{(m + 2)}}\left(\overline{S}_t\left(\bm{w}\right)\right)\overline{f_{j, t}}\left(\bm{w}\right)\overline{f_{j, t}}\left(\bm{w}\right)\prod_{k \in J'}\overline{f_{k, t}}\left(\bm{w}\right)\nonumber\\
    & = \sum_{0 \leq m \leq q}\sum_{\substack{J' \subset [n],\,\left|J'\right| = m\\\bm{l} \in L^{J'},\,\bm{l} = \left(t, \ldots, t\right)\\j \in [n] - J'\\l = l' = t}}\frac{\overline{S}_t\left(\bm{w}\right)^{q - m - 2}}{(q -m - 2)!}\overline{f_{j, t}}\left(\bm{w}\right)\overline{f_{j, t}}\left(\bm{w}\right)\prod_{k \in J'}\overline{f_{k, t}}\left(\bm{w}\right).\label{si:eq:centered_normalized_overlap_noncolliding_expansion_sum_remainder}
\end{align}
Combining expressions Eqns.~\ref{si:eq:centered_normalized_overlap_noncolliding_expansion_main_term}, \ref{si:eq:centered_normalized_overlap_noncolliding_expansion_integral_remainder}, \ref{si:eq:centered_normalized_overlap_noncolliding_expansion_sum_remainder}, Lemma~\ref{si:lemma:remove_collision_terms_nonlinear_function_iterative} provides majorization estimate:
\begin{align}
    & \left(\frac{1}{n}\left\langle \bm{w}^{[p + 1]}, \bm{w}^{[t]} \right\rangle - G_{p + 1, t}\right)^q - \frac{1}{n^q}\sum_{\substack{J' \subset [n],\,\left|J'\right| = q}}\prod_{k \in J'}\left(w_k^{[p + 1]}w_k^{[t]} - G_{p + 1, t}\right)\nonumber\\
    & = \varphi\left(\bm{S}\left(\bm{w}\right)\right) - \sum_{\substack{J' \subset J,\,\left|J'\right| < d\\\bm{l} \in L^{J'}}}\partial_{\bm{l}}\varphi\left(\bm{0}_L\right)\prod_{k \in J'}f_{k, l_k}\left(\bm{w}\right)\nonumber\\
    & \preceq R_{\mathrm{int}, q, t}\left(\bm{w}\right) + R_{\mathrm{sum}, q, t}\left(\bm{w}\right)\nonumber\\
    & = R_{\mathrm{sum}, q, t}\left(\bm{w}\right).
\end{align}
Taking the path integral, the above majorization bound implies
\begin{align}
    & \left|\left\langle \left(\frac{1}{n}\left\langle \bm{z}^{[p + 1]}, \bm{z}^{[t]} \right\rangle - G_{p + 1, t}\right)^q \right\rangle_{\mu^{[p]}\left(\bm{z};\,\xi, [n]\right)} - \left\langle \frac{1}{n^q}\sum_{J' \subset [n],\,\left|J'\right| = q}\prod_{k \in J'}\left(w_k^{[p + 1]}w_k^{[t]} - G_{p + 1, t}\right) \right\rangle_{\mu^{[p]}\left(\bm{z};\,\xi, [n]\right)}\right|\nonumber\\
    & \leq R_{\mathrm{sum}, q, t}\left(\bm{1}_{n \times \mathcal{T}_p}\right)\nonumber\\
    & = \sum_{0 \leq m \leq q}\sum_{\substack{J' \subset [n],\,\left|J'\right| = m\\\bm{l} \in L^{J'},\,\bm{l} = \left(t, \ldots, t\right)\\j \in [n] - J'\\l = l' = t}}\frac{(2/n)^{q - m - 2}}{(q -m - 2)!} \cdot (2/n) \cdot (2/n) \cdot (2/n)^m\nonumber\\
    & = \frac{2^q}{n^q}\sum_{0 \leq m \leq q}\sum_{\substack{J' \subset [n],\,\left|J'\right| = m\\\bm{l} \in L^{J'},\,\bm{l} = \left(t, \ldots, t\right)\\j \in [n] - J'\\l = l' = t}}\frac{1}{(q -m - 2)!}\nonumber\\
    & = \frac{2^q}{n^q}\sum_{0 \leq m \leq q}\binom{n}{m}(n - m)\frac{1}{(q - m - 2)!}\nonumber\\
    & \leq \frac{2^q}{n^q}\sum_{0 \leq m \leq q}\frac{n^{m + 1}}{m!}\frac{1}{(q - m - 2)!}\nonumber\\
    & = \frac{2^q}{n^q}\sum_{0 \leq m \leq q - 2}\frac{n^{m + 1}}{m!}\frac{1}{(q - m - 2)!}\nonumber\\
    & \leq \frac{2^q}{n^q}\sum_{0 \leq m \leq q - 2}\frac{n^{q - 1}}{m!}\frac{1}{(q - m - 2)!}\nonumber\\
    & = \frac{2^{2q - 2}}{(q - 2)!}\frac{1}{n}.\label{si:eq:centered_normalized_overlap_noncolliding_expansion_path_integral_error}
\end{align}
We now bound the expectation of the collision-free centered-monomial product under the path integral:
\begin{align}
    \left|\frac{1}{n^q}\left\langle \sum_{J' \subset [n],\,\left|J'\right| = q}\prod_{k \in J'}\left(w_k^{[p + 1]}w_k^{[t]} - G_{p + 1, t}\right) \right\rangle_{\mu^{[p]}\left(\bm{z};\,\xi, [n]\right)}\right| & = \left|\frac{1}{n^q}\sum_{J' \subset [n],\,\left|J'\right| = q}\left\langle \prod_{k \in J'}\left(w_k^{[p + 1]}w_k^{[t]} - G_{p + 1, t}\right) \right\rangle_{\mu^{[p]}\left(\bm{z};\,\xi, [n]\right)}\right|\nonumber\\
    & = \frac{1}{n^q}\sum_{J' \subset [n],\,\left|J'\right| = q}\left|\left\langle \prod_{k \in J'}\left(w_k^{[p + 1]}w_k^{[t]} - G_{p + 1, t}\right) \right\rangle_{\mu^{[p]}\left(\bm{z};\,\xi, [n]\right)}\right|\nonumber\\
    & \leq \frac{1}{n^q}\sum_{J' \subset [n],\,\left|J'\right| = q}\varepsilon^{[p]}\left(\xi, n, q\right)\nonumber\\
    & = \frac{1}{n^q}\binom{n}{q}\varepsilon^{[p]}\left(\xi, n, q\right)\nonumber\\
    & \leq \frac{1}{q!}\varepsilon^{[p]}\left(\xi, n, q\right)\nonumber\\
    & \leq \frac{1}{q!}\varepsilon^{[p]}_{\mathrm{max}}\left(\xi, n, \mathrm{deg}(\xi)\right).\label{si:eq:centered_normalized_overlap_noncolliding_contribution_bound}
\end{align}
Combining bounds Eqns.~\ref{si:eq:centered_normalized_overlap_noncolliding_expansion_path_integral_error}, \ref{si:eq:centered_normalized_overlap_noncolliding_contribution_bound} yields:
\begin{align}
    \left|\left\langle \varphi_{q, t}\left(\bm{z}\right) \right\rangle_{\mu^{[p]}\left(\bm{z};\,\xi, [n]\right)}\right| & = \left|\left\langle \left(\frac{1}{n}\left\langle \bm{z}^{[p + 1]}, \bm{z}^{[t]} \right\rangle - G_{p + 1, t}\right)^q \right\rangle_{\mu^{[p]}\left(\bm{z};\,\xi, [n]\right)}\right|\nonumber\\
    & \leq \frac{2^{2q - 2}}{(q - 2)!}\frac{1}{n} + \frac{1}{q!}\varepsilon_{\mathrm{max}}^{[p]}\left(\xi, n, \mathrm{deg}\left(\xi\right)\right).
\end{align}
Plugging this bound into exact expression Eq.~\ref{si:eq:finite_size_energy_density_error} for the finite-size error of the energy density, the error can be bounded as:
\begin{align}
    \left|\nu_{\infty} - \nu_n\right| & \leq \sum_{t \in \mathcal{T}_{p - 1}}\left|\Gamma_t\right|\sum_{1 \leq q \leq \mathrm{deg}(\xi)}\xi^{(q)}\left(1\right)\left(\frac{2^{2q - 2}}{(q - 2)!}\frac{1}{n} + \frac{1}{q!}\varepsilon_{\mathrm{max}}^{[p]}\left(\xi, n, \mathrm{deg}(\xi)\right)\right)\nonumber\\
    & \leq 2\gamma p\left(\frac{4}{n}\sum_{1 \leq q \leq \mathrm{deg}(\xi)}\frac{\xi^{(q)}\left(1\right)}{q!}q(q - 1)4^{q - 2} + \varepsilon^{[p]}_{\mathrm{max}}\left(\xi, n, \mathrm{deg}\left(\xi\right)\right)\sum_{1 \leq q \leq \mathrm{deg}\left(\xi\right)}\frac{\xi^{(q)}\left(1\right)}{q!}\right)\nonumber\\
    & \leq 2\gamma p\left(\frac{4\xi''(5)}{n} + \xi(2)\varepsilon^{[p]}_{\mathrm{max}}\left(\xi, n, \mathrm{deg}(\xi)\right)\right).
\end{align}
Then, choosing
\begin{align}
    n & \geq \frac{16\gamma p x}{\varepsilon}
\end{align}
ensures the first term is upper-bounded by $\varepsilon/2$. Next, by Proposition~\ref{si:prop:simplified_threshold_function_bound}, choosing
\begin{align}
    n & \geq 2^{8p^2 + 3}b(x)\left(\frac{2^{\mathrm{deg}(\xi) + 4}\gamma^2p^2x^2}{\varepsilon^2}\right)^{\left(\frac{2e(x)}{\log 2}\right)^{2p}}\left(2\log_2\left(\frac{4\gamma p x}{\varepsilon}\right) + \mathrm{deg}(\xi)\right)^2
\end{align}
ensures the second term is upper-bounded by $\varepsilon/2$. Hence, choosing $n$ as suggested in the statement ensures the energy density's finite-size error is upper-bounded by $\varepsilon$.
\end{proof}
\end{repcorollary}

\subsubsection{Majorization estimates}

In this Section, we prove deferred majorization bounds used in the derivation (Lemma~\ref{si:lemma:centered_moments_maximum_recursion}) of the key recursion in layer index $q$ between maximum centered moments $\varepsilon^{[q]}\left(\xi, s, s''\right)$. These majorization estimates cover three quantities. First, multiplicative factor $\mu^{[p]}_{C, 2}\left(\bm{z}_{S'}, \bm{z}_{S''};\,\xi, S', S''\right)$ in the cavity decomposition of the spin-glass QAOA path integral measure Lemma~\ref{si:lemma:cavity_decomposition_path_integral_measure} is treated in Lemma~\ref{si:lemma:mu_c_2_constant_degree_polynomial_approximation}, making approximation $\mu^{[p]}_{C, 2} \approx 1$ under the path integral fully rigorous. Next, multiplicative factor $\mu^{[p]}_{C, 1}\left(\bm{z}_{S'}, \bm{z}_{S''};\,\xi, S', S''\right)$ in the same cavity decomposition is approximated as a constant-degree polynomial in Lemma~\ref{si:lemma:mu_c_1_constant_degree_polynomial_approximation}, using the non-colliding Taylor expansion results from Section~\ref{si:sec:non_colliding_taylor_expansion}. Finally, Lemma~\ref{si:lemma:spin_boson_measures_ratio_majorization_bound} provides a majorant on the relative density of two spin-boson measures characterized by influence functionals $\bm{K}, \bm{\widetilde{K}}$ respectively (see Definition~\ref{si:def:spin_boson_path_integral_measure}). The latter result is used in the proof of main recursion inequality Lemma~\ref{si:lemma:centered_moments_maximum_recursion} to control the variation of the path integral measure under a shift of the $\bm{G}$ matrix solution to the fixed-point equation, given $\bm{G}$ parametrizes the influence functional in the spin-boson mapping of QAOA (Proposition~\ref{si:prop:g_iteration_spin_boson_multi_states}). A bound on the variation of $\bm{G}$ itself is deferred to Section~\ref{si:sec:perturbation_g_matrix} (Proposition~\ref{si:prop:g_matrix_variation_bound_parametrized_xi} and Corollary~\ref{si:cor:simple_g_matrix_variation_bound_parametrized_xi}).

Associated to the approximation of $\mu^{[p]}_{C, 1}$ in Lemma~\ref{si:lemma:mu_c_1_constant_degree_polynomial_approximation}, two subsidiary estimates are motivated and proven at their points of use below: a bound on the coefficients of the polynomial approximation (Lemma~\ref{si:lemma:mu_c_1p_d_coefficients_series_bound}) and a majorant for $\mu^{[p]}_{C, 1}$ (Lemma~\ref{si:lemma:mu_c_1_majorization_bound}).

We start with an estimate on multiplicative factor $\mu^{[p]}_{C, 2}\left(\bm{z}_{S'}, \bm{z}_{S''};\,\xi, S', S''\right)$ occurring in the cavity decomposition of the path integral measure. In the heuristic discussion of the cavity argument (Section~\ref{si:sec:heuristic_cavity_argument}), we observed this quantity tended to $1$ in the infinite-size limit, pointwise in the spin configuration. As pointwise estimates do not pass to the path integral, we tighten this estimate by providing a majorant to the error in the following Lemma:

\begin{lemma}[Quantifying $\mu_{C, 2} \approx 1$ in path integral measure cavity decomposition]
\label{si:lemma:mu_c_2_constant_degree_polynomial_approximation}
Consider the cavity decomposition of the $p$-level spin-glass QAOA path integral measure introduced in Lemma~\ref{si:lemma:cavity_decomposition_path_integral_measure}. Then, contribution $\mu^{[p]}_{C, 2}\left(\bm{w}_{S'}, \bm{z}_{S''};\,\xi, S', S''\right)$, seen as a function of complex variables $\bm{w}_{S'} \in \mathbb{C}^{S' \times \mathcal{T}_p}$ (with $\bm{z}_{S''} \in \{1, -1\}^{S'' \times \mathcal{T}_p}$ treated as a fixed bit matrix) can be replaced by $1$ up to a precisely quantified majorization error. Namely, denoting by
\begin{align}
    \widetilde{\mu}^{[p]}_{C, 2}\left(\bm{w}_{S'}, \bm{z}_{S''};\,\xi, S', S''\right) := \mu^{[p]}_{C, 2}\left(\bm{w}_{S'}, \bm{z}_{S''};\,\xi, S', S''\right) - 1
\end{align}
the difference, it holds
\begin{align}
     \widetilde{\mu}^{[p]}_{C, 2}\left(\,\cdot\,, \bm{z}_{S''};\,\xi, S', S''\right) & \preceq \overline{\widetilde{\mu}^{[p]}_{C, 2}}\left(\,\cdot\,;\,\xi, S', S''\right),
\end{align}
where
\begin{align}
    \overline{\widetilde{\mu}^{[p]}_{C, 2}}\left(\bm{w}_{S'};\,\xi, S', S''\right) & := \exp\left(\frac{\gamma^2\left|S''\right|^2}{4|S|}\sum_{t, u \in \mathcal{T}_{p - 1}}\xi''\left(\frac{1}{|S|}\left\langle \bm{w}_{S'}^{[t]}, \bm{w}_{S'}^{[u]} \right\rangle + \frac{2\left|S''\right|}{|S|}\right)\right)\nonumber\\
    & \hspace*{20px} \times \frac{\gamma^2\left|S''\right|^2}{4|S|}\sum_{t, u \in \mathcal{T}_{p - 1}}\xi''\left(\frac{1}{|S|}\left\langle \bm{w}^{[t]}_{S'}, \bm{w}^{[u]}_{S'} \right\rangle + \frac{2\left|S''\right|}{|S|}\right).
\end{align}
This majorant can further be bounded as follows at the all-$1$ variables:
\begin{align}
    \overline{\widetilde{\mu}^{[p]}_{C, 2}}\left(\bm{1}_{S' \times \mathcal{T}_{p - 1}};\,\xi, S', S''\right) & \leq \frac{\gamma^2p^2\left|S''\right|^2}{|S|}\xi''\left(1 + \frac{\left|S''\right|}{|S|}\right)\exp\left(\frac{\gamma^2p^2\left|S''\right|}{|S|}\xi''\left(1 + \frac{\left|S''\right|}{|S|}\right)\right)\nonumber\\
    & \leq \frac{\gamma^2p^2\left|S''\right|^2}{|S|}\xi''\left(2\right)\exp\left(\frac{\gamma^2p^2\left|S''\right|^2}{|S|}\xi''(2)\right).
\end{align}
\begin{proof}
Using the elementary majorization bound on the difference between $\exp(z)$ and $1$:
\begin{align}
    \exp\left(z\right) - 1 & \preceq z\exp(z).
\end{align}
and composing majorization bounds, we compute:
\begin{align}
    & \mu^{[p]}_{C, 2}\left(\bm{w}_{S'}, \bm{z}_{S''};\,\xi, S', S''\right) - 1\nonumber\\
    & = \exp\left(-\frac{1}{2}\sum_{t, u \in \mathcal{T}_{p - 1}}\Gamma_t\Gamma_u\sum_{m \geq 2}\frac{1}{m!}\xi^{(m)}\left(\frac{1}{|S|}\left(\left\langle \bm{w}_{S'}^{[t]}, \bm{w}_{S'}^{[u]} \right\rangle + \left\langle \bm{z}_{S''}^{[t]}, \bm{z}_{S''}^{[u]} \right\rangle\right)\right)\frac{1}{|S|^{m - 1}}\left\langle \bm{z}^{[t]}_{S''}, \bm{z}^{[u]}_{S''} \right\rangle^m\right) - 1\nonumber\\
    & \preceq \frac{1}{2}\sum_{t, u \in \mathcal{T}_{p - 1}}\left|\Gamma_t\Gamma_u\right|\sum_{m \geq 2}\frac{1}{m!}\xi^{(m)}\left(\frac{1}{|S|}\left(\left\langle \bm{w}^{[t]}_{S'}, \bm{w}^{[u]}_{S'} \right\rangle + \left|S''\right|\right)\right)\frac{\left|S''\right|^m}{|S|^{m - 1}}\nonumber\\
    & \hspace*{30px} \times \exp\left(\frac{1}{2}\sum_{t, u \in \mathcal{T}_{p - 1}}\left|\Gamma_t\Gamma_u\right|\sum_{m \geq 2}\frac{1}{m!}\xi^{(m)}\left(\frac{1}{|S|}\left(\left\langle \bm{w}^{[t]}_{S'}, \bm{w}^{[u]}_{S'} \right\rangle + \left|S''\right|\right)\right)\frac{\left|S''\right|^m}{|S|^{m - 1}}\right)\nonumber\\
    & \preceq \frac{\gamma^2}{2}\sum_{t, u \in \mathcal{T}_{p - 1}}\sum_{m \geq 2}\frac{1}{m!}\xi^{(m)}\left(\frac{1}{|S|}\left(\left\langle \bm{w}^{[t]}_{S'}, \bm{w}^{[u]}_{S'} \right\rangle + \left|S''\right|\right)\right)\frac{\left|S''\right|^m}{|S|^{m - 1}}\nonumber\\
    & \hspace*{30px} \times \exp\left(\frac{\gamma^2}{2}\sum_{t, u \in \mathcal{T}_{p - 1}}\sum_{m \geq 2}\frac{1}{m!}\xi^{(m)}\left(\frac{1}{|S|}\left(\left\langle \bm{w}^{[t]}_{S'}, \bm{w}^{[u]}_{S'} \right\rangle + \left|S''\right|\right)\right)\frac{\left|S''\right|^m}{|S|^{m - 1}}\right).
\end{align}
We now use majorization bound:
\begin{align}
    \sum_{m \geq 2}\frac{1}{m!}\xi^{(m)}\left(\varphi\left(\bm{w}\right)\right)\frac{\left(s''\right)^m}{|S|^{m - 1}} & = |S|\sum_{m \geq 2}\frac{1}{m!}\xi^{(m)}\left(\varphi\left(\bm{w}\right)\right)\frac{\left(s''\right)^m}{|S|^m}\nonumber\\
    & = |S|\left(\xi\left(\varphi\left(\bm{w}\right) + \frac{s''}{|S|}\right) - \xi\left(\varphi\left(\bm{w}\right)\right) - \xi'\left(\varphi\left(\bm{w}\right)\right)\frac{s''}{|S|}\right)\nonumber\\
    & \preceq |S|\frac{1}{2}\xi''\left(\overline{\varphi}\left(\bm{w}\right) + \frac{s''}{|S|}\right)\left(\frac{s''}{|S|}\right)^2\nonumber\\
    & = \frac{\left(s''\right)^2}{2|S|}\xi''\left(\overline{\varphi}\left(\bm{w}\right) + \frac{s''}{|S|}\right).
\end{align}
valid for any multivariate entire function $\varphi\left(\bm{w}\right)$ majorized by $\overline{\varphi}\left(\bm{w}\right)$. Applying this to $\varphi\left(\bm{w}_{S'}\right) := \left\langle \bm{w}^{[t]}_{S'}, \bm{w}^{[u]}_{S'} \right\rangle/|S| + s''/|S|$, majorized by itself, yields
\begin{align}
    & \mu^{[p]}_{C, 2}\left(\bm{z}_{S'}, \bm{z}_{S''};\,\xi, S', S''\right) - 1\nonumber\\
    & \preceq \frac{\gamma^2}{2}\frac{\left(s''\right)^2}{2|S|}\sum_{t, u \in \mathcal{T}_{p - 1}}\xi''\left(\frac{1}{|S|}\left\langle \bm{w}^{[t]}_{S'}, \bm{w}^{[u]}_{S'} \right\rangle + \frac{2s''}{|S|}\right)\exp\left(\frac{\gamma^2}{2}\frac{\left(s''\right)^2}{2|S|}\sum_{t, u \in \mathcal{T}_{p - 1}}\xi''\left(\frac{1}{|S|}\left\langle \bm{w}^{[t]}_{S'}, \bm{w}^{[u]}_{S'} \right\rangle + \frac{2s''}{|S|}\right)\right).
\end{align}
\end{proof}
\end{lemma}

After obtaining an estimate for multiplicative contribution $\mu^{[p]}_{C, 2}\left(\bm{z}_{S'}, \bm{z}_{S''};\,\xi, S', S''\right)$ in the cavity decomposition of the path integral measure Lemma~\ref{si:lemma:cavity_decomposition_path_integral_measure}, we now turn our attention to multiplicative contribution $\mu^{[p]}_{C, 1}\left(\bm{z}_{S'}, \bm{z}_{S''};\,\xi_{s'/s}, S', S''\right)$. We handle this quantity by expanding it, for each fixed history $\bm{z}_{S''} \in \{1, -1\}^{S'' \times \mathcal{T}_p}$ of the small spin subsystem $S''$, as a polynomial in the centered spin products $z^{[t]}_jz^{[u]}_j - G_{t, u}$ (with $j \in S'$ and $t, u \in \mathcal{T}_p$) from the large spin subsystem.

\begin{lemma}[Constant-degree polynomial approximation of $\mu_{C, 1}$ in path integral measure cavity expansion]
\label{si:lemma:mu_c_1_constant_degree_polynomial_approximation}
Consider the cavity decomposition of the $p$-level spin-glass QAOA path integral measure introduced in Lemma~\ref{si:lemma:cavity_decomposition_path_integral_measure}. Then, for all expansion order parameter $d \geq 0$, contribution $\mu^{[p]}_{C, 1}\left(\bm{z}_{S'}, \bm{z}_{S''};\,\xi_{s'/s}, S', S''\right)$ can be decomposed as:
\begin{align}
    & \mu^{[p]}_{C, 1}\left(\bm{z}_{S'}, \bm{z}_{S''};\,\xi_{s'/s}, S', S''\right)\nonumber\\
    & = \mu^{[p]}_{C, SB}\left(\bm{z}_{S''};\,\xi_{s'/s}, S''\right)\left(1 + \mu^{[p]}_{C, 1', d}\left(\bm{z}_{S'}, \bm{z}_{S''};\,\xi_{s'/s}, S', S''\right) + \mu^{[p]}_{C, 1'', d}\left(\bm{z}_{S'}, \bm{z}_{S''};\,\xi_{s'/s}, S', S''\right)\right),\label{si:eq:mu_c_1_additive_contributions}
\end{align}
where
\begin{align}
    \mu^{[p]}_{C, SB}\left(\bm{z}_{S''};\,\xi_{s'/s}, S''\right) & := \exp\left(-\frac{1}{2}\sum_{t, u \in \mathcal{T}_{p - 1}}\Gamma_t\Gamma_u\xi'_{s'/s}\left(G_{t, u}\right)\left\langle \bm{z}^{[t]}_{S''}, \bm{z}^{[u]}_{S''} \right\rangle\right)\label{si:eq:mu_c_sb_definition}
\end{align}
is the cost contribution to the path integral of QAOA applied to a certain spin-boson Hamiltonian over $S''$ (see Proposition~\ref{si:prop:spin_boson_path_integral_measure_expression}). $\mu^{[p]}_{C, 1', d}\left(\,\cdot\,,\bm{z}_{S''};\,\xi_{s'/s}, S', S''\right)$ and $\mu^{[p]}_{C, 1'', d}\left(\,\cdot\,, \bm{z}_{S''};\,\xi_{s'/s}, S', S''\right)$ extend to entire functions over $\mathbb{C}^{S' \times \mathcal{T}_p}$. $\mu^{[p]}_{C, 1', d}\left(\,\cdot\,,\bm{z}_{S''};\,\xi_{s'/s}, S', S''\right)$ can be expanded as:
\begin{align}
    \mu^{[p]}_{C, 1', d}\left(\bm{w}_{S'}, \bm{z}_{S''};\,\xi_{s'/s}, S', S''\right) & = \sum_{\substack{J' \subset S'\\1 \leq |J'| < d\\\bm{t}, \bm{u} \in \mathcal{T}_{p - 1}^{J'}}}\mu^{[p]}_{C, 1', \bm{t}, \bm{u}}\left(\bm{z}_{S''};\,\xi_{s'/s}, S', S''\right)\prod_{j \in J'}\left(w_j^{[t_j]}w_j^{[u_j]} - G_{t_j, u_j}\right),\label{si:eq:mu_c_1p_d_definition}
\end{align}
for a family of coefficients $\left(\mu^{[p]}_{C, 1', \bm{t}, \bm{u}}\left(\bm{z}_{S''};\,\xi_{s'/s}, S', S''\right)\right)_{J' \subset S',\,1 \leq |J'| \leq d,\,\bm{t}, \bm{u} \in \mathcal{T}_{p - 1}^{J'}}$ bounded as follows:
\begin{align}
    \left|\mu^{[p]}_{C, 1', \bm{t}, \bm{u}}\left(\bm{z}_{S''};\,\xi_{s'/s}, S', S''\right)\right| & \leq \frac{\exp\left(4p^2\gamma^2s''R\xi''\left(1 + 2R\right)\right)}{\left(\left|S'\right|R\right)^{\left|J'\right|}}\prod_{t, u \in \mathcal{T}_{p - 1}}m_{t, u}!,\label{si:eq:mu_c_1p_d_coefficients_definition}
\end{align}
with $m_{t, u} := \left|\left\{j \in J'\,:\,(t_j, u_j) = (t, u)\right\}\right|$ the multiplicity of pair $(t_j, u_j)$ in $J'$-labeled tuples $\bm{t}, \bm{u} \in \mathcal{T}_{p - 1}^{J'}$. $\mu^{[p]}_{C, 1'', d}\left(\,\cdot\,,\bm{z}_{S''};\,\xi_{s'/s}, S', S''\right)$ admits the following majorization bounds:
\begin{align}
    \mu^{[p]}_{C, 1'', d}\left(\,\cdot\,,\bm{z}_{S''};\,\xi_{s'/s}, S', S''\right) & \preceq \overline{\mu^{[p]}_{C, 1'', d}}\left(\,\cdot\,,\bm{z}_{S''};\,\xi_{s'/s}, S', S''\right),\label{si:eq:mu_c_1pp_d_majorization_bound}\\
    \overline{\mu^{[p]}_{C, 1'', d}}\left(\bm{1}_{S' \times \mathcal{T}_{p - 1}},\bm{z}_{S''};\,\xi_{s'/s}, S', S''\right) & \leq 2^{4p^2}\exp\left(4p^2\gamma^2s''R\xi''\left(1 + 2R\right)\right)\nonumber\\
    & \hspace*{10px} \times \left(\left(\frac{4}{R}\right)^d\binom{d + 4p^2 - 1}{4p^2 - 1} + \frac{1}{\left|S'\right|}4p^2\left(4p^2 + 1\right)\frac{(4/R)^2}{\left(1 - 4/R\right)^{4p^2 + 2}}\right),
\end{align}
for all $R > 4$.
\begin{proof}
Starting with the initial expression of $\mu^{[p]}_{C, 1}$ provided in Eq.~\ref{si:eq:cavity_decomposition_path_integral_contribution_1}:
\begin{align}
    \mu^{[p]}_{C, 1}\left(\bm{z}_{S'}, \bm{z}_{S''};\,\xi_{s'/s}, S', S''\right) & = \exp\left(-\frac{1}{2}\sum_{t, u \in \mathcal{T}_{p - 1}}\Gamma_t\Gamma_u\xi'_{s'/s}\left(\frac{1}{\left|S'\right|}\left\langle \bm{z}^{[t]}_{S'}, \bm{z}^{[u]}_{S'} \right\rangle\right)\left\langle \bm{z}^{[t]}_{S''}, \bm{z}^{[u]}_{S''} \right\rangle\right),
\end{align}
We Taylor-expand $\xi'_{s'/s}$ around $G_{t, u}$ to obtain:
\begin{align}
    \mu^{[p]}_{C, 1}\left(\bm{z}_{S'}, \bm{z}_{S''};\,\xi_{s'/s}, S', S''\right) & = \mu_{C, SB}\left(\bm{z}_{S''};\,\xi_{s'/s}, S''\right)\mu_{C, 1'}\left(\bm{z}_{S'}, \bm{z}_{S''};\,\xi_{s'/s}, S', S''\right),
\end{align}
where
\begin{align}
    \mu^{[p]}_{C, SB}\left(\bm{z}_{S''};\,\xi_{s'/s}, S''\right) & := \exp\left(-\frac{1}{2}\sum_{t, u \in \mathcal{T}_{p - 1}}\Gamma_t\Gamma_u\xi_{s'/s}'\left(G_{t, u}\right)\left\langle \bm{z}^{[t]}_{S''}, \bm{z}^{[u]}_{S''} \right\rangle\right),\\
    \mu^{[p]}_{C, 1'}\left(\bm{z}_{S'}, \bm{z}_{S''};\,\xi_{s'/s}, S', S''\right) & := \exp\left(-\frac{1}{2}\sum_{t, u \in \mathcal{T}_{p - 1}}\Gamma_t\Gamma_u\sum_{m \geq 1}\frac{\xi_{s'/s}^{(1 + m)}\left(G_{t, u}\right)}{m!}\left(\frac{1}{\left|S'\right|}\left\langle \bm{z}^{[t]}_{S'}, \bm{z}^{[u]}_{S'} \right\rangle - G_{t, u}\right)^m\left\langle \bm{z}^{[t]}_{S''}, \bm{z}^{[u]}_{S''} \right\rangle\right).
\end{align}
We now obtain a low-degree polynomial approximation of $\mu^{[p]}_{C, 1'}\left(\bm{w}_{S'}, \bm{z}_{S''};\,\xi_{s'/s}, S', S''\right)$, seen as a function of $\bm{w}_{S'} \in \mathbb{C}^{S' \times \mathcal{T}_{p - 1}}$ only ($\bm{z}_{S''}$ frozen), using Proposition~\ref{si:prop:polynomial_approximation_entire_function_path_integral_measure}. Following the notations from the Proposition, initially introduced in Lemma~\ref{si:lemma:remove_collision_terms_nonlinear_function}, we let:
\begin{align}
    D & := S' \times \mathcal{T}_{p - 1},\\
    J & := S',\\
    L & := \mathcal{T}_{p - 1} \times \mathcal{T}_{p - 1},\\
    f_{j,\,(t, u)}: & \left\{\begin{array}{ccc}
         \mathbb{C}^{S' \times \mathcal{T}_{p - 1}} & \longrightarrow & \mathbb{C}\\
         \bm{w}_{S'} & \longmapsto & f_{j,\,(t, u)}\left(\bm{w}_{S'}\right) := \frac{1}{\left|S'\right|}\left(w_j^{[t]}w_j^{[u]} - G_{t, u}\right)
    \end{array}\right.,\\
    \varphi: & \left\{\begin{array}{ccc}
         \mathbb{C}^{\mathcal{T}_{p - 1} \times \mathcal{T}_{p - 1}} & \longrightarrow & \mathbb{C}\\
         \bm{x} = \left(x_{t, u}\right)_{t, u \in \mathcal{T}_{p - 1}} & \longmapsto & \exp\left(-\frac{1}{2}\sum\limits_{t, u \in \mathcal{T}_{p - 1}}\Gamma_t\Gamma_u\sum\limits_{m \geq 1}\frac{\xi_{s'/s}^{(1 + m)}\left(G_{t, u}\right)}{m!}x_{t, u}^{m}\left\langle \bm{z}^{[t]}_{S''}, \bm{z}^{[u]}_{S''} \right\rangle\right) - 1
    \end{array}\right..
\end{align}
Note that $\varphi$ depends on frozen bit matrix $\bm{z}_{S''} \in \{1, -1\}^{S'' \times \mathcal{T}_{p - 1}}$, which is not explicit in the notation for conciseness. Reexpressing $\mu^{[p]}_{C, 1'}$ from these auxiliary definitions:
\begin{align}
    \mu^{[p]}_{C, 1'}\left(\bm{w}_{S'}, \bm{z}_{S''};\,\xi_{s'/s}, S', S''\right) & = 1 + \varphi\left(\bm{S}\left(\bm{w}\right)\right),\\
    \bm{S}\left(\bm{w}\right) & := \left(S_{t, u}\left(\bm{w}\right)\right)_{t, u \in \mathcal{T}_{p - 1}},\\
    S_{t, u}\left(\bm{w}\right) & := \frac{1}{\left|S'\right|}\sum_{j \in S'}\left(w_j^{[t]}w_j^{[u]} - G_{t, u}\right).
\end{align}
We will derive error bounds for polynomial approximation of $\bm{w} \longmapsto \varphi\left(\bm{S}(\bm{w})\right)$ that are uniform in $\bm{z}_{S''}$, only depending on $s'' = |S''|$. $\varphi$ can be simply upper-bounded on any polydisk of radii $R$:
\begin{align}
    M_R & = \sup_{\forall t, u \in \mathcal{T}_{p - 1},\,\left|x_{t, u}\right| \leq R}\left|\varphi\left(\bm{x}\right)\right|\nonumber\\
    & \leq \exp\left(\frac{1}{2}\sum_{t, u \in \mathcal{T}_{p - 1}}\left|\Gamma_t\Gamma_u\right|\sum_{m \geq 1}\frac{\xi_{s'/s}^{(1 + m)}\left(\left|G_{t, u}\right|\right)}{m!}R^ms''\right)\nonumber\\
    & \leq \exp\left(2p^2\gamma^2s''\sum_{m \geq 1}\frac{\xi_{s'/s}^{(1 + m)}\left(1\right)}{m!}R^m\right)\nonumber\\
    & \leq \exp\left(2p^2\gamma^2s''\left(\xi_{s'/s}'\left(1 + R\right) - \xi_{s'/s}'(1)\right)\right)\nonumber\\
    & \leq \exp\left(2p^2\gamma^2s''R\xi_{s'/s}''\left(1 + R\right)\right)\nonumber\\
    & = \exp\left(2p^2\gamma^2s''R\frac{s'}{s}\xi''\left(\frac{s'}{s}\left(1 + R\right)\right)\right)\nonumber\\
    & \leq \exp\left(2p^2\gamma^2s''R\xi''\left(1 + R\right)\right)
\end{align}
Furthermore,
\begin{align}
    f_{j,\,(t, u)}\left(\bm{w}_{S'}\right) & = \frac{1}{\left|S'\right|}\left(w^{[t]}_jw^{[u]}_j - G_{t, u}\right)\nonumber\\
    & \preceq \frac{1}{|S'|}\left(w_j^{[t]}w_j^{[u]} + 1\right)\nonumber\\
    & =: \overline{f_{j,\,(t, u)}}\left(\bm{w}_{S'}\right),
\end{align}
with $\overline{f_{j,\,(t, u)}}\left(\bm{1}_{S' \times \mathcal{T}_{p - 1}}\right) = 2/\left|S'\right|$, so one may take $C = 2$ in Proposition~\ref{si:prop:polynomial_approximation_entire_function_path_integral_measure}. Applying the Proposition, we then define $\mu^{[p]}_{C, 1', d}$ as the truncated Taylor expansion:
\begin{align}
    \mu^{[p]}_{C, 1', d}\left(\bm{w}_{S'}, \bm{z}_{S''};\,\xi_{s'/s}, S', S''\right) & := \sum_{\substack{J' \subset J,\,\left|J'\right| < d\\\bm{l} \in L^{J'}}}\partial_{\bm{l}}\varphi\left(\bm{0}_L\right)\prod_{j \in J'}f_{j, l_j}\left(\bm{w}_{S'}\right)\nonumber\\
    & = \sum_{\substack{J' \subset J,\,1 \leq \left|J'\right| < d\\\bm{l} \in L^{J'}}}\partial_{\bm{l}}\varphi\left(\bm{0}_L\right)\prod_{j \in J'}f_{j, l_j}\left(\bm{w}_{S'}\right)\nonumber\\
    & = \sum_{\substack{J' \subset S',\,1 \leq \left|J'\right| < d\\\bm{t}, \bm{u} \in \mathcal{T}_{p - 1}^{J'}}}\left(\prod_{j \in J'}\frac{\partial}{\partial x_{t_j, u_j}}\varphi\left(\bm{0}\right)\right)\prod_{j \in J'}\frac{1}{\left|S'\right|}\left(w_j^{[t_j]}w_j^{[u_j]} - G_{t_j, u_j}\right)\nonumber\\
    & = \sum_{\substack{J' \subset S'\\1 \leq \left|J'\right| < d\\\bm{t}, \bm{u} \in \mathcal{T}_{p - 1}^{J'}}}\mu^{[p]}_{C, 1', \bm{t}, \bm{u}}\left(\bm{z}_{S''};\,\xi_{s'/s}, S', S''\right)\prod_{j \in J'}\left(w_j^{[t_j]}w_j^{[u_j]} - G_{t_j, u_j}\right),
\end{align}
where in the final line, we let:
\begin{align}
    \mu^{[p]}_{C, 1', \bm{t}, \bm{u}}\left(\bm{z}_{S''};\,\xi_{s'/s}, S', S''\right) & := \frac{1}{\left|S'\right|^{\left|J'\right|}}\prod_{j \in J'}\frac{\partial}{\partial x_{t_j, u_j}}\varphi\left(\bm{0}\right),
\end{align}
restoring all explicit dependences on the left-hand side for rigour. Note this definition only depends on $\bm{t}, \bm{u} \in \mathcal{T}_{p - 1}^{J'}$ through the multisets formed by pairs $\left(t_j, u_j\right)$ ($j \in J'$); in particular, the definition does not depend on the labelling set $J'$. Using Cauchy's estimate on a polydisk (Lemma~\ref{si:lemma:cauchy_estimate_polydisks}) to bound the derivatives of $\varphi$ at $\bm{0}$, these complex numbers can be bounded as:
\begin{align}
    \left|\mu^{[p]}_{C, 1', \bm{t}, \bm{u}}\left(\bm{z}_{S''};\,\xi_{s'/s}, S', S''\right)\right| & \leq \frac{M_{2R}}{\left|S'\right|^{\left|J'\right|}}\prod_{t, u \in \mathcal{T}_{p - 1}}\frac{m_{t, u}!}{R^{m_{t, u}}}\nonumber\\
    & \leq \frac{\exp\left(4p^2\gamma^2s''R\xi''\left(1 + 2R\right)\right)}{\left(\left|S'\right|R\right)^{\left|J'\right|}}\prod_{t, u \in \mathcal{T}_{p - 1}}m_{t, u}!,
\end{align}
for all $R > 2C = 4$, where $m_{t, u} := \left|\left\{j \in J'\,:\,(t_j, u_j) = (t, u)\right\}\right|$ is the multiplicity of pair $(t_j, u_j)$ in $J'$-labeled tuples $\bm{t}, \bm{u}$. Still following the notation of Proposition~\ref{si:prop:polynomial_approximation_entire_function_path_integral_measure}, we define
\begin{align}
    \mu^{[p]}_{C, 1'', d}\left(\bm{w}_{S'}, \bm{z}_{S''};\,\xi_{s'/s}, S', S''\right) & := \psi\left(\bm{w}_{S'}\right)\nonumber\\
    & := \varphi\left(\bm{S}\left(\bm{w}_{S'}\right)\right) - \sum_{\substack{J' \subset J,\,\left|J'\right| < d\\\bm{l} \in L^{J'}}}\partial_{\bm{l}}\varphi\left(\bm{0}\right)\prod_{k \in J'}f_{k, l_k}\left(\bm{w}_{S'}\right)\nonumber\\
    & = \mu^{[p]}_{C, 1'}\left(\bm{w}_{S'}, \bm{z}_{S''};\,\xi_{s'/s}, S', S''\right) - 1\nonumber\\
    & \hspace*{20px} - \sum_{\substack{J' \subset S'\\1 \leq \left|J'\right| < d\\\bm{t}, \bm{u} \in \mathcal{T}_{p - 1}^{J'}}}\mu^{[p]}_{C, 1', \bm{t}, \bm{u}}\left(\bm{z}_{S''};\,\xi_{s'/s}, S', S''\right)\prod_{j \in J'}\left(w_j^{[t_j]}w_j^{[u_j]} - G_{t_j, u_j}\right)
\end{align}
Letting
\begin{align}
    \overline{\mu^{[p]}_{C, 1'', d}}\left(\,\cdot\,, \bm{z}_{S''};\,\xi_{s'/s}, S', S''\right) & := \overline{\psi}\left(\,\cdot\,\right),
\end{align}
Proposition~\ref{si:prop:polynomial_approximation_entire_function_path_integral_measure} gives:
\begin{align}
    \mu^{[p]}_{C, 1'', d}\left(\,\cdot\,, \bm{z}_{S''};\,\xi_{s'/s}, S', S''\right) & \preceq \overline{\mu^{[p]}_{C, 1'', d}}\left(\,\cdot\,, \bm{z}_{S''};\,\xi_{s'/s}, S', S''\right),
\end{align}
with majorant evaluated at all-$1$ variables bounded as follows:
\begin{align}
     \overline{\mu^{[p]}_{C, 1'', d}}\left(\bm{1}_{S' \times \mathcal{T}_{p - 1}}, \bm{z}_{S''} ;\,\xi_{s'/s}, S', S''\right) & \leq 2^{4p^2}\exp\left(4p^2\gamma^2s''R\xi''\left(1 + 2R\right)\right)\nonumber\\
     & \hspace*{20px} \times \left(\left(\frac{4}{R}\right)^d\binom{d + 4p^2 - 1}{4p^2 - 1} + \frac{1}{|S'|}4p^2(4p^2 + 1)\frac{\left(4/R\right)^2}{\left(1 - 4/R\right)^{4p^2 + 2}}\right).
\end{align}
\end{proof}
\end{lemma}

Given the decomposition of contribution $\mu^{[p]}_{C, 1}$ to the path integral measure introduced in Lemma~\ref{si:lemma:mu_c_1_constant_degree_polynomial_approximation}:
\begin{align}
    & \mu^{[p]}_{C, 1}\left(\bm{z}_{S'}, \bm{z}_{S''};\,\xi_{s'/s}, S', S''\right)\nonumber\\
    & = \mu^{[p]}_{C, SB}\left(\bm{z}_{S''};\,\xi_{s'/s}, S''\right)\left(1 + \mu^{[p]}_{C, 1', d}\left(\bm{z}_{S'}, \bm{z}_{S''};\,\xi_{s'/s}, S', S''\right) + \mu^{[p]}_{C, 1'', d}\left(\bm{z}_{S'}, \bm{z}_{S''};\,\xi_{s'/s}, S', S''\right)\right),
\end{align}
and the decomposition of $\mu^{[p]}_{C, 1', d}$ from coefficients $\mu^{[p]}_{C, 1', \bm{t}, \bm{u}}$:
\begin{align}
    \mu^{[p]}_{C, 1', d}\left(\bm{w}_{S'}, \bm{z}_{S''};\,\xi_{s'/s}, S', S''\right) & = \sum_{\substack{J' \subset S'\\\left|J'\right| < d\\\bm{t}, \bm{u} \in \mathcal{T}_{p - 1}^{J'}}}\mu^{[p]}_{C, 1', \bm{t}, \bm{u}}\left(\bm{z}_{S''};\,\xi_{s'/s}, S', S''\right)\prod_{j \in J'}\left(w_j^{[t_j]}w_j^{[u_j]} - G_{t_j, u_j}\right),
\end{align}
we bound the sum of absolute values of these coefficients in the following Lemma:

\begin{lemma}[Bound on absolute series of $\mu^{[p]}_{C, 1', d}$ expansion coefficients]
\label{si:lemma:mu_c_1p_d_coefficients_series_bound}
Consider expansion coefficients $\left(\mu^{[p]}_{C, 1', \bm{t}, \bm{u}}\left(\bm{z}_{S''};\,\xi_{s'/s}, S', S''\right)\right)_{J' \subset S', \bm{t}, \bm{u} \in \mathcal{T}_{p - 1}^{\left|J'\right|}}$ defined in Lemma~\ref{si:lemma:mu_c_1_constant_degree_polynomial_approximation}. For any $\rho > 0$ and parameter $R > \max(4, \rho)$, the following bound holds on the series of these coefficients:
\begin{align}
    \sum_{\substack{J' \subset S'\\\bm{t}, \bm{u} \in \mathcal{T}_{p - 1}^{J'}}}\rho^{\left|J'\right|}\left|\mu^{[p]}_{C, 1', \bm{t}, \bm{u}}\left(\bm{z}_{S''};\,\xi_{s'/s}, S', S''\right)\right| & \leq \frac{\exp\left(4p^2\gamma^2R\xi''\left(1 + 2R\right)\left|S''\right|\right)}{\left(1 - \rho/R\right)^{4p^2}}.
\end{align}
\begin{proof}
We start with bound Eq.~\ref{si:eq:mu_c_1p_d_coefficients_definition} from Lemma~\ref{si:lemma:mu_c_1_constant_degree_polynomial_approximation} on $\left|\mu^{[p]}_{C, 1', \bm{t}, \bm{u}}\left(\bm{z}_{S''};\,\xi_{s'/s}, S', S''\right)\right|$:
\begin{align}
    \sum_{\substack{J' \subset S'\\\bm{t}, \bm{u} \in \mathcal{T}_{p - 1}^{J'}}}\rho^{\left|J'\right|}\left|\mu^{[p]}_{C, 1', \bm{t}, \bm{u}}\left(\bm{z}_{S''};\,\xi_{s'/s}, S', S''\right)\right| & \leq \sum_{\substack{J' \subset S'\\\bm{t}, \bm{u} \in \mathcal{T}_{p - 1}^{J'}}}\rho^{\left|J'\right|}\frac{\exp\left(4p^2\gamma^2s''R\xi''\left(1 + 2R\right)\right)}{\left(\left|S'\right|R\right)^{\left|J'\right|}}\prod_{t, u \in \mathcal{T}_{p - 1}}m_{t, u}!,
\end{align}
where $m_{t, u}$ is an implicit function of $J'$-labeled tuples $\bm{t}, \bm{u}$, counting the multiplicity of pair $(t, u)$ in these joint tuples. Observing the above summand depends only on the size of $J'$, one may sum over this size denoted by $m$, counting $\binom{\left|S'\right|}{m}$ sets $J'$ of size $m$. Given any such set, one may replace the summation over $\bm{t}, \bm{u}$ by summation over multiplicities $\bm{m} := \left(m_{t, u}\right)_{t, u \in \mathcal{T}_{p - 1}}$ summing to $\left|\bm{m}\right| = \sum_{t, u}m_{t, u} = \left|J'\right| = m$; we then count $\binom{m}{\bm{m}}$ tuples for any given multiplicities $\bm{m}$. All in all:
\begin{align}
    \sum_{\substack{J' \subset S'\\\bm{t}, \bm{u} \in \mathcal{T}_{p - 1}^{J'}}}\rho^{\left|J'\right|}\left|\mu^{[p]}_{C, 1', \bm{t}, \bm{u}}\left(\bm{z}_{S''};\,\xi_{s'/s}, S', S''\right)\right| & \leq \sum_{m \geq 1}\binom{\left|S'\right|}{m}\rho^m\frac{\exp\left(4p^2\gamma^2s''R\xi''\left(1 + 2R\right)\right)}{\left(\left|S'\right|R\right)^m}\sum_{\substack{\bm{m} \in \mathbb{N}^{\mathcal{T}_{p - 1} \times \mathcal{T}_{p - 1}}\\\left|\bm{m}\right| = m}}\binom{m}{\bm{m}}\prod_{t, u \in \mathcal{T}_{p - 1}}m_{t, u}!\nonumber\\
    & = \sum_{m \geq 1}\binom{\left|S'\right|}{m}\rho^m\frac{\exp\left(4p^2\gamma^2s''R\xi''\left(1 + 2R\right)\right)}{\left(\left|S'\right|R\right)^m}\sum_{\substack{\bm{m} \in \mathbb{N}^{\mathcal{T}_{p - 1} \times \mathcal{T}_{p - 1}}\\\left|\bm{m}\right| = m}}m!\nonumber\\
    & \leq \sum_{m \geq 1}\rho^m\frac{\exp\left(4p^2\gamma^2s''R\xi''\left(1 + 2R\right)\right)}{R^m}\binom{m + 4p^2 - 1}{4p^2 - 1}\nonumber\\
    & \leq \frac{\exp\left(4p^2\gamma^2s''R\xi''\left(1 + 2R\right)\right)}{\left(1 - \rho/R\right)^{4p^2}}.
\end{align}
\end{proof}
\end{lemma}

Lemma~\ref{si:lemma:mu_c_1_constant_degree_polynomial_approximation} proven above provided a polynomial approximation of multiplicative contribution $\mu^{[p]}_{C, 1}\left(\bm{z}_{S'}, \bm{z}_{S''};\,\xi_{s'/s}, S', S''\right)$ in the cavity decomposition of the path integral measure. The error term in this decomposition is meant to be small in the infinite-size limit. However, we will also need a looser estimate on the same quantity, in the form of a mere majorization bound:

\begin{lemma}[Majorant of $\mu_{C, 1}$]
\label{si:lemma:mu_c_1_majorization_bound}
Consider complex-valued function of bit matrices $\mu^{[p]}_{C, 1}$ defined in Lemma~\ref{si:lemma:cavity_decomposition_path_integral_measure}, Eq.~\ref{si:eq:cavity_decomposition_path_integral_contribution_1}, and consider its extension to matrices of complex numbers:
\begin{align}
    \mu^{[p]}_{C, 1}\left(\,\cdot\,, \bm{z}_{S''};\,\xi_{s'/s}, S', S''\right): \left\{\begin{array}{rcl}
         \mathbb{C}^{S' \times \mathcal{T}_{p - 1}} & \longrightarrow & \mathbb{C}\\
         \bm{w}_{S'} & \longmapsto & \exp\left(-\frac{1}{2}\sum\limits_{t, u \in \mathcal{T}_{p - 1}}\Gamma_t\Gamma_u\xi'_{s'/s}\left(\frac{1}{\left|S'\right|}\left\langle \bm{w}^{[t]}_{S'}, \bm{w}^{[u]}_{S'} \right\rangle\right)\left\langle \bm{z}^{[t]}_{S''}, \bm{z}^{[u]}_{S''} \right\rangle\right)
    \end{array}\right.
\end{align}
This function admits the following majorant:
\begin{align}
    \overline{\mu^{[p]}_{C, 1}}\left(\,\cdot\,;\,\xi_{s'/s}, S', S''\right): \left\{\begin{array}{rcl}
         \mathbb{C}^{S' \times \mathcal{T}_{p - 1}} & \longrightarrow & \mathbb{C}\\
         \bm{w}_{S'} & \longmapsto & \exp\left(\frac{\gamma^2}{2}\sum\limits_{t, u \in \mathcal{T}_{p - 1}}\xi'_{s'/s}\left(\frac{1}{\left|S'\right|}\left\langle \bm{w}^{[t]}_{S'}, \bm{w}^{[u]}_{S'} \right\rangle\right)\left|S''\right|\right)
    \end{array}\right.,
\end{align}
uniformly in bit matrix $\bm{z}_{S''} \in \{1, -1\}^{S'' \times \mathcal{T}_{p - 1}}$. Evaluated at the all-$1$ variables, this majorant can be bounded as:
\begin{align}
    \overline{\mu^{[p]}_{C, 1}}\left(\bm{1}_{S' \times \mathcal{T}_{p - 1}};\,\xi_{s'/s}, S', S''\right) & \leq \exp\left(2p^2\gamma^2\xi'\left(\frac{\left|S'\right|}{\left|S\right|}\right)\left|S''\right|\right)\nonumber\\
    & \leq \exp\left(2p^2\gamma^2\xi'\left(1\right)\left|S''\right|\right).
\end{align}
\begin{proof}
This results from the following estimates, where $\bm{z}_{S''} \in \{1, -1\}^{S'' \times \mathcal{T}_{p - 1}}$ is a fixed bit matrix and one reasons over entire functions of complex variables $\bm{w}_{S'} \in \mathbb{C}^{S' \times \mathcal{T}_{p - 1}}$:
\begin{align}
    \mu^{[p]}_{C, 1}\left(\bm{w}_{S'}, \bm{z}_{S''};\,\xi_{s'/s}, S', S''\right) & \preceq \exp\left(\frac{1}{2}\sum_{t, u \in \mathcal{T}_{p - 1}}\left|\Gamma_t\right|\left|\Gamma_u\right|\xi'_{s'/s}\left(\frac{1}{\left|S'\right|}\left\langle \bm{w}^{[t]}_{S'}, \bm{w}^{[u]}_{S'} \right\rangle\right)\left|\left\langle \bm{z}^{[t]}_{S''}, \bm{z}^{[u]}_{S''} \right\rangle\right|\right)\nonumber\\
    & \preceq \exp\left(\frac{1}{2}\sum_{t, u \in \mathcal{T}_{p - 1}}\gamma^2\xi'_{s'/s}\left(\frac{1}{\left|S'\right|}\left\langle \bm{w}^{[t]}_{S'}, \bm{w}^{[u]}_{S'} \right\rangle\right)\left|S''\right|\right).
\end{align}
The bounds at the all-$1$ variables follows from plugging in definition $\xi_{s'/s}\left(x\right) := \frac{s}{s'}\xi\left(\frac{s'}{s}x\right)$:
\begin{align}
    \overline{\mu^{[p]}_{C, 1}}\left(\bm{1}_{S' \times \mathcal{T}_{p - 1}}, \bm{z}_{S''};\,\xi_{s'/s}, S', S''\right) & \leq \exp\left(2p^2\gamma^2\xi'_{s'/s}\left(1\right)\left|S''\right|\right)\nonumber\\
    & = \exp\left(2p^2\gamma^2\xi'\left(\frac{\left|S'\right|}{|S|}\right)\left|S''\right|\right)\nonumber\\
    & \leq \exp\left(2p^2\gamma^2\xi'(1)\left|S''\right|\right).
\end{align}
\end{proof}
\end{lemma}

This concludes the estimation of multiplicative contributions $\mu^{[p]}_{C, 1}\left(\bm{z}_{S'}, \bm{z}_{S''};\,\xi_{s'/s}, S', S''\right)$ and $\mu^{[p]}_{C, 2}\left(\bm{z}_{S'}, \bm{z}_{S''};\,\xi, S', S''\right)$ in the cavity decomposition of the path integral measure Lemma~\ref{si:lemma:cavity_decomposition_path_integral_measure}. The following Lemma provides the third main majorization estimate required in the proof of recursion result Lemma~\ref{si:lemma:centered_moments_maximum_recursion}, by majorizing the relative density of two spin-boson path integral measures. Intuitively, the change of spin-boson path integral measure is caused by the shift of mixture polynomial in the cavity decomposition of the path integral measure: if the full measure involves polynomial $\xi$, then multiplicative contribution $\mu^{[p]}_{C, 1}$ involves the shifted polynomial $\xi_{s'/s}$, which is associated to a different $\bm{G}$ matrix and hence to a distinct spin-boson measure.

\begin{lemma}[Majorant for ratio of spin-boson path integral measures]
\label{si:lemma:spin_boson_measures_ratio_majorization_bound}
Consider two discrete $p$-layers spin-boson evolutions formed from combining $|S|$ independent spin-boson evolutions with spins labeled by $S$ and $p$ discrete layers. Denote by $\bm{K}, \bm{\widetilde{K}} \in \mathbb{C}^{\mathcal{T}_{p - 1} \times \mathcal{T}_{p - 1}}$ the influence functionals of the two spin-boson evolutions. The path integral measure of the spins-bosons evolution with spins labeled by $S$ is given by:
\begin{align}
    \mu^{[p]}_{SB}\left(\bm{z};\,S\right) & = \mu_B\left(\bm{z};\,S\right)\mu_{C, SB}\left(\bm{z};\,S\right),\\
    \widetilde{\mu}^{[p]}_{SB}\left(\bm{z};\,S\right) & = \mu_B\left(\bm{z};\,S\right)\widetilde{\mu}_{C, SB}\left(\bm{z};\,S\right).
\end{align}
In the above expressions, $\bm{z} = \bm{z}_S\in \{1, -1\}^{S \times \mathcal{T}_p}$ is a bit matrix, and $\mu_B$, common to the two spin-bosons systems, is defined as for the QAOA circuit:
\begin{align}
    \mu^{[p]}_B\left(\bm{z};\,S\right) & := \prod_{j \in S}f\left(\bm{z}_j\right),\\
    f\left(\bm{a}\right) & := \frac{1}{2}\prod_{1 \leq t \leq p}\bra{a_t}e^{i\beta_tX}\ket{a_{t + 1}}\bra{a_{-t - 1}}e^{-i\beta_tX}\ket{a_{-t}},
\end{align}
and reflects the contribution of $X$ rotations. The contribution of displacement operators is reflected in:
\begin{align}
    \mu^{[p]}_{C, SB}\left(\bm{z};\,S\right) & := \exp\left(-\frac{1}{2}\sum_{t, u \in \mathcal{T}_{p - 1}}K_{t, u}\Gamma_t\Gamma_u\left\langle \bm{z}^{[t]}_S, \bm{z}^{[u]}_S \right\rangle\right),\\
    \widetilde{\mu}^{[p]}_{C, SB}\left(\bm{z};\,S\right) & := \exp\left(-\frac{1}{2}\sum_{t, u \in \mathcal{T}_{p - 1}}\widetilde{K}_{t, u}\Gamma_t\Gamma_u\left\langle \bm{z}^{[t]}_S, \bm{z}^{[u]}_S \right\rangle\right).
\end{align}
Extending these cost contributions to matrices of complex numbers $\bm{w} \in \mathbb{C}^{S \times \mathcal{T}_p}$, the following majorization bound holds on their ratio:
\begin{align}
    \frac{\widetilde{\mu}^{[p]}_{SB}\left(\bm{w};\,S\right)}{\mu^{[p]}_{SB}\left(\bm{w};\,S\right)} - 1 & \preceq \frac{\gamma^2}{2}\sum_{t, u \in \mathcal{T}_{p - 1}}\left|\widetilde{K}_{t, u} - K_{t, u}\right|\left\langle \bm{w}_S^{[t]}, \bm{w}_S^{[u]} \right\rangle\exp\left(\frac{\gamma^2}{2}\sum_{t, u \in \mathcal{T}_{p - 1}}\left|\widetilde{K}_{t, u} - K_{t, u}\right|\left\langle \bm{w}^{[t]}_S, \bm{w}^{[u]}_S \right\rangle\right).
\end{align}
\begin{proof}
The ratio of spin-boson measures reduces to the ratio of their cost contributions, and we compute:
\begin{align}
    \frac{\widetilde{\mu}^{[p]}_{SB}\left(\bm{w}_S;\,S\right)}{\mu^{[p]}_{SB}\left(\bm{w}_S;\,S\right)} & = \frac{\exp\left(-\frac{1}{2}\sum\limits_{t, u \in \mathcal{T}_{p - 1}}\widetilde{K}_{t, u}\Gamma_t\Gamma_u\left\langle \bm{w}^{[t]}_S, \bm{w}^{[u]}_S \right\rangle\right)}{\exp\left(-\frac{1}{2}\sum\limits_{t, u \in \mathcal{T}_{p - 1}}K_{t, u}\Gamma_t\Gamma_u\left\langle \bm{w}^{[t]}_S, \bm{w}^{[u]}_S \right\rangle\right)}\nonumber\\
    & = \exp\left(-\frac{1}{2}\sum_{t, u \in \mathcal{T}_{p - 1}}\left(\widetilde{K}_{t, u} - K_{t, u}\right)\Gamma_t\Gamma_u\left\langle \bm{w}^{[t]}_S, \bm{w}^{[u]}_S \right\rangle\right).
\end{align}
From majorization bound
\begin{align}
    e^z - 1 & \preceq ze^z,
\end{align}
and standard composition of majorization bounds:
\begin{align}
    \frac{\widetilde{\mu}_{SB}^{[p]}\left(\bm{w}_S;\,S\right)}{\mu_{SB}^{[p]}\left(\bm{w}_S;\,S\right)} - 1 & \preceq \frac{1}{2}\sum_{t, u \in \mathcal{T}_{p - 1}}\left|\widetilde{K}_{t, u} - K_{t, u}\right|\left|\Gamma_t\right|\left|\Gamma_u\right|\left\langle \bm{w}^{[t]}_S, \bm{w}^{[u]}_S \right\rangle\exp\left(\frac{1}{2}\sum_{t, u \in \mathcal{T}_{p - 1}}\left|\widetilde{K}_{t, u} - K_{t, u}\right|\left|\Gamma_t\right|\left|\Gamma_u\right|\left\langle \bm{w}^{[t]}_S, \bm{w}^{[u]}_S \right\rangle\right)\nonumber\\
    & \preceq \frac{\gamma^2}{2}\sum_{t, u \in \mathcal{T}_{p - 1}}\left|\widetilde{K}_{t, u} - K_{t, u}\right|\left\langle \bm{w}^{[t]}_S, \bm{w}^{[u]}_S \right\rangle\exp\left(\frac{\gamma^2}{2}\sum_{t, u \in \mathcal{T}_{p - 1}}\left|\widetilde{K}_{t, u} - K_{t, u}\right|\left\langle \bm{w}^{[t]}_S, \bm{w}^{[u]}_S \right\rangle\right).
\end{align}
\end{proof}
\end{lemma}

Finally, the following simple Lemma bounds the absolute mass of the mixer path integral measure:

\begin{lemma}[Bounding total weight of $\left|\mu^{[p]}_B\right|$]
\label{si:lemma:path_integral_measure_mixer_contribution_absolute_weight_bound}
Let $S$ a qubit set and $p \geq 0$ a number of QAOA layers and consider the mixer contribution to the path integral measure $\mu_B\left(\bm{z}_S;\,S\right)$ defined in Eq.~\ref{si:eq:path_integral_measure_mixer_contribution}. Then, the total weight of the absolute value of this measure is bounded as:
\begin{align}
    \sum_{\bm{z}_S \in \{1, -1\}^{S \times \mathcal{T}_p}}\left|\mu_B\left(\bm{z}_S;\,S\right)\right| & \leq 2^{p\left|S\right|},
\end{align}
uniformly in mixer angles $\bm{\beta} \in \mathbb{R}^p$.
\begin{proof}
\begin{align}
    \sum_{\bm{z}_S \in \{1, -1\}^{S \times \mathcal{T}_p}}\left|\mu^{[p]}_B\left(\bm{z}_S;\,S\right)\right| & = \sum_{\bm{z}_S \in \{1, -1\}^{S \times \mathcal{T}_p}}\left|\prod_{j \in S}f\left(\bm{z}_j\right)\right|\nonumber\\
    & = \sum_{\bm{z}_S \in \{1, -1\}^{S \times \mathcal{T}_p}}\prod_{j \in S}\left|f\left(\bm{z}_j\right)\right|,
\end{align}
where
\begin{align}
    f\left(\bm{a}\right) & := \frac{1}{2}\bm{1}\left[a_{p + 1} = a_{-p - 1}\right]\prod_{1 \leq t \leq p}\bra{a_t}e^{i\beta_tX}\ket{a_{t + 1}}\bra{a_{-t - 1}}e^{-i\beta_tX}\ket{a_{-t}},\\
    \bm{a} & \in \{1, -1\}^{\mathcal{T}_p}.
\end{align}
Using independence over $j \in S$,
\begin{align}
    \sum_{\bm{z}_S \in \{1, -1\}^{S \times \mathcal{T}_p}}\left|\mu^{[p]}_B\left(\bm{z}_S;\,S\right)\right| & = \prod_{j \in S}\sum_{\bm{z}_j \in \{1, -1\}^{\mathcal{T}_p}}\left|f\left(\bm{z}_j\right)\right|.
\end{align}
Now, expressing bitstring summation as a matrix product (of matrices indexed by bits):
\begin{align}
    \sum_{\bm{a} \in \{1, -1\}^{\mathcal{T}_p}}\left|f(\bm{a})\right| & = \frac{1}{2}\begin{pmatrix}
        1\\
        1
    \end{pmatrix}^T\overrightarrow{\prod_{t = 1}^p}\begin{pmatrix}
        \left|\cos\beta_t\right| & \left|\sin\beta_t\right|\\
        \left|\sin\beta_t\right| & \left|\cos\beta_t\right|
    \end{pmatrix}\overleftarrow{\prod_{t = 1}^p}\begin{pmatrix}
        \left|\cos\beta_t\right| & \left|\sin\beta_t\right|\\
        \left|\sin\beta_t\right| & \left|\cos\beta_t\right|
    \end{pmatrix}\begin{pmatrix}
        1\\
        1
    \end{pmatrix}\nonumber\\
    & = \frac{1}{2}2\prod_{1 \leq t \leq p}\left(\left|\cos\beta_t\right| + \left|\sin\beta_t\right|\right)^2\nonumber\\
    & \leq \frac{1}{2}2\prod_{1 \leq t \leq p}2\nonumber\\
    & \leq 2^p,
\end{align}
where we used bound $\left|\cos\beta\right| + \left|\sin\beta\right| \leq \sqrt{2}$, uniform in $\beta \in \mathbb{R}$. It follows
\begin{align}
     \sum_{\bm{z}_S \in \{1, -1\}^{S \times \mathcal{T}_p}}\left|\mu^{[p]}_B\left(\bm{z}_S;\,S\right)\right| & \leq 2^{p\left|S\right|}.
\end{align}
\end{proof}
\end{lemma}

\subsubsection{Perturbation of \texorpdfstring{$G$}{G} matrix}
\label{si:sec:perturbation_g_matrix}

In this Section, we establish bounds on the variation of the $\bm{G}$ matrix when mixture polynomial $\xi$ varies. Recall from Section~\ref{si:sec:spin_glass_qaoa_g_iteration} that a mixture polynomial is associated to a $\bm{G}$ matrix iteration Eq.~\ref{si:eq:g_iteration}, which is equivalent to a fixed-point equation Eq.~\ref{si:eq:g_fixed_point_equation}. The resulting $\bm{G}$ matrix in turn provides the infinite-size energy density of QAOA on the corresponding spin glass.

Specifically, we consider the $\bm{G}$ matrices $\bm{G}^0$ and $\bm{G}^1$ generated by the $\bm{G}$ iterations defined by two spin glass mixture polynomials $\xi_0, \xi_1$. Explicitly, these matrices are defined recursively for all $\left(t, u\right) \in \mathcal{T}_p \times \mathcal{T}_p$ in $\max\left\{t, u\right\}$ as follows:
\begin{align}
    G^h_{t, u} & := 1 \qquad \textrm{for } |t| = |u| = 1,\\
    G^h_{t, u} & := \sum_{\bm{a} \in \{1, -1\}^{\mathcal{T}_{q - 1}}}a_ta_uf\left(\bm{a}\right)\exp\left(-\frac{1}{2}\sum_{v, w \in \mathcal{T}_{q - 2}}\xi_h'\left(G^h_{v, w}\right)\Gamma_v\Gamma_wa_va_w\right) \qquad \textrm{for } \max\left\{|t|, |u|\right\} = q,\label{si:eq:g_iteration_parametrized_xi}
\end{align}
where $h \in \{0, 1\}$ indexes the relevant $\xi$ polynomial and its corresponding iteratively computed $\bm{G}$ matrix. It can be shown that $\bm{G}^h$ defined recursively in this way satisfies fixed point equation
\begin{align}
    G^h_{t, u} & = \sum_{\bm{a} \in \{1, -1\}^{\mathcal{T}_{q - 1}}}a_ta_uf\left(\bm{a}\right)\exp\left(-\frac{1}{2}\sum_{v, w \in \mathcal{T}_{q - 2}}\xi_h'\left(G^h_{v, w}\right)\Gamma_v\Gamma_wa_va_w\right) \qquad \textrm{for } \max\left\{|t|, |u|\right\} \leq q.\label{si:eq:g_fixed_point_equation_parametrized_xi}
\end{align}
for all layer index $q \in [p + 1]$ (note the inequality on layer indices of $\bm{G}^h$, stronger than the equality from Eq.~\ref{si:eq:g_iteration_parametrized_xi}).
Reinterpreting the above in terms of spin-boson path integral measure, recursion equation Eq.~\ref{si:eq:g_iteration_parametrized_xi} can be rewritten
\begin{align}
    G^h_{t, u} & = \left\langle z_1^{[t]}z_1^{[u]} \right\rangle_{\mu^{[q - 1],h}_{SB}\left(\bm{z}_1;\,\{1\}\right)},\label{si:eq:g_iteration_parametrized_xi_spin_boson_expectation_rephrasing}
\end{align}
where the spin-boson system is over a single spin $1$, and $\mu^{[q - 1],h}_{SB}\left(\,\cdot\,;\,\{1\}\right)$ is the spin path integral measure of the $q$-layers spin-boson system characterized by influence functional $\left(\xi'_h\left(G^h_{v, w}\right)\right)_{v, w \in \mathcal{T}_{q - 2}} \in \mathbb{C}^{\mathcal{T}_{q - 2} \times \mathcal{T}_{q - 2}}$.

\begin{proposition}[$\bm{G}$ matrix perturbation from perturbation of spin glass mixture polynomial]
\label{si:prop:g_matrix_variation_bound_parametrized_xi}
Given the expansion coefficients $\left(c_{h, m}\right)_{m \geq 1} \in \mathbb{R}_+^{\mathbb{N}^*}$ of $\xi'_0, \xi'_1$ in the canonical basis:
\begin{align}
    \xi'_h\left(x\right) & =: \sum_{m \geq 1}mc_{h, m}x^{m - 1},
\end{align}
define
\begin{align}
    \delta_{\xi} & := \left\lVert \xi'_1 - \xi'_0 \right\rVert_1\\
    & := \sum_{m \geq 1}m\left|c_{1, m} - c_{0, m}\right|
\end{align}
as the $1$-norm distance between coefficient vectors. For all $0 \leq q \leq p$, let
\begin{align}
    \delta^{[q]} & := \max_{t, u \in \mathcal{T}_q}\left|G^1_{t, u} - G^0_{t, u}\right|
\end{align}
the maximum variation between $\bm{G}^1$ and $\bm{G}^0$ over entries in $\mathcal{T}_q$. Let $x_1''$ be an upper-bound on $\xi_1''(1)$. Then, the following recursive bound holds on sequence $\left(\delta^{[q]}\right)_{q \geq 0}$:
\begin{align}
    \delta^{[0]} & = 0,\\
    \delta^{[q + 1]} & \leq 2\gamma^2p^2\left(x_1''\delta^{[q]} + \delta_{\xi}\right)\exp\left(2\gamma^2p^2\left(x_1''\delta^{[q]} + \delta_{\xi}\right)\right), \qquad \forall 0 \leq q \leq p - 1.
\end{align}
\begin{proof}
We prove the bound inductively in $q$.

For $q = 0$, this is trivial since $G^0_{t, u} = G^1_{t, u} = 1$ as long as $|t| = |u| = 1$.

Let us assume the bound proven up to level $(q - 1)$ inclusive and let $t, u \in \mathcal{T}_q$ such that $\max\{|t|, |u|\} \leq q + 1$. Using interpretation Eq.~\ref{si:eq:g_iteration_parametrized_xi_spin_boson_expectation_rephrasing} of the $\bm{G}$ fixed-point equation Eq.~\ref{si:eq:g_fixed_point_equation_parametrized_xi} as a spin-boson expectation, and performing a change of path integral measure $\mu^{[q],1}_{SB}\left(\,\cdot\,;\,\{1\}\right) \longrightarrow \mu^{[q],0}_{SB}\left(\,\cdot\,;\,\{1\}\right)$,
\begin{align}
    G^1_{t, u} - G^0_{t, u} & = \left\langle z_1^{[t]}z_1^{[u]} \right\rangle_{\mu_{SB}^{[q],1}\left(\bm{z}_1;\,\{1\}\right)} - \left\langle z_1^{[t]}z_1^{[u]} \right\rangle_{\mu_{SB}^{[q],0}\left(\bm{z}_1;\,\{1\}\right)}\nonumber\\
    & = \left\langle \frac{\mu_{SB}^{[q],1}\left(\bm{z}_1;\,\{1\}\right)}{\mu_{SB}^{[q],0}\left(\bm{z}_1;\,\{1\}\right)}z_1^{[t]}z_1^{[u]} \right\rangle_{\mu_{SB}^{[q],0}\left(\bm{z}_1;\,\{1\}\right)} - \left\langle z_1^{[t]}z_1^{[u]} \right\rangle_{\mu_{SB}^{[q],0}\left(\bm{z}_1;\,\{1\}\right)}\nonumber\\
    & = \left\langle \left(\frac{\mu_{SB}^{[q],1}\left(\bm{z}_1;\,\{1\}\right)}{\mu_{SB}^{[q],0}\left(\bm{z}_1;\,\{1\}\right)} - 1\right)z_1^{[t]}z_1^{[u]} \right\rangle_{\mu_{SB}^{[q],0}\left(\bm{z}_1;\,\{1\}\right)}\nonumber\\
    & =: \left\langle \psi\left(\bm{z}_1\right) \right\rangle_{\mu_{SB}^{[q],0}\left(\bm{z}_1;\,\{1\}\right)}.
\end{align}
By Lemma~\ref{si:lemma:spin_boson_measures_ratio_majorization_bound},
\begin{align}
    \psi\left(\bm{z}_1\right) & \preceq \overline{\psi}\left(\bm{z}_1\right),
\end{align}
where
\begin{align}
    \overline{\psi}\left(\bm{z}_1\right) & = \frac{\gamma^2}{2}\sum_{v, w \in \mathcal{T}_{q - 1}}\left|\xi_1'\left(G^1_{v, w}\right) - \xi'_0\left(G^0_{v, w}\right)\right|z_1^{[v]}z_1^{[w]}\exp\left(\frac{\gamma^2}{2}\sum_{v, w \in \mathcal{T}_{q - 1}}\left|\xi_1'\left(G^1_{v, w}\right) - \xi_0'\left(G^0_{v, w}\right)\right|z_1^{[v]}z_1^{[w]}\right).
\end{align}
By this majorization estimate, it holds:
\begin{align}
    \left|G^1_{t, u} - G^0_{t, u}\right| & \leq \left\langle \overline{\psi}\left(\bm{1}_{\{1\} \times \mathcal{T}_q}\right) \right\rangle_{\mu_{SB}^{[q],0}\left(\bm{z}_1;\,\{1\}\right)}.
\end{align}
To bound $\overline{\psi}\left(\bm{1}_{\{1\} \times \mathcal{T}_q}\right)$, we bound $\xi'_1\left(G^1_{v, w}\right) - \xi'_0\left(G^0_{v, w}\right)$ as follows:
\begin{align}
    \xi'_1\left(G^1_{v, w}\right) - \xi'_0\left(G^0_{v, w}\right) & \leq \left|\xi'_1\left(G^1_{v, w}\right) - \xi'_1\left(G^0_{v, w}\right)\right| + \left|\xi'_1\left(G^0_{v, w}\right) - \xi'_0\left(G^0_{v, w}\right)\right|\nonumber\\
    & \leq \xi_1''\left(1\right)\left|G^1_{v, w} - G^0_{v, w}\right| + \left\lVert \xi_1' - \xi_0' \right\rVert_1\nonumber\\
    & \leq \xi_1''(1)\delta^{[q - 1]} + \delta_{\xi}.
\end{align}
It follows:
\begin{align}
    \overline{\psi}\left(\bm{1}_{\{1\} \times \mathcal{T}_q}\right) & \leq \frac{\gamma^2}{2}\left|\mathcal{T}_{q - 1}\right|^2\left(\xi_1''(1)\delta^{[q - 1]} + \delta_{\xi}\right)\exp\left(\frac{\gamma^2}{2}\left|\mathcal{T}_{q - 1}\right|^2\left(\xi_1''(1)\delta^{[q - 1]} + \delta_{\xi}\right)\right)\nonumber\\
    & = 2\gamma^2q^2\left(\xi_1''(1)\delta^{[q - 1]} + \delta_{\xi}\right)\exp\left(2\gamma^2q^2\left(\xi_1''(1)\delta^{[q - 1]} + \delta_{\xi}\right)\right).
\end{align}
Hence,
\begin{align}
    \delta^{[q]} & \leq 2\gamma^2q^2\left(\xi_1''(1)\delta^{[q - 1]} + \delta_{\xi}\right)\exp\left(2\gamma^2q^2\left(\xi_1''(1)\delta^{[q - 1]} + \delta_{\xi}\right)\right)\nonumber\\
    & \leq 2\gamma^2p^2\left(\xi_1''(1)\delta^{[q - 1]} + \delta_{\xi}\right)\exp\left(2\gamma^2p^2\left(\xi_1''(1)\delta^{[q - 1]} + \delta_{\xi}\right)\right)\nonumber\\
    & \leq 2\gamma^2p^2\left(x_1''\delta^{[q - 1]} + \delta_{\xi}\right)\exp\left(2\gamma^2p^2\left(x_1''\delta^{[q - 1]} + \delta_{\xi}\right)\right)
\end{align}
proving the induction step and the result.
\end{proof}
\end{proposition}

\begin{corollary}[Simple bound on $\bm{G}$ matrix perturbation from perturbation of spin glass mixture polynomial]
\label{si:cor:simple_g_matrix_variation_bound_parametrized_xi}
Recall the setting and notation from Proposition~\ref{si:prop:g_matrix_variation_bound_parametrized_xi}. For simplicity, assume
\begin{align}
    2e\gamma^2p^2x_1''\exp\left(2\gamma^2p^2\delta_{\xi}\right) \geq 2,
\end{align}
increasing upper bound $\gamma$ (or for instance $\delta_{\xi}$) if necessary. Then, the perturbation of $\bm{G}$ entries between spin glass mixture polynomials $\xi_0$ and $\xi_1$ is uniformly bounded as
\begin{align}
    \left|G^1_{t, u} - G^0_{t, u}\right| & \leq 2e\left(2e\gamma^2p^2\right)^p\left(x_1''\right)^{p - 1}\delta_{\xi}, \qquad \forall t, u \in \mathcal{T}_p.
\end{align}
\begin{proof}
By Proposition~\ref{si:prop:g_matrix_variation_bound_parametrized_xi}, $\left(\delta^{[q]}\right)_{q \geq 0}$ satisfies
\begin{align}
    \delta^{[0]} & := 0,\\
    \delta^{[q + 1]} & \leq \left(a\delta^{[q]} + b\right)e^{c\delta^{[q]}}, \qquad \forall 0 \leq q \leq p - 1.
\end{align}
where
\begin{align}
    a & := 2\gamma^2p^2x_1''\exp\left(2\gamma^2p^2\delta_{\xi}\right),\\
    b & := 2\gamma^2p^2\delta_{\xi}\exp\left(2\gamma^2p^2\delta_{\xi}\right),\\
    c & := 2\gamma^2p^2x_1''.
\end{align}
Defining $\overline{a} := \max\{2, ea\}$ by assumption, Lemma~\ref{si:lemma:exponentials_tower_bound} then yields
\begin{align}
    \delta^{[q]} & \leq 2eb\overline{a}^{q - 1} \qquad \forall 0 \leq q \leq p,
\end{align}
as long as
\begin{align}
    -\frac{\log\left(2ebc/\overline{a}\right)}{\log \overline{a}} \geq p,
\end{align}
i.e.
\begin{align}
    b \leq \frac{1}{2ec\overline{a}^{p - 1}}.
\end{align}
We now observe that $ea = 2e\gamma^2p^2x_1''\exp\left(2\gamma^2p^2\delta_{\xi}\right) \geq 2$ by assumption,
hence $\overline{a} = \max\{2, ea\} = ea$, and the last inequality constraint becomes:
\begin{align}
    b & \leq \frac{1}{2ec(ea)^{p - 1}}.
\end{align}
Substituting explicit values of $a$, $c$, this is equivalent to
\begin{align}
    & 2\gamma^2p^2\delta_{\xi}\exp\left(2\gamma^2p^2\delta_{\xi}\right) \leq \frac{1}{2\left(2e\gamma^2p^2x_1''\right)^p\exp\left(2\gamma^2p^2(p - 1)\delta_{\xi}\right)}\nonumber\\
    & \iff \delta_{\xi}\exp\left(2\gamma^2p^3\delta_{\xi}\right) \leq \frac{e}{2\left(2e\gamma^2p^2\right)^{p + 1}\left(x_1''\right)^p}.
\end{align}
which may be satisfied by taking:
\begin{align}
    \delta_{\xi} & \leq \min\left\{\frac{1}{2\gamma^2p^3}, \frac{1}{2\left(2e\gamma^2p^2\right)^{p + 1}\left(x_1''\right)^p}\right\}
\end{align}
\end{proof}
\end{corollary}

\section{Technical tools}

This Section collects technical results used repeatedly in the rigorous version of the cavity analysis of Section~\ref{si:sec:qaoa_cavity_method_analysis}. Section~\ref{si:sec:majorization} dicusses majorization in several-complex-variable calculus. Majorization provides a convenient framework for bounding the error between two functions, since unlike pointwise bounds it is compatible not only with genuine probabilty measures but also with spin path integrals. Section~\ref{si:sec:calculus} collects miscellaneous calculus results, including bounds of complex-variable functions and combinatorial estimates.

\subsection{Majorization}
\label{si:sec:majorization}

In this Section, we introduce majorization, the main technical tool used in this work to quantify errors between functions under spin path integrals (as introduced in Section~\ref{si:sec:unitary_circuit_path_integrals_and_spin_boson_systems}, specifically Definition~\ref{si:def:layered_quantum_circuit_path_integral_measure}). We start with a definition of majorization:

\begin{definition}[Majorization]
\label{si:def:majorization}
Let $f, g: \mathbb{C}^d \longrightarrow \mathbb{C}$ two complex-valued analytic functions of $d$ variables. Given the power series of these two functions:
\begin{align}
    f\left(\bm{w}\right) & = \sum_{m_1, \ldots, m_d \geq 0}f_{m_1, \ldots, m_d}w_1^{m_1}\ldots w_d^{m_d},\\
    g\left(\bm{w}\right) & = \sum_{m_1, \ldots, m_d \geq 0}g_{m_1, \ldots, m_d}w_1^{m_1}\ldots w_d^{m_d},
\end{align}
$f$ is said to be majorized by $g$ is for all $\left(m_1, \ldots, m_d\right) \in \mathbb{N}^d$,
\begin{itemize}
    \item $g_{m_1, \ldots, m_d} \geq 0$,\\
    \item $\left|f_{m_1, \ldots, m_d}\right| \leq g_{m_1, \ldots, m_d}$.
\end{itemize}
\end{definition}

\begin{proposition}[Majorization is transitive]
Let $f: \mathbb{C}^d \longrightarrow \mathbb{C}$ be majorized by $g$, and $g$ be majorized by $h$. Then, $f$ is majorized by $h$.
\end{proposition}

\begin{proposition}[Majorization of monomials]
\label{si:prop:majorization_monomials}
For all dimension $d \geq 0$, complex number $\alpha \in \mathbb{C}$ and natural integer tuple $\left(m_1, \ldots, m_d\right) \in \mathbb{N}^d$, the monomial function:
\begin{align}
    f\left(\bm{z}\right) & := \alpha z_1^{m_1} \ldots z_d^{m_d}
\end{align}
is majorized by:
\begin{align}
    f\left(\bm{z}\right) & := |\alpha| z_1^{m_1} \ldots z_d^{m_d}.
\end{align}
\end{proposition}

\begin{proposition}[Majorization and sum]
\label{si:prop:majorization_sum}
Let $f^1, \ldots, f^n$ analytic functions $\mathbb{C}^d \longrightarrow \mathbb{C}$ respectively majorized by $\overline{f^1}, \ldots, \overline{f^n}$. Then, $f^1 + \ldots + f^n$ is majorized by $\overline{f^1} + \ldots + \overline{f^n}$.
\begin{proof}
Let be given the power series expansions of individual functions and their majorizing counterparts:
\begin{align}
    f^j\left(\bm{w}\right) &:= \sum_{m_1, \ldots, m_d \geq 0}f^j_{m_1, \ldots, m_d}w_1^{m_1}\ldots w_d^{m_d},\\
    \overline{f^j}\left(\bm{w}\right) & = \sum_{m_1, \ldots, m_d \geq 0}\overline{f^j}_{m_1, \ldots, m_d}w_1^{m_1} \ldots w_d^{m_d}.
\end{align}
Then, the power series of sum $f^1$ is given by:
\begin{align}
    \sum_{1 \leq j \leq n}f^j\left(\bm{w}\right) & = \sum_{m_1, \ldots, m_d \geq 0}\left(\sum_{1 \leq j \leq n}f^j_{m_1, \ldots, m_d}\right)w_1^{m_1} \ldots w_d^{m_d}.
\end{align}
The general coefficient of this series is bounded as:
\begin{align}
    \left|\sum_{1 \leq j \leq n}f^j_{m_1, \ldots, m_d}\right| & \leq \sum_{1 \leq j \leq n}\left|f^j_{m_1, \ldots, m_d}\right|\nonumber\\
    & \leq \sum_{1 \leq j \leq n}\overline{f^j}_{m_1, \ldots, m_d}.
\end{align}
The latter bound is non-negative by definition of majorization, and corresponds to the general power series coefficient of $\overline{f^1} + \ldots + \overline{f^n}$. This completes the proof.
\end{proof}
\end{proposition}

\begin{proposition}[Majorization and composition]
\label{si:prop:majorization_composition}
Let $R_f, R_g \in \left(0, +\infty\right]$ and let $f: B\left(0, R_f\right)^d \longrightarrow \mathbb{C}^{d'}$ and $g: B\left(0, R_g\right)^{d'} \longrightarrow \mathbb{C}$ be two analytic functions of $d, d'$ variables respectively, on open polydisks about the origin (the value $+\infty$ recovering entire functions). For $f = \left(f^1, \ldots, f^{d'}\right)$, $f^j: B\left(0, R_f\right)^d \longrightarrow \mathbb{C}$, this means that each of the $d'$ coordinate functions $f^j$ is analytic. Let $f, g$ be majorized by $\overline{f}: B\left(0, R_f\right)^d \longrightarrow \mathbb{C}^{d'}$ and $\overline{g}: B\left(0, R_g\right)^{d'} \longrightarrow \mathbb{C}$; again, for $f$, the statement is meant coordinate-wise. We further assume the majorant composition $\overline{g} \circ \overline{f}$ admits an absolutely convergent power series in a neighbourhood of $\bm{0}$; equivalently, since majorants have non-negative coefficients, that $\left(\overline{f}^j\left(r, \ldots, r\right)\right)_{1 \leq j \leq d'}$ lies within the polydisk of convergence of $\overline{g}$ for some $r > 0$. Then, $g \circ f$ is majorized by $\overline{g} \circ \overline{f}$.
\begin{proof}
Let be given the power series expansions of $f, g, \overline{f}, \overline{g}$:
\begin{align}
    f^j\left(\bm{z}\right) & := \sum_{m_1, \ldots, m_d \geq 0}f^j_{m_1, \ldots, m_d}z_1^{m_1} \ldots z_d^{m_d}, \qquad 1 \leq j \leq d',\\
    g\left(\bm{w}\right) & = \sum_{p_1, \ldots, p_{d'} \geq 0}g_{p_1, \ldots, p_{d'}}w_1^{p_1}\ldots w_{d'}^{p_{d'}},\\
    \overline{f^j}\left(\bm{z}\right) & = \sum_{m_1, \ldots, m_d \geq 0}\overline{f}^j_{m_1, \ldots, m_d}z_1^{m_1}\ldots z_d^{m_d},\\
    \overline{g}\left(\bm{w}\right) & = \sum_{p_1, \ldots, p_{d'} \geq 0}\overline{g}_{p_1, \ldots, p_{d'}}w_1^{p_1} \ldots w_{d'}^{p_{d'}},
\end{align}
where $\bm{z} = \left(z_1, \ldots, z_d\right) \in \mathbb{C}^d$ and $\bm{w} = \left(w_1, \ldots, w_{d'}\right) \in \mathbb{C}^{d'}$. We then compute the composition as:
\begin{align}
    g\left(f\left(\bm{z}\right)\right) & = \sum_{p_1, \ldots, p_{d'} \geq 0}g_{p_1,\ldots, p_{d'}}\prod_{1 \leq j \leq d'}f^j\left(\bm{z}\right)^{p_j}\nonumber\\
    & = \sum_{p_1, \ldots, p_{d'} \geq 0}g_{p_1,\ldots, p_{d'}}\prod_{1 \leq j \leq d'}\left(\sum_{m_1, \ldots, m_d \geq 0}f^j_{m_1, \ldots, m_d}z_1^{m_1}\ldots z_d^{m_d}\right)^{p_j}\nonumber\\
    & = \sum_{p_1, \ldots, p_{d'} \geq 0}g_{p_1, \ldots, p_{d'}}\prod_{1 \leq j \leq d'}\sum_{\substack{m^k_1, \ldots, m^k_d \geq 0\\\forall 1 \leq k \leq p_j}}z_1^{\sum\limits_{1 \leq k \leq p_j}m^k_1}\ldots z_d^{\sum\limits_{1 \leq k \leq p_j}m^k_d}\prod_{1 \leq k \leq p_j}f^j_{m^k_1, \ldots, m^k_d}\nonumber\\
    & = \sum_{p_1, \ldots, p_{d'} \geq 0}g_{p_1, \ldots, p_{d'}}\sum_{\substack{m^{j, k}_1, \ldots, m^{j, k}_d \geq 0\\\forall 1 \leq j \leq d'\\\forall 1 \leq k \leq p_j}}z_1^{\sum\limits_{\substack{1 \leq j \leq d'\\1 \leq k \leq p_j}}m^{j, k}_1}\ldots z_d^{\sum\limits_{\substack{1 \leq j \leq d'\\1 \leq k \leq p_j}}m^{j, k}_d}\prod_{\substack{1 \leq j \leq d'\\1 \leq k \leq p_j}}f^j_{m^{j, k}_1, \ldots, m^{j, k}_d}
\end{align}
By the assumed convergence of the majorant composition, all these series inversions are licit: they are justified by the existence, for some $R > 0$ such that $\left(\overline{f}^j\left(R, \ldots, R\right)\right)_{1 \leq j \leq d'}$ lies within the polydisk of convergence of $\overline{g}$, of a constant $C\left(R\right)$ such that:
\begin{align}
    \left|f^j_{m_1, \ldots, m_d}\right| & \leq C\left(R\right)R^{-m_1 - \ldots - m_d},\\
    \left|g_{p_1, \ldots, p_{d'}}\right| & \leq C\left(R\right)R^{-p_1 - \ldots - p_{d'}}.
\end{align}
for all $m_1, \ldots, m_d \geq 0$ and $p_1, \ldots, p_{d'} \geq 0$. Using Proposition~\ref{si:prop:majorization_monomials} to majorize each monomial
\begin{align}
    z_1^{\sum\limits_{\substack{1 \leq j \leq d'\\1 \leq k \leq p_j}}m_1^{j, k}} \ldots z_d^{\sum\limits_{\substack{1 \leq j \leq d'\\1 \leq k \leq p_j}}m_d^{j, k}},
\end{align}
and Proposition~\ref{si:prop:majorization_sum} to majorize the sum, we obtain that $g \circ f$ is majorized by:
\begin{align}
    \overline{h}\left(\bm{z}\right) & := \sum_{p_1, \ldots, p_{d'}}\overline{g}_{p_1, \ldots, p_{d'}}\sum_{\substack{m_1^{j, k},\,\ldots,\,m_d^{j, k}\\\forall 1 \leq j \leq d'\\1 \leq k \leq p_j}}z_1^{\sum\limits_{\substack{1 \leq j \leq d'\\1 \leq k \leq p_j}}m_1^{j, k}} \ldots z_d^{\sum\limits_{\substack{1 \leq j \leq d'\\1 \leq k \leq p_j}}m_d^{j, k}}\prod_{\substack{1 \leq j \leq d'\\1 \leq k \leq p_j}}\overline{f^j}_{m_1^{j, k}, \ldots, m_d^{j, k}}.
\end{align}
But running the previous calculation backwards, the above function equals $\overline{g} \circ \overline{f}\left(\bm{z}\right)$, concluding the proof.
\end{proof}
\end{proposition}

By combining Proposition~\ref{si:prop:majorization_monomials} and Proposition~\ref{si:prop:majorization_composition}, we obtain:

\begin{proposition}[Majorization and product]
Let $f^1, \ldots, f^n$ be majorized by $\overline{f^1}, \ldots, \overline{f^n}$. Then the product function $f^1 \ldots f^n$ is majorized by the product of majorants $\overline{f^1}\,\ldots\,\overline{f^n}$.
\begin{proof}
The coefficient of a monomial $\bm{z}^{\bm{m}}$ in the product $f^1 \ldots f^n$ is the convolution $\sum_{\bm{m}^1 + \ldots + \bm{m}^n = \bm{m}}\prod_{i}f^i_{\bm{m}^i}$; by the triangle inequality and the coefficient-wise bounds $|f^i_{\bm{m}^i}| \leq \overline{f^i}_{\bm{m}^i}$ (with $\overline{f^i}_{\bm{m}^i} \geq 0$), its modulus is at most $\sum_{\bm{m}^1 + \ldots + \bm{m}^n = \bm{m}}\prod_i \overline{f^i}_{\bm{m}^i}$, which is precisely the non-negative coefficient of $\bm{z}^{\bm{m}}$ in $\overline{f^1}\ldots\overline{f^n}$.
\end{proof}
\end{proposition}

\begin{proposition}[Majorization of exponential partial sums]
\label{si:prop:majorization_exponential}
For all $k \geq 0$, the exponential series partial sum $\exp_{\leq k}$ is majorized by itself, and similarly for the exponential series tail $\exp_{\geq k}$. In particular ($k = 0$), the exponential is majorized by itself.
\begin{proof}
Both $\exp_{\leq k}$ and $\exp_{\geq k}$ have Taylor coefficients $1/m! \geq 0$; a power series with non-negative coefficients is majorized by itself, since each coefficient equals its own modulus.
\end{proof}
\end{proposition}

\subsection{Calculus}
\label{si:sec:calculus}

The following Gaussian integration by parts Lemma is a standard tool in spin glass theory:

\begin{lemma}[Gaussian integration by part]
\label{si:lemma:gaussian_ibp}
Let $\left(\bm{X}, \bm{Y}\right)$, with $\bm{X} \in \mathbb{R}^m, \bm{Y} \in \mathbb{R}^n$ be a jointly centered Gaussian vector. Then, for any differentiable function $\varphi: \mathbb{R}_n \longrightarrow \mathbb{C}$ of suitable (for instance, exponential) growth at infinity:
\begin{align}
    \mathbb{E}_{\left(\bm{X}, \bm{Y}\right)}\left[X_j\varphi\left(\bm{Y}\right)\right] & = \sum_{k \in [n]}\mathbb{E}\left[\mathbb{E}\left[X_jY_k\right]\partial_k\varphi\left(\bm{Y}\right)\right] .
\end{align}
\end{lemma}

In the following derivations, we will frequently truncate the Taylor series of the exponential to a given order; for this purpose, it will prove convenient to introduce specific notation:

\begin{definition}[Partial Taylor sum and Taylor series tail of exponential series]
\label{si:def:exponential_series_partial_sums}
For all natural integer $k \geq 0$, we define the following analytic functions of $z \in \mathbb{C}$:
\begin{align}
    \exp_{\leq k}\left(z\right) & := \sum_{m = 0}^k\frac{z^m}{m!},\\
    \exp_{\geq k}\left(z\right) & := \sum_{m = k}^{+\infty}\frac{z^m}{m!}
\end{align}
\end{definition}

We recall the following standard bound on the tail of the exponential's Taylor series:

\begin{proposition}[Exponential series tail bound]
\label{si:prop:exponential_tail_bound}
For all natural integer $k \geq 0$ and complex number $z \in \mathbb{C}$:
\begin{align}
    \left|\exp_{\geq k}\left(z\right)\right| & \leq \exp\left(|z|\right)\frac{|z|^k}{k!}.
\end{align}
\end{proposition}

\begin{lemma}[Geometric bound for initial terms of tower of exponentials]
\label{si:lemma:exponentials_tower_bound}
Let sequence $\left(x_n\right)_{n \geq 0}$ of non-negative numbers satisfy
\begin{align}
    x_0 \in [0, 1/c]
\end{align}
and inequality
\begin{align}
    x_{n + 1} & \leq \left(ax_n + b\right)e^{cx_n},
\end{align}
for $a, b, c \geq 0$. Then, the initial terms of $\left(x_n\right)_{n \geq 0}$ are bounded as follows:
\begin{align}
    0 \leq x_n \leq \overline{a}^n\left(x_0 + \frac{2eb}{\overline{a}}\right) \qquad \forall n \in [0, n^*],
\end{align}
where
\begin{align}
    n^* := \min\left\{n \geq 0\,:\,x_n > 1/c\right\},
\end{align}
and
\begin{align}
    \overline{a} & := \max\{2, ea\}.
\end{align}
Besides,
\begin{align}
    n^* \geq \left\lceil -\frac{\log\left(\left(x_0 + 2eb/\overline{a}\right)c\right)}{\log\overline{a}} \right\rceil.
\end{align}
In particular, $n^*$ can be made arbitrarily large for sufficiently small $x_0$ and $b$.
\begin{proof}
Define:
\begin{align}
    n^* & := \min\left\{n \geq 0\,:\,x_n > 1/c\right\}.
\end{align}
$1 \leq n^* \leq +\infty$ since $x_0 \leq 1/c$ by assumption. For all $n < n^*$, $x_n \leq 1/c$ and we can write
\begin{align}
    x_{n + 1} & \leq \left(ax_n + b\right)e^{cx_n}\nonumber\\
    & \leq \left(ax_n + b\right)e^{1}\nonumber\\
    & = eax_n + eb.
\end{align}
Hence, for all $n \in [0, n^*]$, $\left(x_n\right)_{n \geq 0}$ is upper-bounded by arithmetic-geometric sequence $\left(y_n\right)_{n \geq 0}$, defined by recursion
\begin{align}
    y_0 & := x_0,\\
    y_{n + 1} & = \overline{a}y_n + eb, \qquad \forall n \geq 0,
\end{align}
where
\begin{align}
    \overline{a} & := \max\left\{2, ea\right\},
\end{align}
and of explicit general term given by
\begin{align}
    y_n & = -\frac{eb}{\overline{a} - 1} + \overline{a}^n\left(y_0 + \frac{eb}{\overline{a} - 1}\right)\nonumber\\
    & \leq \overline{a}^n\left(y_0 + \frac{eb}{\overline{a} - 1}\right).\nonumber\\
    & \leq \overline{a}^n\left(y_0 + \frac{2eb}{\overline{a}}\right)
\end{align}
From inequality $x_n \leq \overline{a}^n\left(y_0 + 2eb/\overline{a}\right)$ for all $n \in [0, n^*]$, we deduce
\begin{align}
    n^* \geq \left\lceil -\frac{\log\left(\left(y_0 + 2eb/\overline{a}\right)c\right)}{\log\overline{a}} \right\rceil.
\end{align}
\end{proof}
\end{lemma}

\begin{lemma}[Binomial coefficient bound]
\label{si:lemma:binomial_coefficient_bound}
For all integers $m, k \geq 0$, the following bound holds:
\begin{align}
    \binom{m + k}{k} \leq 2^{m + k}.
\end{align}
\begin{proof}
This follows from bound $\binom{n}{k} \leq 2^n$ for all $0 \leq k \leq n$, which itself results from the binomial theorem:
\begin{align}
    \binom{n}{k} & \leq \sum_{0 \leq m \leq n}\binom{n}{m}1^m1^{n - m}\nonumber\\
    & = (1 + 1)^n\nonumber\\
    & = 2^n.
\end{align}
\end{proof}
\end{lemma}

\section{Computation of optimal and algorithmic value of spin glass}

In this Section, we detail the method used in this work to compute optimal and algorithmic energy densities of an arbitrary spin glass. The optimal value results from Parisi's formula~\cite{PhysRevLett.43.1754,talagrand_parisi_formula} (see also \cite{Panchenko2013} for a modern and very general presentation). The algorithmic value results from a variational principle of El Alaoui, Montanari and Sellke~\cite{optimization_mean_field_spin_glasses}.

A mixed even/odd spin glass Hamiltonian is
specified by its covariance or mixture polynomial
\begin{equation}
    \xi(q) = \sum_{p \geq 2} w_p q^p,
\end{equation}
where the weights \(w_p \geq 0\) determine the mixture.  The pure \(p\)-spin Hamiltonian is a centered Gaussian process indexed by spin
configurations. With the normalization used in this paper, one may write
\begin{equation}
    H_{N,p}(\sigma)
    =
    \frac{1}{N^{(p-1)/2}}
    \sum_{i_1,\ldots,i_p=1}^N
    J^{(p)}_{i_1,\ldots,i_p}
    \sigma_{i_1}\cdots\sigma_{i_p}, \quad \sigma \in \{\pm 1\}^N
\end{equation}
with independent $\mathcal{N}(0, 1)$ couplings. The pure model is the special case $\xi(q)=w_p q^p$. In this paper, we  choose the convention that $w_p = \frac1p$. The mixed-spin Hamiltonian can be written as
\begin{equation}
    H_N(\sigma)
    =
    \sum_{p \geq 2} \sqrt{w_p}\, H_{N,p}(\sigma).
\end{equation}

Let \(\gamma:[0,1)\to \mathbb{R}_{\geq 0}\) be an integrable function called a \emph{candidate order parameter}. Given
\(\gamma\), define \(\Phi_\gamma(t,x)\) as the solution of the \emph{Parisi PDE}:
\begin{equation}
    \partial_t \Phi_\gamma(t,x)
    =
    -\frac{\xi''(t)}{2}
    \left(
        \partial_{xx}\Phi_\gamma(t,x)
        +
        \gamma(t)\left(\partial_x\Phi_\gamma(t,x)\right)^2
    \right),
    \qquad 0\leq t<1,
\end{equation}
with terminal condition
\begin{equation}
    \Phi_\gamma(1,x)=|x|.
\end{equation}
The corresponding \emph{Parisi functional} is
\begin{equation}
    \mathcal{P}(\gamma)
    =
    \Phi_\gamma(0,0)
    -
    \frac{1}{2}\int_0^1 t\xi''(t)\gamma(t)\,dt.
\end{equation}
The Parisi formula identifies the infimum of this functional with the limiting
zero-temperature energy. In the usual finite-temperature formulation,
one first computes the limiting free energy
\begin{equation}
    F_N(\beta)
    =
    \frac{1}{N}\log \sum_{\sigma\in\{\pm1\}^N}
    \exp\left(\beta H_N(\sigma)\right),
\end{equation}
and then sends \(N\to\infty\). The zero-temperature limit is obtained by scaling
by \(\beta\) and letting \(\beta\to\infty\):
\begin{equation}
    \lim_{\beta\to\infty}\frac{1}{\beta}
    \lim_{N\to\infty} F_N(\beta)
    =
    \lim_{N\to\infty}\frac{1}{N}\max_{\sigma}H_N(\sigma) =: \textup{OPT}(\xi).
\end{equation}
In this zero-temperature setting, the Parisi functional above is the variational
problem whose minimum gives the limiting ground-state/free-energy value under the
chosen normalization.

The connection between OPT and the Parisi functional is called the zero-temperature Parisi formula:
\begin{equation}
\label{si:eqn:opt_parisi_formula}
    \mathrm{OPT}(\xi)
    =
    \inf_{\gamma \in \mathcal{A}_{\mathrm{OPT}}}\mathcal{P}(\gamma), \tag{OPT Problem}
\end{equation}
where $\mathcal{A}_{\mathrm{OPT}}$ is the set of all candidate order parameters $\gamma$ satisfying a monotonicity condition:
\begin{equation}
    \gamma(s)\leq \gamma(t)
    \qquad\text{whenever }s\leq t.
\end{equation}
The minimizer $\gamma^{\star}$ is called \emph{the order parameter} associated with $\xi(q)$ (in the zero temperature phase).

The algorithmic threshold ALG is defined the following relaxation of the above optimization problem:
\begin{equation}
\label{si:eqn:alg_parisi_formula}
    \mathrm{ALG}(\xi)
    =
    \inf_{\gamma \in \mathcal{A}_{\mathrm{ALG}}}\mathcal{P}(\gamma), \tag{ALG Problem}
\end{equation}
where \(\mathcal{A}_{\mathrm{ALG}}\) is the set of all candidate order parameters $\gamma$ such that  $\xi''(t)\gamma(t)$ is integrable and also has finite total variation over any compact subset of $[0,1)$ \cite{optimization_mean_field_spin_glasses}. Since a non-decreasing $\gamma$ has finite total variation over any compact subset of $[0, 1)$, we have $\mathcal{A}_{\mathrm{OPT}}\subset \mathcal{A}_{\mathrm{ALG}}$, and thus  \eqref{si:eqn:alg_parisi_formula} is a proper relaxation of \eqref{si:eqn:opt_parisi_formula}, i.e.  ALG $\leq $ OPT always. Particularly, a spin glass optimization problem associated with mixture $\xi(q)$ has the  \emph{overlap gap property} (OGP) \cite{Gamarnik_2021_ogp_survey} when the minimier of \eqref{si:eqn:alg_parisi_formula} is not feasible for \eqref{si:eqn:opt_parisi_formula}, in other words ALG $\neq$ OPT.

To solve the above two problems numerically, we utilize a piecewise constant discretization of the elements of $\mathcal{A}_{\textup{ALG}}, \mathcal{A}_{\textup{OPT}}$. We refer to the number of steps as the \emph{number of atoms}. Particularly,  the numerics approximate $\gamma$ by a nonnegative array
\begin{equation}
    \gamma = (\gamma_0,\ldots,\gamma_{m-1}),
\end{equation}
where \(m=\texttt{num\_atoms}\). This array is interpreted as a step function on
the grid
\begin{equation}
    0 = q_0 < q_1 < \cdots < q_m = 1,
    \qquad q_i = i/m,
\end{equation}
by setting \(\gamma(t)=\gamma_i\) for \(t\in[q_i,q_{i+1})\). The stepwise nature of the candidate solution enables the Parisi PDE to be efficiently solved.

For each order parameter segment, the Parisi PDE can be solved by a Cole--Hopf transform
on each time interval. The numerics implement this update as a Gaussian convolution:
\begin{equation}
    \Phi_i(x)
    =
    \frac{1}{\gamma_i}
    \log \mathbb{E}
    \exp\left(\gamma_i \Phi_{i+1}\left(x+\sqrt{\Delta_i}\,Z\right)\right),
\end{equation}
where \(Z \sim \mathcal{N}(0,1)
\) and
\begin{equation}
    \Delta_i = \xi'(q_{i+1}) - \xi'(q_i).
\end{equation}
The derivatives of $\Phi_i$ are computed using the following to differentiate a convolution:
\begin{align*}
    \partial_x^{j}[f \star g](x) = \int_{\mathbb{R}}f(y)\partial_x ^{j}g(x-y)dy,
\end{align*}
where $g$ is a Gaussian pdf and hence $\partial_x^{j}g(x-y)$ is known in closed form. Hence the derivatives can also be computed efficiently using optimized convolution subroutines.

In the implementation, the Gaussian expectation is evaluated by discrete
convolution on a uniform spatial grid
\begin{equation}
    x \in [-\texttt{trunc},\texttt{trunc}],
    \qquad \texttt{nx}+1 \text{ grid points}.
\end{equation}
For each step, the integral in $\mathcal{P}(\gamma)$ is evaluated as a dot product
\begin{equation}
    \int_0^1 t\xi''(t)\gamma(t)\,dt
    =
    \sum_{i=0}^{m-1} L_i \gamma_i,
\end{equation}
where
\begin{equation}
    L_i
    =
    \int_{q_i}^{q_{i+1}} t\xi''(t)\,dt,
\end{equation}
can be computed in closed form. The above provides a procedure to compute the cost function $\mathcal{P}$ for both \eqref{si:eqn:alg_parisi_formula} and \eqref{si:eqn:opt_parisi_formula}, an approximate minimizer is found utilizing off-the-shelf, constrained optimizers. Particularly for \eqref{si:eqn:opt_parisi_formula}, we tell the optimizer to enforce the non-negativity constraint on $\gamma$. In the continuum limit, both problems are known to be convex (for \eqref{si:eqn:opt_parisi_formula} see  \cite[Theorem 20]{Jagannath_2015} and for \eqref{si:eqn:alg_parisi_formula} see \cite[Proof of Lemma 6.15]{optimization_mean_field_spin_glasses}) .

The above was applied to compute ALG and OPT for a mixture that interpolates from a pure $p=2$ to a pure $p=8$:
\begin{equation}
    \xi_\alpha(q)
    =
    (1-\alpha)\frac{q^2}{2}
    +
    \alpha\frac{q^8}{8}, \alpha \in [0, 1]
\end{equation}
used in the numerical experiments.

\end{document}